\documentclass[a4paper,11pt]{article}
\pdfoutput=1 

\usepackage{jheppub} 

\usepackage[T1]{fontenc} 
\usepackage{appendix}
\usepackage{comment}

\newtheorem{thm}{Theorem}
\newtheorem{zadacha}[thm]{Exercise}

\title{\boldmath Basic introduction to higher-spin theories}

\author[a,b]{Dmitry Ponomarev}

\affiliation[a]{Institute for Theoretical and Mathematical Physics,\\
Lomonosov Moscow State University, Lomonosovsky avenue, Moscow, 119991, Russia}
\affiliation[b]{I.E. Tamm Theory Department, Lebedev Physical Institute,\\
 Leninsky avenue, Moscow, 119991, Russia}

\emailAdd{ponomarev@lpi.ru}

\abstract{This is a  collection of my lecture notes on the higher-spin theory course given for students at the Institute for Theoretical and Mathematical Physics, Lomonosov Moscow State University. The goal of these lectures is to give an introduction to higher-spin theories accessible to master level students which would enable them to read the higher-spin literature. I start by introducing basic relevant notions of representation theory and the associated field-theoretic descriptions. Focusing on massless symmetric fields I review different approaches to interactions as well as the no-go results. I end the lectures by reviewing some of the currently available positive results on interactions of massless higher-spin fields, namely, holographic, Chern-Simons and chiral higher-spin theories.}

\begin{document} 
\maketitle
\flushbottom

\section{Introduction}

Higher-spin theories  is a fascinating topic of modern theoretical physics, which has deep connections to other important areas, such as holography, various forms of bootstrap and string theory. Broadly speaking, with the ultimate goal being the construction of quantum gravity models,  higher-spin program attempts to chart the landscape of consistent field theories scrutinising and, possibly, expanding the consistency requirements along the way. For a recent review on the current status of higher-spin theories, their connection to other areas of theoretical physics and for discussions on possible future directions we refer the reader to \cite{Bekaert:2022poo}.

When approaching this problem, one quickly discovers the following striking feature: although consistent field theories are abundant at free level, all known interacting theories are, essentially, just variants of a handful of theories -- such as scalar theories and theories of spin-$\frac{1}{2}$ fermions, the Yang-Mills theory or General Relativity -- and involve only very limited types of lower-spin fields. A notable exception to this conclusion is provided by string theory, which, however, strongly relies on the world-sheet description, while its reformulation as a field theory constitutes a separate research direction. The fact that we were not able to construct interacting theories beyond a couple of known ones -- which, moreover, involve a very limited set of fields -- either requires an explanation or implies that there is something that we are crucially missing about field theories.

Although the construction of any interacting higher-spin theory would constitute an important progress, more interest is usually drawn to  theories involving massless fields. The main reason for that is that massless fields with spin greater than $\frac{1}{2}$ inevitably require symmetries to be present in their description. Symmetries, in turn, have always played a central role in physics and mathematics: while in physics they serve as an important constructive principle and the  basis for unifications, they are responsible for improved quantum properties, etc, in mathematics they inspired many algebraic and geometric constructions. In particular, the Yang-Mills theory and General Relativity -- massless spin-1 and spin-2 theories with gauge algebras given by local internal symmetries and diffeomorphisms respectively -- constitute an essential part of our current description of nature.

A particular class of massless fields is given by symmetric fields. These are simpler and better explored. One reason that makes symmetric massless fields simpler is that they lead to symmetries of a conventional type, while massless mixed-symmetry fields require more unusual reducible gauge symmetries. Besides that, as we will see shortly, only symmetric massless fields propagate in a four-dimensional space-time, which is a setup of most physical relevance.

In these lectures, we will primarily focus on symmetric massless higher-spin fields and the problem of their interactions. In this context ''higher-spin'' refers to spin greater than two, as for these fields even a free field description let alone introducing interactions becomes notoriously difficult. 
In particular, serious difficulties with interactions of higher-spin gauge fields are 
reflected  in the famous no-go theorems from 60's and 70's, which prove that for some natural assumptions massless higher-spin fields cannot interact.

 Most of these results are based on the $S$-matrix theory, which is directly applicable only to scattering in flat space. 
To avoid the negative results of the no-go theorems it was  suggested that massless higher-spin theories may exist, but they do not admit the Minkowski background as a solution, instead, these are more naturally defined around the (A)dS background. This idea lead to an important progress and, in particular, a concrete higher-spin model -- the Vasiliev theory -- was put forward. The existence of higher-spin gauge theories with the (A)dS space as a favourable background received solid support from holography later. 

Besides that, currently, higher-spin theories have been constructed explicitly in certain simplified setups, in which they appear as very natural generalisations of the associated lower-spin theories. In particular, using the frame-work of the Chern-Simons theories, one can straightforwardly extend 3d gravity to the higher-spin case. In a similar manner, chiral higher-spin theories can be regarded as natural higher-spin generalisations of the self-dual Yang-Mills theory and self-dual gravity. In these lectures we give a basic introduction to these and other results, focusing on symmetric higher-spin gauge fields and their interactions.

This lecture course is organised as follows. We will start by reviewing the basics of the free field theory both in the Minkowski and in the AdS backgrounds. We will then review the Flato-Fronsdal theorem and higher-spin algebras, which will be used later. Next, we discuss various approaches to interactions, both at the Lagrangian and the $S$-matrix level. We then review some of the no-go theorems and explain how they constrain interactions of higher-spin gauge fields. We end the lectures by reviewing more recent results, that bypass the standard no-go theorems in one way or another. These include higher-spin theories constructed holographically, Chern-Simons theories and chiral higher-spin theories.

Before proceeding to the main text, we would like to make a couple of remarks. Firstly, we will often be satisfied with the physical level of rigour, some of the statements  will be given without the proof, especially, if these proofs are technical and can be easily found elsewhere. We will try to be clear each time this happens and give references to more comprehensive discussions. Secondly,
the higher-spin literature by now is quite extensive and reviewing all the important results in a clear and consistent manner appears to be an arduous task. We, therefore, focus on simpler and more accessible topics. The results that were left out are briefly discussed at the end of the lectures. Thirdly, we would like to mention other reviews --
 see, in particular, \cite{Sorokin:2004ie,Bekaert:2006py,Vasilievutrecht,Rahman:2015pzl,Kessel:2017mxa} -- which give an accessible  presentation of some topics that we cover here, supplementing them with further details. We refer the reader to these  for a different perspective and a more complete picture.  Other references are given throughout the text. Finally, we assume basic familiarity of the reader with Lie groups and Lie algebras, their representation theory, basic knowledge of General Relativity as well as classical and quantum field theories is also assumed.


\section{Free fields in flat space and UIR's of the Poincare group}
\label{sec:2}

In this section we explain the relation between free fields in flat space and unitary irreducible representations of the Poincare group. We then proceed to the classification of these representations. 
This discussion is rather standard and  follows the steps of section 2 of  \cite{Weinberg}. For technical details and extensions, we refer the reader to \cite{Bekaert:2006py}.

\subsection{From free fields to UIR's of the isometry algebra}

The basic experimental fact is that in our Universe gravitational fields are of order unity and they approximately describe a maximally symmetric background, -- the Minkowski or the (A)dS spaces being the examples of the most symmetric backgrounds --  while the remaining fields are vanishingly small. This motivates a perturbative approach to field theories, that is when one starts from the background and then considers corrections -- order by order -- in small fields. Note that from this perspective fluctuations of the gravitational field over its background are treated on the same footing with other small fields.

In the present section we will focus on the flat space case, which means that at the leading order we have only the background gravitational field, given by the d-dimensional Minkowski metric
\begin{equation}
\label{23mar1}
\eta= {\rm diag}(-,+,\dots,+).
\end{equation}
 This metric is \emph{invariant} with respect to the action of the vector fields, generating the Poincare group $ISO(d-1,1)$ -- the isometries of the Minkowski space. The associated Lie algebra has two types of generators: translation $P_\mu$ and Lorentz transformations $J_{\mu\nu}$, which commute as follows
\begin{equation}
\label{1x22sep1}
\begin{split}
i[J^{\mu\nu},J^{\rho\sigma}]&=\eta^{\nu\rho}J^{\mu\sigma}-\eta^{\mu\rho}J^{\nu\sigma}-\eta^{\sigma\mu}J^{\rho\nu}+\eta^{\sigma\nu}J^{\rho\mu},\\
 i[P^\mu,J^{\rho\sigma}]&=\eta^{\mu\rho}P^\sigma - \eta^{\mu\sigma}P^\rho,\\
 i[P^\mu,P^\rho]&=0.
 \end{split}
\end{equation}

Another basic experimental fact is that these symmetries survive when fluctuating fields are added. This symmetry, however, works in a slightly different way: fluctuations themselves are no longer invariant, instead, they transform under symmetries, while \emph{the equations of motion} they satisfy \emph{remain invariant}. We will now consider the implications of this statement at the first order in fluctuations -- the order of free theories. 

At this order fields -- which we denote by  $\varphi$ -- satisfy linear equations, schematically, given by
\begin{equation}
\label{22sep1x2}
L(x,\partial_x)\varphi(x) =0,
\end{equation}
where $L$ is some linear operator. Terms of the form $\varphi^2$, $\varphi^3$, etc. are absent in (\ref{22sep1x2}) as these are of higher orders in perturbation theory. Due to the fact that (\ref{22sep1x2}) is linear in $\varphi$, its solutions form a linear space. Moreover, it is required that the solution space is invariant with respect to the action of the Poincare group. 
This means that, there exists a way to define an action of the Poincare group elements $G$ on $\varphi$
\begin{equation}
\label{24mar1}
\delta_G \varphi = U(G)\varphi
\end{equation}
with $U(G)$ being  a linear operator,
 and this action leaves (\ref{22sep1x2}) invariant. Consistency requires that for any pair of elements of the Poincare group $G_1$ and $G_2$ one has
 \begin{equation}
 \label{23mar3}
 U(G_1) U(G_2)=U(G_1 G_2),
 \end{equation}
 where on the left-hand side we have a composition of linear maps on the field space, while on the right-hand side $G_1$ and $G_2$ are multiplied with the Poincare group product.
Mathematically speaking, this implies that solutions of (\ref{22sep1x2}) realise a \emph{representation} of the Poincare group\footnote{Strictly speaking, it is required that free fields realise projective representations, that is representations up to a phase. This extension is important if one needs to deal with fermions.   Inclusion of fermions in the higher-spin context is relatively straightforward and, usually, does not add much except technical difficulties. Fermions will not be discussed here. For the discussion on fermions in the context of representation theory, see e.g. \cite{Weinberg}.}.

Group elements, which are infinitesimally close to the group unity can be related to the Lie algebra generators via
\begin{equation}
\label{1x23sep3}
G=1+i\varepsilon g.
\end{equation}
This allows one to induce a Lie  algebra action on fields from (\ref{24mar1})
\begin{equation}
\label{23mar2}
\delta_P \varphi =U(P)\varphi, \qquad \delta_J\varphi = U(J)\varphi
\end{equation}  

The simplest example of this setup is the free scalar field
\begin{equation}
\label{1x23sep1}
(\Box-m^2)\varphi=0.
\end{equation}
In this case the action of the Poincare generators is defined as 
\begin{equation}
\label{1x23sep2}
U(P_\mu)\varphi = -i\partial_\mu \varphi, \qquad U(J_{\mu\nu})\varphi=-i(x_\mu \partial_\nu-x_{\nu}\partial_\mu)\varphi.
\end{equation}

\begin{zadacha}
Check that (\ref{1x23sep2}) satisfy (\ref{1x22sep1}).
\end{zadacha}

\begin{zadacha}
Check that (\ref{1x23sep2}) is a symmetry of (\ref{1x23sep1}).
\end{zadacha}

Another consistency requirement which is relevant at the linear order comes from the quantum theory and it is the requirement of \emph{unitarity}. To start, one requires that there is a positive definite sesquilinear form on the on-shell states of the theory, that is
\begin{equation}
\label{15mar1}
\begin{split}
(\varphi,\psi)&=(\psi,\varphi)^*,\\
(\varphi,\xi_1 \psi_1+\xi_2\psi_2)&=\xi_1 (\varphi,\psi_1)+\xi_2 (\varphi,\psi_2),\\
(\eta_1\varphi_1+\eta_2\varphi_2,\psi)&=\eta_1^* (\varphi_1,\psi)+\eta_2^*(\varphi_2,\psi),\\
(\psi,\psi)&>0,\qquad \text{for} \qquad \psi \ne 0,
\end{split}
\end{equation}
where $\varphi$ and $\psi$ are states in the representations space. 
The role of this form is to compute probability amplitudes in quantum theory and the above requirements, in particular, are necessary to ensure that probabilities are non-negative.

Besides that, it is required that  norm (\ref{15mar1}) is invariant with respect to the action of the isometry group 
\begin{equation}
\label{15mar2}
(U(G)\varphi,U(G)\psi)=(\varphi,\psi).
\end{equation} 
Then, the Lie algebra generators $g$,   (\ref{1x23sep3}), act in a self-adjoint way
\begin{equation}
\label{15mar2x1}
(U(g)\varphi,\psi)=(\varphi,U(g)\psi).
\end{equation}
For the Poincare algebra we thus have
\begin{equation}
\label{1x23sep4}
(U(P)\varphi,\psi)=(\varphi,U(P)\psi), \qquad (U(J)\varphi,\psi)=(\varphi,U(J)\psi).
\end{equation}

Finally, if a representation is a 
direct sum of other representations, then in quantum theory it receives the interpretation of multi-particle states. Instead, one-particle states or elementary particles are described by \emph{irreducible} representations of the isometry algebra. 

Summing up this discussion, we conclude that \emph{free elementary fields in the Minkowski space can be identified with  unitary irreducible representations (UIR's) of the Poincare group}.

Before proceeding to the classification of such representations, we would like to anticipate one its important feature. Namely, there is a theorem stating that unitary representations of simple non-compact groups may be either trivial or infinite-dimensional.\footnote{See, e.g. \cite{Bekaert:2006py} for a more precise statement and for the proof of this theorem.} Since the Poincare group -- as well as isometries of other most symmetric backgrounds such as the (A)dS space isometries -- is non-compact, we infer that on-shell fields should necessarily be infinite-dimensional as vector spaces. 
The Poincare group is not simple, though, its Lorentz subgroup, which is expected to be realised non-trivially, is still non-compact, hence, 
the conclusion of the theorem also applies to fields in flat space.
This is why in field theory fields are inevitably described by functions on some space: this is the standard way to realise an infinite-dimensional vector space.

\subsection{Wigner's method of induced representations for UIR's of the Poincare group}

In the previous section we saw that free fields in the Minkowski space can be identified with UIR's of the Poincare group. Thus, in order to classify all possible free field theories in flat space, we need to classify UIR's of the Poincare group. Despite the fact that there exist other approaches to the problem, Wigner's technique of induced representation is the most standard approach, which gives an exhaustive classification of UIR's of the Poincare group. We will now review it.

To start, one notes that due to the fact that translations commute, one can choose a basis in the representation space so that translations act diagonally\footnote{Note that this is only true for unitary representations. Indeed, if unitarity is not imposed, the best one can achieve when diagonalising even  a single matrix is the Jordan canonical form.}
\begin{equation}
\label{1x23sep5}
U(P_\mu) \varphi_{p,\sigma}=p_\mu\varphi_{p,\sigma}.
\end{equation}
Here $p$ is the label of the representation space basis which refers to the eigenvalue of $P$. Since the eigenspace with fixed $p$ does not have to be one-dimensional, we supply $\varphi$ with an extra label $\sigma$. 

We have already specified how translations act in our basis and now we move on to the Lorentz transformations. To this end, we consider a finite Lorentz transformation, defined by a pseudo-orthogonal  matrix $\Lambda^{\mu}{}_\nu$. The associated action in a given representation will be denoted $U(G(\Lambda,0))$ with $0$ referring to the fact that translations are not performed. We would like to apply this transformation to the state with definite momentum $p$ and find what is the momentum of the resulting state. We have
\begin{equation}
\label{23mar4}
\begin{split}
U(P^\mu) U(G(\Lambda,0))\varphi_{p,\sigma}&=U(G(\Lambda,0))\big[U^{-1}(G(\Lambda,0))U(P^\mu) U(G(\Lambda,0))\big]\varphi_{p,\sigma}\\
&=U(G(\Lambda,0))U(\Lambda^{-1}{}_\rho{}^\mu P^\rho)\varphi_{p,\sigma}
\end{split}
\end{equation}
To obtain the second equality, one has to use commutation relations of the Poincare algebra together with (\ref{23mar3}). In essence, (\ref{23mar4}) just says that $P$ transforms as a vector.
 For details we refer the reader to \cite{Weinberg}. By using that  $(\Lambda^{-1})_\rho{}^\mu=\Lambda^\mu{}_\rho$ and linearity of $g\to U(g)$, we find
\begin{equation}
\label{1x23sep6}
\begin{split}
U(P^\mu) U(G(\Lambda,0))\varphi_{p,\sigma}
=\Lambda^{\mu}{}_\rho p^\rho U(G(\Lambda,0))\varphi_{p,\sigma}.
\end{split}
\end{equation}
Therefore, $U(G(\Lambda, 0))\varphi_{p,\sigma}$ is a state with definite momentum equal to $\Lambda^\mu{}_\rho p^\rho$ and it should be expressible as a linear combination of states $\varphi_{\Lambda p,\sigma}$
\begin{equation}
\label{1x23sep7}
U(G(\Lambda,0))\varphi_{p,\sigma}=\sum_{\sigma'}C_{\sigma'\sigma}(\Lambda,p)\varphi_{\Lambda p,\sigma'}.
\end{equation}
This implies that for any $p$ in the spectrum, besides states $\varphi_{p,\sigma}$, the representation space should contain all states with momenta $\Lambda p$, where $\Lambda$ is a matrix of an arbitrary Lorentz transformation. The resulting manifold in the momentum space is called \emph{the orbit} of $p$.

By now, we have already settled with the way the Lorentz transformations act on $p$ labels. 
To specify the representations of the Poincare group we are after, it remains to define how the Lorentz transformations act on $\sigma$ labels.

 To this end, on a given orbit in $p$ space, we choose any convenient momentum $p_0$. Next, for every $p$ on the orbit, we pick some convenient Lorentz transformation, such that 
\begin{equation}
\label{1x23sep8}
p^\mu = L^{\mu}{}_\nu (p)(p_0)^\nu.
\end{equation}
By the very definition of an orbit such $L$ exists, though, it does not have to be unique. The precise choice of $p_0$ and $L$ is immaterial for the following construction. Finally, we choose some basis $\varphi_{p_0,\sigma}$ for states with momentum $p_0$, while for other momenta  the basis is chosen, so that $L$ does not act on $\sigma$ labels, that is
\begin{equation}
\label{1x23sep9}
\varphi_{p,\sigma}\equiv N(p)U(G(L(p),0))\varphi_{p_0,\sigma}.
\end{equation}
Here again, $N(p)$ is some arbitrary factor, that can be adjusted in any convenient way. 

We have already fixed the basis in the representation space completely. In a given basis, we will now find how general Lorentz transformations act on $\sigma$ labels of states with  arbitrary momenta on the orbit of $p_0$.

 We, thus,  act on (\ref{1x23sep9}) with an arbitrary Lorentz transformation, which gives
\begin{equation}
\label{1x23sep10}
\begin{split}
U(G(\Lambda,0))\varphi_{p,\sigma}&=N(p)U(G(\Lambda L(p),0))\varphi_{p_0,\sigma}\\
&=N(p)U(G(L(\Lambda p),0))U(G(L^{-1}(\Lambda p)\Lambda L(p),0))\varphi_{p_0,\sigma}.
\end{split}
\end{equation}
In the last line $L^{-1}(\Lambda p)\Lambda L(p)$ leaves $p_0$ invariant, therefore,  $U(G(L(\Lambda p),0))$ acts on  states with momentum $p_0$. The latter action has already been specified in (\ref{1x23sep9}). We, therefore, find that to define the left-hand side of (\ref{1x23sep10}) it remains to specify the action of Lorentz transformations  $W$, that leave $p_0$ invariant 
\begin{equation}
\label{1x23sep11}
p^\mu_0 = W^\mu{}_\nu p^\nu_0.
\end{equation}
A subgroup  of the Lorentz group that leaves a given vector $p_0$ invariant
is called \emph{the stability subgroup} of $p_0$ or \emph{the Wigner little group}. 
Since $W$ leaves $p_0$ invariant, it only acts on the $\sigma$ labels of $\varphi_{p_0,\sigma}$. 

Obviously, once an UIR of the Poincare group is known, it uniquely defines an UIR of its subgroup -- the Wigner little group. 
One can also go in the opposite direction: starting from an UIR of the Wigner little group one can induce a representation of the Poincare group using (\ref{1x23sep5}), (\ref{1x23sep9}), (\ref{1x23sep10}). Obviously, the resulting induced representation is also irreducible. Let us show that it is also unitary.

To see that we note that 
hermiticity of $P$ implies that the inner product is vanishing  for states with different momenta 
\begin{equation}
\label{1x23sep13}
(\varphi_{p_1,\sigma_1},\varphi_{p_2,\sigma_2}) =0 ,\qquad  p_1\ne p_2.
\end{equation}
For states with equal momenta, by Lorentz invariance, the inner product can be induced from the inner product of states with momentum $p_0$. The letter is defined by the inner product of the Wigner little group, which is positive definite, since the representation of the Wigner little group we start with is unitary. It is not hard to see that the inner product for  states $\varphi_{p,\sigma}$ constructed in this way is manifestly positive definite and Poincare invariant. Thus, the induced representation of the Poincare group is, indeed, unitary.

In summary, we have found that UIR's of the Poincare group and of the Wigner little group are in one-to-one correspondence. Moreover,  the former can be induced from the latter by the construction that we presented above, which is called \emph{Wigner's method of induced representations}. As a corollary of this construction, we find that \emph{UIR's of the Poincare group are classified by:} 
\begin{itemize}
\item[i)] orbits of momenta $p$ under Lorentz transformations;
\item[ii)] UIR's of the stability subgroup of some fixed momentum on a chosen  orbit.
\end{itemize}
In the following we will discuss different orbits, their stability subgroups and UIR's of the latter.

\subsection{Classification of UIR's of the Poincare group}
\label{sec:wig}

In this section we discuss different classes of orbits and the associated stability subgroups. We then move on to the classification  of representations for these subgroups.  

\subsubsection{Classes of orbits}

It is easily seen that there are six types of orbits of vectors with respect to the Lorentz transformations.

Clearly, the vanishing momentum, $p=0$,  forms an orbit.
 For non-zero vectors, the only invariant one can construct out of $p$ is $p^2=\eta_{\mu\nu}p^\mu p^\nu$. Therefore, different values of $m^2\equiv -p^2$ label different orbits.  There are three classes of $m^2$ that lead to different types of physics.
  For $m^2>0$ the momentum orbit is a two-sheeted hyperboloid. One of the sheets has $p^0>0$, while the other one has $p^0 <0$. These are different orbits associated with positive and negative energy modes of \emph{massive} particles\footnote{By the Poincare group we mean its component, which is continuously connected to the unity.  For the whole Poincare group -- with the time reversal and the  parity transformation included -- positive and negative modes belong to the same orbit.}. 
  For $m^2=0$ the orbit is given by $p^2=0$, which describes  a cone in  momentum space. Similarly, its parts with $p^0>0$ and $p^0<0$ form different orbits. The associated representations are called positive and negative energy \emph{massless} representations.
  Finally, $m^2<0$ corresponds to a one-sheeted hyperboloid.  It is a single orbit and the associated representations are called \emph{tachyonic}.
 
Now we proceed to the discussion of  the associated stability subgroups. 
 For a vanishing vector the stability subgroup is, clearly, the whole Lorentz group.
  By going to the frame in which $p_0=(m,0,\dots, 0)$, it is trivial to see that the stability subgroup in the massive case is $SO(d-1)$. In a similar manner, one can see that the stability subgroup in the tachyonic case is $SO(d-2,1)$.
 
 The stability subgroup in the massless case is less obvious. We will explore it at the level of the Lie algebra.
To start, we use Lorentz transformations to bring  momentum to the form
\begin{equation}
\label{15sep5}
(p_+, p_-,p_i)=(p_+,0,0),
\end{equation}
where the light-cone coordinates are defined by
\begin{equation}
\label{26mar1}
p_{\pm}=\frac{1}{\sqrt{2}}(p_{d-1}\pm p_0), \qquad p_i=\{ p_1,\dots, p_{d-2}\}.
\end{equation}
In these coordinates, the metric has the following non-vanishing components
\begin{equation}
\label{14sep5}
\eta^{+-}=\eta^{-+}=1, \qquad \eta^{ij}=\delta^{ij}.
\end{equation}

We are looking for the subalgebra of the Lorentz algebra, that leaves (\ref{15sep5}) invariant. This amounts to solving
\begin{equation}
\label{15sep6}
J_{\mu\nu}p^\nu=0,
\end{equation}
which gives
\begin{equation}
\label{15sep7}
J_{\mu -}=0.
\end{equation}
The generators that are not set to zero by (\ref{15sep7}) are $J_{ij}$ and $J_{i+}$. Clearly, $J_{ij}$ generate $SO(d-2)$. It is trivial to see that $J_{i+}$ is a vector with respect to $J_{ij}$. Moreover, using the commutation relations of the  Lorentz algebra (\ref{1x22sep1}), one easily finds that $J_{i+}$ commute with each other. Summing up, we found that the stability subalgebra of (\ref{15sep5}) is  $iso(d-2)$.

Zero-momentum and tachyonic representations have non-compact stability subgroups. The theorem that we quoted above implies that the associated unitary representations are infinite-dimensional. This means that the associated particles for each momentum should carry infinitely many degrees of freedom. Though, there is nothing wrong with that in the context of the previous analysis, this is not what we observe in nature. Moreover, for tachyonic representations energy is not bounded from below, which is another requirement that one typically adds to have consistent physics. These types of representations will not be discussed below. 

To proceed, we need to study UIR's of the Wigner little group in each remaining case -- massive and massless fields. For massive fields the Wigner little group  is  the orthogonal group $SO(d-1)$ and its UIR's are well-known. These will be discussed in the following section. 

For massless fields the stability subgroup is $ISO(d-2)$ and the analysis is a bit more complicated.
 Its UIR's  can be induced from the stability subgroup of ''momentum'' $\pi_i \equiv J_{i+}$ in the same way as for the Poincare group. The only difference is that now the counterpart of the Lorentz group is the orthogonal group $SO(d-2)$. The latter has only two orbits: $\pi^2>0$ and $\pi =0$. For the orbit $\pi=0$  the inhomogeneous part of $ISO(d-2)$ is realised trivially, so, in effect, the analysis boils down to the classification of UIR's of $SO(d-2)$.
 By a shift of dimension, this classification reduces to the one in the massive case. The associated representations are often called \emph{helicity} representation or, with some abuse of terminology, just \emph{massless} representations. 
 
 Another class of massless representations -- those with $\pi^2>0$ -- is called \emph{continuous spin} or \emph{infinite spin} representations. These also feature infinitely many degrees of freedom at fixed momentum -- labelled by points on a sphere of constant $\pi^2$. We will not discuss continuous spin fields in the following.

\subsubsection{Representations of the orthogonal algebra}
\label{sect:232}

As we  saw  the problem of classification of fields of most physical relevance -- massless helicity and massive fields -- reduces to the problem of classification of unitary irreducible representations of the orthogonal group $SO(n)$. Below we will consider a couple of suggestive examples and then give the end result for this classification. For a more comprehensive analysis and further details see \cite{Bekaert:2006py,Vasilievutrecht} and references therein.

An obvious representation of $SO(n)$ is the scalar one -- with the representation space being one-dimensional and transforming trivially under $SO(n)$. The inner product 
\begin{equation}
\label{1x29oct1}
(\varphi,\psi)\equiv \bar\varphi\psi
\end{equation}
is, obviously, positive definite and satisfies the remaining requirements in (\ref{15mar1}).

The second most trivial example is given by the vector representation. To form an invariant norm, one utilises an $so(n)$-invariant metric $\delta_{ij}$
\begin{equation}
\label{1x29oct2}
(\varphi,\psi)\equiv \bar\varphi^i\psi^j\delta_{ij}.
\end{equation}
Again, it is straightforward to see that it satisfies all requirements in (\ref{15mar1}).

Once the vector representation is available, we proceed to tensor products thereof, which results into tensor representations of $SO(d)$. For the simplest case of a tensor product of two vectors, we have a rank-two tensor $\varphi^{ij}$. It is then easy  to see that decomposition of $\varphi$
\begin{equation}
\label{1x29oct3}
\varphi^{ij}=\varphi_S^{ij}+\varphi_A^{ij}+\delta^{ij}\varphi_T
\end{equation}
into its symmetric traceless, antisymmetric  and pure trace parts 
\begin{equation}
\label{1x29oct4}
\varphi_S^{ij}=\varphi_S^{ji}, \qquad \varphi_S^{ij}\delta_{ij}=0,\qquad
\varphi_A^{ij}= -\varphi_A^{ji}
\end{equation}
is invariant under the action of $SO(d)$. It turns out that each component $\varphi_S$, $\varphi_A$ and $\varphi_T$ is irreducible and gives a unitary representation. Moreover, $\varphi_T$ transforms as a scalar.

\begin{zadacha}
Show that the property that the rank-2 tensor is symmetric is invariant with respect to general linear transformations.
\end{zadacha}

\begin{zadacha}
Show that the norm
\begin{equation}
\label{1x29oct2x1}
(\varphi_S,\psi_S)\equiv \bar\varphi_S^{ik}\psi_S^{jl}\delta_{ij}\delta_{kl}
\end{equation}
is positive definite in $d=2$.
\end{zadacha}

From this example we learn that tensor representations of $SO(n)$ are, in general, reducible. This happens due to the fact that symmetry properties as well as contractions with the invariant metric remain invariant with respect to the action of $SO(n)$. A non-trivial question that we need to answer is what precise symmetry and trace constraints lead to irreducible representations. Below, we present the answer to this question.

To characterise a symmetry of a tensor it is essential  to use
\emph{Young diagrams}. A Young diagram is a diagram consisting of boxes 
arranged into rows of non-increasing lengths. For example, a
Young diagram
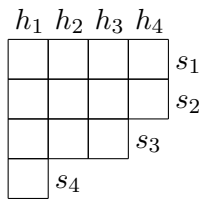
\begin{eqnarray*}
\begin{picture}(75,75)%
\put(0,45){\begin{picture}(15,15)%
\put(0,0){\line(1,0){15}}
\put(0,15){\line(1,0){15}}
\put(0,0){\line(0,1){15}}
\put(15,0){\line(0,1){15}}
\end{picture}}
\put(15,45){\begin{picture}(15,15)%
\put(0,0){\line(1,0){15}}
\put(0,15){\line(1,0){15}}
\put(0,0){\line(0,1){15}}
\put(15,0){\line(0,1){15}}
\end{picture}}
\put(30,45){\begin{picture}(15,15)%
\put(0,0){\line(1,0){15}}
\put(0,15){\line(1,0){15}}
\put(0,0){\line(0,1){15}}
\put(15,0){\line(0,1){15}}
\end{picture}}
\put(45,45){\begin{picture}(15,15)%
\put(0,0){\line(1,0){15}}
\put(0,15){\line(1,0){15}}
\put(0,0){\line(0,1){15}}
\put(15,0){\line(0,1){15}}
\end{picture}}
\put(0,30){\begin{picture}(15,15)%
\put(0,0){\line(1,0){15}}
\put(0,15){\line(1,0){15}}
\put(0,0){\line(0,1){15}}
\put(15,0){\line(0,1){15}}
\end{picture}}
\put(15,30){\begin{picture}(15,15)%
\put(0,0){\line(1,0){15}}
\put(0,15){\line(1,0){15}}
\put(0,0){\line(0,1){15}}
\put(15,0){\line(0,1){15}}
\end{picture}}
\put(30,30){\begin{picture}(15,15)%
\put(0,0){\line(1,0){15}}
\put(0,15){\line(1,0){15}}
\put(0,0){\line(0,1){15}}
\put(15,0){\line(0,1){15}}
\end{picture}}
\put(45,30){\begin{picture}(15,15)%
\put(0,0){\line(1,0){15}}
\put(0,15){\line(1,0){15}}
\put(0,0){\line(0,1){15}}
\put(15,0){\line(0,1){15}}
\end{picture}}
\put(0,15){\begin{picture}(15,15)%
\put(0,0){\line(1,0){15}}
\put(0,15){\line(1,0){15}}
\put(0,0){\line(0,1){15}}
\put(15,0){\line(0,1){15}}
\end{picture}}
\put(15,15){\begin{picture}(15,15)%
\put(0,0){\line(1,0){15}}
\put(0,15){\line(1,0){15}}
\put(0,0){\line(0,1){15}}
\put(15,0){\line(0,1){15}}
\end{picture}}
\put(30,15){\begin{picture}(15,15)%
\put(0,0){\line(1,0){15}}
\put(0,15){\line(1,0){15}}
\put(0,0){\line(0,1){15}}
\put(15,0){\line(0,1){15}}
\end{picture}}
\put(0,0){\begin{picture}(15,15)%
\put(0,0){\line(1,0){15}}
\put(0,15){\line(1,0){15}}
\put(0,0){\line(0,1){15}}
\put(15,0){\line(0,1){15}}
\end{picture}}
\put(2.5,64){$h_1$}
\put(17.5,64){$h_2$}
\put(32.5,64){${h_3}$}
\put(47.5,64){${h_4}$}
\put(62.5,49){${s_1}$}
\put(62.5,34){${s_2}$}
\put(47.5,19){${s_3}$}
\put(17.5,4){${s_4}$}
\end{picture}
\nonumber
\end{eqnarray*}
consists of four rows of lengths $s_1=4$, $s_2=4$, $s_3=3$ and $s_4=1$. 
A Young diagram with lengths of rows equal to $s_1$, $s_2$, $\dots$ 
will be denoted as $\mathbf{Y}(s_1,s_2,\dots)$.
Equivalently, a Young diagram can be defined by listing heights of 
its columns $h_1$, $h_2$, $\dots$. Then it will be denoted  
as $\mathbf{Y}[h_1,h_2,\dots]$. 

There are two equivalent ways of  assigning a symmetry 
of a Young diagram 
\begin{equation}
\label{1x29oct5}
\mathbf{Y}(s_1,s_2,\dots,s_{h_1})
\end{equation}
to a rank-$k$ tensor $\varphi$ with $k=\sum_i{s_i}=\sum_i{h_i}$. First, one should associate 
each index of $\varphi$ with a box of the Young diagram $\mathbf{Y}$. 
Then the tensor $\varphi$ is said to possess a symmetry of $\mathbf{Y}$ 
in \emph{symmetric convention} iff:
\begin{itemize}
\item[i)]  tensor $\varphi$ is symmetric with respect 
to permutations of indices associated with the same row of $\mathbf{Y}$,
\item[ii)] symmetrization of $\varphi$ in all indices associated 
 with  the $i$th row of $\mathbf{Y}$ with any index 
 from  the $j$th row with $j> i$ gives zero identically.
\end{itemize}
Alternatively, one says that a tensor $\varphi$  possesses a symmetry of $\mathbf{Y}$ 
in \emph{antisymmetric convention} iff:
\begin{itemize}
\item[i)] tensor $\varphi$ is antisymmetric with respect 
to permutations of indices associated with the same column of $\mathbf{Y}$,
\item[ii)] antisymmetrization of $\varphi$ in all indices associated 
 with the $i$th column of $\mathbf{Y}$ with any index from the 
 $j$th column with $j> i$ gives zero identically.
\end{itemize}

If a tensor $\varphi$ possesses a symmetry of a Young diagram $\mathbf{Y}$ we
will often say that $\varphi$ is of shape $\mathbf{Y}$.
A $GL(n)$ tensor  cannot be of shape that has more than $n$ rows, otherwise, the tensor vanishes identically. This is seen the easiest for Young diagrams with one column: non-vanishing components of a totally antisymmetric rank-$k$
tensor have all their indices different, which is impossible for $k>n$.

The important fact is that \emph{tensors possessing symmetries of Young diagrams with no more than $n$ rows
cover all finite-dimensional irreducible representations of $GL(n)$}.
For example, totally symmetric (of shape $\mathbf{Y}(s)$) and 
totally antisymmetric (of shape $\mathbf{Y}[h]$) tensors are irreducible 
under $GL(n)$. Other tensors are said to be of \emph{mixed-symmetry}.

We will mostly use the symmetric convention for the Young symmetry of tensors. For tensor $\varphi$ with symmetry $\mathbf{Y}(s_1,s_2,\dots)$ we will use notation
\begin{equation}
\label{1x29oct6}
\varphi^{i(s_1),i(s_2),\dots}.
\end{equation}
For conventions on symmetric indices and symmetrisation, see appendix \ref{appa:conventions}. For tensors with symmetry given in the antisymmetric convention we will use
\begin{equation}
\label{1x29oct6x1}
\varphi^{i[h_1],i[h_2],\dots}.
\end{equation}

The following exercises illustrate some aspects of the above stated results.

\begin{zadacha}
\label{z1}
 Show that if a rank-3 tensor is symmetric in the first two indices and antisymmetric in the first and the third indices, then it is identically zero.
\end{zadacha}

\begin{zadacha}
 Show that if a rank-3 tensor $\varphi^{i_1i_2,j}$ has the Young symmetry $\mathbf{Y}(2,1)$ in the symmetric convention, then 
\begin{equation}
\label{1x29oct7}
\tilde\varphi^{i_1i_2j}=\varphi^{i_1i_2,j}-\varphi^{ji_2,i_1}
\end{equation}
has symmetry $\mathbf{Y}[2,1]$ in the antisymmetric convention.
\end{zadacha}

\begin{zadacha}
 Show that a rank-3 tensor of symmetry $\mathbf{Y}(1,2)$ is identically zero.
 \end{zadacha}

\begin{zadacha}
 Show that a tensor of symmetry $\mathbf{Y}(1,1,\dots, 1)$ is totally antisymmetric.
 \end{zadacha}

To make tensors irreducible under $SO(m,k)$\footnote{The above discussion on irreducibility applies to any signature. Though, keep in mind that tensorial representations for non-Euclidean signature are non-unitary.}
in addition to the Young symmetry conditions 
some trace constraints have to be imposed.
The simplest trace constraint making $\mathbf{Y}$-shaped tensor 
irreducible under (pseudo)orthogonal group is 
just the requirement of the tensor to be traceless 
with respect to the $SO(m,k)$-invariant metric on all pairs of indices.
\emph{To indicate that a tensor, in addition to the Young symmetry is also traceless, we will use} $\mathbb{Y}$. Clearly, antisymmetric tensor are automatically traceless, so for them this distinction is immaterial. 

It is important to stress that whenever a traceless tensor $\varphi$ 
is of some shape $\mathbb{Y}$ with more than $n=m+k$ boxes  
in the first two columns, then  $\varphi$ is identically zero.
Young diagrams such that the sum of the heights of the 
first two columns does not exceed $n$ are said to be \emph{allowed}.

\begin{zadacha}
Show that a traceless tensor of shape $\mathbb{Y}(2,1)$ is identically zero in $d=2$.
\end{zadacha}

It remains to note that in addition to $\delta_{ij}$, $SO(m,k)$ features another invariant tensor: the totally antisymmetric rank-$n$ Levi-Civita tensor $\epsilon^{i[n]}$, see appendix \ref{appa:conventions} for conventions. By contracting it with all indices of a totally antisymmetric tensor of rank $r$, we obtain an equivalent representation, given by a tensor with symmetry $\mathbf{Y}[n-r]$. For $n$ even and for a tensor of shape $\mathbf{Y}[\frac{n}{2}]$, one can impose an additional irreducibility condition that the tensor is (anti)-self-dual. Similar considerations apply to traceless tensors of allowed shapes and to dualisations with respect to the indices of the first column. One can show that the resulting dual tensors are still traceless and have the Young symmetry characterised by diagrams of  allowed shapes. The same is not true for dualisations on other columns, in particular, dual tensors are no longer traceless.

A natural way to obtain tensors with the Young symmetry is to start from a tensor possessing no symmetry and act on it with symmetry projectors as we did in the rank-2 case in (\ref{1x29oct3}). For a tensor of arbitrary rank it works as follows. First, one needs to associate each index of a tensor with a box in a Young diagram of a suitable size. Then, by carrying out antisymmetrization on indices in each column and then symmetrising the result on indices in every row we obtain a tensor possessing symmetry of the given Young diagram in the symmetric convention. Similarly, if we first symmetrise indices in each row and then antisymmetrize them in every column we obtain a tensor with the Young symmetry in the antisymmetric convention. It is worth keeping in mind that by distributing indices in the very same Young diagram at the first step of the procedure in different ways and carrying out (anti)-symmetrisations we end up getting different tensors, which, may not even be related by permutations of indices. 

To illustrate this last phenomenon, we consider a tensor $T_{ijk}$ and first antisymmetrize it on $i$ and $k$ and then symmetrise it on $i$ and $j$. As a result, we obtain a tensor
\begin{equation}
\label{5xnov1}
R_{ij,k}=\frac{1}{4}(T_{ijk}+T_{jik}-T_{kji}-T_{kij}),
\end{equation}
which has the symmetry type $\mathbf{Y}(2,1)$ with $k$ being in the second row. By using the Young symmetry, it is easy to see that there are only two independent permutations of indices of $R$. These can be taken to be $R_{ij,k}$ and $R_{ik,j}$. Besides that, one can construct another tensor by first antisymmetrizing $T$ on $i$ and $j$ and then symmetrising it on $i$ and $k$. As a result we obtain
\begin{equation}
\label{5xnov2}
P_{ik,j}=\frac{1}{4}(T_{ijk}+T_{kji}-T_{jik}-T_{jki}),
\end{equation}
which has the same symmetry type $\mathbf{Y}(2,1)$, but  with $j$ being in the second row. Again, it has two non-trivial permutations, which can be taken to be $P_{ij,k}$ and $P_{ik,j}$. It is not hard to check that all tensors  $R_{ij,k}$, $R_{ik,j}$,  $P_{ij,k}$ and $P_{ik,j}$ are linearly independent. In particular, $R_{ij,k}$ and $P_{ij,k}$, despite having the same symmetry type and locations of indices in the Young diagram, are not proportional.

Above we explained how by imposing symmetry, trace and duality constraints one can make a tensor representation of $SO(m,k)$ irreducible. Tensors with these irreducibility conditions imposed  \emph{give a complete list of finite-dimensional irreducible representations of $SO(n)$}. An important result in the context of our field-theoretic discussion is that in the Euclidean $SO(n)$ case \emph{all these representations are unitary}.

\section{From massless UIR's to the Fronsdal action}
\label{sec:3}

In the previous section we saw that massless UIR's of the Poincare group  can be induced from UIR's of the Wigner little group, $ISO(d-2)$, associated with light-like momenta $p^2=0$. We will focus here, more specifically, on helicity representations, for which the inhomogeneous part of the Wigner little group is realised trivially, so, in effect, the little group reduces to $SO(d-2)$.  In this section our goal will be to rephrase this representation theory result in a more conventional field-theoretic manner. From now on, we will mostly  focus on the symmetric representations of the Wigner little group.

 This section gives a slightly rearranged material from section 2.1 of \cite{Didenko:2014dwa}. Note that we use different symmetrization conventions compared to \cite{Didenko:2014dwa}. For our conventions on dealing with symmetric indices, see appendix \ref{appa:conventions}.

\subsection{Lorentz-covariant form of massless representations}
\label{sec:3.1}

After the success of the Special Relativity, physicists acknowledged the important role of Lorentz symmetries as symmetries of nature. In field theory it is typical to make these symmetries manifest and to do that the standard way is to employ tensor fields. This is not, strictly speaking, necessary. One can successfully deal with massless field theories e. g. in the light-cone gauge, which will be discussed in section \ref{sec:13}.  In the present section we will show how the representation theory analysis of the previous section can be rephrased into a manifestly Lorentz-covariant form.

The solution to this problem in the massless case is given by equations 
\begin{equation}
\label{14sep1}
\begin{split}
\Box\varphi^{\mu(s)}=0,\\
\partial_{\nu} \varphi^{\nu \mu(s-1)}=0,\\
\varphi_\nu{}^{\nu \mu(s-2)}=0,
\end{split}
\end{equation}
modulo pure gauge solutions of the form
\begin{equation}
\label{14sep2}
\delta \varphi^{\mu(s)}=\partial^\mu\xi^{\mu(s-1)},
\end{equation}
where the gauge parameter satisfies
\begin{equation}
\label{14sep3}
\begin{split}
\Box\xi^{\mu(s-1)}=0,\\
\partial_\nu \xi^{\nu \mu(s-2)}=0,\\
\xi_\nu{}^{\nu \mu(s-3)}=0.
\end{split}
\end{equation}
In other words, \emph{massless} bosonic fields of spin $s\ge 1$ \emph{are gauge fields} (for $s=0$ (\ref{14sep2}) does not make sense and gauge symmetry is absent). 
Let us demonstrate, that system (\ref{14sep1})-(\ref{14sep3}), indeed, describes massless UIR's as constructed using Wigner's approach. 

To start, we recall that the very definition of tensor fields implies that they transform appropriately under diffeomorphisms, which, in particular, include transformations from the Poincare group. With the proper normalisation, for translations one has
\begin{equation}
\label{15apr1x1}
U(P_\mu)\varphi =-i\partial_\mu\varphi.
\end{equation}
Thus, the $p$ labels of Wigner's approach of induced representations are nothing but momenta in the Fourier space.

Then, it is easy to see that  the first equation in (\ref{14sep1}) implies that momenta carried by solutions to (\ref{14sep1})-(\ref{14sep3}) are, indeed, light-like, $p^2=0$. As in the Wigner's procedure, we will now focus on a solution with a well-defined momentum and then, using Lorentz symmetry transform it to some convenient form. 

To this end,  it will by helpful to use the light-cone coordinates (\ref{26mar1}) and pick
 the standard light-like momentum as in (\ref{15sep5}).
Then, the second equation in (\ref{14sep1}) implies that 
\begin{equation}
\label{14sep7}
\varphi^{+ \mu(s-1)}=0,
\end{equation}
that is $\varphi$ with at least one index being ''$+$'' vanishes. Next, gauge symmetry (\ref{14sep2}) implies that 
\begin{equation}
\label{14sep8}
\varphi^{- \mu(s-1)}\sim 0,
\end{equation}
that is all $\varphi$ with at least one index being ''$-$'' are pure gauge. Taking (\ref{14sep7}), (\ref{14sep8}) together, we are left with $\varphi$'s with all indices in the set $2,\dots,d-1$. These are exactly the tensor indices of the Wigner little group $SO(d-2)$ discussed below (\ref{15sep7}).
 The third equation in (\ref{14sep1}) then implies that, in addition, $\varphi$ is traceless
\begin{equation}
\label{14sep9}
\varphi_j{}^{j i(s-2)}=0.
\end{equation}
The last equation describes a tensor representation of the Wigner little group of shape $\mathbb{Y}(s)$, as we intended to show.
Accordingly, (\ref{14sep1})-(\ref{14sep3}) describe a massless spin-s field.

\begin{zadacha}
 Show that the rank-two antisymmetric field $\varphi$
 \begin{equation}
\label{27mmar1}
\begin{split}
\Box\varphi^{\mu[2]}=0,\\
\partial_\nu \varphi^{\nu \mu}=0,
\end{split}
\end{equation}
with gauge symmetry 
\begin{equation}
\label{27mar2}
\delta \varphi^{\mu[2]}=\partial^{[\mu}\xi^{\mu]},
\end{equation}
where the parameter satisfies 
\begin{equation}
\label{27mar3}
\partial_{\mu}\xi^{\mu}=0, \qquad \Box \xi^\mu=0
\end{equation}
and, the gauge symmetry is redundant in the sense that gauge transformations with parameters
\begin{equation}
\label{27mar4}
\delta\xi^\mu =\partial^\mu \lambda, \qquad \Box \lambda=0
\end{equation}
do not act on $\varphi$, describes the massless helicity representation characterised by tensor $\mathbf{Y}[2]$. This gives a simplest example of a non-symmetric massless field and, as we can see, these may require redundant gauge symmetries also known as gauge for gauge transformations.
 \end{zadacha}

\subsection{Fronsdal's action}

For many purposes, e. g. for quantisation, it is convenient to have an action for a field theory.
In our case, we are looking for an action, that would produce equations (\ref{14sep1}) together with symmetries (\ref{14sep2}), (\ref{14sep3}).

In field theory one, usually,  treats differential and algebraic equations differently. Na\-me\-ly, while it is considered acceptable to impose algebraic equations by hand, one requires that differential equations result from the variational principle. Similarly, it is not typical to require that gauge parameters satisfy any differential constraints off-shell. Thus, for the system (\ref{14sep1}), we would like to derive the first two equations from action variation, while the last equation, the requirement of tracelessness, can be set by hand. Similarly, while the last equation in (\ref{14sep3}) can be imposed as an off-shell constraint, we demand that the first two equations come out dynamically, e.g. from partial  gauge fixing.

It is typical that a system of equations of motion cannot be derived from a Lagrangian unless \emph{auxiliary fields} are added. Indeed, Lagrangian systems may only contain as many equations of motion as they contain fields, so if the number of dynamical fields does not match  the number of equations of motion, auxiliary fields need to be added. This argument becomes  more complex for gauge theories, as some equations may result from gauge fixing. Still, adding auxiliary fields is typically a necessity.  The role of these fields is to produce the missing equations, while still not contributing to the degrees of freedom count by virtue of being either pure gauge or being expressed in terms of the physical fields, once the equations of motion are taken into account. The quest for the right set of auxiliary fields, that allows one to write an action leading to the desired equations of motion is  often a tedious and indirect task. Moreover, the result of this procedure is not unique: different off-shell systems may result into the same equations of motion.

The \emph{minimal} set of auxiliary fields suitable for the description of system (\ref{14sep1})-(\ref{14sep3}) was found by Fronsdal \cite{Fronsdal:1978rb}\footnote{A systematic search of the minimal set of traceless symmetric fields that would be sufficient to write a Lagrangian for a free massive spin-$s$ field was carried out in \cite{Singh:1974qz}. The Fronsdal action was obtained as the massless limit of \cite{Singh:1974qz}. The Fronsdal action for a massless spin-$s$ field features traceless symmetric tensors of ranks $s$ and $s-2$, see (\ref{14sep10}), (\ref{30dec2}). The rank-$s$ traceless field is, clearly, necessary as it features (\ref{14sep1}) explicitly. The rank-$(s-2)$ traceless field is auxiliary. It is needed because for $\xi$ free of any differential constraints gauge transformation  (\ref{14sep2}) violates tracelessness of $\varphi$: even for $\xi$ traceless, $\varphi$ should have at least one non-trivial trace originating from the divergence of $\xi$. Thus, at least one auxiliary symmetric and traceless  rank-$(s-2)$ field is necessary  to accommodate the trace of $\varphi$. This proves that Fronsdal's set of off-shell fields is, indeed, minimal provided we do not allow  differential constraints on gauge parameters imposed off-shell.}. In this approach, the off-shell field is required to be double-traceless
\begin{equation}
\label{14sep10}
\varphi_{\mu \nu}{}^{\mu \nu \rho(s-4)}=0.
\end{equation}
It can be interpreted as a combination of two traceless fields
\begin{equation}
\label{30dec2}
\varphi^{\mu(s)}=\psi^{\mu(s)}+\eta^{\mu\mu}\chi^{\mu(s-2)}, \qquad \psi_{\nu}{}^{\nu \mu(s-2)}=0, \qquad \chi_{\nu}{}^{\nu \mu(s-4)}=0,
\end{equation}
where $\psi$ is the dynamical field, which will  be identified with $\varphi$ in (\ref{14sep1}), while $\chi$ is auxiliary.

The symmetries of the action read
\begin{equation}
\label{14sep11}
\delta\varphi^{\mu(s)}=\partial^\mu \xi^{\mu(s-1)},
\end{equation}
where 
\begin{equation}
\label{14sep12}
\xi_{\mu}{}^{\mu \nu(s-3)}=0.
\end{equation}
The action itself is given by
\begin{equation}
\label{14sep13}
\begin{split}
S=&-\frac{1}{2}\int d^dx \left( \partial_\mu\varphi^{\nu(s)}\partial^\mu \varphi_{\nu(s)}-\frac{s(s-1)}{2}\partial_\mu \varphi^{\nu}{}_{\nu}{}^{\rho(s-2)}\partial^\mu \varphi_\sigma{}^\sigma{}_{\rho(s-2)}\right.\\
&\qquad\qquad\quad+ s(s-1) \partial^\mu \varphi^\nu{}_\nu{}^{\rho(s-2)}\partial^\lambda \varphi_{\lambda \mu \rho(s-2)}-s\partial_\mu\varphi^{\mu \nu(s-1)} \partial^\rho\varphi_{\rho \nu(s-1)}\\
&\qquad \qquad\qquad\qquad\qquad\qquad\quad \left.-\frac{s(s-1)(s-2)}{4} \partial_\mu\varphi^\nu{}_\nu{}^{\mu \rho(s-3)}\partial^\lambda\varphi_\lambda{}^\sigma{}_{\sigma \rho(s-3)}\right).
\end{split}
\end{equation}
It is fixed up to an overall factor by the requirement of gauge invariance (\ref{14sep11}), which everywhere in  these lectures is understood up to boundary terms.
Below our goal will be to reproduce the system (\ref{14sep1})-(\ref{14sep3}) from (\ref{14sep13}).

To start, we note that integrating by parts (\ref{14sep13}) can be brought to the form
\begin{equation}
\label{14sep14}
S=\frac{1}{2}\int d^dx \varphi_{\mu(s)}G^{\mu(s)}[\varphi],
\end{equation}
where
\begin{equation}
\label{14sep15}
G^{\mu(s)} \equiv F^{\mu(s)}-\frac{s(s-1)}{4}\eta^{\mu\mu}F^{\mu(s-2)\nu}{}_\nu,
\end{equation}
\begin{equation}
\label{14sep16}
F^{\mu(s)} \equiv \Box \varphi^{\mu(s)}-s\partial^\mu D^{\mu (s-1)},
\end{equation}
\begin{equation}
\label{14sep17}
D^{\mu(s-1)}\equiv \partial_\nu \varphi^{\nu \mu(s-1)}-\frac{s-1}{2}\partial^\mu\varphi^{\mu(s-2)\nu}{}_\nu.
\end{equation}
This means that equations of motion read
\begin{equation}
\label{14sep18}
G^{\mu(s)}\approx 0.
\end{equation}
To arrive to this conclusion, one needs to exploit the following two facts. Firstly, $G$ is self-adjoint, when regarded as an operator acting on $\varphi$. This ensures that by varying (\ref{14sep14}) with respect to $\varphi$ which is written explicitly in (\ref{14sep14}) we get the same contribution to the equations of motion as by varying  with respect to the second $\varphi$ contained in $G$. Secondly, considering that $\varphi$ is double-traceless, the variational principle only entails that the double-traceless part of $G$ is vanishing. However, as can be checked, $G$ as written in (\ref{14sep15}) is already double-traceless
\begin{equation}
\label{15apr2x1}
G^{\mu(s-4)\nu\rho}{}_{\nu\rho} =0, \qquad F^{\mu(s-4)\nu\rho}{}_{\nu\rho} =0,
\end{equation}
 so equations of motion, indeed, are given by (\ref{14sep18}).

Equations of motion can be rewritten in the equivalent form 
\begin{equation}
\label{14sep19}
F^{\mu(s)}\approx 0.
\end{equation}
To see this, we take the trace of (\ref{14sep18}) with (\ref{14sep15}) and (\ref{15apr2x1}) taken into account. This leads to
\begin{equation}
\label{14sep20}
F^{\mu(s-2)\nu}{}_\nu\approx 0.
\end{equation}
It is then easy to see that (\ref{14sep18}) implies (\ref{14sep19}). In a similar way one shows that (\ref{14sep19}) entails (\ref{14sep18}), so these are equivalent\footnote{Relation between $F$ and $G$ is the higher-spin analogue of that between $R_{\mu\nu}$ and $G_{\mu\nu}\equiv R_{\mu\nu}-\frac{1}{2}Rg_{\mu\nu}$ in General Relativity.}.

We will now focus on (\ref{14sep19}). It is convenient to impose the de Donder gauge
\begin{equation}
\label{14sep21}
D^{\mu(s-1)}=0.
\end{equation}
The residual gauge symmetry is given by (\ref{14sep11}), where the gauge parameter is constrained 
\begin{equation}
\label{14sep22}
\Box \xi^{\mu(s-1)}=0.
\end{equation}
The equations of motion in this gauge give
\begin{equation}
\label{14sep23}
\Box \varphi^{\mu(s)}\approx 0.
\end{equation}

Solutions to (\ref{14sep22}) are non-trivial, so the gauge can be fixed further.
Consider variation 
\begin{equation}
\label{14sep24}
\delta \varphi^{\mu(s-2)\nu}{}_\nu=\frac{2}{s}\partial_\nu\xi^{\nu \mu(s-2)}.
\end{equation}
Taking into account (\ref{14sep22}) and (\ref{14sep23}), we can see from (\ref{14sep24}) that traces of $\varphi$ can be gauged away
\begin{equation}
\label{14sep25}
\varphi^{\mu(s-2)\nu}{}_\nu=0.
\end{equation}
Gauge parameter of the residual symmetry then satisfies
\begin{equation}
\label{15sep2}
\partial_\mu\xi^{\mu \nu(s-2)}=0.
\end{equation}
Finally, considering definition (\ref{14sep17}) and gauge conditions (\ref{14sep21}), (\ref{14sep25}), we find 
\begin{equation}
\label{15sep1}
\partial_\mu \varphi^{\mu \nu(s-1)}=0.
\end{equation}

Summing up, we found that the Fronsdal action results into (\ref{14sep23}), (\ref{14sep25}), (\ref{15sep1}) for fields and into (\ref{14sep12}), (\ref{14sep22}), (\ref{15sep2}) for gauge parameters, which gives exactly the system (\ref{14sep1})-(\ref{14sep3}), that we intended to describe.

Before finishing this section, let us note that the fact that the action (\ref{14sep14}) is gauge invariant, implies
\begin{equation}
\label{15sep3}
\delta S=\int d^dx \partial_\mu \xi_{\mu(s-1)}G^{\mu(s)}=-\int d^dx  \xi_{\mu(s-1)}\partial_\nu G^{\nu \mu(s-1)}=0.
\end{equation}
This requires
\begin{equation}
\label{15sep4}
\partial_\mu G^{\mu \nu(s-1)}=\eta^{\nu\nu}H^{\nu(s-3)},
\end{equation}
for $H$ being any tensor. Indeed, considering that $\xi$ is traceless, the form of $\partial G$ in (\ref{15sep4}) is necessary to ensure that (\ref{15sep3}) vanishes. Equation (\ref{15sep4}) is a higher-spin version of the familiar \emph{Bianchi identities}.

\begin{zadacha}
 Check that the Fronsdal tensor $F$ given in (\ref{14sep16}) is gauge invariant with respect to (\ref{14sep11}).
 \end{zadacha}

\begin{zadacha}
 Show that the residual symmetry in the de Donder gauge (\ref{14sep21}) is given by (\ref{14sep22}).
 \end{zadacha}

\begin{zadacha}
 Verify the Bianchi identity (\ref{15sep4}) (a bit more technical).
 \end{zadacha}

\begin{zadacha}
 Check that for $s=1$ the Fronsdal action reproduces the action of Maxwell's theory.
 \end{zadacha}
 
 \begin{zadacha}
Linearise the action of General Relativity in terms of metric fluctuations. Check that it agrees with the $s=2$ case of the Fronsdal action.
 \end{zadacha}
 
 The last problem explains why one typically refers to approaches employing fields of Fronsdal's action as \emph{metric-like}.

As a side remark, we mention that Maxwell's field strength and the linearised Riemann tensor can be straightforwardly generalised to higher spins. Namely, the higher-spin field strength is defined as
\begin{equation}
\label{5xnov3}
F_{\mu_1\nu_1,\mu_2\nu_2,\dots, \mu_s\nu_s}\equiv \frac{1}{2^s}(\partial_{\mu_1}\partial_{\mu_2}\dots \partial_{\mu_s}\varphi_{\nu(s)}-
\partial_{\nu_1}\partial_{\mu_2}\dots \partial_{\mu_s}\varphi_{\mu_1\nu(s-1)}+\dots ),
\end{equation}
where the implicit terms antisymmetrize the expression on any pair $(\mu_i,\nu_i)$. Then, $F$ has symmetry of a rectangular two-row Young diagram of length $s$ in the antisymmetric convention. Considering that derivatives commute, $F$ is gauge invariant with respect to the Fronsdal-type gauge transformations (\ref{14sep11}). Note that this argument does not rely on trace properties of $\varphi$ and $\xi$. Analogously, one can see that $F$ satisfies the Bianchi identity
\begin{equation}
\label{5xnov4}
\partial_{[\rho}F_{\mu_1\nu_1],\mu_2\nu_2,\dots, \mu_s\nu_s}=0
\end{equation}
and other properties similar to those of the Maxwell field strength and the linearised Riemann tensor.

\subsection{Massive fields}

Despite this goes away from the main goal of our course, for completeness, we briefly mention the analogous construction for massive fields. 

To start, the Lorentz-covariant way to present massive UIR's induced by Wigner's approach is 
\begin{equation}
\label{16sep1}
\begin{split}
(\Box-m^2)\varphi^{\mu(s)}=0,\\
\partial_\nu \varphi^{\nu \mu(s-1)}=0,\\
\varphi_\nu{}^{\nu \mu(s-2)}=0.
\end{split}
\end{equation}
By picking the standard massive momentum as
\begin{equation}
\label{16sep2}
p_0=m \qquad \text{and} \qquad  p_\mu=0, \quad \mu\ne 0,
\end{equation}
it is straightforward to see that the representation of the little group $SO(d-1)$ defined by (\ref{16sep1}) is given by traceless rank-$s$ tensors, which correspond to massive spin-$s$ fields.  

As for massless fields, writing an action that would result in (\ref{16sep1}) is not easy. Again, the problem is that equations (\ref{16sep1}) for general $s$ are not naturally Lagrangian in the sense that there are more fields than equations. 

In the simplest spinning case the action can be written as
\begin{equation}
\label{16sep3}
S=\int d^dx \left(-\frac{1}{4}F_{\mu \nu}F^{\mu\nu}-\frac{1}{2}\varphi_\mu m^2 \varphi^\mu \right),
\end{equation}
where 
\begin{equation}
\label{16sep4}
F_{\mu\nu}=\partial_\mu \varphi_\nu-\partial_\nu \varphi_\mu.
\end{equation}
By varying (\ref{16sep3}), we get
\begin{equation}
\label{16sep5}
\Box \varphi^\mu-\partial^\mu \partial_\nu \varphi^\nu-m^2 \varphi^\mu\approx 0.
\end{equation}
Taking its divergence, we obtain a lower-derivative consequence
\begin{equation}
\label{16sep6}
\partial_\mu\varphi^\mu\approx 0.
\end{equation}
Plugging it back into (\ref{16sep5}), we find 
\begin{equation}
\label{16sep7}
(\Box -m^2) \varphi^\mu\approx 0.
\end{equation}
Equations (\ref{16sep6}), (\ref{16sep7}) are just (\ref{16sep1}) in the spin-1 case. 
Thereby, we find that, despite the apparent mismatch between the number of equations and the number of fields in the massive spin-1 theory, the necessary equations of motion can  still be made Lagrangian. This is achieved by designing the action in such a way that the resulting equations of motion have a lower-derivative consequence. Action (\ref{16sep3}) is known as the Proca theory.

In a similar manner one can construct a Lagrangian that would produce (\ref{16sep1}) in the spin-2 case. The resulting action is known as the Fierz-Pauli theory. It requires a single traceful rank-2 symmetric, $\mathbf{Y}(2)$, off-shell field or, equivalently, one traceless symmetric rank-2 field and one scalar, $\mathbb{Y}(2)\oplus\mathbb{Y}(0)$.
  For spins $s\ge 3$ equations (\ref{16sep1}) cannot be derived from a Lagrangian unless more complicated sets of auxiliary fields are introduced. A solution to this problem with a \emph{minimal} set of auxiliary fields was given in \cite{Singh:1974qz}. There also exist other formulations that involve additional auxiliary fields. In particular, it can be beneficial to add auxiliary fields so that the theory does not have lower-derivative consequence such as (\ref{16sep6}), instead featuring gauge symmetries. This is typically achieved by virtue of symmetries that act algebraically, which are also known as \emph{Stueckelberg} symmetries. The following example illustrates how this works for the massive spin-1 case. 

\begin{zadacha}
 Consider a theory 
\begin{equation}
\label{30dec1}
S=\int d^dx \left(-\frac{1}{4}F_{\mu\nu}F^{\mu\nu}-\frac{1}{2}\varphi_\mu m^2 \varphi^\mu -\frac{1}{2}\partial_\mu\varphi \partial^\mu \varphi+m\partial_\mu\varphi \varphi^\mu\right),
\end{equation}
where $\varphi^\mu$ and $\varphi$ are independent dynamical fields. Show its equivalence to (\ref{16sep3}). Such formulations for massive fields naturally result from the dimensional reduction of massless theories. 
\end{zadacha}

\subsection{Why irreducible multiplets?}

In the previous sections we explained how irreducible massless and massive spinning  fields can be described at the Lagrangian level. To this end we needed some specially tailored constructions, the main role of which was to ensure that the system does not describe anything else but the irreducible representation we intend to describe. One may wonder whether we can do something simpler if we relax the requirement that the system describes an irreducible representation. The standard lore of higher-spin theories is that
unless special care is taken, the theory involves pathological degrees of freedom in the sense that these either violate unitarity, give ghosts or tachyons. Below we will demonstrate this with a simple example. 

To this end, we consider an example of a massless field with one Lorentz index
\begin{equation}
\label{15sep8}
\Box \varphi^\mu=0
\end{equation}
and see what are the roles of its components from the point of view of representation theory and why it is necessary to eliminate two degrees of freedom to make the associated theory consistent.

A somewhat oversimplified version of the argument goes as follows. Equation (\ref{15sep8}) comes from the action
\begin{equation}
\label{16sep8}
S=\frac{1}{2}\int d^dx \varphi_\mu \Box \varphi^\mu.
\end{equation}
There is the standard procedure to induce an inner product once the action is known. This procedure results in the norm, which is negative for $\varphi^0$ and positive for $\varphi^\mu$ with $\mu$ spatial or vice versa. In other words, the inner product defined this way is not positive definite and, hence, the theory is not unitary. Alternatively, one can compute the stress-energy tensor for (\ref{16sep8}) and see that the energy is not positive definite. 
This, again, is related to the fact that spatial and time components of $\varphi^\mu$ contribute to energy with opposite signs.

Below we will give another version of this argument, which relates to the previous discussion better. Namely, we would like to make the connection with the construction of UIR's using the approach of Wigner. 
Clearly,  the fact that (\ref{15sep8}) cannot realise a unitary representation of $ISO(d-2)$ already follows from $ISO(d-2)$ being non-compact and $\varphi^\mu$ having finitely many components. Nevertheless, it is instructive to  demonstrate the absence of the invariant and positive-definite inner product for solutions of (\ref{15sep8}) more explicitly.

To start, it is clear that the representation associated with (\ref{15sep8}) is massless in the sense of the relevant orbit being $p^2=0$. Therefore, we again pick the special momentum as in (\ref{15sep5}). Then, with respect to the action of the Wigner little group, detailed in section \ref{sec:wig}, $\varphi^\mu$ naturally splits into 
\begin{equation}
\label{15sep9}
\varphi^\mu=\{\varphi^+,\varphi^-,\varphi^i \}.
\end{equation}
Clearly, each component in this split is invariant under $SO(d-2)$. As for the inhomogeneous part of the group, which is generated by $\pi_i\equiv J_{i+}$, we have, 
\begin{equation}
\begin{split}
\label{15sep10}
J_{i+}[\varphi^+]=-i\varphi^i, \qquad J_{i+}[\varphi^j]=i\delta_i{}^j \varphi^-, \qquad J_{i+}[\varphi^-]=0.
\end{split}
\end{equation}
To derive (\ref{15sep10}) one needs to use the standard formula for the action of the Lorentz algebra on vectors 
\begin{equation}
\label{15sep10xx1}
J^{\mu\nu}[\varphi^\rho]=-i\eta^{\rho\nu}\varphi^\mu+i \eta^{\rho\mu}\varphi^\nu
\end{equation}
in the light-cone coordinates. Equation (\ref{15sep10}) shows that $\pi^i$ acts in a nilpotent manner
\begin{equation}
\label{15sep11}
\varphi^+ \quad \to \quad \varphi^i \quad \to \quad \varphi^- \quad \to \quad 0,
\end{equation}
which already suggests that it cannot be represented by a unitary operator. We will study this more carefully later.

Before that, we would like to compare (\ref{15sep8}) with the spin-1 helicity representation, described by (\ref{14sep1})-(\ref{14sep3}) with $s=1$. At the level of the Wigner little group, to arrive to the spin-1 helicity representation from (\ref{15sep8}), we need to make two steps. First, we should note that the states spanned by $\{\varphi^i,\varphi^-\}$ form a subrepresentation in the original representation  carried by $\varphi^\mu$, see (\ref{15sep11}). Thus, we can consistently drop $\varphi^+$ states and focus on the representation realised on $\{\varphi^i,\varphi^-\}$. In the covariant language, this is achieved by requiring that the field is divergenceless, $\partial_\mu\varphi^\mu=0$. Next, (\ref{15sep11}) also implies that the resulting representation has a subrepresentation with states $\varphi^-$. These can be quotiented out, which in the covariant language is achieved due to gauge invariance. As a result, we obtain a representation realised on the space $\varphi^i$, on which $\pi$ acts trivially -- just as required for the helicity representations.

We now return to the representation of the Wigner little group associated with (\ref{15sep8}), which has a non-trivial action of $\pi$, (\ref{15sep10}).
 According to our general logic, we should figure out  whether this representation is unitary, that is whether it admits a positive-definite invariant norm, which we will now do.
 
To start, demanding invariance  with respect to $SO(d-2)$, we conclude that the inner product  is of the form
\begin{equation}
\label{16sep9}
(\varphi,\psi)=\alpha \bar\varphi^+ \psi^++\beta \bar\varphi^-\psi^-+\gamma(\bar\varphi^- \psi^++\bar\varphi^+ \psi^-)+\delta \bar\varphi^i\psi_i.
\end{equation}
Next, for the norm to be positive we need to require $\delta > 0$. In particular, $\delta \ne 0$ and, as a result, $(\pi^i[\varphi^+],\psi^j)\ne 0$.  Self-adjointness of $\pi$ then entails $(\varphi^+,\pi^i [\psi^j])\ne 0$. By considering how $\pi$ acts on $\psi^j$, (\ref{15sep10}), this entails $\gamma \ne 0$. Besides that, $(\pi^i[\varphi^j],\psi^-)=(\varphi^j,\pi^i[\psi^-])=0$, where the last equality is due to the fact that $\pi$ annihilates $\psi^-$. This leads to  $\beta=0$. Summing up, we find that 
for the $SO(d-2)$ scalar modes, the inner product is of the form
\begin{equation}
\label{16sep10}
\alpha \bar\varphi^+ \psi^++\gamma(\bar\varphi^- \psi^++\bar\varphi^+ \psi^-), \qquad \gamma\ne 0.
\end{equation}
It is trivial to see that this inner product cannot be positive definite.
In fact, $\beta=0$ alone implies that $\varphi^-$ are zero-norm states.
 As a result, we infer that the representation described by (\ref{15sep8}) is non-unitary. 

From this example we can draw a conclusion that by randomly dropping constraints and gauge symmetries that we need to describe a unitary irreducible representation we are likely to arrive at a representation that is not only reducible, but also non-unitary. Typically this also gives theories with energies not bounded below. This, certainly, does not mean that we should only construct theories by putting healthy irreducible fields together. Instead, one may come up with a heuristic construction which, for one or another reason, gives rise to a reducible, but not pathological spectrum. This, for instance, happens for string theory. Still, as we would like to be as general as possible and have the consistency conditions on the spectrum under control, we will stick to the approach in which the theory is built out of fields, each describing a UIR of the Poincare group.

\subsection{Further reading}

Some authors find the double-trace constraint of  Fronsdal's fields unattractive. It is possible to remove these in different ways. In particular, the approach of \cite{Francia:2002aa,Francia:2002pt} features unconstrained symmetric Lorentz tensors. A similar effect is achieved in the formulation of \cite{Segal:2001qq}. There are also other approaches that further extend the spectrum of the off-shell fields. For instance, the triplet  approach \cite{Sagnotti:2003qa} features a pair of auxiliary tensors motived by string theory. Alternatively, one can try to achieve simplifications of the description of the massless higher-spin fields by constraining the off-shell spectrum. In particular, a very simple formulation involving only symmetric traceless tensors was suggested in \cite{Skvortsov:2007kz,Campoleoni:2012th}. It, however, requires gauge parameters to be divergence-free, which may appear somewhat unnatural\footnote{In \cite{Francia:2013sca} it was shown that the divergence-free constraint can be solved in terms of gauge parameters without differential constraints, which leads to a more complex pattern of reducible gauge transformations.}. 

Despite our main focus in the present course is on massless symmetric fields, we briefly mention that free actions are known in other cases and within different approaches. In particular, Fronsdal's approach can be extended to massless fermionic fields \cite{Fang:1978wz} and to massless mixed-symmetry fields \cite{Labastida:1987kw}.
In the latter case, the theory develops a set of gauge symmetries, which are, in general, reducible.

 The minimal approach to massive symmetric fields of \cite{Singh:1974qz} can be reformulated in the Stueckelberg form \cite{Zinoviev:2001dt}. The action for symmetric continuous spin fields was found only recently \cite{Schuster:2014hca}. To the best of or knowledge, actions for arbitrary massive mixed-symmetry fields are not known. For a recent review on the state of the art in this area, see e.g. \cite{Bekaert:2017khg}.

\section{Fields in AdS as lowest-weight representations}

In the same way as we analysed field perturbations around the Minkowski space, one may consider expansions in small fields around other backgrounds. In this context, the central role is played by the most symmetric backgrounds, for which the rules of quantum field theory are better understood. This, in particular, is related to the fact that to quantise a theory one should  be able to find its complete set of solutions, which is hard to do once one goes away from the most symmetric setup.
 Besides that, it is also important to have a time-like Killing vector, so that one is able to define energy and then split solutions into positive and negative energy modes. 
 
 The group-theoretic approach that we used in previous sections also crucially relies on the presence of rich symmetries. In particular, the very notion of spin in the Minkowski space was defined based on the field transformation properties under the Minkowski space isometries. For general backgrounds the necessary isometries are absent and spin cannot be defined. 
 
 In the present section we will discuss the anti-de Sitter space -- the space of constant negative curvature. In particular, we will construct UIR's of its isometry group, which will then be identified with fields propagating in the anti-de Sitter space. 
 In a similar manner one can also study field theories in de Sitter space, which is the space of constant positive curvature. The representation theory of the de Sitter isometry group, however, has one unattractive property: all its UIR's have energy unbounded from below. This is why, one usually prefers to deal with the anti-de Sitter space instead. We will not discuss the de Sitter case in detail here, though, some references will be given at the end of the section.

\subsection{AdS space}

To start, we review what AdS${}_d$ space is. The AdS space is a space of constant curvature, which can be conveniently described by a hyperboloid embedded into a space of one dimension larger, referred to as the \emph{ambient space},
 \begin{equation}
 \label{7apr5}
 X^M X_M = -l^2 < 0,
 \end{equation}
 where $X^M$ are ambient coordinates, $l$ is the AdS space radius and the ambient metric is
 \begin{equation}
 \label{7apr6}
 \eta_{MN} = \rm{diag}(-,+,\dots,+,-).
 \end{equation}
 The ambient space indices run over $d+1$ value, $M=0,1,\dots, d$.
 The cosmological constant is given by
  \begin{equation}
 \label{7apr7}
 \Lambda \equiv -\frac{(d-1)(d-2)}{2 l^2}
 \end{equation}
and it is defined so that the metric induced on the hyperboloid satisfies
 \begin{equation}
 \label{7apr8}
{  R}_{\mu\nu}-\frac{1}{2}g_{\mu\nu}{  R}+\Lambda g_{\mu\nu}=0,
 \end{equation}
 which, in turn, comes from the Lagrangian density\footnote{Unlike most of the higher-spin literature, we define the cosmological constant with the same factors that are typically used in General Relativity. For example, in  \cite{Didenko:2014dwa}  the cosmological constant is defined as $\Lambda\equiv-1/l^2$.}
 \begin{equation}
 \label{7apr9}
 {\cal L}={  R}-2\Lambda,
 \end{equation}
 where ${ R}$ is the scalar curvature and ${ R}_{\mu\nu}$ is the Ricci tensor. The AdS Riemann tensor satisfies
 \begin{equation}
 \label{21sep1}
 {  R}_{\mu\nu\lambda\rho}=-\frac{1}{l^2}(g_{\mu\lambda}g_{\nu\rho}-g_{\mu\rho}g_{\nu\lambda}),
 \end{equation}
 which, in particular,  implies
 \begin{equation}
 \label{21sep2}
 [\nabla_\mu,\nabla_\nu]v^\rho=-\frac{1}{l^2}\left(\delta^\rho{}_\mu v_\nu-\delta^\rho{}_\nu v_\mu\right).
 \end{equation}
 
 The key advantage of the ambient space construction is that it makes the $AdS$ isometries manifest. These are 
 \begin{equation}
 \label{7apr10}
J^{MN}\equiv - i (X^M \partial^N - X^N \partial^M).
 \end{equation}
 In these terms one defines the AdS deformed translations, also called transvections, by
  \begin{equation}
 \label{7apr11}
 P^\mu\equiv l^{-1} J^{d\mu},
 \end{equation}
 while the Lorentz generators are just the components of $J^{MN}$ with $M$ and $N$ in the set from $0$ to $d-1$.
 The commutator of transvections gives
  \begin{equation}
 \label{7apr12}
 i[P^\mu,P^\nu]=\frac{1}{l^2}J^{\mu\nu},
 \end{equation}
 so unlike in flat space, these do not commute and cannot take definite values simultaneously.
 In the flat space  limit, $l\to \infty$, one reproduces the commutation relations of the Poincare algebra. 
 The energy is defined by
 \begin{equation}
 \label{7apr13}
 H\equiv P^0 = -P_0 =l^{-1}J^{d0}.
 \end{equation}

 The AdS space defined as above is a one-sheeted hyperboloid with a periodic time direction, which leads to the problem of closed time-like curves.  In particular, map $e^{-2\pi i J^{d0}}$ rotates the AdS space along the time direction by angle $2\pi$ and eventually carries every its point to itself. To avoid this issue, one considers the AdS covering space -- a space with countably many copies of the original hyperboloid -- so that $e^{-2\pi i J^{d0}}$ maps a point of copy number $n$ to a point of copy number $n+1$ with the same coordinates. It is this covering space that we will refer to as the AdS space in the following. Some further details on the AdS space geometry can be found in \cite{Fronsdal:1974ew}.

\subsection{UIR's as the lowest-weight modules}
\label{sec:4.2}

By the same logic that we used in flat space,  free fields  in AdS${}_d$ can be identified with unitary irreducible representations of the AdS${}_d$ isometry algebra, $SO(d-1,2)$. Here we will review how these are constructed. This analysis can be found, e.g. in section 1.1 of \cite{Vasiliev:2004cm}.

\subsubsection{General construction}

Group $SO(d-1,2)$ is non-compact, hence, its unitary representations are infinite-di\-men\-sional. To construct these representations, one considers their decomposition into representations of the maximal compact subgroup of $SO(d-1,2)$, which is  $SO(2)\oplus SO(d-1)$.
Considering that the latter group is compact, its unitary representations are finite-dimensional. We already know how to construct UIR's of the orthogonal group from section \ref{sect:232}. For the direct product of two orthogonal groups UIR's are constructed by taking tensor products of UIR's of each factor. Accordingly, UIR's of $SO(2)\oplus SO(d-1)$ are given by tensor products of  UIR's of $SO(d-1)$ -- characterised by some Young diagrams $\mathbb{Y}$ -- and UIR's of $SO(2)$.

As for the $SO(2)$ factor, we can proceed by constructing its UIR's as $SO(2)$ tensors, which is what the previous discussion suggests. It should be noted, however, that this $SO(2)$ is the group of time translations and once we replace the AdS hyperboloid  with its covering space, $SO(2)$ should also be replaced with its universal covering group $\mathbb{R}$, the additive group of real numbers.
 All irreducible representations of the latter group are one-dimensional and can be characterised by a single number, which is the eigenvalue of the single group generator  
 \begin{equation}
 \label{8apr1}
 E\equiv  J^{d0}.
 \end{equation}
 This generator is a dimensionless version of the energy generator (\ref{7apr13}). Unitarity requires that its eigenvalue is real. If 
 we considered tensor representation of $SO(2)$ instead, $E$ could have only assumed  integer values. 
 
Accordingly, decomposition of a representation of $SO(d-1,2)$ into representations of its compact subgroup can be schematically written as
 \begin{equation}
 \label{30mar1}
\varphi = \oplus_i |E_i,\mathbb{Y}_i\rangle,
 \end{equation}
 where $E_i$ are some real numbers, $\mathbb{Y}_i$ are shapes of tensor representations of $SO(d-1)$ and the sum can be infinite. Equation (\ref{30mar1}) specifies the action of the compact subgroup generators on the states of the $SO(d-1,2)$ representation we are after.
 
  It remains to define how the non-compact generators act.
 These can be conveniently organised into 
  \begin{equation}
 \label{8apr2}
 J^{+m}\equiv J^{0m}+ i J^{dm}, \qquad J^{-m}\equiv J^{0m}-iJ^{dm},
 \end{equation}
 $m= 1,\dots , d-1$. They act as raising and lowering operators with respect to energy
 \begin{equation}
 \label{8apr3}
 [E,J^{+m} ]= J^{+m}, \qquad [E,J^{-m}]= - J^{-m}.
 \end{equation}
For a representation with energy bounded from below,  there should be a term in the decomposition (\ref{30mar1}), which is annihilated by the energy-lowering operators\footnote{Somewhat speculatively, the lowest-energy space for $SO(d-1,2)$ representations can be compared with $\varphi_{p,\sigma}$ space with $p=(m,0,\dots,0)$ for massive fields in flat space. In particular, for the latter states energy, indeed, acquires minimal value $E_0=m$ and these also furnish a representation of $SO(d-1)$, which is the  Wigner little  group in the massive case. Yet, there are, major differences between these two setups. Most importantly, transvections do not commute, so, only energy takes a definite value on $|E_0,\mathbb{Y}_0\rangle$. Still, this analogy was used fruitfully to construct UIR's of $SO(d-1,2)$ in a form which is very intuitive from the flat space perspective \cite{Fronsdal:1974ew}.}
  \begin{equation}
 \label{22sep2}
 J^{-m}|E_0,\mathbb{Y}_0\rangle =0.
 \end{equation}
The states $|E_0,\mathbb{Y}_0\rangle$ are then the lowest-weight states in the representation.  The remaining states in (\ref{30mar1})  are generated by the raising operators
 \begin{equation}
 \label{22sep3}
  J^{+m_1}\dots J^{+m_n} |E_0,\mathbb{Y}_0\rangle.
 \end{equation}
   
From (\ref{8apr3}) one finds that the value of energy for (\ref{22sep3}) is $E_0+n$. In turn, representations of $SO(d-1)$ carried by (\ref{22sep3}) can be found by evaluating a tensor product of $\mathbb{Y}_0$ with $\mathbf{Y}(n)$, the latter representation being carried by $n$ commuting  $J^{+}$'s.
 The $SO(d-1,2)$ representation thus defined will be referred to as the lowest-energy or the lowest-weight representation and will be denoted $V(E_0,\mathbb{Y}_0)$. These representations are also often referred to as the Verma modules.
 
 Above we assumed that only a single irreducible representation of the compact subgroup satisfies the lowest-weight condition (\ref{22sep2}). As it is not hard to see, if, instead, the lowest-weight space transforms as a reducible representation of $SO(2)\oplus SO(d-1)$, then the lowest-weight representation of $SO(d-1,2)$ constructed out of it is not irreducible. As we are ultimately interested in irreducible representations, we will assume that the lowest-weight space (\ref{22sep2}) is unique.

Next, we would like to turn our attention to unitarity of $V(E_0,\mathbb{Y}_0)$. 
 Recall that for $SO(d-1,2)$ generators unitarity implies $(J^{AB})^\dagger=J^{AB}$, which due to $i$'s in definition (\ref{8apr2}) entails
 \begin{equation}
 \label{22sep4}
 (J^{+m})^\dagger = J^{-m}, \qquad   (J^{-m})^\dagger = J^{+m}.
 \end{equation}
 Then, the inner product for two  vectors (\ref{22sep3}) is
 \begin{equation}
 \label{22sep5}
 \begin{split}
 &(  J^{+p_1}\dots J^{+p_k} |E_0,\mathbb{Y}_0\rangle,  J^{+m_1}\dots J^{+m_n} |E_0,\mathbb{Y}_0\rangle)\\
 & \qquad \qquad \qquad =
 \langle E_0,\mathbb{Y}_0| J^{-p_k}  \dots J^{-p_1}J^{+m_1}\dots J^{+m_n} |E_0,\mathbb{Y}_0\rangle.
 \end{split}
 \end{equation}
 Here we used the bra-ket notation
 \begin{equation}
 \label{22sep6}
 (u,v)\equiv \langle u| v\rangle
 \end{equation}
 and (\ref{22sep4}) to move $J^+$ through the inner product, converting them to $J^-$.  
 
 It is easily seen that self-adjointness of energy implies that the inner product is vanishing  between the states with $k\ne n$. For states with $k=n$ we can use the commutation relation
   \begin{equation}
 \label{8apr4}
 [J^{-m},J^{+n}]=2(\delta^{mn}E - i J^{mn})
 \end{equation}
 to pull all $J^-$'s to the right and all $J^+$'s to the left. Now, recall that $|E_0,\mathbb{Y}_0\rangle$ is the lowest-weight state, so $J^-$'s applied to it give zero. Similarly, $J^+$'s annihilate $\langle E_0,\mathbb{Y}_0|$. Therefore, after commuting $J^-$'s to the right and $J^+$'s to the left in (\ref{22sep5}),
 the only non-vanishing contributions come from commutators of $J^+$ and $J^-$.
 These can be evaluated with   (\ref{8apr4}) and, thus, the inner product (\ref{22sep5}) for $k=n$ gets expressed, schematically, as
 \begin{equation}
 \label{30mar2}
  \langle E_0,\mathbb{Y}_0| f(J^{mn}, E) |E_0,\mathbb{Y}_0\rangle
 \end{equation}
 with some polynomial function $f$. 
 
 Considering that $E$ and $J^{mn}$ are the compact group generators, $f(J^{mn}, E) |E_0,\mathbb{Y}_0\rangle$ still belongs to the lowest-energy space. Therefore, we can see that the inner product  for any two states in the $SO(d-1,2)$ representation with $k=n$ can be induced from the inner product of the lowest-weight states, while it is vanishing for states with $k\ne n$.
 It is now straightforward to implement the requirement of unitarity. Namely, the norm for the lowest-energy states, as well as the norms for all states with $k=n$ that are induced from it should be positive-definite.

 \subsubsection{Spin-1 example}
 
To better understand how this construction works, we will illustrate it with a simple example.
 Consider a representation $|E_0,\mathbb{Y}(1)\rangle$ with the lowest-energy space of energy $E_0$ and carrying a vector representation of $SO(d-1)$.
 As will be shown later, this representation can be identified with a  spin-1 field in the AdS space.
 
  To make the $SO(d-1)$ symmetry manifest, we will write its tensor indices explicitly. In particular, the lowest-energy space will be denoted  $|E_0\rangle^a$. The vector representation of $SO(d-1)$ is unitary and up to an irrelevant positive number, its inner product is fixed to be
\begin{equation}
\label{22sep7}
\langle E_0|^a |E_0\rangle^b = \delta^{ab}.
\end{equation}

  Next we consider the inner product of states involving one raising operator. Employing (\ref{8apr4}), we obtain
 \begin{equation}
 \label{22sep8}
 \langle E_0|^d J^{-a} J^{+b}|E_0\rangle^c = 
2  \langle E_0|^d (\delta^{ab} E-i J^{ab}) |E_0\rangle^c.
 \end{equation}
Since $|E_0\rangle^c$ is the $SO(d-1)$ vector, we have
 \begin{equation}
 \label{22sep9}
 J^{ab} |E_0\rangle^c=-i \delta^{bc} |E_0\rangle^a +i \delta^{ac} |E_0\rangle^b.
 \end{equation}
 Together with (\ref{22sep7}) this leads to
   \begin{equation}
 \label{22sep10}
 \langle E_0|^d J^{-a} J^{+b}|E_0\rangle^c = 
2\left(E_0\delta^{ab}\delta^{cd}-\delta^{bc}\delta^{ad}+\delta^{ac}\delta^{bd} \right).
 \end{equation}
 
States $J^{+b}|E_0\rangle^c$ carry two $SO(d-1)$ indices and can be split into three irreducible representations of $SO(d-1)$: the symmetric traceless rank-2 tensor, the antisymmetric rank-2 tensor and the scalar representation.
 With respect to this decomposition the inner product (\ref{22sep10}) is diagonal and 
for each of the representations up to an overall factor it is given by the associated positive-definite inner product of $SO(d-1)$. Requiring these factors to be positive, we find three different constraints on $E_0$, which are necessary to ensure unitary of the representation. 
 
 We will consider the simplest constraint that comes from the inner product of scalars. To this end, we take 
 traces of (\ref{22sep10}) with  $\delta_{da}$ and $\delta_{bc}$. This gives
   \begin{equation}
 \label{22sep11}
\delta_{da} \delta_{bc}\langle E_0|^d J^{-a} J^{+b}|E_0\rangle^c = 
2(d-1)\left(E_0-d+2 \right).
 \end{equation}
Hence, for dimensions  $d>1$, this inner product  is positive for
 \begin{equation}
 \label{22sep12}
 E_0>d-2.
 \end{equation}
 
Equation (\ref{22sep12}) gives a single constraint that comes from  positivity of the inner product for some states appearing at level $k=n=1$. To ensure that the lowest-weight representation $V(E_0,\mathbb{Y}_0)$ is unitary, we should consider norms for all other states (\ref{22sep3}) and require their positivity in a similar manner. Performing this analysis explicitly is a tedious task, so we will not carry it out  here. As the end result one finds that  (\ref{22sep12}) is, actually, the strongest positivity constraint,
thereby, representations generated from $|E_0\rangle^a$ with $E_0$ satisfying 
(\ref{22sep12})
are unitary. 

The case of
 \begin{equation}
 \label{22sep13}
 E_0=d-2
 \end{equation}
 requires a separate analysis.
 From (\ref{22sep11})
 we find 
  that  $J^{+a}|E_0\rangle_a$ has the vanishing inner product with itself, so it is a null state.
  One can also see that
 \begin{equation}
 \label{22sep14}
 J^{-a}J^{+b}|E_0\rangle_b = 2(E_0-d+2)|E_0\rangle^a,
 \end{equation}
 which means that with (\ref{22sep13}) satisfied $J^{+b}|E_0\rangle_b$ is a lowest-energy vector in the sense (\ref{22sep2}). It can be used to build a lowest-weight representation $V(d-1,\mathbb{Y}(0))$  in the same way  we did with $|E_0\rangle^a$. 
It is then straightforward to see  that, similarly to  $J^{+a}|E_0\rangle_a$, all  states in $V(d-1,\mathbb{Y}(0))$ have vanishing norms.
At the same time,  (\ref{22sep14}) ensures that $V(d-1,\mathbb{Y}(0))$ is a subrepresentation of the original representation
 $V(d-2,\mathbb{Y}(1))$, so  $V(d-1,\mathbb{Y}(0))$ can be consistently quotiented out of 
 $V(d-2,\mathbb{Y}(1))$, leaving us with 
  a unitary representation. The $SO(d-1,2)$ representation that results from this quotienting will be denoted
 \begin{equation}
 \label{31mar1}
 D(d-2,\mathbb{Y}(1))\equiv V(d-2,\mathbb{Y}(1))\; / \; V(d-1,\mathbb{Y}(0)).
 \end{equation}
 
 Finally, for
  \begin{equation}
 \label{22sep15}
 E_0<d-2
 \end{equation}
 the lowest-energy representation generated from $|E_0\rangle^a$ is, clearly, non-unitary, as it features negative norm states.

\begin{zadacha}
\label{exc30mar}
 Consider a representation with the lowest weight being scalar with respect to $SO(d-1)$. Then apply the raising operators twice and focus on the scalar representation of $SO(d-1)$ in that sector. By requiring that the norm of this state is positive, find the constraint on the energy of the lowest weight state $E_0$.
 \end{zadacha}

  \subsubsection{Unitarity bounds for symmetric fields}
  \label{sec:323}
  
   The situation illustrated in the previous example is generic. Namely, for high enough $E_0$,  representation $V(E_0,\mathbb{Y}_0)$ is unitary. By decreasing $E_0$, at some point we arrive at its value $E_0(\mathbb{Y}_0)$, for which some of the states have vanishing norms. These states, however, form a subrepresentation and can be consistently quotiented out, which eventually gives a unitary representation $D(E_0(\mathbb{Y}_0),\mathbb{Y}_0)$.  For even lower values of $E_0$, the lowest-weight representation becomes non-unitary. \emph{This gives the classification of lowest-weight UIR's of $SO(d-1,2)$} we were after. The critical value $E_0(\mathbb{Y}_0)$ is well-known for general  shapes of Young diagrams and can be found, e.g. in \cite{Metsaev:1995re}. Below we will discuss  the case of $\mathbb{Y}_0=\mathbb{Y}(s)$ in more detail, as it is relevant for symmetric higher-spin fields.

  As in the spin-1 case, one can start from the lowest-energy state  $|E_0\rangle^{a(s)}$, which carries $\mathbb{Y}(s)$ representation of $SO(d-1)$. By studying the inner product for $J_a^+|E_0\rangle^{a(s)}$ with itself, one finds a bound 
    \begin{equation}
 \label{17dec1}
 E_0\ge s+d-3, \qquad s\ge 1.
 \end{equation}
   It turns out that constraints from positivity of the inner product for other states are milder, so (\ref{17dec1}), actually, gives the unitarity bounds for these lowest-energy representations. For $E_0$ above the unitarity bound, these representations will be called \emph{massive} spin-$s$ fields. In turn, for $E_0$ on the unitarity bound, these representations will be called \emph{massless} spin-$s$ fields.  The phenomenon of shortening of the lowest-weight representation with $E_0$ on the unitarity bound, will be later related to the occurrence of gauge invariance in field theory. 
This explains why it is natural to regard fields with $E_0$ on the unitarity bound as the AdS counterparts of massless fields in flat space.
 
 In turn, in the scalar case the unitarity bound is given by 
  \begin{equation}
 \label{17dec2}
 E_0\ge \frac{d-3}{2}, \qquad s_0 =0.
 \end{equation}
 It can be found by solving exercise \ref{exc30mar}. For the critical value  $E_0=\frac{d-3}{2}$ this representation also acquires zero-norm states, which should be quotiented out. It may be tempting to interpret this phenomenon as gauge invariance as well. 
 Scalar fields, however, never acquire gauge invariance in flat space, so it seems surprising that something reminiscent of gauge invariance occurs for scalar fields in AdS. 
  As we will see later, despite, for general values of $E_0$ the associated lowest-energy representations can, indeed, be interpreted as scalar fields in the AdS space, for the critical value of $E_0$ the shortened representation has too little degrees of freedom to be interpreted this way. Instead, the critical case corresponds to the scalar field in the $d-1$-dimensional Minkowski space
 \begin{equation}
 \label{17dec3}
 \Box \varphi =0
 \end{equation}
 with $SO(d-1,2)$ playing the role of the $d-1$-dimensional conformal group. This phenomenon plays an important role in the context of higher-spin holography, which will be discussed later.

\subsection{The Casimir operators}  
 \label{sec:4.3}

 In the lowest-weight approach  representations of $SO(d-1,2)$ are generated from the lowest-weight vectors $|E_0, \mathbb{Y}_0\rangle$, which, in turn, are specified by their $SO(2)\oplus SO(d-1)$ labels $E_0$ and $\mathbb{Y}_0$. Thus, UIR's of $SO(d-1,2)$ admit a natural labelling with $E_0$ and $\mathbb{Y}_0$, which, moreover,  uniquely defines these representations. Alternatively, representations of $SO(d-1,2)$ can be labelled by the values of the $SO(d-1,2)$  Casimir operators. Below we will compute the value of the quadratic Casimir operator for the lowest-energy representations. This result will be useful in the following section for establishing the connection between the lowest-energy representations and their field-theoretical realisations. 
  
 We remind the reader that the quadratic Casimir operator is defined by
 \begin{equation}
 \label{8apr5}
 {\cal C}_2(so(d-1,2)) = \frac{1}{2}J^{AB}J_{AB}.
 \end{equation}
 It acts diagonally on all states in the representation space of an irreducible representation with the same eigenvalue. This eigenvalue is, thus, characteristic of a given representation. To extract it, we will evaluate ${\cal C}_2(so(d-1,2))$
 on the lowest-energy state. 
 With some algebra, which involves
 \begin{equation}
 \label{8apr6}
 J^{0m}J^{0m}+J^{dm}J^{dm}=(d-1)E+J^{+m}J^{-m},
 \end{equation}
 we find
 \begin{equation}
 \label{8apr7}
 {\cal C}_2(so(d-1,2))|E_0,\mathbb{Y}_0\rangle = \left( E_0(E_0-d+1)+{\cal C}_2(so(d-1))\right)|E_0,\mathbb{Y}_0\rangle,
 \end{equation}
 where
 \begin{equation}
 \label{8apr8}
 {\cal C}_2(so(d-1)) \equiv \frac{1}{2}J^{mn}J^{mn}
 \end{equation}
 is the quadratic Casimir operator of $so(d-1)$. The value of  $ {\cal C}_2(so(d-1))$ for symmetric representation of $so(d-1)$ will be found below.
 
 \begin{zadacha}
Reproduce (\ref{8apr6}) and (\ref{8apr7}).
 \end{zadacha}
 
 In a similar manner one can express higher Casimir operators in terms of the lowest-weight labels. However, for our purposes it will be sufficient to know only the value of the quadratic Casimir operator. To be more precise,  in the field theory description,  a lowest-energy representation $V(E_0, \mathbb{Y}_0 )$ will be realised as a divergence-less  tensor field of  shape $\mathbb{Y}_0$\footnote{The fact that the shape $\mathbb{Y}_0$ carries over 
to  the field-theory description in such a straightforward manner is not entirely obvious.
This can be shown, for example, by comparing the values of higher Casimir operators or, more directly,  one can find the lowest-energy state on the field theory side and identify the associated $E_0$ and $\mathbb{Y}_0$.
 We will not do that, as our goal, anyway, is not to provide a mathematically strict proof of the equivalence of the two representations, but rather support this statement with some evidence and familiarise the reader with the frequently used tools.}. 
 The associated equations of motion have only one free parameter -- the value of mass squared in the mass term. The mass squared will then be related to $E_0$ and to do that it will be sufficient 
  to compare a single Casimir operator computed  in the field theory and in the lowest-energy realisations.

\subsection{Further reading}

An accessible overview of the UIR's of $SO(d-1,2)$ can be found in \cite{Minwalla:1997ka}.
All lowest-energy  unitary irreducible representations of $SO(d-1,2)$ are available, see \cite{doi:10.1063/1.1705183,Mack:1975je,Siegel:1988gd,Metsaev:1995re,Metsaev:1997nj,Metsaev:1998xg}. In particular, unitarity bounds on energies are known for any shape $\mathbb{Y}_0$. More mathematically advanced discussions on the subject can be found in \cite{Enright1983,Ferrara:2000nu,Bourget:2017kik}. For completeness, we mention that the classification of UIR's of the de Sitter space isometry group, $SO(d,1)$, can be found in \cite{1962258,doi:10.1063/1.1665471,dobrev1977harmonic}, see also introductory part of \cite{Basile:2016aen}.

\section{Massless fields in AdS in the Lorentz-covariant form}

We will start from presenting the manifestly Lorentz-covariant description for symmetric higher-spin fields in the AdS space and then we will establish its connection with the UIR's of $SO(d-1,2)$ constructed in the previous section. More details on massless fields in AdS can be found in section 2.2 of \cite{Didenko:2014dwa}, section 3.1 of \cite{Mikhailov:2002bp} and section 6 of \cite{Kessel:2017mxa} and references therein.

\subsection{Covariant form of equations of motion}
\label{sec:51}
Equations of motion for massive fields read
\begin{equation}
\label{22sep16}
\begin{split}
\left(\Box-\frac{1}{l^2}[(s-2)(s+d-3)-s] -m^2\right)\varphi^{\mu(s)}=0,\\
\nabla_\nu \varphi^{\nu \mu(s-1)}=0,\\
\varphi_\nu{}^{\nu \mu(s-2)}=0.
\end{split}
\end{equation}
One important point that we would like to stress immediately is that the notion of mass for fields in the AdS space does not have the  meaning that it has in flat space. Indeed, by definition, $m^2$ in flat space is minus the value of momentum squared $P^2$, which can be evaluated on any state in the UIR, and thus serves as one of the representation labels.  Equivalently, $P^2$ is the quadratic Casimir operator of the Poincare algebra and $-m^2$ is its value for a given UIR. Naive attempts to extend this notion to the AdS space fail. In particular, replacing translations with transvections in $P^2$ does not work, as $P^2$ is not the Casimir operator of $so(d-1,2)$. 

Still, we need to write equations of motion for fields in the AdS space and these inevitably feature a mass term, that is the term without derivatives. Similarity of these equations with the flat space ones suggests to write the coefficient of this term as $m^2$ with, possibly, some corrections that vanish in the flat-space limit.  There is no unique convention how to do that. 
 In (\ref{22sep16}) we adopted a convention for which $m^2=0$ corresponds to the point at which the system acquires gauge invariance. In this way, massless fields in the AdS space are gauge fields, as in flat space.

Setting $m^2=0$ in (\ref{22sep16}), we find
\begin{equation}
\label{22sep17}
\begin{split}
\left(\Box-\frac{1}{l^2}[(s-2)(s+d-3)-s]\right)\varphi^{\mu(s)}=0,\\
\nabla_\nu \varphi^{\nu \mu(s-1)}=0,\\
\varphi_\nu{}^{\nu \mu(s-2)}=0.
\end{split}
\end{equation}
These equations are invariant with respect to gauge transformations
\begin{equation}
\label{22sep18}
\delta \varphi^{\mu(s)}=\nabla^\mu\xi^{\mu(s-1)},
\end{equation}
where the gauge parameter satisfies
\begin{equation}
\label{22sep19}
\begin{split}
\left(\Box-\frac{1}{l^2}(s-1)(s+d-3) \right)\xi^{\mu(s-1)}=0,\\
\nabla_\nu \xi^{\nu \mu(s-2)}=0,\\
\xi_\nu{}^{\nu \mu(s-3)}=0.
\end{split}
\end{equation}

\begin{zadacha}
 Check that consistency of $\nabla \cdot \varphi =0$ with (\ref{22sep18}) requires the mass term for $\xi$ as indicated in (\ref{22sep19}).
 \end{zadacha}

\begin{zadacha}
 Show that the wave equation for the gauge parameter in  (\ref{22sep19}) entails the mass term for $\varphi$ as indicated in  (\ref{22sep17}).
 \end{zadacha}
 
 These exercises imply that $m^2=0$ is, indeed, the unique value for which (\ref{22sep16}) develops gauge invariance (\ref{22sep18}).
 Similarly, the mass term for the gauge parameter (\ref{22sep19}) is fixed uniquely by consistency with other conditions. 
  From this perspective, system (\ref{22sep17})-(\ref{22sep19}) can be regarded as the unique deformation of the massless flat-space system (\ref{14sep1})-(\ref{14sep3}) to AdS.

\vspace{0.5cm}

\subsection{The Fronsdal action}

The action that gives rise to (\ref{22sep17})-(\ref{22sep19}) was also found by Fronsdal \cite{Fronsdal:1978vb}. It is structurally identical to its flat space counterpart and its analysis follows the same lines.  The only differences that the AdS case brings is additional mass terms and covariant derivatives, which do not commute. Let us review how it works in more detail.

The Fronsdal action is given by 
\begin{equation}
\label{22sep20}
S=\frac{1}{2}\int d^dx\sqrt{-g} \varphi_{\mu(s)}G^{\mu(s)},
\end{equation}
where
\begin{equation}
\label{22sep21}
G^{\mu(s)}\equiv F^{\mu(s)}-\frac{s(s-1)}{4}g^{\mu\mu}F^{\mu(s-2)\nu}{}_\nu,
\end{equation}
\begin{equation}
\label{22sep22}
\begin{split}
F^{\mu(s)}&\equiv \Box \varphi^{\mu(s)}-s\nabla^\mu D^{\mu(s-1)}\\
&-\frac{1}{l^2}[(s-2)(d+s-3)-s]\varphi^{\mu(s)}-\frac{1}{l^2}s(s-1)g^{\mu\mu}\varphi^\nu{}_\nu{}^{\mu(s-2)},
\end{split}
\end{equation}
\begin{equation}
\label{22sep23}
D^{\mu(s-1)}\equiv \nabla_\nu \varphi^{\nu \mu(s-1)}-\frac{s-1}{2}\nabla^\mu\varphi^{\mu(s-2)\nu}{}_\nu.
\end{equation}

As in flat space, free equations of motion are
\begin{equation}
\label{22sep24}
F^{\mu(s)}\approx 0.
\end{equation}
Next, we impose the de Donder gauge
\begin{equation}
\label{22sep25}
D^{\mu(s-1)}=0
\end{equation}
and (\ref{22sep24}) becomes
\begin{equation}
\label{22sep26}
\Box \varphi^{\mu(s)}-\frac{1}{l^2}[(s-2)(d+s-3)-s]\varphi^{\mu(s)}-\frac{1}{l^2}s(s-1)g^{\mu\mu}\varphi^\nu{}_\nu{}^{\mu(s-2)}\approx 0.
\end{equation}
The residual gauge parameter satisfies the first equation in (\ref{22sep19}).

\begin{zadacha}
Check this.
 \end{zadacha}

Next, we consider variation
\begin{equation}
\label{22sep27}
\delta \varphi^{\mu(s-2)\nu}{}_\nu=\frac{2}{s}\nabla_\nu\xi^{\nu \mu(s-2)}.
\end{equation}
To be able to use this equation to gauge away the trace part of $\varphi$, one should first make sure that both sides of (\ref{22sep27}) satisfy the same wave equation. From (\ref{22sep26}) one has
\begin{equation}
\label{22sep28}
\Box \varphi^{\mu(s-2)\nu}{}_\nu -\frac{1}{l^2}(s^2+(d-2)s-2) \varphi^{\mu(s-2)\nu}{}_\nu\approx 0.
\end{equation}
One then checks that the first equation in (\ref{22sep19}) entails the same mass term for the wave equation for $\nabla\cdot \xi$.

\begin{zadacha}
Check this.
 \end{zadacha}

Therefore, we find that (\ref{22sep27}) can, indeed, be used to set
\begin{equation}
\label{22sep29}
\varphi^{\mu(s-2)\nu}{}_\nu=0,
\end{equation}
which leaves the gauge parameter divergenceless
\begin{equation}
\label{22sep30}
\nabla_\nu\xi^{\nu \mu(s-2)}=0.
\end{equation}
Finally, substituting (\ref{22sep29}) into (\ref{22sep25}), we find
\begin{equation}
\label{22sep31}
\nabla_\nu\varphi^{\nu \mu(s-1)}=0.
\end{equation}
Thus, we have shown that (\ref{22sep20}) upon variation and fixing gauges, gives 
 (\ref{22sep17})-(\ref{22sep19}), as we intended to show.
 
 Finally, we note that, as in flat space,  gauge invariance of the action (\ref{22sep20}) leads to  the Bianchi identity
 \begin{equation}
\label{22sep32}
\nabla_\nu G^{\nu \mu(s-1)}=g^{\mu\mu}H^{\mu(s-3)}.
\end{equation}

\subsection{Quadratic Casimir operator for $so(k,l)$}

In section \ref{sec:4.3} we computed the quadratic Casimir operator  for the lowest-weight representations of $so(d-1,2)$. The result is expressed in terms of the quadratic Casimir operator of the orthogonal algebra $so(d-1)$ evaluated on the lowest-weight space.
We will now compute the latter Casimir operator for traceless symmetric fields. 
In order to do that we will use  the approach of \emph{generating functions}, which is frequently used in the higher-spin literature to facilitate manipulations with tensor indices.

To be more general, we consider the case of $so(k,l)$ with $k+l=n$.  For dealing with symmetric tensors one introduces a single auxiliary $so(k,l)$ vector $u^a$, called polarisation vector. Then, by contracting tensor indices of a given tensor with $u$, we obtain a generating function, which looks like a scalar function of one auxiliary variable. 
For definiteness, we will use normalisation
\begin{equation}
\label{14oct1}
\varphi(u)\equiv \sum_m \frac{1}{m!} \varphi^{a(m)}u_{a_1} \dots u_{a_m}.
\end{equation}
As we can see, generating function $\varphi(u)$ allows us to deal with an infinite set of tensors simultaneously. 

All standard operations with tensors can be conveniently recast into the language of generating functions. In particular, it is straightforward to see that tracelessness of $\varphi^{a(m)}$ translates into 
\begin{equation}
\label{28sep2}
\frac{\partial}{\partial u^a}\frac{\partial}{\partial u_a}\varphi=0
\end{equation}
for the associated generating function. Similarly, one can easily see that 
\begin{equation}
\label{28sep2x1}
u^a\frac{\partial}{\partial u^a}\varphi=s \varphi
\end{equation}
implies that there is only rank-$s$ tensor in the sum (\ref{14oct1}).

The language of generating functions also allows one to deal with $so(k,l)$ transformations efficiently. Namely, under $so(k,l)$ the generating function transforms as
\begin{equation}
\label{28sep1}
J^{ab} \varphi (u)=-i \left(u^a \frac{\partial}{\partial u_b}-u^b \frac{\partial}{\partial u_a}\right) \varphi (u).
\end{equation}
In these terms, it is straightforward to evaluate the quadratic Casimir operator
\begin{equation}
\label{28sep3}
 {\cal C}_2(so(k,l))\varphi(u)  \equiv \frac{1}{2}J^{ab}J_{ab}\varphi(u)=s(s+n-2)\varphi(u).
\end{equation}
To obtain the last equality we had  to commute $\partial_u$ through $u$ so that they combine into $u\cdot\partial_u$. Then we used (\ref{28sep2x1}), the fact that the trace vanishes (\ref{28sep2}) and $n$ comes from $\delta^a{}_{a}$.

Similarly, for tensors with families of symmetric indices, we can introduce several auxiliary polarisation vectors. For example, to encode two-row Young diagrams it is convenient to introduce 
\begin{equation}
\label{14oct2}
\varphi(u,v)\equiv \sum_{m,p} \frac{1}{m!p!} \varphi^{a(m),b(p)}u_{a_1} \dots u_{a_m} v_{b_1} \dots v_{b_p}.
\end{equation}
Then, as it is not hard to see, the Young symmetry condition can be written as
\begin{equation}
\label{14oct3}
u^a\frac{\partial}{\partial v^a} \varphi(u,v)=0,
\end{equation}
 where it is understood that $a(m)$ correspond to the first row of the Young diagram, while $b(p)$ corresponds to the second one.

\begin{zadacha}
 Find the value of the quadratic Casimir operator for a tensor representation of $so(n)$ of shape $\mathbb{Y}(s_1,s_2)$. 
 \end{zadacha}

\subsection{Ambient-space formalism for tensor fields in  AdS}

Previously we defined the AdS space as a hyperboloid embedded into the ambient space via (\ref{7apr5}). The key idea behind the ambient space formalism is that it allows one to make the $so(d-1,2)$ symmetry of the AdS space manifest at the expense of introducing one redundant space-time coordinate. Below we will explain how this approach extends to tensor fields in the AdS space. 
This can be done in slightly different ways, see e.g. \cite{Bekaert:2010hk,Sleight:2016hyl} for more details.

As could have been anticipated, to describe a tensor field in AdS, we will use a tensor field in the ambient space instead. Once a symmetric tensor field in the ambient space $\varphi_{A(s)}(X)$ is given, the AdS field can be defined 
 via the pull-back
\begin{equation}
\label{2apr1}
\varphi_{A(s)}(X) \quad \to \quad \varphi_{\mu(s)}(x) = \frac{\partial X^{A_{1}}(x) }{\partial x^{\mu_1}}\dots 
 \frac{\partial X^{A_{s}}(x) }{\partial x^{\mu_s}} \varphi_{A(s)}(X(x)),
\end{equation}
where $x$ are some intrinsic AdS coordinates or, equivalently, $X(x)$ solves (\ref{7apr5}) identically. This map is surjective, but not injective for two reasons. Firstly, the ambient field on the right-hand side of (\ref{2apr1}) is evaluated on the AdS hyperboloid, so the values the field takes away from the AdS hyperboloid are irrelevant. Secondly, $\partial X/\partial x$ is a $(d+1)\times d$ matrix, which projects ambient space vectors to the tangent space of the AdS submanifold\footnote{Due to the presence of the ambient metric we do not distinguish vectors and 1-forms. Moreover, the ambient metric allows us to unambiguously decompose ambient space vectors at $X^2=-l^2$ into AdS tangent and AdS transverse parts, which are, moreover,  orthogonal to each other. Namely, tangent vectors $V^A$ are defined by the condition that these are annihilated by 1-form $d(X^2+l^2)$, that is, $V^AX_A=0$. Its orthogonal complement, spanned by vectors proportional to $X^A$, can be identified as the AdS transverse vector space. By lowering indices, we get the analogous decomposition for 1-forms.}. This means that  components of the ambient tensor which are transverse to AdS drop out from the right-hand side of (\ref{2apr1}).

There are different ways to proceed from here. First approach is to impose constraints on the ambient field, so that the map (\ref{2apr1}) becomes injective. In this case, for every tensor field in the AdS space one has a unique ambient space representative. Then any operation one has to perform with the AdS field can be reformulated in terms of its ambient space counterpart. For example, it is convenient to require that the ambient space field has fixed homogeneity degree in the radial direction
\begin{equation}
\label{2apr2}
X^B\frac{\partial}{\partial X^B} \varphi_{A(s)}(X)= \kappa \varphi_{A(s)}(X),
\end{equation}
which allows one to extend it away from the AdS hyperboloid in the unique manner. The homogeneity degree $\kappa$ can be any, though, there are some choices, which are more convenient than other.
Besides that one imposes
\begin{equation}
\label{2apr3}
X^A\varphi_{A(s)}(X)= 0,
\end{equation}
which implies that $\varphi_{A(s)}$ does not have components in the direction transverse to AdS, while the remaining components are uniquely defined by the AdS tensor. 

The second approach is to accept that a tensor field in AdS does not have a unique ambient space representative. Then one has to make sure that the  redundant components of the ambient tensor drop out from the analysis. In particular, one has to ensure that the ambient tensor is always evaluated on (\ref{7apr5}) and, moreover, the radial derivative
\begin{equation}
\label{30sep3}
\hat \kappa \equiv X^A\frac{\partial}{\partial X^A}
\end{equation}
drops out from all intrinsic AdS objects. Unlike in  (\ref{2apr2}) here we do not assume that $\hat \kappa$ takes a definite value.  Similarly, to ensure that the redundant components of the ambient tensor do not contribute, all its indices should be contracted with the projector to the AdS tangent space 
\begin{equation}
\label{28sep8x1}
P^A{}_B \equiv \delta^A{}_B+\frac{X^AX_B}{l^2}, \qquad X^AP_{A}{}^B=0, \qquad P^2=P.
\end{equation}
It is straightforward to see that due to the second equation in (\ref{28sep8x1}), the projected tensor, indeed, satisfies (\ref{2apr3}).

These two approaches are completely equivalent. For definiteness, we will use the first one. Then the ambient space representative of the AdS metric reads 
\begin{equation}
\label{28sep8}
 \qquad g_{AB} \equiv \eta_{AB}+\frac{X_AX_B}{l^2}.
\end{equation}
In turn, the AdS covariant derivative is simply given by the ambient partial derivative, which is
then followed by the projector that brings the result back to the space of tensors that satisfy
 (\ref{2apr3}). For
 symmetric tensors this gives
\begin{equation}
\label{30sep4}
\nabla_B\varphi_{A(s)} = P_B{}^D P_{A_1}{}^{C_1}\dots P_{A_s}{}^{C_s}\partial_D \varphi_{C(s)}, 
\end{equation}
where $\partial$ denotes the partial derivative with respect to $X$. To see that this is, indeed, the AdS covariant derivative, one can check that it is torsion-free, metric compatible and satisfies the remaining properties of the covariant derivative\footnote{For general embeddings, the ambient space Levi-Civita connection induces a Levi-Civita connection onto the embedded submanifold. The ambient and the induced geometric objects are connected by the Gauss-Codazzi equations.}.

The result (\ref{30sep4}) can be simplified by 
 pulling all $X$ through $\partial$ and using (\ref{2apr3})
\begin{equation}
\label{30sep5}
\nabla_B\varphi_{A(s)} = \partial_B \varphi_{A(s)}+\frac{X_B}{l^2} \kappa \varphi_{A(s)}-s\frac{X_A}{l^2}\varphi_{BA(s-1)}.
\end{equation}
This implies
\begin{equation}
\label{30sep6}
\nabla^A\varphi_{A(s)}=0 \quad \Leftrightarrow \quad   \partial^A\varphi_{A(s)}=0,
\end{equation}
that is the condition that the divergence of the AdS field  vanishes corresponds to the vanishing of the ambient space divergence. 
Similarly, one can compute the AdS Laplacian
\begin{equation}
\label{30sep7}
\nabla \cdot\nabla \varphi_{A(s)}=\partial\cdot \partial \varphi_{A(s)}+\frac{1}{l^2}[\kappa (\kappa+d-1)-s]\varphi_{A(s)},
\end{equation}
where
\begin{equation}
\label{30sep8}
\partial \cdot \partial \equiv \frac{\partial}{\partial X^A}\frac{\partial}{\partial X_A}.
\end{equation}
Note the appearance of the radial derivative $\kappa$ in (\ref{30sep7}). 
This happens due to the fact that the ambient space Laplacian involves derivatives in the AdS transverse direction and these have to be subtracted to match it with the AdS intrinsic Laplacian appearing on the other side of the equation.

\begin{zadacha}
Derive (\ref{30sep5}).
 \end{zadacha}
 
 \begin{zadacha}
Derive (\ref{30sep7}).
 \end{zadacha}

It will be convenient to deal with generating functions for ambient tensors. 
 By contracting them with the ambient space polarization vector $U$ in the same way as in (\ref{14oct1}), we obtain a generating function $\varphi(X;U)$. 
 In these terms, the condition that selects the AdS tensors (\ref{2apr3}) reads
 \begin{equation}
\label{28sep7}
X^A\frac{\partial}{\partial U^A}\varphi(X;U)=0,
\end{equation}
 while the AdS tracelessness and vanishing divergence amount to
 \begin{equation}
\label{28sep9x1}
\frac{\partial}{\partial U^A}\frac{\partial}{\partial U_A}\varphi (X;U)=0, 
\qquad
\frac{\partial}{\partial U_A}\frac{\partial}{\partial X^A}\varphi(X;U)=0.
\end{equation}
 Finally, the action of $so(d-1,2)$ isometries on the generating functions is given by
\begin{equation}
\label{28sep6}
\begin{split}
J^{AB}\varphi&=L^{AB}\varphi+M^{AB}\varphi,\\
 L^{AB}\varphi&=-i \left(X^A \frac{\partial}{\partial X_B}-X^B \frac{\partial}{\partial X_A}\right)\varphi,\\
  M^{AB}\varphi&=-i \left(U^A \frac{\partial}{\partial U_B}-U^B \frac{\partial}{\partial U_A}\right)\varphi.
\end{split}
\end{equation}

\subsection{Wave operator in terms of lowest-weight labels}

We will now use the machinery of the ambient space formalism to  compute the quadratic Casimir operator of $so(d-1,2)$  and to relate it to the wave equations of section \ref{sec:51}.
By definition for the quadratic Casimir operator of $so(d-1,2)$ we get
\begin{equation}
\label{28sep9}
 {\cal C}_2(so(d-1,2))\varphi(X;U)=
 \left(\frac{1}{2}L^{AB}L_{AB}+L^{AB}M_{AB}+\frac{1}{2}M^{AB}M_{AB}\right) \varphi(X;U).
\end{equation}
Now we will proceed with each term one by one. 

The last term is the easiest. Taking into account (\ref{28sep9x1}), we can use the result (\ref{28sep3}) except that now we are dealing with $so(d-1,2)$, not $so(k,l)$, so we need to replace $n$ with $d+1$ in (\ref{28sep3}) to account for a different contribution from taking the trace. This gives
\begin{equation}
\label{28sep10}
\frac{1}{2}M^{AB}M_{AB} \varphi(X;U) = s(s+d-1)\varphi(X;U)
\end{equation}
for a rank-$s$ ambient tensor. 
As for the rest, we have
\begin{equation}
\label{30sep1}
L^{AB}M_{AB}=-2X^AU_A\frac{\partial}{\partial X_B} \frac{\partial}{\partial U^B}-2U^A\frac{\partial}{\partial U^A}+2U^A \frac{\partial}{\partial X^A}X^B\frac{\partial}{\partial U^B},
\end{equation}
\begin{equation}
\label{30sep2}
L^{AB}L_{AB}=2\left(l^2 \frac{\partial}{\partial X^A}\frac{\partial}{\partial X_A}+
\kappa (\kappa+d-1)
\right).
\end{equation}
The first and the last terms in (\ref{30sep1}) vanish due to $\varphi$ being divergence-free and tangential, while the second terms is just $-2s$.
Putting everything together we find
\begin{equation}
\label{30sep9}
 {\cal C}_2(so(d-1,2))\varphi=
 (l^2 \partial\cdot \partial +\kappa (\kappa+d-1) +s(s+d-3))\varphi.
\end{equation}

Now we would like to rewrite the right-hand side of (\ref{30sep9}) in terms of the AdS Laplacian. To this end we eliminate $\partial\cdot\partial$ by employing (\ref{30sep7}). This gives
\begin{equation}
\label{30sep10}
 {\cal C}_2(so(d-1,2))\varphi=
 (l^2 \nabla\cdot \nabla +s(s+d-2))\varphi.
\end{equation}
Substituting here the value of the quadratic Casimir operator for general $E_0$, see (\ref{8apr7}), (\ref{28sep3}), we get
\begin{equation}
\label{30ssep11}
(l^2 \nabla\cdot\nabla - [E_0 (E_0-d+1)-s])\varphi=0.
\end{equation}
This is the wave equation we were seeking: it allows us to identify the parameter $E_0$ in the lowest-weight module construction with the mass-like term in the wave equation.

To be more precise, matching (\ref{30ssep11}) with (\ref{22sep16}), we find
\begin{equation}
\label{23sep1}
l^2m^2=E_0 (E_0-d+1)-(s-2)(d+s-3).
\end{equation}
As a test, we can see that for the unitarity bound value $E_0 = s+d-3$ -- which was argued to correspond to the massless case in section \ref{sec:323} --
we find $m^2=0$, the latter being the value of the mass term for which wave equations become gauge invariant.

Finally, as another consistency check,  we note that the radial derivative $\kappa$ dropped out form the end result. This was anticipated, as we were relating two intrinsic AdS quantities -- the AdS Laplacian and the value of the  Casimir operator evaluated on  the AdS field -- so any dependence on the way the field was extended away from the AdS hypersurface had to cancel out.

\begin{zadacha}
Consider the on-shell scalar field $(\Box-M^2)\varphi =0$ in AdS. By solving $J^{-m}\varphi=0$ find its lowest weight state. Find the associated values of $E_0$ and $s$.
 \end{zadacha}

\subsection*{Further reading}

The Fronsdal action for massless fields in the anti-de Sitter space was obtained in \cite{Fronsdal:1978vb} both in the ambient space formalism and in intrinsic terms. It is equally applicable for the description of gauge fields in de Sitter space. This construction was extended to fermions in \cite{Fang:1979hq}. Alternative descriptions for these cases can be found, e.g. in \cite{Buchbinder:2001bs,Buchbinder:2007vq}. Discussion of massless mixed-symmetry fields at the level of equations of motion can be found in \cite{Metsaev:1995re,Metsaev:1997nj,Metsaev:1998xg}.
Actions for massless mixed-symmetry fields in AdS are not known beyond special cases. This is related to the fact that these fields feature less gauge symmetries than their flat space counterparts \cite{Brink:2000ag}, so many of the technical difficulties characteristic to
flat space
 massive fields, in the AdS space appear already for massless fields. Another interesting phenomenon of the anti-de Sitter case is that for special values of mass massive fields develop gauge symmetries with the massless case being only one of such values. For the remaining special values gauge symmetry is smaller than in the massless case, so these fields are called \emph{partially-massless}. Discussions on partially massless fields can be found, e.g. in \cite{Deser:2001us}.
A more detailed review of the ambient-space formalism can be found in \cite{Bekaert:2010hk}.
For a recent discussion of general fields in de Sitter space and the associated representation theory aspects, see \cite{Basile:2016aen}.

\section{Singletons and higher-spin algebra}

In  section \ref{sec:323} we encountered a peculiar scalar representation of $so(d-1,2)$ with a shortened spectrum.  Representations of this type were discovered for AdS${}_4$ in \cite{Dirac:1963ta} and are known as singletons. Singleton representations play an important role in higher-spin theories. In particular, the Flato-Fronsdal theorem says that the tensor product of two singleton representations gives an infinite tower of massless higher-spin fields. Besides that,  higher-spin algebras can be identified as symmetry algebras of singleton representations. Both these statements serve as the basis of the higher-spin holography.

In this section we will prove the Flato-Fronsdal theorem and define the higher-spin algebra. We will focus on the AdS${}_4$ case, as 
due to the isomorphism $so(3,2)\sim sp(4,\mathbb{R})$ the relevant analysis is technically  simpler.
 Along the way, we will introduce some standard higher-spin background, such as the Weyl ordering and the star product.

\subsection{Flato-Fronsdal theorem from characters}

In this section we will prove the Flato-Fronsdal theorem \cite{Flato:1978qz} for $so(3,2)$ using characters.
For  additional details we refer the reader to \cite{Bae:2016rgm}.

To start, we summarise the relevant information on the lowest-weight representations of $so(3,2)$. 
In this case the compact subalgebra is $so(2)\oplus so(3)$. UIR's of $so(3)$ are classified by tensors of shapes $\mathbb{Y}(J)$ and $\mathbb{Y}(J,1)$, which are allowed Young diagrams in three dimensions\footnote{As usual, we ignore fermions in our analysis.}. Tensors of shape $\mathbb{Y}(J,1)$ can be dualised to symmetric ones by employing the Levi-Civita tensor $\epsilon^{abc}$. Thus, all inequivalent UIR's of $so(3)$ are given by traceless symmetric tensors of shapes $\mathbb{Y}(J)$. Accordingly, to characterise a representation of $so(2)\oplus so(3)$ one has to give a pair of numbers $(E,J)$, which will be referred to as compact weights. For brevity, we will also drop $\mathbb{Y}$ in our notation for the lowest-weight representations, e.g. the singleton representation in $d=4$ will be denoted $D(\frac{1}{2},0)$.

The Flato-Fronsdal theorem features the singleton representation as well as massless symmetric representations. To prove the theorem using characters, we will need the spectrum of compact weights for each of these representations. 
The scalar singleton has the following compact weights
\begin{equation}
\label{3oct1}
 D\left(\tfrac{1}{2},0\right): \qquad E=J+\frac{1}{2}, \qquad J=0,1,2,\dots.
\end{equation}
The term ''singleton'' refers to the fact that $ D(\frac{1}{2},0)$ has a single value of $E$ for each $J$, thus pairs $(E,J)$ form a single line in the compact weight space.
For massless fields of $s>0$ one has
\begin{equation}
\label{3oct2}
D(s+1,s): \qquad E=J+1,J+2,\dots, \qquad J=s,s+1,s+2,\dots.
\end{equation}
For massless $s=0$ one has instead 
\begin{equation}
\label{3oct3}
D(1,0): \qquad E=J+1,J+3,\dots, \qquad J=0,1,2,\dots.
\end{equation}
Multiplicity of each weight space is one in all cases. Let us note that unlike for singletons and massless spinning fields, there is no representation shortening for $D(1,0)$. The latter is a scalar representation with a specific value of $E_0$, which can be obtained by extension of the relation between $E_0$ and spin, $E_0=s+1$, for massless spinning fields to the scalar case. This is the only reason why we refer to this scalar field as massless.

We will not give a detailed derivation of (\ref{3oct1})-(\ref{3oct3}). They key features of these formulae are rather simple to understand by solving the following exercises.
 
\begin{zadacha}
 The lowest weight vector in (\ref{3oct2}) has $E=s+1$ and $J=s$. Explain why  states with $E=s+2$ and $J=s-1$ are absent in the spectrum. Explain how states with $E=s+2$ and $J=s+1$ are obtained from the lowest-weight space.
 Explain how states with $E=s+2$ and $J=s$ are obtained from the lowest-weight space.
 \end{zadacha}
 
\begin{zadacha}
 The lowest-weight vector in (\ref{3oct3}) has $E=1$ and $J=0$. Explain how  states with $E=2$ and $J=1$ are obtained from the lowest-weight vector. Explain why the state with $E=2$ and $J=0$, which is present for general spin (\ref{3oct2}) is now absent.
  Explain why  states with $E=3$ and $J=1$, which are present for general spin (\ref{3oct2}) are now absent from the spectrum.
  \end{zadacha}
 
  \begin{zadacha}
   The lowest-weight vector in (\ref{3oct1}) has $E=\frac{1}{2}$ and $J=0$. Explain how  state with $E=\frac{3}{2}$ and $J=1$ are obtained from the lowest-weight vector.   
   Explain why the state with $E=\frac{5}{2}$ and $J=0$ is absent in the spectrum. 
   \end{zadacha}

The idea behind the method of characters is as follows. Above we saw that representations of $so(3,2)$  decompose into representations of $so(2)\oplus so(3)$, labelled by $E$ and $J$, and the spectrum of compact weights uniquely defines the associated $so(3,2)$ representation. Instead of keeping track of this list of weight spaces, it is more convenient to introduce a function, which unambiguously records this information. To be more precise, one defines the character as
\begin{equation}
\label{3oct4}
\chi(\alpha,\beta) \equiv \sum_{E,J} n(E,J)\alpha^{2E}\chi_J(\beta),
\end{equation}
where 
\begin{equation}
\label{3oct5}
\chi_J(\beta)\equiv \sum_{m=-J}^J \beta^{2m}=\frac{\beta^{2J+1}-\beta^{-2J-1}}{\beta -\beta^{-1}}
\end{equation}
and $n(E,J)$ is the multiplicity with which the weight $(E,J)$ appears in the representation. Putting it differently, one chooses the Cartan subalgebra of $so(3,2)$ as $J_{05}\oplus J_{12}$, then each vector with  weights $E$ and $m$ in the representation space contributes $\alpha^{2E}\beta^{2m}$ to the character. Conversely, once the character is known one can decompose it into the power series, thus identifying the weights $E$ and $m$ that contribute. As spin-$J$ representation of $so(3)$ has weights with respect to $J_{12}$ ranging form $-J$ to $J$, we find it convenient to group the result in $\chi_{J}$'s.

Let us now proceed to the actual computation of characters for (\ref{3oct1})-(\ref{3oct3}). It is pretty straightforward and gives
\begin{equation}
\label{3oct6}
\chi\left(D\left(\tfrac{1}{2},0\right)\right)=\frac{\alpha+\alpha^{-1}}{(\alpha^{-1}\beta^{-1}-\alpha\beta)(\beta \alpha^{-1}-\alpha\beta^{-1})},
\end{equation}
\begin{equation}
\label{3oct7}
\chi\left(D(s+1,s)\right)=\frac{\frac{(\alpha\beta)^{2s}}{\alpha\beta-(\alpha\beta)^{-1}}+\frac{(\alpha/\beta)^{2s}}{\beta/\alpha-\alpha/\beta}}{(\beta-\beta^{-1})(\alpha-\alpha^{-1})}, \qquad s>0,
\end{equation}
\begin{equation}
\label{3oct8}
\chi\left(D(1,0)\right)=\frac{1}{\alpha}\frac{\frac{1}{\alpha\beta-(\alpha\beta)^{-1}}+\frac{1}{\beta/\alpha-\alpha/\beta}}{(\beta-\beta^{-1})(\alpha^2-\alpha^{-2})}.
\end{equation}

\begin{zadacha}
Derive (\ref{3oct6})-(\ref{3oct8}).
\end{zadacha}

Then one can check that
\begin{equation}
\label{3oct9}
\sum_{s=0}^{\infty} \chi\left(D(s+1,s)\right)= \chi\left(D\left(\tfrac{1}{2},0\right)\right) \chi \left(D\left(\tfrac{1}{2},0\right)\right).
\end{equation}
Considering that the character of the tensor product is the product of characters -- which can be easily derived from definitions of the character and the tensor product -- one concludes that
\begin{equation}
\label{3oct10}
D\left(\tfrac{1}{2},0\right) \otimes D\left(\tfrac{1}{2},0\right) = \sum_{s=0}^{\infty} D(s+1,s).
\end{equation}
In other words, \emph{the tensor product of two scalar singletons is the direct sum of massless representations} of $so(3,2)$.

\begin{zadacha}
Using the definitions of the character and of the tensor product of representations, show that the character of the tensor product of two representations equals the product of characters of the representations being multiplied.
\end{zadacha}

For completeness, we mention that there is also a fermionic singleton representation $D(1,\tfrac{1}{2})$. To acknowledge Dirac for the discovery of singleton representations, $D(1,\tfrac{1}{2})$ and $D(\tfrac{1}{2},0)$ are often denoted $Di$ and $Rac$ respectively.

\begin{zadacha}
Using characters, find the spectrum of the tensor product of two fermionic singletons. 
\end{zadacha}

\subsection{Flato-Fronsdal theorem from oscillators}

It turns out that singletons of $so(3,2)$ admit a simple oscillator realisation \cite{Dirac:1963ta}, which we will now review.
We will follow the  review part of \cite{Bae:2016rgm}.

\subsubsection{Oscillator realisation of singletons}

The AdS${}_4$ isometry algebra $so(3,2)$ can be realised in terms of two pairs of oscillators
\begin{equation}
\label{5oct1}
[a,a^\dagger]=1, \qquad [b, b^\dagger]=1.
\end{equation}
In these terms the $so(3,2)$ generators read\footnote{This happens due to the isomorphism $so(3,2)\sim sp(4,\mathbb{R})$.
It may be instructive to consider a $4d$ symplectic manifold and Hamiltonians  quadratic in the Darboux coordinates $z=\{q^1,q^2,p^1,p^2 \}$. Via the Poisson bracket these generate vector fields $\xi^i(z)\partial_i$ that preserve the symplectic form and, moreover, being linear in the Darboux coordinates, $\xi \propto z$, preserve the origin $z=0$. Then, it is not hard to see that these vector fields induce linear transformations on the tangent bundle at the origin, which preserve the symplectic form at this point. Accordingly, these generate  $ sp(4,\mathbb{R})$ and $\partial_j \xi^i(z)$ are  the associated $sp(4,\mathbb{R})$ matrices. Representation (\ref{5oct2}) gives a version of this construction, in which the Poisson bracket is replaced with the commutator.}
\begin{equation}
\label{5oct2}
\begin{split}
E&=\frac{1}{2}(a^\dagger a+b^\dagger b +1), \\
J^{-1}&=\frac{1}{2}(a^2+b^2), \qquad J^{-2}=-\frac{i}{2}(a^2-b^2), \qquad J^{-3} = iab,\\
J^{+1}&=(J^{-1})^\dagger=\frac{1}{2}((a^\dagger)^2+(b^\dagger)^2),\\
 J^{+2}&=(J^{-2})^\dagger=\frac{i}{2}((a^\dagger)^2-(b^\dagger)^2),
  \qquad J^{+3}=(J^{-3})^\dagger=-i a^\dagger b^\dagger,\\
  J^{12}&=\frac{1}{2}(b^\dagger b-a^\dagger a), \qquad
  J^{13}=\frac{1}{2}(b^\dagger a+a^\dagger b),
  \qquad
  J^{23}=-\frac{i}{2}(b^\dagger a-a^\dagger b).
\end{split}
\end{equation}

\begin{zadacha}
  Check that $[J^{-1},J^{+1}]=2E$ and $[J^{-1},J^{+2}]=-2iJ^{12}$, as required by (\ref{8apr4}).
  \end{zadacha}

The oscillator realisation of $Rac$ works as follows. We start from a state, that satisfies
\begin{equation}
\label{5oct3}
a|0\rangle =0, \qquad b|0\rangle =0.
\end{equation}
It is  annihilated by all lowering operators $J^{-i}$, (\ref{5oct2}), so it is the lowest-weight vector. Moreover, one easily sees that
\begin{equation}
\label{5oct4}
E|0\rangle = \frac{1}{2}|0\rangle, \qquad J^{ij}|0\rangle =0,
\end{equation}
so the lowest-weight vector has energy equal to $1/2$ and spin equal to zero. Therefore, $|0\rangle$ is the lowest-weight vector of the singleton representation of $so(3,2)$. The remaining states are generated, as usual, by applying the raising operators. As a result, the states of the singleton module are spanned by
even functions of the creation operators
\begin{equation}
\label{5oct5}
Rac:\qquad f(a^\dagger,b^\dagger)|0\rangle, \qquad f(-a^\dagger,-b^\dagger)=f(a^\dagger,b^\dagger).
\end{equation}
Note that since $a^\dagger$ and $b^\dagger$ commute, expression $ f(a^\dagger,b^\dagger)$ is defined unambiguously. 

The norm on singleton states given by functions $f_1$ and $f_2$ is defined as
\begin{equation}
\label{22jun1}
(f_1,f_2)\equiv \langle 0| f_1^*(a,b)f_2(a^\dagger,b^\dagger)|0\rangle.
\end{equation}
Accordingly, Hermitian conjugation amounts to $a\leftrightarrow a^\dagger$, $b\leftrightarrow b^\dagger$, reversing the order of operators and making the complex conjugation.

\begin{zadacha}
Check that $E^\dagger = E$, $(J^{ij})^\dagger = J^{ij}$ and $(J^{-m})^\dagger = J^{+m}$, thereby, representation (\ref{5oct2}) is unitary.
  \end{zadacha}

\begin{zadacha}
  Consider the Di representation, generated from lowest-weight vectors $a^\dagger|0 \rangle$ and $b^\dagger|0\rangle$. Find the associated values of energy and spin.
  \end{zadacha}

\subsubsection{Tensor product of two singletons}

To proceed with the Flato-Fronsdal theorem, we consider the tensor product of two scalar singletons. To this end we need to double the set of oscillators
\begin{equation}
\label{5oct6}
[a_1,a_1^\dagger]=1, \qquad [b_1, b_1^\dagger]=1, \qquad 
[a_2,a_2^\dagger]=1, \qquad [b_2, b_2^\dagger]=1.
\end{equation}
Then, the states of the tensor product are generated by 
\begin{equation}
\label{5oct6x0}
\begin{split}
Rac\otimes Rac:\qquad &c(a_1^\dagger,b_1^\dagger,a_2^\dagger,b_2^\dagger)|0\rangle_1 \otimes |0\rangle_2, \\
&c(-a_1^\dagger,-b_1^\dagger,a_2^\dagger,b_2^\dagger)=c(a_1^\dagger,b_1^\dagger,-a_2^\dagger,-b_2^\dagger)=c(a_1^\dagger,b_1^\dagger,a_2^\dagger,b_2^\dagger).
\end{split}
\end{equation}

It will be convenient to introduce
\begin{equation}
\label{5oct6x1}
A_i =\{a_1^\dagger,a_2^\dagger \}, \qquad B_i =\{b_1^\dagger,b_2^\dagger \}, \qquad i=1,2.
\end{equation}
Then $c$ can be regarded as a generating function of $gl(2)$-tensors with two groups of symmetric indices associated with $A$ and $B$.
By the definition of the tensor product the generators of $so(3,2)$ act  as
\begin{equation}
\label{5oct7}
J (V_1\otimes V_2) = J V_1\otimes V_2 + V_1 \otimes J V_2, \qquad V_i \in Rac_i.
\end{equation}

Our goal is to identify irreducible $so(3,2)$ representations in this tensor product. To this end, we will first find its lowest-energy states. This will give a reducible representation of $so(3)\oplus so(2)$. Our next goal will be to identify $(E,J)$ of its irreducible components.
 In order to do that, we will use that each irreducible representation of $so(3)\oplus so(2)$ with weights $(E,J)$ has the unique lowest-$J^{12}$ vector.
Moreover, this vector  has the energy eigenvalue equal $E$ and the $J^{12}$ eigenvalue equal $-J$ . Thus, by applying $E$ and $J^{12}$ to the lowest-energy and lowest-$J^{12}$ states, we will be able to find energies and spins of lowest-weight representations $D(E,J)$ in the tensor product of singletons.

To find the lowest-$J^{12}$ states, we will need the $J^{12}$-lowering operator
\begin{equation}
\label{5oct8}
{\cal J}_-\equiv a^\dagger b=-\frac{1}{2}(J^{23}+iJ^{13}).
\end{equation}
It is easy to see that
\begin{equation}
\label{5oct9}
[J^{12},{\cal J}_-]=-{\cal J}_-,
\end{equation}
so, ${\cal J}_-$, indeed, lowers $J^{12}$.

Having explained our strategy, let us proceed to its first step.
For $c$ defined in (\ref{5oct6x0}),  the lowest-energy condition gives
\begin{equation}
\label{5oct10}
\begin{split}
\delta^{ij}\frac{\partial^2}{\partial A_i \partial A_j}c^{\rm lw}=0,\\
\delta^{ij}\frac{\partial^2}{\partial A_i \partial B_j}c^{\rm lw}=0,\\
\delta^{ij}\frac{\partial^2}{\partial B_i \partial B_j}c^{\rm lw}=0.\\
\end{split}
\end{equation}
In other words, we find that tensors encoded by $c^{\rm lw}$ are traceless with respect to the $so(2)$-invariant metric on every pair of indices. 

\begin{zadacha}
  Compute  $a(a^\dagger)^n|0\rangle$ by commuting $a$ to the right until it annihilates the vacuum. Find the general formula for $af(a^\dagger)|0\rangle$ with polynomial $f$.
\end{zadacha}

\begin{zadacha}
 Using the results of the previous exercise derive (\ref{5oct10}).
\end{zadacha}

Next, we impose the requirement of being the lowest weight with respect to $J^{12}$. This gives
\begin{equation}
\label{5oct11}
\delta^{ij}A_i \frac{\partial}{\partial B_j}c^{\rm lw}=0.
\end{equation}
This means that tensors encoded by $c^{\rm lw}$ have the symmetry of Young diagrams with $B$ in the second row. 

Summing up, we find that the lowest-energy and lowest-$J_{12}$ states in $Rac\otimes Rac$ can be classified by traceless tensors with  the symmetry of two-row Young diagrams in two dimensions. As discussed in section \ref{sect:232}, all traceless $so(d)$ tensors with the number of boxes in first two columns that exceeds  $d$ are vanishing. Therefore, the only non-trivial traceless tensors in $d=2$ are the rank-two antisymmetric tensor and all symmetric tensors of ranks starting from zero. 
Thus, denoting
\begin{equation}
\label{5oct12}
mc^{\rm lw}_{m,n} = A^i \frac{\partial}{\partial A^i}c^{\rm lw}_{m,n}, \qquad n c^{\rm lw}_{m,n} = B^i \frac{\partial}{\partial B^i}c^{\rm lw}_{m,n},
\end{equation}
we find that the only possible solutions to (\ref{5oct10}), (\ref{5oct11}) are of the form
\begin{equation}
\label{5oct13}
(m,n)=(1,1), \qquad (m,n)=(m,0).
\end{equation}

Proceeding further, we note that the only rank-2 antisymmetric tensor in $d=2$ is the Levi-Civita one, so the $(m,n)=(1,1)$ solution is of the form
\begin{equation}
\label{5oct14}
c^{\rm lw}_{1,1}=\epsilon^{ij}A_i B_j = A_1B_2-A_2B_1 = a_1^\dagger b_2^\dagger - a_2^\dagger b_1^\dagger.
\end{equation}
Clearly, it does not have the right parity (\ref{5oct6x0}) with respect to the change of signs of oscillators for each singleton separately. 
The remaining solutions have $n=0$, so $c$ only involves $A_i$.  By requiring symmetry with respect to the change of the signs of all oscillators, we find that $m$ should be even, $m=2k$. 

Next, we would like to solve the trace constraints (\ref{5oct10}) explicitly. Since $c$ that we are left with is $B$-independent, the last two equations are satisfied automatically. To solve the first one, it is convenient to use the light-cone coordinates
\begin{equation}
\label{5oct15}
A_+ = A_1+iA_2, \qquad A_- = A_1-iA_2.
\end{equation}
The two independent polynomial solutions at homogeneity degree $2k$ are
\begin{equation}
\label{5oct16}
(A_1+i A_2)^{2k}, \qquad (A_1-iA_2)^{2k}.
\end{equation}
Finally, we recall that (\ref{5oct6x0}) implies that the solutions we are looking for are even on $A_1$ and on $A_2$ separately. This leaves us with 
\begin{equation}
\label{5oct17}
c^{\rm lw}_{2k,0}=\frac{1}{2}[(A_1+i A_2)^{2k}+ (A_1-iA_2)^{2k}].
\end{equation}

Summarising our analysis, we have found that the lowest-energy and lowest-$J^{12}$ states in the tensor product of two singletons can be characterised by integer $k\ge 0$ and are given by 
\begin{equation}
\label{5oct18}
|V_k\rangle \equiv \frac{1}{2}[(a^\dagger_1+i a^\dagger_2)^{2k}+ (a^\dagger_1-ia^\dagger_2)^{2k}]|0\rangle \otimes |0\rangle.
\end{equation}
It is straightforward to compute that
\begin{equation}
\label{5oct19}
\begin{split}
E|V_k\rangle &= \frac{1}{2}(a_1^\dagger a_1+b_1^\dagger b_1+a_2^\dagger a_2+b_2^\dagger b_2 +2)
|V_k\rangle = (k+1)|V_k\rangle,\\
J^{12}|V_k\rangle &= \frac{1}{2}(b_1^\dagger b_1-a_1^\dagger a_1+b_2^\dagger b_2-a_2^\dagger a_2)
|V_k\rangle = -k |V_k\rangle.
\end{split}
\end{equation}
Therefore, the tensor product of two singletons decomposes into the lowest-weight representations $D(k+1,k)$, $k\ge 0$, which are the right weights for massless fields. 

In a similar manner one can analyze the tensor product of spinor singletons, the tensor product of the spinor and the scalar singletons as well as the symmetric tensor products of the scalar and the spinor singletons.

\begin{zadacha}
 By modifying the analysis above, decompose the tensor product of spinor singletons into irreducible representations.
\end{zadacha}

\subsection{Higher-spin algebra}

In this section, we will define  the higher-spin algebra for the AdS${}_4$ case. Its relation to massless higher-spin fields will become clear later. Before proceeding to the higher-spin algebra, we will introduce one auxiliary tool, -- the star product -- which is extensively used in the higher-spin literature.

\subsubsection{Wigner-Weyl map and the star product}

In the previous section we encountered  oscillator variables (\ref{5oct1}). What distinguishes them from usual variables is that oscillators do not commute. Because of that, when dealing with oscillator variables, one runs into difficulties related to order ambiguities. More precisely, for a function of oscillator variables, by rearranging the order of oscillators in it, one can bring it to different forms, which are not manifestly equal to the original expression. Certainly, this non-uniqueness of a representation for the same oscillatorial function is unattractive. 
To remedy this problem, one may require that oscillators always appear in a certain ordering, which prevents their further non-trivial reordering and thus, every expression receives its unique representation.
 One example of such an ordering is the normal ordering, that is when $a$ and $b$ appear to the right from $a^\dagger$ and $b^\dagger$. This ordering may be convenient  if we want to act with our oscillatorial expressions  on the vacuum (\ref{5oct3}).

For our purposes it will be more convenient to use the symmetric ordering which is also known as \emph{the Weyl ordering}. Before going to its definition, we will change notation by adding hats to the oscillators $a\to \hat a$ and $a^\dagger \to \hat a^\dagger$. This is done to distinguish them from their counterparts, to be introduced soon, which will commute trivially. 

According to the \emph{definition}, all linear combinations of 
\begin{equation}
\label{9oct1}
(\alpha \hat a +\beta \hat a^\dagger)^n,
\end{equation}
with arbitrary commuting numbers $\alpha$ and $\beta$, are Weyl ordered. In particular, $\hat a^2$, $\hat a\hat a^\dagger + \hat a^\dagger \hat a$ and $(\hat a^\dagger)^2$ are all Weyl ordered. In turn, $\hat a\hat a^\dagger$ is not, unless it appears in combination $\hat a\hat a^\dagger + \hat a^\dagger \hat a$. Similarly,
\begin{equation}
\label{9oct2}
\hat a^2 (\hat a^\dagger)^2+ \hat a \hat a^\dagger \hat a \hat a^\dagger +\hat a  (\hat a^\dagger)^2 \hat a +\hat a^\dagger \hat a^2 \hat a^\dagger + \hat a^\dagger \hat a \hat a^\dagger \hat a +
(\hat a^\dagger)^2 \hat a^2
\end{equation}
is Weyl ordered. 
As it is not hard to see,  any polynomial expression of $\hat a$ and $\hat a^\dagger$ 
can be brought to a unique Weyl ordered form
by properly commuting $\hat a$ and $\hat a^\dagger$.

Next, once a Weyl ordered function of oscillators  is given, we can associate with it a function of  \emph{commuting} variables simply  by replacing $\hat a\to a$, $\hat a^\dagger\to a^\dagger$.
For the example above, we have
\begin{equation}
\label{9oct3}
\hat a^2 (\hat a^\dagger)^2+ \hat a \hat a^\dagger \hat a \hat a^\dagger +\hat a  (\hat a^\dagger)^2 \hat a +\hat a^\dagger \hat a^2 \hat a^\dagger + \hat a^\dagger \hat a \hat a^\dagger \hat a +
(\hat a^\dagger)^2 \hat a^2\quad  \leftrightarrow \quad  6 a^2 (a^\dagger)^2.
\end{equation}
The map between functions of oscillators and functions of commuting variables as described above is called \emph{the Wigner-Weyl map}. Let us denote it by $W$
\begin{equation}
\label{9oct4}
f(\hat a,\hat a^\dagger) \leftrightarrow W[f](a,a^\dagger).
\end{equation}
One often calls $W[f]$ the \emph{symbol} of $f$.
We emphasize again, that for $f$ in the Weyl ordered form map (\ref{9oct4}) amounts to $\hat a\to a$, $\hat a^\dagger\to a^\dagger$, while for $f$ not in the Weyl ordered form, one should first manipulate it by performing commutators so that it becomes Weyl ordered and only then replace $\hat a\to a$, $\hat a^\dagger\to a^\dagger$. 

\begin{zadacha}
  Find the Weyl ordered form of $\hat a^2 \hat a^\dagger$.
\end{zadacha}

An important part of the oscillator algebra is that \emph{any two functions of oscillators can be multiplied}. For two Weyl ordered functions, their product, in general, is not Weyl ordered. This implies that the usual operator product of oscillator functions corresponds to some non-trivial operation with their symbols. This operation is called \emph{the star product} and it is \emph{defined} by 
\begin{equation}
\label{9oct5}
W[f_1]\star W[f_2] \equiv W[f_1 f_2].
\end{equation}
Here, on the right-hand side functions $f_1$ and $f_2$ are multiplied with the usual operator product, which is then followed by the 
Wigner-Weyl map. In turn, on the left-hand side, one first maps $f_1$ and $f_2$ to their symbols, which are then multiplied with the star product, with the latter defined so that the two sides of (\ref{9oct5}) agree.

For the case of two pairs of oscillators, which will be relevant in the following,  the star product  reads 
\begin{equation}
\label{6oct1}
\begin{split}
&\big(f\star g\big)(a,a^\dagger,b,b^\dagger) \\
&\qquad\quad = f(a,a^\dagger,b,b^\dagger) \exp \frac{1}{2}\left( \frac{\overleftarrow\partial}{\partial a}
{\frac{\overrightarrow\partial}{\partial a^\dagger}}
-
{\frac{\overrightarrow\partial}{\partial a}}
{\frac{\overleftarrow\partial}{\partial a^\dagger}}+
{\frac{\overleftarrow\partial}{\partial b}}
{\frac{\overrightarrow\partial}{\partial b^\dagger}}
-
{\frac{\overrightarrow\partial}{\partial b}}
{\frac{\overleftarrow\partial}{\partial b^\dagger}}
\right)
g(a,a^\dagger,b,b^\dagger).
\end{split}
\end{equation}
It admits an alternative integral representation
\begin{equation}
\label{6oct2}
\begin{split}
f\star g =&\frac{1}{\pi^4}\int ds ds^\dagger dt dt^\dagger du du^\dagger dv dv^\dagger
f(a+s,a^\dagger+s^\dagger,b+t,b^\dagger+t^\dagger)\\
&\qquad \qquad g(a+u,a^\dagger+u^\dagger,b+v,b^\dagger+v^\dagger)
\exp \left(-2[su^\dagger +tv^\dagger-s^\dagger u-t^\dagger v] \right).
\end{split}
\end{equation}
While the first formula is more convenient when we want to compute the star product of two polynomials, the second one is more handy to deal with non-polynomial functions of oscillators, such as exponentials. Derivation of (\ref{6oct1}) and (\ref{6oct2}) can be found, e.g. in \cite{BSH}.

\begin{zadacha}
  Find $(a^\dagger)^2a\star a^2 a^\dagger$ using (\ref{6oct1}).
\end{zadacha}

\begin{zadacha}
  Compute $e^{\alpha a^\dagger a}\star e^{\beta a^\dagger a}$ where $\alpha$ and $\beta$ are numbers using  (\ref{6oct2}).
\end{zadacha}

\subsubsection{Back to higher-spin algebra}
\label{sec:632}

We start by considering the associative algebra of linear maps of the $Rac$ representation space to itself. With the oscillator realisation of the $Rac$ representation as in (\ref{5oct5}) this algebra is generated by operators of the form
\begin{equation}
\label{6oct3}
v(\hat a,\hat a^\dagger,\hat b,\hat b^\dagger)
\end{equation}
with the additional requirement that the total homogeneity degree of $h$  in all oscillators is even\footnote{The fact that any operator on the Hilbert space of states generated from the vacuum by raising operators is expressible in terms of creation and annihilation operators is rather standard and can be found e.g. in \cite{Weinberg}.}. If $Rac$ was a finite-dimensional linear space $V$, the counterpart of this associative algebra would have  been given by ${\rm End}(V)$ -- the algebra of all linear maps in the $V$ space. 
Putting it differently, the states of the singleton representation are spanned by functions of two variables with the additional parity requirement. Then, (\ref{6oct3}) are just the most general linear differential operators that preserve the aforementioned parity condition.

The product of two linear transformations in the $Rac$ space  is given by the composition of these transformations. For two transformations $v_1$ and $v_2$, their composition is generated by their operator product $v_2 v_1$. This product, is, clearly, associative. \emph{The higher-spin algebra} in AdS${}_4$ is the Lie algebra obtained from this associative algebra, by taking
the commutator of generators as the Lie bracket
\begin{equation}
\label{5apr1}
[v_1,v_2](\hat a,\hat a^\dagger,\hat b,\hat b^\dagger) \equiv v_1(\hat a,\hat a^\dagger,\hat b,\hat b^\dagger) v_2(\hat a,\hat a^\dagger,\hat b,\hat b^\dagger) - v_2(\hat a,\hat a^\dagger,\hat b,\hat b^\dagger) v_1(\hat a,\hat a^\dagger,\hat b,\hat b^\dagger).
\end{equation}
By using the Weyl-Wigner map, $w_i \equiv W[v_i]$, the higher-spin algebra can be realised on functions of commuting variables 
\begin{equation}
\label{5apr2}
\begin{split}
&[w_1,w_2](a, a^\dagger, b, b^\dagger) \equiv  [w_1,w_2]_{\star}(a, a^\dagger, b, b^\dagger)\\
&\qquad\qquad \equiv 
w_1(a, a^\dagger, b, b^\dagger) \star w_2(a, a^\dagger, b, b^\dagger) -w_2(a, a^\dagger, b, b^\dagger) \star w_1(a, a^\dagger, b, b^\dagger).
\end{split}
\end{equation}
The total homogeneity degree for symbols $w$ remains even
\begin{equation}
\label{5apr3}
\left( a \frac{\partial}{\partial a}+ a^\dagger \frac{\partial}{\partial  a^\dagger}
+ b \frac{\partial}{\partial b}+ b^\dagger \frac{\partial}{\partial  b^\dagger}
 \right)w(a, a^\dagger, b, b^\dagger) = 2k w (a, a^\dagger, b, b^\dagger), \qquad k \in \mathbb{N}\cup \{ 0\}.
\end{equation}
We denote the higher-spin algebra so defined by $hs(3,2)$.

Let us now derive some important properties of the higher-spin algebra. To start, it is straightforward to see that symbols for $so(3,2)$ generators (\ref{5oct2}) are all quadratic in the oscillators. Moreover, the commutator of two $so(3,2)$ generators gives again the $so(3,2)$ generator, so $so(3,2)$ forms a subalgebra of the higher-spin algebra, $so(3,2)\subset hs(3,2)$.

Next, let us consider the adjoint action of $so(3,2)$ on $hs(3,2)$
\begin{equation}
\label{6oct4}
g [w] \equiv [g,w]_{\star}, \qquad g\in so(3,2), \quad w\in hs(3,2).
\end{equation}
The commutator $[g,w]_{\star}$ is linear in $w$, moreover, it gives an element of the higher-spin algebra as a result. Together with other obvious properties of (\ref{6oct4}) this means that with respect to the adjoint action the higher-spin algebra forms a representation of $so(3,2)$. Below, our goal will be to decompose this representation into irreducible ones. 

To this end, we will explicitly evaluate the commutator (\ref{6oct4}). Since $g$ is polynomial, it is more convenient to use representation (\ref{6oct1}) for the star-product. As $g$ is quadratic in oscillators, only the first three terms in the expansion of the exponent (\ref{6oct1}) are non-trivial. Moreover, the first and the last terms are symmetric in $w\leftrightarrow g$, so these drop out from the commutator. We are left with 
\begin{equation}
\label{6oct7}
[g, w]_\star = g \left( {\frac{\overleftarrow\partial}{\partial a}}
{\frac{\overrightarrow\partial}{\partial a^\dagger}}
-
{\frac{\overrightarrow\partial}{\partial a}}
{\frac{\overleftarrow\partial}{\partial a^\dagger}}+
{\frac{\overleftarrow\partial}{\partial b}}
{\frac{\overrightarrow\partial}{\partial b^\dagger}}
-
{\frac{\overrightarrow\partial}{\partial b}}
{\frac{\overleftarrow\partial}{\partial b^\dagger}}
\right)
w.
\end{equation}

It is straightforward to see that $[g, w]_\star$ has the same homogeneity degree in oscillators as $w$ itself. Therefore the representation of $so(3,2)$ carried by $w$ splits into representations with fixed homogeneity degrees in oscillators. Each such representation is, obviously, finite-dimensional, so it should be equivalent to some tensor representation of $so(3,2)$\footnote{This analysis can be streamlined
if one groups the oscillators into an $sp(4,\mathbb{R})$ vector $Y^A=\{a,a^\dagger,b,b^\dagger \}$. In these terms it is not hard to see that the higher-spin algebra decomposes into symmetric tensors of $sp(4,\mathbb{R})$.}.

To identify these tensor representations,  we will follow the familiar route.
As before, we will be searching for the lowest-weight vectors of energy and $J^{12}$. Evaluating the lowest energy and lowest $J^{12}$ constraints in terms of symbols we find
\begin{equation}
\label{6oct8}
\begin{split}
[a^2,w^{\rm lw}]_\star&=2a\frac{\partial}{\partial a^\dagger}w^{\rm lw}=0,\\
[b^2,w^{\rm lw}]_\star&=2b\frac{\partial}{\partial b^\dagger}w^{\rm lw}=0,\\
[ab,w^{\rm lw}]_\star&=b\frac{\partial}{\partial a^\dagger}w^{\rm lw}+a\frac{\partial}{\partial b^\dagger}w^{\rm lw}=0,\\
[a^\dagger b,w^{\rm lw}]_\star&=-b\frac{\partial}{\partial a}w^{\rm lw}+a^\dagger\frac{\partial}{\partial b^\dagger}w^{\rm lw}=0.
\end{split}
\end{equation}

\begin{zadacha}
 Derive (\ref{6oct8}).
\end{zadacha}

From the first two equations we learn that $w^{\rm lw}$ is independent of $a^\dagger$ and $b^\dagger$. The third equation is then trivially satisfied, while the last equation implies that $w^{\rm lw}$ is also independent of $a$.
In other words, the lowest-weight vectors are of the form
\begin{equation}
\label{6oct9}
w^{\rm lw}=b^{2k},
\end{equation}
where we also took into account (\ref{5apr3}).
It is then straightforward to find their weights
\begin{equation}
\label{6oct10}
\begin{split}
[E,b^{2k}]_\star=\frac{1}{2}[a^\dagger a + b^\dagger b,b^{2k}]_\star=-kb^{2k},\\
[J^{12},b^{2k}]_\star=\frac{1}{2}[  b^\dagger b-a^\dagger a,b^{2k}]_\star=-kb^{2k}.
\end{split}
\end{equation}

Finally, let us identify them with tensorial representations. We claim that these representations are given by 
$\mathbb{Y}(k,k)$-shaped tensors of $so(3,2)$. This can be verified in the following way. If we use $U^M$ and $V^M$ as polarisation vectors for the first and the second rows of the aforementioned tensor 
then the generating function for its lowest-weight vector is given by
\begin{equation}
\label{6oct11}
[(iU^1+U^2)(iV^0+V^4)-(iV^1+V^2)(iU^0+U^4)]^k.
\end{equation}
It can be checked that (\ref{6oct11}) is, indeed, the lowest weight with respect to $E$ and $J^{12}$, moreover, it is traceless, has the desired Young symmetry and eigenvalues of $E$ and $J^{12}$ are both $-k$. Thus, the $\mathbb{Y}(k,k)$-shaped tensors of $so(3,2)$, indeed, carries the same lowest-weight representation as (\ref{6oct10}).

\begin{zadacha}
  Check this.
\end{zadacha}

\subsection{Further reading}

The Flato-Fronsdal theorem can be analogously derived for the $Di$ representation. It can also be extended to higher dimensions \cite{Vasiliev:2004cm} and to tensor products of more complicated representations, see e.g. \cite{Basile:2014wua}. Higher-spin algebra in AdS${}_4$ was constructed as a candidate symmetry algebra underlying massless higher-spin theories \cite{Fradkin:1986ka}. It was later realised that it can be alternatively defined as an algebra of endomorphisms of the singleton representation and extended to any dimension \cite{Eastwood:2002su}, see also
\cite{Vasiliev:2003ev}. 
For various extensions of these results and further references, see \cite{Iazeolla:2008ix,Govil:2013uta,Joung:2014qya,Basile:2018dzi}.

\section{Perturbative approach to interactions in gauge theories}
\label{section:7}

In the previous sections we studied the consistency conditions for \emph{free} theories in the Minkowski and  the AdS spaces, focusing mainly on massless fields. We found that formulation of massless dynamics in the manifestly Lorentz covariant approach requires gauge invariance. In this section we switch our attention to theories of \emph{interacting} massless fields. These are also required to be gauge theories, which is the key constraint a consistent interacting massless theory has to satisfy. 
Below, we will start by discussing the structures underlying interacting gauge theories and then consider the associated constraints expanded in the coupling constant. 
Except for minor rearrangements and extensions, the content of  this section  follows \cite{Berends:1984rq}. Discussions on global symmetries can be found in \cite{Joung:2013nma}.

\subsection{Preliminary remarks}

As we already mentioned, formulations of massless higher-spin theories in terms of tensor fields -- such as in the Fronsdal theory -- inevitably require gauge invariance. Gauge invariance is necessary for the reason that tensor fields have more components than needed to describe the correct physical degrees of freedom, so the excessive components have to be factored out via gauge invariance.
 When going to interactions we are still interested in making only physical degrees of freedom interact, so the non-linear action should remain gauge invariant. Besides deforming the action at non-linear level, we may also need to deform the free theory gauge transformations with non-linear corrections. This happens, for example, for the Yang-Mills theory and for General Relativity and we should allow for this possibility when studying gauge theories in general.
  
\emph{Gauge invariance of the action}
\begin{equation}
\label{27oct3}
\delta^\xi S=0
\end{equation}
\emph{is the key requirement that we are going to impose}. This consistency condition will be expanded in powers of the coupling constant and solved order by order. The approach towards  construction of theories based on solving general consistency conditions is often referred to as bootstrap, so our perturbative approach to interactions in gauge theories can be regarded as a  bootstrap procedure. Note, however, that unlike in the $S$-matrix bootstrap or in the conformal bootstrap, we assume that the theory has an action and the consistency conditions are formulated in its terms. The perturbative procedure that we are going to review now is often referred to as \emph{the Noether procedure}.

Every gauge invariant theory \emph{automatically} has \emph{gauge symmetries} that \emph{close}. This means that the commutator of infinitesimal gauge transformations is again a gauge transformation
\begin{equation}
\label{28oct1}
[\delta^{\xi_1},\delta^{\xi_2}]\varphi= \delta^{\xi'}\varphi,
\end{equation}
up to, possibly, on-shell trivial terms. In (\ref{28oct1}) $\xi'$, clearly, depends on gauge parameters $\xi_1$ and $\xi_2$. Besides that, in general, it may also depend on fields. Despite the gauge symmetry closure is a consequence of the action being gauge invariant, usually, the former condition is much easier to analyze.

A yet simpler consequence of (\ref{27oct3}) comes from the fact that 
any symmetry transformations are maps from the field space to itself labelled by gauge parameters. A combination of symmetry transformations is just a composition of the associated maps,  therefore, \emph{the action of symmetries is always associative}. This, in turn, entails \emph{the Jacobi identity} for commutators
\begin{equation}
\label{29oct1}
\sum_{{\rm cylic}}[\delta^{\xi_1},[\delta^{\xi_2},\delta^{\xi_3}]]\varphi=0.
\end{equation}
This condition is always automatically satisfied if (\ref{28oct1}) holds, but (\ref{29oct1}) is easier to deal with in practice.

Instead of considering the constraints following from \emph{gauge} invariance of the action, one can consider analogous constraints imposed by invariance with respect to \emph{global} symmetries. In the present context global symmetries are understood as a subsector of gauge symmetries, which leave the vacuum solution -- the Minkowski  space (or the (A)dS space) with other fields vanishing -- invariant. Particular examples of global symmetries are given by internal $su(N)$ global transformations and Poincare transformations,  which are associated with the Yang-Mills theory and gravity respectively. 
 The global symmetry constraints are weaker than those following from gauge invariance, but are still very stringent and, at the same time, are often much easier to solve.

In the reminder of this section we will expand the aforementioned constraints in the coupling constant and discuss various subtleties related to the ensuing consistency conditions. 
In particular, we will find out that locality and field redefinitions play an important role in the Noether procedure. 

\subsection{Gauge invariance of the action}

The expansion of the non-linear action in powers of the coupling constant reads
\begin{equation}
\label{27oct1}
S=S_2+gS_3+g^2 S_4+\dots.
\end{equation}
Here $S_2$ refers to the quadratic part of the action, the Fronsdal action. The term $S_3$ is cubic in fields, $S_4$ is quartic and so on. In a similar manner we can expand the gauge transformation
\begin{equation}
\label{27oct2}
\delta^\xi\varphi = \delta^\xi_0\varphi+g \delta^\xi_1\varphi+g^2 \delta^\xi_2\varphi+\dots.
\end{equation}
Here $\delta_0$ refers to the gauge transformation in the Fronsdal theory, while $\delta_1$ and $\delta_2$ are corrections to it, which are linear and quadratic in $\varphi$ respectively. Without loss of generality, one can keep gauge transformations linear in $\xi$, because non-linear terms can be eliminated by redefinitions of the gauge parameter.

Next, we require that the complete non-linear action is gauge invariant (\ref{27oct3}).
 This condition can be expanded in powers of the coupling constant, leading to a series of constraints,  each of which has to be satisfied separately. This leads to 
 \begin{equation}
 \label{27oct4}
 \begin{split}
 \delta^\xi_0S_2 &=0, \\
  \delta^\xi_1S_2+\delta^\xi_0S_3& =0, \\
    \delta^\xi_2S_2+\delta^\xi_1S_3+  \delta^\xi_0S_4& =0
 \end{split}
 \end{equation}
and so on. 

The first condition in (\ref{27oct4}) is satisfied trivially, because Fronsdal's action is gauge invariant under Fronsdal's gauge symmetry.

As for the second condition, one quickly notices that the second term on the left-hand side is trivial on the free shell
\begin{equation}
\label{27oct5}
\delta^\xi_1S_2=\delta^\xi_1\varphi \frac{\delta S_2}{\delta\varphi} \approx 0.
\end{equation}
Here we use $\approx$ to denote equality that holds once free equations of motion are taken into account. The free equation of motion is just
\begin{equation}
\label{27oct6}
\frac{\delta S_2}{\delta\varphi} \approx 0,
\end{equation}
which explains (\ref{27oct5}).
Accordingly, the second equation in (\ref{27oct4}) is equivalent to
\begin{equation}
\label{27oct7}
\delta^\xi_0S_3=\delta^\xi_0\varphi\frac{\delta S_3}{\delta\varphi}\approx 0.
\end{equation}
In this form the consistency condition boils down to the constraint imposed on a single unknown, the cubic vertex $S_3$. Once $S_3$ is known, one plugs it into the second equation in (\ref{27oct4}), which allows one to recover $\delta_1$.

Upon carrying out this analysis, the following subtleties should be taken into account. First is that among possible interactions $S_3$ there are some \emph{fake} ones, that can be \emph{obtained by field redefinitions}
\begin{equation}
\label{27oct8x0}
\varphi \to \varphi+g f(\varphi)
\end{equation}
 of the quadratic action. For example, starting from 
\begin{equation}
\label{27oct8}
S_2= \frac{1}{2}\int d^dx \big(  \varphi_1\Box \varphi_1 + \varphi_2\Box \varphi_2 + \varphi_3\Box \varphi_3  \big),
\end{equation}
 field redefinition 
\begin{equation}
\label{27oct9}
\varphi_1 \to \varphi_1+g \varphi_2 \varphi_3
\end{equation}
 produces a fake cubic interaction
\begin{equation}
\label{27oct10}
S^{\rm f} = S_3^{\rm f}+\dots, \qquad S^{\rm f}_3 = \int d^dx \varphi_2 \varphi_3\Box \varphi_1,
\end{equation}
where we omitted higher-order terms. Despite (\ref{27oct10}) looks as a cubic interaction vertex just by virtue of being cubic in fields, the theory (\ref{27oct8}), (\ref{27oct10}) is actually free, which can be seen by going to the original field frame. Since, we are interested in constructing genuinely interacting theories, we need a machinery to systematically detect fake interactions and factor them out. 

A systematic approach to diagnose fake interactions is based on their property, which can be easily seen from the example above.
Namely, fake interaction  vanish when lower-order equations of motion are taken into account. In the example above $S_3^{\rm f}\approx 0$ since $\Box \varphi_1 \approx 0$, the latter being the equation of motion in the free theory.
It is not hard to show that whenever two derivatives in the cubic vertex are contracted with each other, by employing integration by parts one can rewrite this vertex in the form in which $\Box$'s act on one of the fields, so such vertices give fake interactions in massless theories. This gives an easy way to spot fake cubic interactions and remove them. This argument can be straightforwardly extended to higher-order vertices.

Another subtlety refers to a step at which one extracts $\delta_1$ once $S_3$ is known, employing the second equation in (\ref{27oct4}). This equation has the form of an inhomogeneous linear equation in $\delta_1$, therefore, its general solution is given as a sum of a particular solution of (\ref{27oct4}) and of the general solution of the homogeneous equation 
\begin{equation}
\label{27oct5x1}
(\delta^\xi_1)^{\rm h}S_2=(\delta^\xi_1)^{\rm h}\varphi \frac{\delta S_2}{\delta\varphi} = 0.
\end{equation}
Considering that the quadratic action is invariant with respect to undeformed transformations, the general solution of (\ref{27oct5x1}) for $(\delta^\xi_1)^{\rm h}\varphi$ has the form $\delta_0^{\tau}\varphi$ with $\tau$ linear in $\varphi$.
 These solutions correspond to the ambiguity in redefinitions of gauge parameters
\begin{equation}
\label{27oct11}
\xi \to \xi + g f(\xi,\varphi),
\end{equation}
where at this order $f$ is linear in both arguments. Accordingly, terms of this form in $\delta^\xi \varphi$ can be dropped without loss of generality.

Before finishing the order $g$ analysis, let us note that the relevant consistency condition is linear in deformations $S_3$ and $\delta_1$. This means that any linear combination of solutions for $S_3$ and $\delta_1$ is still a solution. In the field theory terms this implies that \emph{at the cubic order all consistent cubic vertices can come with arbitrary coupling constants}. 

 At the next order, $g^2$, we are lead to consider the last equation in (\ref{27oct4}). Again, it features two unknowns: $\delta_2$ and $S_4$. As for the order-$g$ analysis, one notices that $\delta_2S_2\approx 0$, which allows one to obtain a condition on $S_4$ and cubic coupling constants. Once $S_4$ is known, $\delta_2$ can be extracted. 

By following this procedure one can, in principle, reconstruct the complete nonlinear theory order by order. 

\subsection{Closure of gauge transformations}

As mentioned, a simple \emph{consequence of} the existence of a \emph{gauge invariant action} (\ref{27oct3}) is that the associated \emph{gauge transformations close}, (\ref{28oct1}), with a parameter $\xi'$, which is, in general, field-dependent
\begin{equation}
\label{28oct2}
\xi'\equiv [\xi_1,\xi_2]=C(\varphi,\xi_1,\xi_2).
\end{equation}
In other words, consistent gauge theories require the existence of some bracket $[\xi_1,\xi_2]$ for gauge parameters. From this perspective, $C$ plays a role of the structure constant, which, in general, can be field-dependent. Gauge theories with non-vanishing $C$ are referred to 
as \emph{non-Abelian}.

In fact, a more general possibility exist. For a set of fields $\varphi^i$ with $i$ labelling different fields in the set,  one may have
\begin{equation}
\label{28oct2x1}
[\delta^{\xi_1},\delta^{\xi_2}]\varphi^i= \delta^{\xi'}\varphi^i+\delta^Q\varphi^i,
\qquad
\delta^Q\varphi^i\equiv Q^{i,j}(\varphi,\xi_1,\xi_2)\frac{\delta L}{\delta \varphi^j}
\end{equation}
with $Q^{i,j}=-Q^{j,i}$. Variation of the $\delta^Q$ type is a trivial symmetry of the action. Indeed, 
\begin{equation}
\label{28oct2x2}
\delta^Q L = Q^{i,j}(\varphi,\xi_1,\xi_2)\frac{\delta L}{\delta \varphi^i}\frac{\delta L}{\delta \varphi^j}=0.
\end{equation}
We will not consider the possibility of transformation closure as in (\ref{28oct2x1}) in what follows.

In (\ref{28oct2}) we encountered a field-dependent gauge parameter $\xi'$. In the following section, when discussing the Jacobi identity, we will need to deal with the commutators of such parameters. To prepare ourselves for that analysis, we compute here the commutator of field-dependent gauge parameters. To highlight that these parameters depend on fields, we introduce a new notation $\pi (x,\varphi(x))$. Whenever dependence of parameters on fields will be irrelevant, we will replace $\pi$  with $\xi$ back.

Let the gauge transformation be defined by function $T$
\begin{equation}
\label{28oct3}
\varphi \to \varphi' =\varphi + \delta^\pi \varphi =\varphi+T(\varphi,\pi(x,\varphi(x))).
\end{equation}
As mentioned above, without loss of generality,  we can assume that $T$ is linear in $\pi$.
Commuting two such transformations we obtain
\begin{equation}
\label{28oct4}
\begin{split}
[\delta^{\pi_1},\delta^{\pi_2}]\varphi &=\frac{\delta T(\varphi,\pi_1)}{\delta \varphi}T(\varphi,\pi_2)-
\frac{\delta T(\varphi,\pi_2)}{\delta \varphi}T(\varphi,\pi_1)
\\
&+T(\varphi,\frac{\delta \pi_1}{\delta \varphi}T(\varphi,\pi_2))-
T(\varphi,\frac{\delta \pi_2}{\delta \varphi}T(\varphi,\pi_1)).
\end{split}
\end{equation}
To derive (\ref{28oct4}) linearity of $T$  on the gauge parameter was used. 

\begin{zadacha}
Derive this. Be sure that when computing, say, $\delta^1 \delta^2\varphi$ you apply $\delta^1$ to $\delta^2\varphi$ and not substitute $\delta^1\varphi$ instead of $\varphi$ in $\delta^2\varphi$, because otherwise you will get a wrong sign in (\ref{28oct4}).
\end{zadacha}

Focusing  on field-independent parameters,  -- then the last line in (\ref{28oct4}) drops out -- we find
\begin{equation}
\label{28oct5}
\frac{\delta T(\varphi,\xi_1)}{\delta \varphi}T(\varphi,\xi_2)-
\frac{\delta T(\varphi,\xi_2)}{\delta \varphi}T(\varphi,\xi_1)
=T(\varphi,C(\varphi,\xi_1,\xi_2)).
\end{equation}
This gives a  non-trivial constraint we were after: for general $T$ the left-hand side of (\ref{28oct5}) cannot be presented in the form of the right-hand side with some $C$. It is  worth noting that redefinitions (\ref{27oct11}) affect $C$.

Returning to field-dependent parameters, we employ (\ref{28oct5}) to eliminate the first two terms on the right-hand side of (\ref{28oct4}). We, thus, find the commutator in the field-dependent case
\begin{equation}
\label{28oct6}
[\delta^{\pi_1},\delta^{\pi_2}]\varphi \equiv \delta^{[\pi_1,\pi_2]}\varphi 
\end{equation}
where we defined
\begin{equation}
\label{28oct7}
[\pi_1,\pi_2]\equiv C(\varphi,\pi_1,\pi_2)+\frac{\delta \pi_1}{\delta \varphi}T(\varphi,\pi_2)
-\frac{\delta \pi_2}{\delta \varphi}T(\varphi,\pi_1).
\end{equation}

Having settled with the generalities of the algebra closure, let us analyze the associated constraints perturbatively.
As before, we expand the gauge transformations in powers of fields
\begin{equation}
\label{29oct4}
T(\varphi,\xi)=T_0(\varphi,\xi)+g T_1(\varphi,\xi)+ g^2 T_2(\varphi,\xi)+\dots.
\end{equation}
The leading non-vanishing term on the left-hand side of (\ref{28oct5}) is
\begin{equation}
\label{29oct5}
g\frac{\delta T_1(\varphi,\xi_1)}{\delta \varphi}T_0(\varphi,\xi_2)-
g\frac{\delta T_1(\varphi,\xi_2)}{\delta \varphi}T_0(\varphi,\xi_1)
\end{equation}
and this expression is field-independent.
Hence, the leading term on the right-hand side of (\ref{28oct5})  should also be of order $g$ in the coupling constant and it should also be field-independent.
This implies that the perturbative expansion of the structure constants goes as 
\begin{equation}
\label{29oct6}
C(\varphi,\xi_1,\xi_2)= gC_0(\varphi,\xi_1,\xi_2)+g^2C_1(\varphi,\xi_1,\xi_2)+\dots.
\end{equation}
As usual, here the lower index refers to the homogeneity degree of a function in fields. 

In these terms, the lowest order constraint that follows from 
(\ref{28oct5}) reads
\begin{equation}
\label{29oct7}
{ T_1(T_0(\varphi,\xi_2),\xi_1)}-
{ T_1(T_0(\varphi,\xi_1),\xi_2)}=
 T_0(\varphi, C_0(\varphi,\xi_1,\xi_2)).
\end{equation}
To simplify the left-hand side, we used that $T_1$ is linear in fields.

At the next order, $g^2$, we find
\begin{equation}
\label{29oct8}
\begin{split}
{ T_1(T_1(\varphi,\xi_2),\xi_1)}-
{ T_1(T_1(\varphi,\xi_1),\xi_2)}
+\frac{\delta T_2(\varphi,\xi_1)}{\delta \varphi}T_0(\varphi,\xi_2)-
\frac{\delta T_2(\varphi,\xi_2)}{\delta \varphi}T_0(\varphi,\xi_1)\\
=T_0(\varphi,C_1(\varphi,\xi_1,\xi_2))+T_1(\varphi,C_0(\varphi,\xi_1,\xi_2)).
\end{split}
\end{equation}
In a similar way one proceeds to higher orders.

\subsection{The Jacobi identity for gauge transformations}

As mentioned before, \emph{for any symmetry that closes the Jacobi identity} (\ref{29oct1}) {is trivially satisfied}. 
However, when rewritten in terms of the gauge algebra bracket, namely, as
\begin{equation}
\label{29oct2}
\sum_{{\rm cylic}}[\xi_1,[\xi_2,\xi_3]]=0
\end{equation}
which is then expressed in terms of the associated  structure constants $C$, it presents a \emph{non-trivial constraint}
\begin{equation}
\label{29oct3}
\sum_{{\rm cylic}} \left(C\left(\varphi, \xi_1,C(\varphi,\xi_2,\xi_3)\right)-\frac{\delta C(\varphi,\xi_2,\xi_3)}{\delta \varphi}T(\varphi,\xi_1)\right)=0.
\end{equation}
To derive it, we used (\ref{28oct7}).
Equation (\ref{29oct3}) can be regarded as a generalised version of the Jacobi identity for the structure constants $C$, which is corrected due to $C$ being field-dependent.

\begin{zadacha}
Derive (\ref{29oct3}).
\end{zadacha}

Proceeding to the perturbative analysis, it is not hard to see that the leading order contribution to (\ref{29oct3}) is of order $g^2$. At this order one finds
\begin{equation}
\label{29oct9}
\sum_{{\rm cylic}} \left(C_0\left(\varphi, \xi_1,C_0(\varphi,\xi_2,\xi_3)\right)-{ C_1(T_0(\varphi,\xi_1),\xi_2,\xi_3)}\right)=0.
\end{equation}
Contributions at higher orders in $g$ can be obtained analogously.

\subsection{Global symmetries}
\label{sec:7.4}

Finally, we would like to focus on a small sector of gauge parameters associated with \emph{global symmetries}. As we will see now, constraints from invariance under global symmetries are much simpler than those resulting from complete gauge invariance. At the same time, these are strong enough to rule out many potential gauge theories.  

Namely, we focus our attention on gauge parameters $\hat\xi$ that satisfy
\begin{equation}
\label{29oct10}
\delta^{\hat \xi}_0\varphi=0.
\end{equation}
One sometimes calls transformations associated with $\hat \xi$ $\;$''symmetries of the vacuum'' as these leave the vacuum solution $\varphi=0$ invariant.  For the Yang-Mills theory and gravity such parameters generate global internal symmetries and the Poincare algebra respectively, which are the global symmetries of these theories. Accordingly, we will also refer to the algebra generated by $\hat \xi$, which will be constructed below for general gauge theories, as the global symmetry algebra. 

For the Fronsdal theory in the Minkowski space (\ref{29oct10})  gives
\begin{equation}
\label{29oct11}
\partial^a\hat\xi^{a(s-1)}=0,
\end{equation}
which implies that $\hat\xi$ is a \emph{traceless Killing tensor} for the Minkowski metric.  
When put in the form of a generating function with an auxiliary variable $u$ (\ref{29oct11}) gives
\begin{equation}
\label{29oct11x1}
u^\mu \frac{\partial}{\partial x^\mu}\hat \xi (u,x)=0,
\end{equation}
which means that  $\hat \xi$ defines a two-row Young diagram with $s-1$ indices in the upper row. Moreover, tracelessness on the indices of the upper row implies that this tensor is traceless on any pair of indices. In a similar manner, one can show that the solution to (\ref{29oct10}) for the Fronsdal theory in the AdS space is given by the AdS space traceless Killing tensors.

\begin{zadacha}
Show that if a tensor is traceless with respect to contractions on indices in the first row of the Young diagram, then it is traceless on any pair of indices.
\end{zadacha}

Despite $\hat \xi$ does not act on fields in the free theory, in general, it does so at  full non-linear level
\begin{equation}
\label{29oct12}
\delta^{\hat \xi}\varphi = g \delta^{\hat \xi}_1\varphi+g^2 \delta^{\hat \xi}_2\varphi+\dots.
\end{equation}
The first term in this expansion defines linear maps $\delta_1^{\hat \xi}$  in the field space labelled with parameters $\hat\xi$. Below we will study the properties of this map using the machinery developed above.

First, we consider the gauge algebra closure condition at the leading order (\ref{29oct7}). Taking into account (\ref{29oct10}), we find 
\begin{equation}
\label{29oct13}
 T_0(\varphi, C_0(\varphi,\hat\xi_1,\hat\xi_2))=0.
\end{equation}
It implies that 
  the bracket defined by $C_0$ maps Killing tensors to Killing tensors.
  At the next order, the algebra closure condition (\ref{29oct8}) restricted to Killing tensors leads to
\begin{equation}
\label{29oct14}
\begin{split}
{ T_1(T_1(\varphi,\hat\xi_2),\hat\xi_1)}-
{ T_1(T_1(\varphi,\hat\xi_1),\hat\xi_2)}
=T_0(\varphi,C_1(\varphi,\hat\xi_1,\hat\xi_2))+T_1(\varphi,C_0(\varphi,\hat\xi_1,\hat\xi_2)).
\end{split}
\end{equation}
Taking into account that the first term on the right-hand side of (\ref{29oct14}) is pure gauge in free theory, one concludes that \emph{$T_1$ realize a representation of the bracket $C_0$ on the physical degrees of freedom}\footnote{Strictly speaking, this argument shows that the representation is realised on the off-shell fields quotiented by pure gauge degrees of freedom. When quotienting out pure gauge degrees of freedom, \emph{off-shell} gauge transformations and \emph{off-shell} gauge parameters are used.} of the theory we are dealing with. Then, the Jacobi identity (\ref{29oct9}) for Killing tensors gives
\begin{equation}
\label{29oct15}
\sum_{{\rm cylic}} C_0\left(\varphi, \hat\xi_1,C_0(\varphi,\hat\xi_2,\hat\xi_3)\right)=0
\end{equation}
that is \emph{$C_0$ restricted to Killing tensors defines the commutator of a usual Lie algebra}. 

It is also instructive to consider constraints (\ref{27oct4}) and their consequences for Killing tensors. The second equation gives
 \begin{equation}
 \label{29oct16}
  \delta^{\hat \xi}_1S_2=0.
 \end{equation}
It implies that \emph{the free action is invariant with respect to the global symmetries}. The same refers to free equations of motion. Note that if global symmetries mix fields of different spins, (\ref{29oct16})  relates prefactors of the associated free actions. Already such a simple and basic requirement that these coefficients are non-singular and of the same sign  constrains admissible global symmetry algebras. 

Similarly, one finds that 
\begin{equation}
 \label{29oct17}
   \delta^{\hat \xi}_1S_3\approx 0,
 \end{equation}
so the cubic action is invariant with respect to global symmetries up to terms that vanish on-free-shell.  Equivalently, cubic amplitudes are invariant with respect to the global symmetries. In a similar manner one can show that \emph{the tree-level $S$-matrix of a consistent gauge theory is invariant with respect to the global symmetry transformations}. Amplitudes in gauge theories will be discussed in more detail in section \ref{sec:9}.

Let us summarise our findings on global symmetries in gauge theories. Firstly, there should be a Lie algebra realised on the vector space, spanned by the direct sum of traceless Killing tensors. Secondly, it is necessary that this algebra has a representation, with the representation space being  the direct sum of off-shell gauge fields modulo pure  gauge degrees of freedom. Thirdly, it is required that the free action as well as the tree-level $S$-matrix is invariant with respect to the global symmetries. 

What is remarkable about global symmetries is that the associated constraints are very simple and relatively easy to access. For example, the structure constants $C_0$ for the global Lie algebra can be extracted already at order $g$, which is the first non-trivial order of the deformation procedure. Besides that, unlike the general formula (\ref{29oct9}), the Jacobi identity for Killing tensors  (\ref{29oct15}) involves  $C_0$ only and thus closes at this order. At the same time, it turns out that the global symmetry constraints are rather stringent. In particular, as we will discuss below, in the higher-spin case even satisfying the Jacobi identity non-trivially is a problem. Besides that, there are well-known theorems, that constrain possible symmetries of the $S$-matrix -- which will be also discussed below -- and, thereby, rule out the associated gauge theories. Both the simplicity and the power of the global symmetry constraints make them an indispensable tool in the analysis of gauge theories. 

Finally, we remark that  considering the way we arrived to global symmetry constraints, these  may appear as artefacts of the approach, based on tensors fields. Indeed, from the point of view of the consistency of the free theory, $\hat\xi$ plays no role: it is introduced as a natural partner of other components of $\xi$, the role of the latter being to eliminate the unphysical degrees of freedom, while $\hat\xi$ itself does not act on fields at free level at all. From this perspective, the presence of $\hat\xi$ is not required by the representation theory considerations, but it is, rather, a consequence of the approach that we used to embed the associated representations into fields. Alternatively, one can, for instance, consider a free theory in the light-cone gauge, see section \ref{sec:13}. Then both $\xi$ and $\hat\xi$ are absent from the outset, thus, if constraints analogous to the global symmetry ones are also present in the light-cone gauge approach, their justification should be very different from that in the covariant formalism.

\section{Applications of the deformation procedure}
\label{sec:8}

To illustrate the general procedure we reviewed in the previous section,  we will consider examples of self-interacting theories of spin-1 and spin-3 fields and then we will give the classification of cubic higher-spin vertices.

\subsection{Derivation of the Yang-Mills theory}

In this section we apply the general analysis that we just presented to the case of spin-1 self-interactions. In particular, we will construct the Yang-Mills theory relying just on general principles of the consistency of interactions of gauge fields. This analysis also shows that the Yang-Mills theory is the unique interacting theory of spin-1 fields with no more than two derivatives\footnote{Strictly speaking, there is a possibility that the first non-trivial vertex is quartic or higher order in fields, which we do not consider.}.

\subsubsection{The leading order analysis}

We start from the free theory of massless spin-1 field
\begin{equation}
\label{YM1}
S_2=-\int d^dx F^{\mu\nu}F_{\mu\nu}, \qquad F_{\mu\nu}=\frac{1}{2}(\partial_{\mu}A_{\nu}-\partial_{\nu}A_{\mu}),
\end{equation}
which is gauge invariant with respect to
 \begin{equation}
\label{YM3}
\delta_0A_{\mu}=\partial_{\mu}\xi.
\end{equation}

A cubic vertex then involves three Lorentz indices carried by fields, which, to maintain Lorentz invariance, should be either contracted with each other, or with derivatives acting on these fields. This implies that we should have an odd number of derivatives in a vertex. Obviously,  if the number of derivatives is higher than three, derivatives should be contracted with each other. As was discussed before, the associated vertices give fake interactions. Therefore, we conclude that there are only two classes of possible non-trivial cubic interactions of spin-1 gauge fields: those with 1 and those with 3 derivatives respectively. 

The next observation is that, due to the fact that gauge transformation (\ref{YM3}) contains a single term with a fixed number of derivatives, condition (\ref{27oct7}) should be satisfied separately for 1- and for  3-derivative vertices. 

Our main goal is to reconstruct the Yang-Mills theory, still, for completeness, we briefly mention
the results of the application of the Noether procedure  to the 3-derivative case. The only non-trivial cubic vertex of this type is
\begin{equation}
\label{YM4}
S^{BI}_3=\int d^dxf_{abc} F^{a\mu\nu}F^{b}{}_{\mu\lambda}F^{c}{}_{\nu}{}^{\lambda},
\end{equation}
where we added internal indices and $f_{abc}$ is a totally antisymmetric coupling constant. Internal indices were added to make this vertex non-trivial. Indeed, it is not hard to see that $F^{3}$ in (\ref{YM4}) is totally antisymmetric in color indices, so for a single spin-1 field this vertex vanishes. This also explains why $f_{abc}$ is totally antisymmetric: $f$ with other symmetries produce the vanishing  contraction with $F^3$.
Since the  vertex (\ref{YM4}) is constructed out of the field strength, it is identically gauge invariant with respect to transformations of the free theory. Therefore, it does not require any deformation of gauge transformations or completion with higher order vertices.
Speaking differently, vertex (\ref{YM4})  together with the free action (\ref{YM1}) presents a consistent theory. The Born-Infeld theory is a famous example of a theory, that involves only vertices built out of the field strength. Due to that, in the higher-spin literature one often refers to this type of vertices as \emph{the Born-Infeld type}. 
Vertices of this type are also referred to as \emph{non-minimal}, to contrast them with minimal interactions obtained by replacing partial derivatives with covariant ones, as well as to emphasise that they have more derivatives than minimal interactions.

We now focus our attention on the 1-derivative case. We will immediately consider the coloured case, because, similarly to (\ref{YM4}),  we will not be able to construct any consistent interactions with a single field.   Accordingly, we start from a set of fields $A_\mu{}^a$, with $a$ being an internal index, each of them having the Maxwell-type action and gauge symmetries
\begin{equation}
\label{YM5}
S_2=-\int d^dx F_{a}{}^{\mu\nu}F^{a}{}_{\mu\nu},
\qquad 
F^a{}_{\mu\nu}=\frac{1}{2}(\partial_{\mu}A^a_{\nu}-\partial_{\nu}A^a_{\mu}),
\qquad
\delta_0A^a_{\mu}=\partial_{\mu}\xi^a.
\end{equation}
Here and in what follows internal indices are raised and lowered with $\delta_{ab}$.

Next, we make the most general ansatz for a cubic vertex with one derivative, which leads to
\begin{equation}
\label{YM6}
S_3=\int d^dx (f_{abc}A_{\mu}^a\partial^{\mu}A^b_{\nu}A^{c\nu}+
h_{abc}\partial^{\mu}A_{\mu}^aA_{\nu}^bA^{c\nu}).
\end{equation}
Taking into account the possibility to integrate by parts, we can always achieve 
\begin{equation}
\label{YM7}
f_{abc}=-f_{acb}.
\end{equation}
Besides that, we notice that $A_{\nu}^bA^{c\nu}$ is symmetric in $b$ and $c$. This means that $h_{a[bc]}$ does not contribute to the action, which, in turn, implies that without loss of generality one can require
\begin{equation}
\label{YM8}
h_{abc}=h_{acb}.
\end{equation}

We now proceed with the requirement that this vertex is gauge invariant on the free shell, (\ref{27oct7}). After a somewhat lengthy computation, we find that
\begin{equation}
\label{2nov1}
\begin{split}
\delta_0 S_3 &\approx \int d^dx \Big[ f_{abc}\big( -\xi^a \partial_\nu \partial^\mu A_\mu^b A^{c\nu}+\xi^b \partial^\mu \partial_\nu A_\mu^a A^{c\nu}+\xi^b\partial^\mu \partial_\nu A_\mu^c A^{a\nu}\\
&\qquad \qquad \qquad \quad+2\xi^b\partial_\nu A_\mu^a \partial^\mu A^{\nu c}+\xi^b \partial^\mu A_{\mu}^a\partial_\nu A^{c\nu}
+\xi^b A_\mu^a \partial^\mu \partial_\nu A^{c\nu}\Big)\\
&+2h_{abc}\Big(\xi^a \partial_\nu \partial^\mu A_\mu^b A^{c\nu}+\xi^a \partial_\mu A_\nu^b \partial^\mu A^{c\nu}-\xi^b\partial_\nu\partial^\mu A_\mu^b A^{c\nu}-\xi^b \partial^\mu A_\mu^a\partial_\nu A^{c\nu}\big) \Big].
\end{split}
\end{equation}
To achieve this, we used the symmetry properties (\ref{YM7}), (\ref{YM8}), integration by parts and the free equations of motion
\begin{equation}
\label{2nov2}
\Box A_\mu^a \approx \partial^\nu\partial_\mu A_\nu^a
\end{equation}
to eliminate d'Alembertians. For (\ref{27oct7}) to be satisfied, we need to set the prefactor of $\xi$ to zero identically (we cannot use equations of motion any more as this will reintroduce d'Alembertians). 

\begin{zadacha}
Reproduce (\ref{2nov1}).
\end{zadacha}

In (\ref{2nov1}) there are several types of terms, in which derivatives and Lorentz indices are contracted in different ways. Each such group of terms should cancel separately. 
Focusing on a term of a particular type -- those that involve $\partial_\mu A_\nu^b \partial^\mu A^{c\nu}$ -- one notices that it appears only as
\begin{equation}
\label{2nov3}
\xi^a h_{abc} \partial_\mu A_\nu^b \partial^\mu A^{c\nu}.
\end{equation}
To make sure that it is zero, we need to require 
\begin{equation}
\label{2nov4}
h_{abc}=-h_{acb},
\end{equation}
which together with (\ref{YM8}) leads to 
\begin{equation}
\label{2nov5}
h_{abc}=0.
\end{equation}
Therefore, the $h$ term in (\ref{YM6}) does not lead to consistent interactions.

Similarly, having set $h$ to zero and focusing on, say,  terms that involve $\partial^\mu A_{\mu}^a\partial_\nu A^{c\nu}$ we find that these appear only as
\begin{equation}
\label{2nov6}
\xi^b f_{abc}\partial^\mu A_{\mu}^a\partial_\nu A^{c\nu}.
\end{equation}
To make it vanish, we should impose
\begin{equation}
\label{2nov7}
f_{abc}=-f_{cba}.
\end{equation}
Together with (\ref{YM7}) it entails that $f$ is totally antisymmetric 
\begin{equation}
\label{2nov8}
f_{[abc]}=f_{abc}.
\end{equation}

This implies that non-trivial 1-derivative interactions  can be constructed once one has at least three massless spin-1 fields, otherwise, $f$ is vanishing. 
It can be also seen that with (\ref{2nov5}) and (\ref{2nov8}) satisfied all terms in (\ref{2nov1}) cancel out. Thus, the resulting cubic vertex is consistent at least at the leading order in the coupling constant.

Next, we switch to the reconstruction of the first deformation to the gauge transformation. To this end, we need to solve
\begin{equation}
\label{2nov9}
\delta_1 S_2+\delta_0 S_3=0.
\end{equation}
Evaluating each term explicitly, we find
\begin{equation}
\label{2nov10}
\delta_0 S_3 =-\int d^d xg\xi^a f_{abc}(\Box A^b_\nu-\partial_\nu \partial^\mu A_\mu^b)A^{c\nu}
\end{equation}
and
\begin{equation}
\label{2nov11}
\delta_1 S_2 = \int d^dx \delta_1 A^{a\nu}(\Box A_{a\nu}-\partial^\mu \partial_\nu A_{a\mu}).
\end{equation}
Then, (\ref{2nov9}) leads to 
\begin{equation}
\label{2nov12}
\delta_1 A_\mu^a =- f^{a}{}_{bc}\xi^b A^c_\mu.
\end{equation}
Thus, we managed to reconstruct the action and gauge transformations of the Yang-Mills theory to the leading order in interactions.

\begin{zadacha}
Reproduce (\ref{2nov10}).
\end{zadacha}

\subsubsection{The second order analysis}

To proceed further, in practice, it is more convenient to move to the analysis of the symmetry closure and the Jacobi identity for gauge transformations, see below. The resulting constraints are easier and may already indicate that the deformation procedure is obstructed or, on the contrary, can be stopped at some order in non-linearities.
 However, we find it instructive to demonstrate that the deformation procedure at the level of gauge invariance of the action can be completed on its own.
 
Focusing on  the invariance of the action, at the next order we impose
\begin{equation}
\label{2nov13}
\delta_0 S_4+\delta_1 S_3 +\delta_2 S_2=0.
\end{equation}
Since, $\delta_1 S_3\ne 0$, either $S_4$ or $\delta_2$ or both of them have to be non-vanishing. One can notice, however, that $\delta_1S_3$ has one derivative, while for local $\delta_2$  variation $\delta_2 S_2$ has at least two derivatives due to the fact that the free action is of the second order. This implies that it is only $\delta_0 S_4$ that can compensate for non-vanishing $\delta_1S_3$ in (\ref{2nov13}), while we can safely set 
\begin{equation}
\label{2nov13x1}
\delta_2=0.
\end{equation}

Then, due to the fact that $\delta_0$ already carries one derivative, $S_4$ should not involve any derivatives. Thus, we make a general ansatz for $S_4$ in the form
\begin{equation}
\label{2nov14}
S_4 =  \int d^dx e_{abcd}A_\mu^a A^{b\mu} A_\nu^c A^{d\nu},
\end{equation}
where $e$ is a yet undetermined set of parameters. By evaluating $\delta_0 S_4$ and $\delta_1 S_3$ explicitly, after a tedious, but straightforward computation that we do not detail here, we find that (\ref{2nov13}) can be fulfilled only when the Jacobi identity for $f$ is satisfied
\begin{equation}
\label{YM13}
f^a{}_{bc}f^{c}{}_{de}+f^a{}_{dc}f^{c}{}_{eb}+f^a{}_{ec}f^{c}{}_{bd}=0.
\end{equation}
Moreover, one then finds
\begin{equation}
\label{2nov15}
e_{abcd}=-\frac{1}{4}f^{e}{}_{ac}f_{ebd}.
\end{equation}

Finally, at  order $g^3$ the consistency condition leads to
\begin{equation}
\label{2nov16}
\delta_0 S_5 +\delta_1 S_4+\delta_2 S_3+ \delta_3 S_2=0.
\end{equation}
By checking 
\begin{equation}
\label{2nov17}
\delta_1 S_4=0,
\end{equation}
we find that it is consistent to set all $S_i$ with $i>4$ and $\delta_j$ with $j>1$ to zero. In other words, the deformation procedure can be terminated with 
\begin{equation}
\label{2nov18}
S=S_2+gS_3+g^2S_4, \qquad \delta =\delta_0+g\delta_1.
\end{equation}
This gives the Yang-Mills theory and the associated gauge symmetries. 

\subsubsection{Transformations closure and the Jacobi identity}

As was mentioned, once the first deformation of gauge transformations is known, it may be useful to consider the constraint
resulting from the requirement of their closure instead of analysing invariance of the action.  At the first non-linear order one has to impose
(\ref{29oct7}). Evaluating the left-hand side explicitly with $T_0$ and $T_1$ given in (\ref{YM5}), (\ref{2nov12}), we find
\begin{equation}
\label{2nov18x1}
{ T_1(T_0(\varphi,\xi_2),\xi_1)}-
{ T_1(T_0(\varphi,\xi_1),\xi_2)}=
-\partial_\mu \big(f^{a}{}_{bc}\xi_1^b \xi_2^c \big).
\end{equation}
Considering that the gauge transformation in the free theory is  given by the gradient, (\ref{29oct7}) can, indeed, be solved and
\begin{equation}
\label{2nov19}
C_0^a(\varphi,\xi_1,\xi_2)=-f^{a}{}_{bc}\xi_1^b \xi_2^c.
\end{equation}

Next, we look at the consistency condition at the next order (\ref{29oct8}).
To start, we check whether we can get away without introducing further deformations, that is with $T_2=0$ and $C_1=0$.
Then, (\ref{29oct8}) reduces to
\begin{equation}
\label{2nov20}
\begin{split}
{ T_1(T_1(\varphi,\xi_2),\xi_1)}-
{ T_1(T_1(\varphi,\xi_1),\xi_2)}
=T_1(\varphi,C_0(\varphi,\xi_1,\xi_2)).
\end{split}
\end{equation}
In a given case, (\ref{2nov20}) leads to the Jacobi identity (\ref{YM13}). Therefore, gauge transformations can, indeed, be truncated at order $T_1$, which together with $T_0$ reproduces the gauge transformations of the Yang-Mills theory. Besides that, we found that the structure constants $C$ are field-independent and $f$'s satisfy the Jacobi identity. These results are consistent with what was found before. At the same time, as one can see, at  a technical level, the analysis of the closure of gauge symmetries is simpler than the full-fledged analysis of the gauge invariance of the action.

Finally, we can also study this problem from the perspective of the Jacobi identity for gauge symmetries. Once $C_0$ is available, (\ref{2nov19}), we consider the first order condition (\ref{29oct9}) and check whether we can get away without introducing $C_1$. In our case we find
\begin{equation}
\label{2nov21}
\sum_{{\rm cylic}} C_0\left(\varphi, \xi_1,C_0(\varphi,\xi_2,\xi_3)\right)=
\sum_{{\rm cylic}}  f^a{}_{bc}\xi_1^b f^c{}_{de}\xi_2^d\xi_3^e,
\end{equation}
which, again, vanishes, provided the Jacobi identity for $f$ (\ref{YM13}) is fulfilled. Therefore, from this perspective, we find again, that deformations of $C$ can be stopped at the first non-trivial order $C_0$ in a yet simpler way.

With the example of the Yang-Mills theory we, hopefully, managed to illustrate some of the general features of the perturbative approach towards the construction of gauge theories. In particular, eventually one is interested in the construction of the gauge invariant action, which can be done directly by imposing gauge invariance order by order.  However, it may be instructive to consider the consequences of the gauge invariance of the action -- the closure of gauge transformations and the requirement that the commutator of gauge transformations satisfies the Jacobi identity. These latter conditions are easier to analyse and this analysis may already indicate that some deformations of the action cannot lead to a consistent theory or, on the contrary, that a consistent deformation exists.  It is worth keeping in mind, however, that the existence of a consistent gauge algebra does not guarantee that an action with this gauge symmetry actually exists, see e.g. \cite{Boulanger:2006gr,Bekaert:2010hp}.

Another scenario for which the analysis of symmetries can be fruitful is that of General Relativity. Namely, due to the fact that the Einstein-Hilbert action involves the inverse metric, once expanded around the Minkowski background in small fields, it leads to an infinite series of vertices. Because of that, perturbative reconstruction of the action in closed form  does not seem feasible in this case. 
At the same time, gauge transformations of General Relativity -- diffeomorphisms --  truncate at the first non-trivial order of the deformation procedure, therefore, these can be accessed perturbatively. The knowledge of the exact symmetry of the theory can then be used to reconstruct the associated action.

\subsection{Spin-3 self-interactions}
\label{sec:8.2}

In this section we review the results of the application of the deformation procedure to the case of spin-3 self-interactions. As the analysis becomes far more technical  than in the spin-1 case, we will just present the key features of it. Some additional technical details can be found in \cite{Berends:1984rq}.

As usual, one starts from the Fronsdal action for a spin-3 field. At the next step one considers all possible cubic deformations $S_3$ and requires that the action is gauge invariant on the free shell, $\delta_0S_3\approx 0$. One finds that in general dimension there are four independent non-fake cubic vertices satisfying this constraint. These have from 3 to 9 derivatives, with the number of derivatives being odd\footnote{We will derive this and analogous statements for general spins in the next section.}.
The highest-derivative vertex is analogous to the Born-Infeld one in the spin-1 case (\ref{YM4}) -- it is identically gauge invariant with respect to the gauge symmetry of the free theory, hence, at this order the deformation procedure can be stopped. One is more interested, however, in cases which lead to non-trivial deformations of gauge transformations and the associated algebra. 
The three-derivative vertex is of this type and we will discuss it below. 

As in the spin-1 case,  non-trivial  cubic interactions can be constructed only when we are dealing with several species of the fields. From the leading order analysis one finds that the three-derivative vertex is of the schematic form 
\begin{equation}
\label{3nov1}
S_3=\int d^dx f_{abc}\varphi_{\mu(3)}^a\partial^{\mu}\partial^{\mu}\partial^{\mu}\varphi^b_{\nu(3)}\varphi^{c\nu(3)}+\dots,
\end{equation}
where $f$ is totally antisymmetric similarly to the Yang-Mills case.
Substituting it into the second equation in (\ref{27oct4}), we find the associated deformation of the gauge transformation
\begin{equation}
\label{3nov2}
\delta_1\varphi^a_{\mu(3)}=- f^{abc}\partial^\nu \partial^\nu\varphi^b_{\mu(3)}\xi^c_{\nu(2)}+\dots.
\end{equation}
Next, we require that the algebra closes to the leading order (\ref{29oct7}) and, thus, extract $C_0$
\begin{equation}
\label{3nov3}
C^a_{0|\mu(2)}(\varphi, \xi_1,\xi_2)=-f^{abc}\partial^{\nu}\partial^{\nu}\xi^b_{1|\mu(2)}\xi^c_{2|\nu(2)}+\dots.
\end{equation}
Then, we look at the requirement of the closure of the gauge algebra at the next order (\ref{29oct8}). The problematic term comes from 
\begin{equation}
\label{3nov4}
\begin{split}
&{ T_1(T_1(\varphi,\xi_2),\xi_1)}-
{ T_1(T_1(\varphi,\xi_1),\xi_2)}
\\
& \qquad\qquad=(f^{abc}f_{bd}{}^{e}-f^{abe}f_{bd}{}^{c})\partial^\nu\partial^\nu\partial^\rho\partial^\rho\varphi^d_{\mu(3)}\xi_{1|\rho(2)}\xi_{2|\nu(2)}+\dots.
\end{split}
\end{equation}
The prefactor made of $f$ can only vanish if $f$ is zero\footnote{Indeed, if 
\begin{equation}
\label{3nov5}
f^{abc}f_{bd}{}^{e}=f^{abe}f_{bd}{}^{c}
\end{equation}
then
\begin{equation}
\label{3nov5x1}
f^{abc}f_{bd}{}^{e}=f^{abe}f_{bd}{}^{c}=-f^{eba}f_{bd}{}^{c}=-f^{ebc}f_{bd}{}^{a}=f^{cbe}f_{bd}{}^{a}=
f^{cba}f_{bd}{}^{e}=-f^{abc}f_{bd}{}^{e},
\end{equation}
which entails
\begin{equation}
\label{3nov6}
f^{abc}f_{bd}{}^{e}=0.
\end{equation}
Taking $u$ and $v$ any auxiliary vectors, we can construct 
\begin{equation}
\label{3nov7}
F^a\equiv f^{abc}u_bv_c.
\end{equation}
Then (\ref{3nov6}) implies $F^2=0$. Recalling that the internal space metric is $\delta_{ab}$, this entails $F=0$. Vanishing of $F$ for any $u$ and $v$ implies that $f$ equals zero.}. Hence, if we want to have  non-trivial interactions, the term that we wrote out on the right hand side of (\ref{3nov4}) is inevitably present.

A closer analysis shows that this contribution cannot be cancelled by any other terms present in (\ref{29oct8}). To be more specific, (\ref{3nov2}) has at most two derivatives acting on $\varphi$, so
\begin{equation}
\label{3nov8}
T_1(\varphi,C_0(\varphi,\xi_1,\xi_2))
\end{equation}
cannot cancel (\ref{3nov4}). Analogously, 
\begin{equation}
\label{3nov9}
\frac{\delta T_2(\varphi,\xi_1)}{\delta \varphi}T_0(\varphi,\xi_2)-
\frac{\delta T_2(\varphi,\xi_2)}{\delta \varphi}T_0(\varphi,\xi_1)
\end{equation}
has at least one derivative that acts on $\xi$ (it comes from $T_0$), so it cannot cancel (\ref{3nov4}) neither. Finally, in
\begin{equation}
\label{3nov10}
T_0(\varphi,C_1(\varphi,\xi_1,\xi_2))
\end{equation}
one $\mu$ index, that comes from $T_0$, is carried by a derivative, not by $\varphi$ as in (\ref{3nov4}), so this term is not helpful neither. 

We, thus, found that the problematic term (\ref{3nov4}) cannot be cancelled, therefore, (\ref{29oct8}) cannot be satisfied.
This gives  one of the many no-go results for higher-spin interactions. In the given case we can conclude that unlike in the Yang-Mills theory, cubic vertex (\ref{3nov1}) cannot be completed to a consistent interacting theory, unless we include other cubic vertices, which may involve fields of other spins. A more complete analysis \cite{Bekaert:2010hp} shows that this problem cannot be cured even if fields of other spins are introduced. In other words, at least within the manifestly covariant framework that we used here,  the vertex (\ref{3nov1}) cannot be present in a consistent higher-spin theory.

\subsection{Classification of cubic higher-spin vertices}
\label{sec:8.3}

Here we present another result of the application of the deformation procedure of section \ref{section:7} -- a complete classification of cubic vertices for massless higher-spin fields. 
Our presentation of this result closely follows section 2 of \cite{Joung:2011ww}.

To simplify the analysis, we will consider fields in the traceless transverse gauge. Using the approach of generating functions, this implies that for the master field $\varphi(x,u)$ one has
\begin{equation}
\label{3nov11}
\partial^2_u\varphi =0, \qquad \partial_u\cdot\partial_x\varphi =0, \qquad \Box \varphi \approx 0.
\end{equation}
Gauge transformations then read
\begin{equation}
\label{3nov12}
\delta_0 \varphi = u\cdot \partial_x \xi
\end{equation}
with $\xi$ satisfying the same constraints as $\varphi$ (\ref{3nov11}). As we know from section \ref{sec:3.1}, this correctly describes massless degrees of freedom.

We will now proceed following the standard steps of the deformation procedure of section \ref{section:7} in this simplified setup. It is not immediately clear that the result of this deformation procedure is equivalent to the one that starts from the Fronsdal action, though, it can be shown, that it is, indeed, the case. Relevant references can be found at the end of this section.

The general  ansatz for cubic interactions reads
\begin{equation}
\label{3nov13}
\begin{split}
S_3 &= \frac{1}{3!}\int d^dx B_{a_1a_2a_3}(\partial_{x_1},\partial_{x_2},\partial_{x_3},\partial_{u_1},\partial_{u_2},\partial_{u_3})\\
& \qquad \qquad\qquad\qquad\qquad\qquad\qquad\varphi^{a_1}(x_1,u_1)\varphi^{a_2}(x_2,u_2)\varphi^{a_3}(x_3,u_3)\Big|_{x_i=x,u_i=0}.
\end{split}
\end{equation}
Here we used the standard trick: fields are initially put at different points, so that distinguishing derivatives acting on different fields is made easier, while after derivatives are evaluated, one sets $x_i=x$, thus, putting all fields to the same point, as required for a local interaction. Besides that, the action (\ref{3nov13}) is evaluated at $u_i=0$, which implies that all tensor indices carried by fields are contracted by means of operators inside $B$.

Note that due to the Bose symmetry
\begin{equation}
\label{19aprn1}
\varphi^{a_i}(x_i,u_i)\varphi^{a_j}(x_j,u_j)= \varphi^{a_j}(x_j,u_j)\varphi^{a_i}(x_i,u_i)
\end{equation}
only the totally symmetric in permutations part of $B$ 
\begin{equation}
\label{5nov0}
B(2,1,3)=B(1,3,2)=B(1,2,3)
\end{equation}
contributes to (\ref{3nov13}).

Next, Lorentz invariance implies that all Lorentz indices of $\partial_x$ and $\partial_u$ are contracted covariantly. Part of these contractions vanish due to conditions (\ref{3nov11}). Besides that, contractions $\partial_{x_i}\cdot \partial_{x_j}$ lead to fake interactions, as was explained before. 
The first non-trivial type of contractions is   $\partial_{u_i}\cdot \partial_{x_j}$. Fixing the freedom of integrating $\partial_{x_j}$ in such expressions by parts, we are lead to the following independent contractions
\begin{equation}
\label{5nov1}
{ Y}_1\equiv \partial_{u_1}\cdot \partial_{x_{23}}, \qquad 
{ Y}_2\equiv \partial_{u_2}\cdot \partial_{x_{31}}, \qquad 
{ Y}_3\equiv \partial_{u_3}\cdot \partial_{x_{12}}, 
\end{equation}
where
\begin{equation}
\label{5nov2}
\partial_{x_{ij}}\equiv \partial_{x_i}-\partial_{x_j}.
\end{equation}
Furthermore, there are three independent contractions of tensor indices carried by fields
\begin{equation}
\label{5nov3}
{ Z}_1\equiv \partial_{u_2}\cdot \partial_{u_3}, \qquad 
{ Z}_2\equiv \partial_{u_3}\cdot \partial_{u_1}, \qquad 
{ Z}_3\equiv \partial_{u_1}\cdot \partial_{u_2}. 
\end{equation}
Thus, keeping only non-trivial contractions we are lead to a more refined ansatz
\begin{equation}
\label{5nov4}
B_{a_1a_2a_3}(\partial_{x_1},\partial_{x_2},\partial_{x_3},\partial_{u_1},\partial_{u_2},\partial_{u_3})=
C_{a_1a_2a_3}({ Y}_1,{ Y}_2,{ Y}_3,{ Z}_1,{ Z}_2,{ Z}_3).
\end{equation}

Next, we require that gauge variation of (\ref{3nov13}) vanishes on the free shell up to total derivatives. In these notations variation of the action goes as follows. First, we proceed as usual
\begin{equation}
\label{5nov5}
\begin{split}
\delta_0S_3 &= \frac{1}{2}\int d^dx C_{a_1a_2a_3}({ Y}_i,{ Z}_i)
\delta\varphi^{a_1}(x_1,u_1)\varphi^{a_2}(x_2,u_2)\varphi^{a_3}(x_3,u_3)\Big|_{x_i=x,u_i=0}\\
&=\frac{1}{2}\int d^dx C_{a_1a_2a_3}({ Y}_i,{ Z}_i)
u_1\cdot\partial_{x_1}\xi^{a_1}(x_1,u_1)\varphi^{a_2}(x_2,u_2)\varphi^{a_3}(x_3,u_3)\Big|_{x_i=x,u_i=0}.
\end{split}
\end{equation}
Then we commute $u_1\cdot \partial_{x_1}$ through $C$, thus getting
\begin{equation}
\label{5nov6}
\begin{split}
\delta_0S_3 &=\frac{1}{2}\int d^dx [C_{a_1a_2a_3},
u_1\cdot\partial_{x_1}]\xi^{a_1}(x_1,u_1)\varphi^{a_2}(x_2,u_2)\varphi^{a_3}(x_3,u_3)\Big|_{x_i=x,u_i=0}\\
&+
\frac{1}{2}\int d^dx u_1\cdot\partial_{x_1} C_{a_1a_2a_3}
\xi^{a_1}(x_1,u_1)\varphi^{a_2}(x_2,u_2)\varphi^{a_3}(x_3,u_3)\Big|_{x_i=x,u_i=0}.
\end{split}
\end{equation}
The last term vanishes. Indeed, its integrand is a polynomial expression in $u_i$ and by $u_i=0$ we are instructed to take its constant part. However, $u_1\cdot \partial_{x_1}$ contributes one power of $u_1$,
so $C$ has one more $\partial_{u_1}$'s than $\xi$ has $u_1$'s, which means that $C$ annihilates $\xi$.
Therefore, in (\ref{5nov6}) only the first term is non-trivial. By requiring that it vanishes on the free shell for any 
 $\varphi$ and $\xi$, we find
\begin{equation}
\label{5nov7}
[C_{a_1a_2a_3},u_1\cdot \partial_{x_1}]\approx 0.
\end{equation}
Employing (\ref{5nov0}), we also obtain
\begin{equation}
\label{5nov8}
[C_{a_1a_2a_3},u_2\cdot \partial_{x_2}]\approx 0, \qquad 
[C_{a_1a_2a_3},u_3\cdot \partial_{x_3}]\approx 0.
\end{equation}

Explicit computation gives\footnote{There is no summation over $j$ here.}
\begin{equation}
\label{6nov1}
[{ Y}_i,u_j \cdot \partial_{x_j}]\approx 0, \qquad 
[{ Z}_i,u_j\cdot\partial_{x_j}]=-\frac{1}{2}\epsilon_{ijk}{ Y}_k.
\end{equation}
As usual, these equalities are understood up to total derivatives. 
Then, 
\begin{equation}
\label{6nov2}
[C_{a_1a_2a_3}({ Y}_i,{ Z}_i),u_j\cdot \partial_{x_j}]\approx[{ Z}_i,u_j\cdot\partial_{x_j}]\partial_{{ Z}_i}C_{a_1a_2a_3}({ Y}_i,{ Z}_i),
\end{equation}
so (\ref{5nov7}), (\ref{5nov8}) amounts to
\begin{equation}
\label{6nov3}
({ Y}_1\partial_{{ Z}_2} - { Y}_2\partial_{{ Z}_1})C_{a_1a_2a_3}({ Y}_i,{ Z}_i)=0
\end{equation}
and its cyclic permutations in fields' indices. Equation (\ref{6nov3}) implies that ${ Z}_1$ and ${ Z}_2$ can enter $C$ only via ${ Z}_1{ Y}_1+{ Z}_2{ Y}_2$. Together with the remaining equations we find that the general solution for $C$ is given by
\begin{equation}
\label{6nov4}
C_{a_1a_2a_3}({ Y}_i,{ Z}_i) = 
K_{a_1a_2a_3}({ Y}_i,{ G}),
\end{equation}
where 
\begin{equation}
\label{6nov5}
\begin{split}
{ G}&\equiv { Z}_1{ Y}_1+{ Z}_2{ Y}_2+{ Z}_3{ Y}_3\\
&\qquad \qquad=
\partial_{u_2}\cdot \partial_{u_3} \partial_{u_1}\cdot \partial_{x_{23}}+
\partial_{u_3}\cdot \partial_{u_1} \partial_{u_2}\cdot \partial_{x_{31}}+
\partial_{u_1}\cdot \partial_{u_2} \partial_{u_3}\cdot \partial_{x_{12}}.
\end{split}
\end{equation}

To obtain vertices from (\ref{6nov4}) we need to focus on $K$'s, which are polynomial in $Y$ and $G$.
This is a natural requirement since for local theories the number of the space-time derivatives should be a non-negative integer. Similarly, the number of index contractions should be a non-negative integer.
Then, expanding $K$ into the power series in $Y$ and $G$, we obtain the classification of cubic interactions for massless fields in the form
\begin{equation}
\label{6nov6}
K_{a_1a_2a_3}({ Y}_i,{ G})=
\sum_{s_1,s_2,s_3=0}^\infty\sum_{n=0}^{{\rm min}\{s_1,s_2,s_3\}} g^{s_1s_2s_3n}_{a_1a_2a_3}{ G}^n { Y}_1^{s_1-n}
{ Y}_2^{s_2-n}
{ Y}_3^{s_3-n},
\end{equation}
where $g$ are independent coupling constants.
Here spins $s_i$ were identified from  the number of $\partial_{u_i}$ that $K$ features.

Before ending this section, let us make two corollaries of the classification (\ref{6nov6}).
Since each $Y$ and $G$ contribute one space-time derivative, from (\ref{6nov6}) one finds that the total number of  derivatives entering the vertex is 
\begin{equation}
\label{6nov7}
N(\partial_x)=s_1+s_2+s_3-2n
\end{equation}
and it takes values in the range
\begin{equation}
\label{6nov8}
s_1+s_2+s_3-2{\rm min}\{s_1,s_2,s_3\} \le N(\partial_x)\le s_1+s_2+s_3.
\end{equation}

Besides that, symmetry (\ref{5nov0}) entails 
\begin{equation}
\label{6nov9}
g^{s_2s_1s_3n}_{a_2a_1a_3} = g^{s_1s_3s_2n}_{a_1a_3a_2} =(-1)^{s_1+s_2+s_3} g^{s_1s_2s_3n}_{a_1a_2a_3}.
\end{equation}
In particular, in the colorless case this implies that coupling constants $g^{s_2s_1s_3n}$ are totally symmetric functions of  spins for the total spin even, while they are totally antisymmetric in spins otherwise. In the latter case, clearly, the coupling constants vanish when at least two of the fields have the same spin.
 We have already encountered this phenomenon for spin-1 and spin-3 self-interactions. 

\subsection{Remarks}

Below we will  make a couple of  general remarks that can be derived from the results we presented, as well as mention some issues that we did not discuss in depth.

\paragraph{Peculiarities of higher-spin interactions.} As one can see from (\ref{6nov8}), whenever a cubic vertex involves spin higher than 2, it inevitably has more than two derivatives. In particular, one can see that there are no two-derivative interactions of the type $s-s-2$. In lower-spin theories, interactions of this type appear when one \emph{minimally couples} a theory to gravity, more specifically, they come from covariantization of the two-derivative kinetic terms. 
In other words, in higher-spin theories minimal interactions with gravity are not possible. 
A somewhat different version of this statement is known as the Aragone-Deser no-go theorem \cite{Aragone:1979hx}.
As we will see later, this is a feature of the Fronsdal approach, while in the light-cone gauge minimal interactions with gravity do exist.
The minimal gravitational coupling of higher-spin gauge fields can be also constructed in the spinor-helicity formalism for amplitudes as well as using the twistor-space techniques. 

\paragraph{The role of locality.} In the perturbative approach to interactions a significant role is played by locality. To illustrate this, consider, the second equation in (\ref{27oct4}). To solve it, we used the fact that $\delta_1 S_2\approx 0$. This is true, however, only if $\delta_1\varphi$ does not cancel the wave operator arising from $\delta S_2/\delta \varphi$. Of course, this may only happen if $\delta_1\varphi$ has non-local $\Box^{-1}$-type factors. The presence of factors of this type trivialises the deformation procedure. In particular, in the example above one can make any $S_3$ consistent by properly fitting non-local $\delta_1\varphi$. 
Similarly, our no-go conclusions for spin-3 self-interactions were based on the fact that different terms in the consistency condition had derivatives distributed differently. If one was able to undo derivatives, which amounts to using non-local operations, this argument would have not been applicable any more. 
A rigorous discussion of this issue can be found in \cite{Barnich:1993vg}.

\paragraph{Appearance of obstructions.} Above we found that it is not a problem to construct consistent cubic vertices for massless higher-spin fields. All these are independent and give actions, which are consistent to the leading order in the coupling constant $g$. Difficulties with the construction of consistent higher-spin interactions typically appear at the next, quadratic in $g$, order.
This was the case for the spin-3 self-interaction that we considered. Below we will discuss the no-go theorems and the obstructions to interactions highlighted by these theorems first occur at order $g^2$.

\paragraph{Parity-odd interactions.} One more possibility to construct vertices, which are manifestly invariant with respect to the  Lorentz algebra transformations, is to use the totally antisymmetric tensor $\epsilon_{a[d]}$.
The Levi-Civita tensor
changes the sign under time reversal  and under spatial parity transformations. Accordingly, vertices that involve  $\epsilon_{a[d]}$ are 
called parity-odd or parity-breaking, or parity-violating see e.g.  \cite{Boulanger:2005br,Boulanger:2013zza,Conde:2016izb,Kessel:2018ugi} for construction of such vertices.

\paragraph{Dimension-dependent identities.} In dimension $d$ one has
\begin{equation}
\label{8nov1}
\delta^{[\nu_1}{}_{\mu_1}\dots \delta^{\nu_{d+1}]}{}_{\mu_{d+1}}=0
\end{equation}
due to the fact that indices take only $d$ values. This leads to trivialisation of certain interactions in lower dimensions. In particular, in four dimensions
\begin{equation}
\label{8nov2}
{ G}{ Y}_1{ Y}_2{ Y}_3 \propto 
\delta^{[\nu_1}{}_{\mu_1}\dots \delta^{\nu_{5}]}{}_{\mu_{5}} \frac{\partial}{\partial u_{1\mu_1}} \frac{\partial}{\partial u_{2\mu_2}} \frac{\partial}{\partial u_{3\mu_3}} \frac{\partial}{\partial x_{1\mu_4}} \frac{\partial}{\partial x_{2\mu_5}}
\frac{\partial}{\partial u_{1}^{\nu_1}}
\frac{\partial}{\partial u_{2}^{\nu_2}}
\frac{\partial}{\partial u_{3}^{\nu_3}}
\frac{\partial}{\partial x_{1}^{\nu_4}}
\frac{\partial}{\partial x_{2}^{\nu_5}},
\end{equation}
 hence, it vanishes. This entails that in (\ref{6nov6}) only vertices with $n=0$ and $n={\rm min}\{s_1,s_2,s_3 \}$ are non-trivial in $d=4$.

\paragraph{Higher derivatives and instabilities.} Since higher-spin interactions involve higher derivatives already at cubic order, one may wonder whether higher-spin theories suffer from Ostrogradsky's type instabilities. We recall that these instabilities may occur in theories which posses second and higher time derivatives of the dynamical variables. It is worth keeping in mind, however, that in the same way as we dealt with fake interactions, second time derivatives can be always converted to second spatial derivatives via a non-covariant field redefinition. In this way, one avoids higher time derivatives and the associated consequences of Ostrogradsky's theorem.

Moreover, this theorem does not apply directly to gauge theories, see \cite{Woodard:2015zca} for review. One way to proceed with the analysis of stability in this case is to fix a gauge, which can be done in different ways. In section \ref{sec:13} we will construct cubic vertices for massless higher-spin fields in the light-cone gauge. These do not involve any time derivatives, which is a peculiar feature of interactions in the light-cone formalism.  Therefore, at least, a naive application of Ostrogradsky's theorem does not signal any instabilities. 
For further discussions on instability in the higher-spin context we refer the reader to \cite{Kaparulin:2014vpa}.

\subsection{Further reading}

When discussing the general perturbative approach towards interactions in gauge theories in section \ref{section:7} we skipped some subtleties and extensions, such
as the possibility of the gauge algebra to close on-shell as well as the case of reducible gauge symmetries. These can be treated in a similar manner or streamlined in the BRST framework, see \cite{Barnich:1994db,Barnich:1994mt,LucenaGomez:2015eei} for review on the BRST approach in this context and for applications to low-spin examples. 

The literature on the applications of the Noether procedure and, in particular, on cubic vertices in higher-spin theories   is very extensive, so we will  only quote some results. 
 In the Fronsdal formalism the complete cubic vertices -- those that do not rely on the traceless-transverse gauge -- were given in \cite{Manvelyan:2010jr,Sagnotti:2010at}. Cubic vertices were also constructed within other off-shell frameworks for massless symmetric fields, see e.g. \cite{Buchbinder:2006eq,Fotopoulos:2010ay,Francia:2016weg} and references therein. For certain mixed symmetry fields cubic vertices were constructed in \cite{Boulanger:2011qt}. These techniques were also used to show that interactions of certain mixed-symmetry fields do not exist \cite{Bekaert:2002uh,Bekaert:2004dz} and that multi-graviton theories are inconsistent \cite{Boulanger:2000rq}.

\section{Consistency of massless tree-level scattering}
\label{sec:9}

The main consistency condition that we explored in section \ref{section:7} was the requirement that the action for massless higher-spin fields remains gauge invariant at the interacting level. Once the action of the theory is known, one can compute the associated amplitudes employing the standard Feynman rules. Then, the condition of gauge invariance naturally translates into the consistency conditions -- known as the Ward identities -- on the scattering amplitudes.

In the context of the perturbative approach to interactions, studying amplitudes instead of the action can be beneficial, at least, for two reasons. Firstly, on the external lines amplitudes feature states which are on the free shell, which allows one to avoid 
 ambiguities and complications related to off-shell extensions of the theory. Secondly, amplitudes in gauge theories are invariant with respect to gauge symmetries of the free theory. Thus, when dealing with amplitudes, one does not need to worry about the deformation of the gauge symmetry, which, eventually, simplifies the analysis.

 Below we will review the amplitude approach towards interactions in gauge theories and illustrate how it works with the Yang-Mills example. Most of this section is a somewhat rearranged standard lore that can be found in books on quantum field theory and on the $S$-matrix \cite{Weinberg,Peskin:257493,schwartz2014quantum,tHooft:1973wag}.

\subsection{General consistency conditions}

 Amplitudes are functions of the scattering data, that is of labels that parametrise on-shell particles on the external lines. As these are on-shell, it is convenient to impose the transverse-traceless gauge, which brings us to (\ref{14sep1}), while the associated gauge symmetries are given by (\ref{14sep2}), (\ref{14sep3}). When dealing with amplitudes, it is more convenient to use the momentum representation. Therefore, the external lines will be represented by tensors
 \begin{equation}
\label{9nov1}
\varphi^{\mu(s)}(p),
\end{equation}
 which are traceless, on the free shell, 
 \begin{equation}
 \label{11mar1x1}
 p^2\varphi^{\mu(s)}(p)=0
 \end{equation}
 and transverse
  \begin{equation}
 \label{11mar2x1}
 p_\mu\varphi^{\mu(s)}(p)=0.
 \end{equation}
  The probability amplitude for the scattering process involving $\varphi_i$ on external lines is then a linear functional on $\varphi_i$
\begin{equation}
\label{9nov2}
\begin{split}
A_n&(\varphi_1^{\mu_1(s_1)}(p_1),\dots, \varphi_n^{\mu_n(s_n)}(p_n))
\\
&\qquad\equiv
\int d^dp_1\dots d^dp_n\varphi_1^{\mu_1(s_1)}(p_1) \dots \varphi_n^{\mu_n(s_n)}(p_n) A_{n|\mu_1(s_1),\dots,\mu_n(s_n)}(p_1,\dots,p_n)
\end{split}
\end{equation}
with values in complex numbers. In the following, we will discuss the consistency requirements for the scattering amplitude $A_{n|\mu_1(s_1),\dots,\mu_n(s_n)}(p_1,\dots,p_n)$. Moreover, we will consider only the connected tree-level contribution to $A$.

The key consistency requirement is  that the amplitude is invariant with respect to transformations from the Poincare algebra. Invariance with respect to translations implies that $A$ has a momentum-conserving delta function as a factor
\begin{equation}
\label{9nov3}
A_{n|\mu_1(s_1),\dots,\mu_n(s_n)}(p_1,\dots,p_n)=
M_{n|\mu_1(s_1),\dots,\mu_n(s_n)}(p_1,\dots,p_n)\delta^d(p_1+\dots+p_n).
\end{equation}
In turn, Lorentz invariance  can be ensured if all Lorentz indices are contracted in the manifestly covariant way.

Besides that, for gauge fields one demands that the amplitude is gauge invariant with respect to the residual symmetry  (\ref{14sep2}), (\ref{14sep3}), which acts on the external lines of the amplitude as in the \emph{free} theory\footnote{This is related to the fact that amplitudes are obtained from the correlators which, in turn, are computed in the interaction picture, in which the external lines evolve with the free Hamiltonian.}
\begin{equation}
\label{9nov4}
\delta \varphi^{\mu(s)}(p)= p^\mu \xi^{\mu(s-1)}(p).
\end{equation} 
 Then, gauge invariance of the amplitude implies
\begin{equation}
\label{9nov5}
p_{i}^{\mu_i}\xi_i^{\mu_i(s_i-1)}(p_i) M_{n|\mu_1(s_1),\dots,\mu_n(s_n)}(p_1,\dots,p_n)=0 ,\qquad \forall i, \; \xi_i.
\end{equation}
This, in turn, leads to
\begin{equation}
\label{9nov5x1}
p_{i}^{\mu_i} M_{n|\mu_1(s_1),\dots,\mu_n(s_n)}(p_1,\dots,p_n)=0 ,\qquad \forall i, 
\end{equation}
which is known as \emph{the Ward identity} for scattering amplitudes. 

\subsubsection{Feynman rules}

In order to be able to compute amplitudes from the action, one needs to know all the ingredients entering the Feynman rules: propagators and vertices. Vertices will be fixed in the course of the deformation procedure, while propagators are defined, as usual,  as the Green functions for the free equations of motion with the properly set boundary conditions. Due to the presence of the source in the equation for the propagator, it does not satisfy the free equations of motion identically and, hence, the traceless-transverse gauge cannot be imposed on the propagator.
 However, one can still impose the de Donder gauge (\ref{14sep21}), as it does not require the field to be on-shell. 

Despite we will not use massless propagators beyond well-known lower-spin cases, we will present the general result for completeness. 
The generating function for the massless higher-spin propagator in the de Donder gauge is given by
\begin{equation}
\label{10nov1}
{\cal D}_s^d(u_1,u_2;p)=-i\frac{{\cal P}_s^{d-2}(u_1,u_2)}{p^2-i\epsilon},
\end{equation}
where ${\cal P}_s^d$ is the standard traceless projector in $d$ dimensions\footnote{For a symmetric rank-$s$ tensor $\varphi^{\mu(s)}$ its traceless part is given by an expression of the form
\begin{equation}
\label{IJTP1}
\varphi_{0}^{\mu(s)}\equiv \varphi^{\mu(s)}+\alpha_1 \eta^{\mu\mu}\varphi_\nu{}^{\nu\mu(s-2)}+\alpha_2 \eta^{\mu\mu}\eta^{\mu\mu}\varphi_{\nu\rho}{}^{\nu\rho\mu(s-4)}+\dots,
\end{equation}
where coefficients $\alpha_1$, $\alpha_2$, $\dots$ are defined from the requirement that $\varphi_0$ is traceless. Clearly, the operation defined by (\ref{IJTP1})
is a projection: for $\varphi$ traceless one has $\varphi_0=\varphi$. The traceless projection (\ref{IJTP1}) can be alternatively implemented via contraction with the traceless projector ${\cal P}_s^d$
\begin{equation}
\label{IJTP2}
\varphi_{0}^{\mu(s)}\equiv ({\cal P}_s^d)^{\mu(s)}{}_{\nu(s)}\varphi^{\nu(s)} = (\delta^\mu{}_\nu\dots \delta^\mu{}_\nu + \alpha_1\eta^{\mu\mu}\eta_{\nu\nu}\delta^\mu{}_\nu\dots \delta^\mu{}_\nu+\dots ) \varphi^{\nu(s)} .
\end{equation}
Tensor ${\cal P}_s^d$ is, clearly, symmetric and traceless on each group of indices. Employing polarisation vectors $u_1$ and $u_2$ we can turn it into a generating function that satisfies (\ref{10nov2}).
}
\begin{equation}
\label{10nov2}
{\cal P}_s^d(u_1,u_2) = \frac{1}{(s!)^2}(u_1\cdot u_2)^s +\dots, \qquad
\partial_{u_1}^2{\cal P}_s^d(u_1,u_2)=
\partial_{u_2}^2{\cal P}_s^d(u_1,u_2)=0.
\end{equation}
Explicitly one has
\begin{equation}
\label{10nov3}
{\cal P}_s^d(u_1,u_2) = \frac{1}{(s!)^2}\sum_{k=0}^{[s/2]}t_{s,k}^d(u_1^2)^k (u_2^2)^k (u_1\cdot u_2)^{s-2k},
\end{equation}
where
\begin{equation}
\label{10nov4}
t^{d}_{s,k}=\frac{(-1)^k s!}{4^kk!(s-2k)!(\frac{d}{2}-1+s-k)_k}, \qquad (a)_k=\frac{\Gamma(a+k)}{\Gamma(a)}.
\end{equation}

\begin{zadacha}
Derive $t_{s,1}^d$ by requiring (\ref{10nov2}).
\end{zadacha}

For a pedagogical derivation of this result we refer the reader to \cite{Ponomarev:2016jqk}, see also \cite{Francia:2007qt,Bekaert:2009ud}. Note that the upper label of ${\cal P}$ in (\ref{10nov1}) is equal to $d-2$, not to $d$, so the propagator is in a way traceless in $d-2$ dimension. This property can be traced to the tracelessness of the tensor of the little group $so(d-2)$, which characterises the massless representation that the propagator is associated with. In the following we will only deal with  tree-level diagrams, so  ghost propagators will not be necessary.

Usually, one is interested in the $S$-matrix, to which amplitudes contribute with the factor of $i$. When computing $iA$, the Feynman rules work as follows. As we have just discussed, for the internal lines one uses the propagator (\ref{10nov1}). Vertices are then read off the action, with an extra factor of $i$, symmetry factor $n!$ for the $n$-point vertex and the Bose symmetrisation over all legs in the vertex. For more details, we refer the reader to  quantum field theory courses quoted above. For our purposes it will be natural to compute $A$ directly and at tree-level only. Then one can drop $-i$ from (\ref{10nov1}) as well as factors of $i$ contributed by vertices -- these anyway cancel out. Finally, we will not be interested in the $i\epsilon$ issues of the propagator, so $i\epsilon$ will be systematically omitted as well.

\subsection{Perturbative analysis}
\label{sec:9.2}
Now we are ready to proceed to the perturbative construction of interacting gauge theories in terms of amplitudes. We will first describe how this procedure works conceptually and then illustrate it with the example of the Yang-Mills theory.

 At the first non-trivial order we are looking for cubic vertices for which the amplitudes satisfy the Ward identity (\ref{9nov5}) with $n=3$.
In practice, this amounts  to the analysis of section \ref{sec:8.3}, performed in the Fourier space instead of the position space. Therefore, we can use our previous results and immediately proceed to the next order of the perturbative procedure.

Tree-level four-point amplitudes contain two types of contributions: exchanges in the three channels and the contact diagram
\begin{equation}
\label{9nov7}
M_4 = M^{\rm exch}_{4}+M^{\rm cont}_{4}, \qquad  
M^{\rm exch}_{4}=M^{\rm exch}_{4|s}+M^{\rm exch}_{4|t}+M^{\rm exch}_{4|u}.
\end{equation}
 Once 3-pt amplitudes are known, cubic vertices can be reconstructed only up to on-shell trivial terms, see discussions below. Fixing these in some convenient manner -- e. g. with the prescription that derivatives never get contracted with each other, as we did before -- we can compute the exchanges. 
 Then the four-point vertex is derived from the requirement that the total four-point amplitude satisfies the Ward identity.

 In a similar manner, inductively, one  proceeds to higher orders. Namely, let us assume that we have fixed all vertices up to order $n-1$. This allows us to compute all $n$-point exchange diagrams. Then, the contribution of the contact $n$-point diagram is derived from the requirement that the total $n$-point amplitude satisfies the Ward identity. This, in turn, allows us to find the $n$-point vertex.

This outlines the amplitude approach to the perturbative construction of gauge theories. Before applying it, it is worth clarifying two important points.

\subsubsection{Locality}

The language of amplitudes is very visual in demonstrating the role of locality in perturbative approaches towards the  construction of interacting gauge theories. Namely, as we mentioned in the previous section, by relaxing locality the perturbative approach to interactions trivialises in the sense that constraints imposed by gauge invariance, become, essentially, empty. We will now see the amplitude counterpart of this phenomenon.

The constraints that we considered above imply that any set of Poincare invariant $n$-point amplitudes that satisfiy the Ward identity is an equally valid one. Indeed,  at a given order, by subtracting from an  amplitude all exchange diagrams, which are fixed at lower orders of the perturbative procedure, one finds the contribution associated with a contact diagram.
The latter, in turn,  defines a vertex in the action at that order. Therefore, for any set of  amplitudes that satisfy the Ward identity, one can write a gauge invariant action, which reproduces them via the Feynman rules. 
 In reality, massless theories do not have this arbitrariness. It is removed by additionally imposing the requirement of locality. 

To see how locality affects this analysis, let us return to the four-point amplitude that we constructed above and demonstrate that, typically, it corresponds to a non-local four-point vertex.
Indeed, exchange amplitudes have poles -- $1/s$ for the s-channel exchange and similarly for other channels. At the same time, for local four-point vertices, the associated $M^{\rm cont}_{4}$ is an analytic function of the Mandelstam variables\footnote{To make this statement more precise, we should be more clear on what we mean by locality. For example, one can define that local vertices have finitely many derivatives. Then, the associated amplitudes are polynomial in the Mandelstam variables. One can consider a relaxed version of locality for which local amplitudes should not have singularities, that is these should be given by entire functions. These are two natural options 
for the analyticity properties required from local  contact amplitudes. }. This means that for local theories singularities of $M_4$ should match exactly those of exchanges. This is a crucial requirement that locality adds at this order\footnote{In the amplitude literature a somewhat different terminology is used. The property that the amplitude's singularities match those contributed by exchanges is called unitarity, while locality refers to the property that amplitudes may only have single poles, no other singularities at tree level are allowed.}.  In practice, one finds that it
 significantly reduces the number of consistent $M_4$'s and, moreover, usually, one can only solve for $M_4$  for particularly chosen $M_3$'s. A similar pattern holds for higher-point amplitudes.

\subsubsection{Field redefinitions and on-shell trivial terms} 

The perturbative procedure outlined in section \ref{sec:9.2} involved a step at which one needs to promote an amplitude for a contact diagram to the associated vertex. 
This step is ambiguous. Here we would like to clarify the nature of this ambiguity and better explore its manifestations in the context of our procedure.

Obviously, due to the fact that the external lines of the amplitude are put on shell, once a contact amplitude is known, the associated vertex can be reconstructed only up to on-shell trivial terms\footnote{We emphasise, that by on-shell trivial terms we mean those that vanish on the \emph{free} equations of motion.}. Previously, we already encountered vertices of this type. It was shown that on-shell trivial vertices give fake interactions in the sense that these can be produced by field redefinitions of the free action. Thus, despite a vertex associated with a given contact amplitude is not unique, this non-uniqueness is fully accounted for by fake interactions. One possible way to proceed with this freedom is to fix it, e.g. as we did  in section \ref{section:7}.

It is worth pointing out, that the ambiguity of adding on-shell trivial terms to a vertex at a given order via field redefinitions, results into ambiguities in vertices at higher orders, which do not simply reduce to on-shell trivial terms. This can be easily seen from simple examples. 
Still, the effect of field redefinitions is very easy to deal with using  the $S$-matrix language. Namely,  the well-known result from the $S$-matrix literature states that \emph{the complete $S$-matrix is invariant under local  field redefinitions} \cite{tHooft:1973wag}. This means that despite field redefinitions may change contributions of individual diagrams, the total amplitudes remains intact. The aforementioned property of the $S$-matrix is very convenient in the context of the perturbative deformation procedure as it automatically quotients out the field redefinition ambiguity.

In summary, we conclude that the non-uniqueness of the off-shell extension of contact amplitudes to  vertices has no physical significance in the sense that 
different choices do not result in 
 different theories.

\begin{zadacha}
Consider a theory of a  free massless scalar. Make a field redefinition $\varphi \to \varphi + g\varphi^2$. Convince yourself that the quartic vertex in this theory is not on-shell trivial. Compute the connected tree-level four-point amplitude and show that it vanishes.
\end{zadacha}

 Considering that, in the perturbative procedure  we are anyway dealing with amplitudes, which remain invariant under field redefinitions, it seems natural to change the procedure so that it does not require the action at intermediate steps and, thus, does not suffer from field redefinition ambiguities. This leads us to the so-called \emph{on-shell methods}. The basic idea behind the on-shell methods is as follows. All singularities of tree-level diagrams are poles contributed by exchanges. At these poles, the propagators in exchanges become singular, while  the external lines of the subdiagrams that these propagators connect go on-shell.
  Therefore, the residues at these poles are given by products of amplitudes for subdiagrams, resulting from cutting the singular propagators in the original diagram. In this way, the singular part of an $n$-point amplitude can be defined in terms of lower-point amplitudes. In turn, the regular part of the amplitude is adjusted so that the total amplitude satisfies the Ward identity.
   The on-shell methods is a powerful tool, which is widely used for computing amplitudes, see \cite{Elvang:2013cua} for review.

\subsection{Derivation of the Yang-Mills theory}
\label{sec:9.3}
To illustrate the above approach, let us derive the leading order amplitudes and the action of the Yang-Mills theory from general principles. 

As we mentioned before, at the 3-point amplitude level, the analysis of the Ward identity is equivalent to that of section \ref{sec:8.3}, though, carried out in the momentum space.
We will not repeat the derivation here, instead, we will check that the resulting amplitude does satisfy the Ward identity.
  
The 3-point amplitude, associated with the cubic Yang-Mills vertex -- see first term in (\ref{YM6}) -- reads
\begin{equation}
\label{15nov1x1}
\begin{split}
&A_{3|a_1a_2a_3}^{\lambda_1\lambda_2\lambda_3}(p_1,p_2,p_3) = 
\\
&\quad f_{a_1a_2a_3}
[(p_2-p_3)^{\lambda_1}\eta^{\lambda_2\lambda_3}+(p_3-p_1)^{\lambda_2}\eta^{\lambda_3\lambda_1}+(p_1-p_2)^{\lambda_3}\eta^{\lambda_1\lambda_2}]\delta^d(p_1+p_2+p_3).
\end{split}
\end{equation}
Note the extra factor of $3!$ compared to the vertex followed by the Bose symmetrisation over all six permutations of the external lines. The left-hand side of the Ward identity with respect to the first external line can be obtained from (\ref{15nov1x1}) by contracting $A$ with $p_{1|\lambda_1}$. By employing  momentum conservation, one can see that $p_1\cdot p_2$ and $p_1\cdot p_3$ vanish as a consequence of $p_i^2=0$. The remaining terms, using momentum conservation, 
can be shown to vanish due to the fact that $p_{i|\lambda_i}\varphi^{\lambda_i}=0$, see (\ref{11mar2x1})\footnote{Here $\varphi^\lambda$ is a vector potential against which the amplitude is supposed to be integrated, (\ref{9nov2}). We change notation from $A$ to $\varphi$ not to confuse it with the amplitude itself.}.
The Ward identities for other legs of the amplitude hold due to its Bose symmetry.

\begin{zadacha}
Show that (\ref{15nov1x1}), indeed, satisfies the Ward identity.
\end{zadacha}

At the next order, we are supposed to compute exchanges. The s-channel exchange reads
\begin{equation}
\label{15nov2}
\begin{split}
&A_{4|s|a_1a_2a_3a_4}^{\rm exch|\lambda_1\lambda_2\lambda_3\lambda_4} =
\\
&\quad\int d^dp_e d^dp_{e'} f_{a_1a_2a_e}[(p_2-p_e)^{\lambda_1}\eta^{\lambda_2\lambda_e}+(p_e-p_1)^{\lambda_2}\eta^{\lambda_e\lambda_1}+(p_1-p_2)^{\lambda_e}\eta^{\lambda_1\lambda_2}]\\
&\qquad\qquad\qquad\qquad\qquad\frac{\delta^{a_ea_{e'}}\eta_{\lambda_e\lambda_{e'}}}{p_e^2}\delta^d(p_e+p_{e'})\delta^d(p_1+p_2+p_e)\delta^d(p_3+p_4+p_{e'})\\
&\qquad\qquad\qquad f_{a_3a_4a_{e'}}[(p_4-p_{e'})^{\lambda_3}\eta^{\lambda_4\lambda_{e'}}+(p_{e'}-p_3)^{\lambda_4}\eta^{\lambda_{e'}\lambda_3}+(p_3-p_4)^{\lambda_{e'}}\eta^{\lambda_3\lambda_4}],
\end{split}
\end{equation}
where $e$ and $e'$ label two ends of the propagator. Internal momenta $p_e$ and $p_{e'}$ are easily integrated out due to the presence of delta functions. Contracting (\ref{15nov2})
with $p_{1|\lambda_1}$, after some manipulations, we get
\begin{equation}
\label{15nov3}
\begin{split}
\delta_1 A_{4|s}^{\rm exch}=
f_{a_1a_2}{}^{a_e}f_{a_3a_4a_e}
(2\eta^{\lambda_2\lambda_4}p_4^{\lambda_3}-2\eta^{\lambda_2\lambda_3}p_3^{\lambda_4}+(p_3-p_4)^{\lambda_2}\eta^{\lambda_3\lambda_4})\delta^d(p_1+\dots+p_4).
\end{split}
\end{equation}

\begin{zadacha}
Check this.
\end{zadacha}

Note that $p_e^2$ got cancelled in the denominator. This is a general phenomenon and it can be easily explained. Namely, exchanges involve cubic vertices, for which the associated amplitudes satisfy the Ward identity. When one leg, $e$, of such a vertex goes off-shell, it is no longer supposed to satisfy the Ward identity, though, the failure of this amplitude to satisfy the Ward identity should involve a factor $p_e^2$: otherwise, it will not go to zero when this leg returns on-shell. This $p_e^2$ factor cancels the analogous factor in the propagator. Note also that if this cancellation did not occur, one would not be able to obtain a gauge invariant four-point amplitude, without invoking non-local four-point interactions.

The four-point vertex that is capable of fixing the failure of the exchange amplitude to be gauge invariant of the form (\ref{15nov3}) should be free of derivatives. Indeed, when going from the amplitude to the Ward identity, we get an extra power of momentum, while (\ref{15nov3}) is already linear in momenta, so there should be no derivatives in the four-point vertex.
It should be also quadratic in $f$, so the general ansatz for such a vertex is
\begin{equation}
\label{15nov4}
S_4=\alpha\int d^dx f_{a_1a_2}{}^e f_{a_3a_4e}A_{\lambda_1}^{a_1}A_{\lambda_2}^{a_2}A_{\lambda_3}^{a_3}A_{\lambda_4}^{a_4}\eta^{\lambda_1\lambda_3}\eta^{\lambda_2\lambda_4}.
\end{equation}

The associated contact amplitude contains 24 terms. Here we  write out 8 of them, which have the s-channel exchange colour structure
\begin{equation}
\label{15nov5}
\begin{split}
A_{4|s}^{\rm cont}=
\; 4\alpha\; f_{a_1a_2}{}^ef_{a_3a_4e}(\eta^{\lambda_1\lambda_3}\eta^{\lambda_2\lambda_4}-\eta^{\lambda_1\lambda_4}\eta^{\lambda_2\lambda_3})\delta^d(p_1+\dots+p_4).
\end{split}
\end{equation}
Its failure to satisfy the Ward identity on the first leg reads
\begin{equation}
\label{15nov6}
\begin{split}
\delta_1A_{4|s}^{\rm cont}=
 4\alpha\; f_{a_1a_2}{}^ef_{a_3a_4e}(p^{1|\lambda_3}\eta^{\lambda_2\lambda_4}-p^{1|\lambda_4}\eta^{\lambda_2\lambda_3})\delta^d(p_1+\dots+p_4).
\end{split}
\end{equation}

The total gauge variation of the four-point amplitude is obtained by summing (\ref{15nov3}) and (\ref{15nov6}) and then adding the contributions from other channels, which can be found by replacements $2\leftrightarrow 3$ and $2\leftrightarrow 4$ in (\ref{15nov6}). Eliminating $p_1$ via momentum conservation and requiring the result to vanish, we reproduce the Jacobi identity (\ref{YM13}), as well as fix $\alpha=1/4$.

\begin{zadacha}
Check this.
\end{zadacha}

In summary, we managed to reproduce the cubic and the quartic vertices of the Yang-Mills theory with all the associated constraints, such as the Jacobi identity, from the requirement 
that three- and four-point amplitudes satisfy the Ward identity. Strictly speaking, this is not yet the end of the story, as one needs to make sure that all higher-point amplitudes satisfy the Ward identity as well. In practice, this, essentially, amounts to computing all tree-level amplitudes in the Yang-Mills theory, which is a notoriously complicated task. Thus, unlike the perturbative approach based on the gauge invariance of the action, which in the Yang-Mills case terminates after first few steps as the action is at most quartic in fields, the perturbative approach based on the analysis of amplitudes never terminates. This makes the latter approach somewhat disadvantageous compared to the former one, especially, when constructing theories which end up having finitely many terms in the action. Still, as discussed above, the key consistency conditions for massless theories occur at the order $g^2$ and these can be easily accessed using both methods.

As a preparation for the following discussion, let us note that $\delta_1A_{4}^{\rm cont}$ vanishes in the limit $p_1\to 0$, as it manifestly has a factor of $p_1$. In turn, this is not true for $\delta_1 A_{4}^{\rm exch}$, see (\ref{15nov3}). This implies that for $p_1=0$ gauge variation of exchanges should vanish separately. Focusing, say, on contributions from exchanges that involve $\eta^{\lambda_3\lambda_4}$ and trading all $p^{i\lambda_2}$ for $p^{3\lambda_2}$ (this can be done by using momentum conservation, $p_1=0$ and $p_2\cdot \varphi_2=0$), we arrive to the Jacobi identity. In other words, the Jacobi identity for the structure constants can be inferred solely from the so-called \emph{soft limit} $p_1\to 0$, in which contact contributions are irrelevant. The soft limit also plays a key role in the no-go theorem by Weinberg, which will be discussed in the next section.

\subsection{Further reading}

Here we used the framework in which the scattering data was represented by momenta and polarization tensors. In this language, four-point amplitudes involving higher-spin fields were studied e. g. in  \cite{Taronna:2017wbx,Roiban:2017iqg}, see also references therein. Besides this representation, 
there exist other approaches to amplitudes, which 
employ other ways to encode the scattering data.  In particular, the spinor-helicity formalism is especially efficient for massless scattering in $4d$, see \cite{Elvang:2013cua} and references therein. This formalism allows one to easily classify three-point spinning amplitudes as well as to see obstructions to consistent interactions at higher orders of the deformations procedure \cite{Benincasa:2007xk,Benincasa:2011pg}. Quite remarkably, the results of the cubic order analysis in the spinor-helicity formalism are not equivalent to those obtained within the framework that relies on Lorentz tensors. This issue will be  discussed later in the context of chiral higher-spin theories.

\section{No-go theorems}

In previous sections we explored the landscape of gauge theories. We did that trying to stay as general as possible: we imposed a very limited set of natural field theory requirements -- such as unitarity, absence of ghosts and locality -- and searched  all gauge theories that satisfy these requirements. As a result, we managed to construct the known gauge theories, such as the Yang-Mills theory. At the same time, we found that interactions among higher-spin gauge fields are problematic. In particular, we found that a theory of interacting gauge fields of spin-3, analogous to the Yang-Mills theory, does not exist. By proceeding in a similar manner, as a result of a rather tedious analysis, one can show that interacting higher-spin theories of the most interesting and still very general class --  those with a non-Abelian gauge algebra and with higher-spin symmetries entering more non-trivially than as internal symmetries -- do not exist in flat space. 

One can avoid this negative conclusion by arguing that, despite we tried to be as general as possible,  some of the requirements that we imposed -- implicitly or explicitly -- are still too strong -- these can be either relaxed or lifted -- and with the appropriate set of requirements interacting massless higher-spin theories do exist. Of course, one cannot relax any assumptions without control as this may lead to physical or mathematical inconsistency of the theory. Keeping this in mind, in the higher-spin context, one usually explores the following possibilities to relax the standard assumptions or to extend the setup in a wider sense: consider higher-spin fields in the AdS space, explore different sets of off-shell fields, 
discard the necessity of the Lagrangian description, relax the analyticity requirement for the $S$-matrix, etc.  As we will see later, there are indications that these changes, indeed, lead to new possibilities for interacting higher-spin theories. 

Besides the no-go result that we sketch below, there exists a wide spectrum of other no-go theorems, which use different approaches and different sets of assumptions. Getting familiar with all these results will help us  understanding 
how each assumption leads to obstructions to higher-spin interactions and, accordingly, which of these need to be relaxed in an interacting higher-spin theory. The complete list of no-go theorems that constrain interactions of higher-spin gauge fields is rather extensive. Below we will review two powerful no-go results, which played an important role historically. We will also complete the no-go argument based on the deformation approach of the previous sections. 
For a more comprehensive list of analogous results we refer the reader to \cite{Bekaert:2010hw}.

\subsection{Weinberg's no-go theorem}
\label{sec:10.1}

Probably, the first and at the same time a very powerful no-go theorem for interactions of massless higher-spin fields was found by Weinberg in \cite{Weinberg:1964ew}. A detailed discussion of the Weinberg no-go theorem is given in textbook \cite{schwartz2014quantum}. Here we will review it in a somewhat adapted version. 

For simplicity, we will focus on scalar theories, interacting with a massless higher-spin field of spin $s$ via a vertex
\begin{equation}
\label{16nov1}
S_3=g_s\int d^dx (\partial_{u_1}\cdot \partial_{x_{23}})^s \varphi(x_1,u_1)\varphi(x_2)\varphi(x_3)\Big|_{x_i=x, u_1=0}
\end{equation}
and, possibly, higher order interactions.
In section \ref{sec:8.3} we showed that (\ref{16nov1}) gives the unique consistent cubic vertex to the leading  order in interactions when the scalar field is massless, though, it is also consistent for massive scalar fields.

\begin{zadacha}
Show that (\ref{16nov1}) is the unique gauge invariant vertex  when $\varphi(x)$ is a massive field.
\end{zadacha}

Let us consider a tree level scattering process, which involves $n$ scalar fields of, possibly, different masses and one massless higher-spin field.
As discussed in the previous section, it is described by a scattering amplitude 
\begin{equation}
\label{16nov2}
M_{\mu(s)}(p_1,\dots,p_n;q),
\end{equation}
where $q$ is the momentum of the spin-$s$ field.
In the soft limit $q\to 0$ this amplitude is dominated by diagrams, in which the gauge field is attached to one of the external lines $i$ via vertex (\ref{16nov1}). This happens because the propagator that connects this vertex with the remaining part of the diagram has a pole\footnote{Here ''$\approx$'' denotes an approximate equality.} 
\begin{equation}
\label{16nov3}
\frac{1}{p_e^2 +m_i^2}= \frac{1}{(p_i+q)^2 +m_i^2}\approx\frac{1}{p^2_i+2p_i q  +m_i^2}=\frac{1}{2p_i q},
\end{equation}
where $p_e$ is the exchanged momentum and we used that $p_i^2+m_i^2=0$, since the external line is on-shell. In the limit $q\to 0$ (\ref{16nov3}) becomes infinite, while other types of diagrams remain finite. 

Then, in the soft limit, the residue of the amplitude factorises
\begin{equation}
\label{16nov4}
\begin{split}
M_{\mu(s)}(p_1,\dots,p_n;q) &\approx \sum_i g^i_s M'(p_1,\dots,p_n) \frac{1}{2p_i q}(p_i-p_e)_{\mu_1} \dots (p_i-p_e)_{\mu_s} \\
&\approx \sum_i g_s^i{M}'(p_1,\dots,p_n) \frac{2^s}{2p_i q}p_{i|\mu_1}\dots p_{i|\mu_s} ,
\end{split}
\end{equation}
where $M'$ is an amplitude for the diagram obtained from the original one by removing the singular propagator together with the cubic vertex that involves the higher-spin gauge field. 
To obtain (\ref{16nov4}) we used the explicit expression for the cubic vertex (\ref{16nov1}) and the fact that $p_e+p_i\to 0$ in the soft limit due to momentum conservation. Note that the $p_i$ argument of $M'$ in the soft limit is on-shell, which means that $M'$ is a consistent scalar $n$-point scattering amplitude.

The total amplitude should satisfy the Ward identity, that is
\begin{equation}
\label{16nov6}
q^\mu{M}_{\mu(s)}(p_1,\dots,p_n;q)=0.
\end{equation}
Employing (\ref{16nov4}) we obtain
\begin{equation}
\label{16nov7}
\begin{split}
q^\mu{ M}_{\mu(s)}(p_1,\dots,p_n;q)\approx
2^{s-1}  \sum_i g^i_s{ M}(p_1,\dots,p_n) p_{i|\mu_1}\dots p_{i|\mu_{s-1}}.
\end{split}
\end{equation}
Contributions from other diagrams are at least linear in $q$, so these do not survive in the soft limit, see section \ref{sec:9.3} for the illustration in the Yang-Mills theory.  This implies that unless the amplitude for purely scalar scattering is vanishing, we need to demand
\begin{equation}
\label{16nov8}
\sum_i g_s^i p_{i|\mu_1}\dots p_{i|\mu_{s-1}} =0.
\end{equation}
This is the formula we were aiming to derive. Let us now study its consequences for particular values of spins. 

For $s=0$ the Ward identity is irrelevant, so we start with $s=1$. Then (\ref{16nov8}) implies
\begin{equation}
\label{16nov9}
\sum_i g_1^i=0,
\end{equation}
which can be identified as the charge conservation. For $s=2$ we obtain
\begin{equation}
\label{16nov10}
\sum_i g_2^ip_{i|\mu}=0.
\end{equation}
Considering that momentum is conserved, for otherwise general momenta configurations (\ref{16nov10}) can only be fulfilled if
\begin{equation}
\label{16nov11}
g_2^i = g_2.
\end{equation}
 This condition implies that gravity couples to all fields with the same coupling constant. It can be regarded as  the $S$-matrix version of the equivalence principle. Finally, considering spins higher than 2, we find that (\ref{16nov8}) cannot be satisfied together with momentum conservation. 
 
 The Weinberg theorem presented in the current form rules out interactions of massless higher-spin fields with massive scalars via the cubic vertex (\ref{16nov1}). For massless scalars, diagrams with massless higher-spin fields exchanged also contribute to the soft limit, hence, the above analysis should be revisited.
 The conclusion of this analysis remains the same: no matter how coupling constants of vertices are adjusted, the Ward identity for the scattering amplitude cannot be satisfied. 
  This can be derived from the results of \cite{Taronna:2017wbx,Roiban:2017iqg}, see also references therein. One often summarises the Weinberg no-go theorem by saying that higher-spin particles may exist, but they cannot have couplings that survive in the limit of low energy, thus, they cannot mediate long-range interactions.

\subsection{The Coleman-Mandula theorem}
\label{sec:10.2}

The Coleman-Mandula theorem is another powerful no-go theorem which played an important historical role.
It focuses on the symmetries of the $S$-matrix and shows that once some mild assumptions are made these are severely constrained. 
Below we will state the theorem. Its proof is rather long and intricate. It also uses tools which are slightly away from the general flow of this course, so we do not present the proof of the Coleman-Mandula theorem in the main text.
 An interested reader can find it in appendix \ref{app:cm}.

The theorem is formulated as follows.
Assuming that the amplitude is analytic and non-vanishing except certain isolated points in the Mandelstam plane and that the spectrum contains finitely many particles with mass lower than any fixed value, the Coleman-Mandula theorem  proves that the $S$-matrix symmetry algebra can only be a direct product of the Poincare algebra and some internal symmetry algebra\footnote{Here, ''internal'' refers to the fact that the internal algebra generators commute with the Poincare algebra. Equivalently, one can say that internal symmetries do not act neither on momenta nor on spin labels. In particular, according to this definition the $U(1)$ associated with the Maxwell theory is regarded as internal symmetry.}. 
One also makes the standard assumption  that symmetries act on multi-particle states by acting on each single-particle state separately -- that is representations carried by multi-particle states are the tensor products of single-particle representations -- and that these leave the norm invariant in the sense that  the Lie algebra generators are represented by Hermitian operators. The proof of the theorem deals with the 2-to-2 scattering, but it applies to higher-point amplitudes as well.

\subsection{Constraints on higher-spin symmetries induced by vertices}
\label{sec:10.3}

In section \ref{sec:7.4} we found that consistency of massless higher-spin theories at the Lagrangian level requires the existence of a Lie algebra realised on  the space of Killing tensors. Moreover, the structure constants of this Lie algebra cannot be arbitrary: it should be possible to \emph{induce} them from the deformation of the action.
By this we mean that $C_0$ that define these structure constants can be obtained from (\ref{29oct7}), in which $T_1$ should be consistent with some cubic action in the sense that $T_1$ should go together with some $S_3$ and this pair should solve the second line of (\ref{27oct4}).
A systematic analysis of this problem is technically tedious, so here we only present  its qualitative conclusions in a schematic form. For more details we refer the reader to \cite{Joung:2013nma}, see also \cite{Bekaert:2010hp}.

With a slight abuse of notation, we will denote generators of spin-$s$ global symmetries -- those associated with Killing tensors $\hat\xi^{s-1}$ -- with $T_s$. Then,
the analysis shows that 
 for induced higher-spin global symmetries one has
\begin{equation}
\label{17jan12}
[T_s,T_2]=0, \qquad \forall s>2.
\end{equation}
Below we will explore the consequences of this result. More specifically, we are interested to see whether the higher-spin algebra can be non-Abelian and whether it can contain the Poincare algebra more non-trivially than a factor in the direct product. 

To this end, let us assume that higher-spin generators commute non-trivially and that their commutators produce spin-2 generators
\begin{equation}
\label{17jan13}
[T_s,T_{s'}] = T_2 + \sum_{s''>2}T_{s''}.
\end{equation}
 Then, the Jacobi identity leads to
\begin{equation}
\label{17jan14}
[[T_s,T_{s'}],T_2]=-[[T_{s'},T_2],T_s]-[[T_2,T_s],T_{s'}]=0,
\end{equation}
where the right-hand side is vanishing due to (\ref{17jan12}). Alternatively, by evaluating the left-hand side of (\ref{17jan14}) with (\ref{17jan13}), we find
\begin{equation}
\label{17jan15}
[[T_s,T_{s'}],T_2]= [T_2,T_2]+ \sum_{s''>2} [T_{s''},T_2] = [T_2,T_2].
\end{equation}
Consistency of (\ref{17jan14}) and (\ref{17jan15}) requires
\begin{equation}
\label{17jan16}
[T_2,T_2]=0.
\end{equation}

Therefore, we find that the $T_2$ generator can appear in the commutator of two higher-spin generators (\ref{17jan13}) only when global symmetries of the spin-2 field are Abelian. In this case, however, the spin-2 field cannot be identified with graviton. 
 On the contrary, if we demand that $[T_2,T_2]=T_2$, as is the case of the Poincare algebra, we find that $T_2$ cannot appear on the right-hand side of (\ref{17jan13}), which, together with (\ref{17jan12}) implies that higher-spin generators correspond to internal symmetries. 
 
 This conclusion is similar to that of the Coleman-Mandula theorem. Note, however, that it does not rely on any assumptions such as the finiteness of the spectrum below any given mass squared. At the same time, it relies on the existence of the Lagrangian, which is written in terms of the Fronsdal   fields.

 It is also worth remarking that (\ref{17jan12}) alone is not quite satisfactory. Indeed, it implies that parameters $\hat\xi^{s-1}$ of global higher-spin symmetries do not transform under the action of the Poincare algebra. This result is in stark conflict with the expectation that $\hat\xi^{s-1}$ should transform as the respective Killing tensor.

\subsection{What is not ruled out}
\label{sec:10.4}

At the end of this section, let us briefly spell out various possibilities for interacting higher-spin gauge theories, which are not ruled out by the above no-go results. 

In section \ref{sec:8} we already encountered the Born-Infeld type vertices, that is vertices constructed out of field strengths. Vertices of this type can be added at any order of the deformation procedure and lead to theories, which are gauge invariant with respect to the undeformed gauge symmetry to all orders. The Born-Infeld type higher-spin gauge theories are not ruled out by the Weinberg no-go theorem as these do not involve vertex  (\ref{16nov1}). In turn, the Born-Infeld-type interactions do not contradict the Coleman-Mandula theorem, as the associated higher-spin algebra is Abelian and, therefore, internal\footnote{We remind the reader that we use ''internal'' in the sense that the associated symmetries commute with the Poincare algebra. Abelian symmetries commute with all generators, hence, these are internal.}. The same refers to the arguments of section \ref{sec:10.3}. In a similar manner, one cannot rule out slightly more non-trivial theories: those for which the free theory gauge transformations get deformed, but the algebra still remains Abelian.
For discussions on vertices of this type see \cite{Bekaert:2010hw} and references therein.

Speaking of higher-spin gauge theories with symmetries that non-trivially extend the Poincare algebra, these appear to be severely constrained. The argument of section \ref{sec:10.3} rules out the possibility that these theories may have a local Lagrangian description in terms of Fronsdal-like fields. To avoid this conclusion, one may try to explore different off-shell descriptions of higher-spin fields. 
In particular, as we will see in section \ref{sec:13}, in the light-cone gauge approach new possibilities for higher-spin interactions occur. Another interesting illustration of the phenomenon that the existence of an interacting theory may depend on the field choice is provided by the dual formulations of gravity: while General Relativity is a classically consistent interacting theory of massless spin-2, interactions of the dual graviton are obstructed \cite{Bekaert:2002uh}.

Irrespectively of the field choice and even of the existence of the Lagrangian description, higher-spin  theories with symmetries that non-trivially extend the Poincare algebra  are still severely constrained by the Coleman-Mandula theorem. The most obvious way to avoid its conclusions is to allow an infinite spectrum of massless higher-spin fields. 
The assumption of a finite spectrum below any value of mass crucially features  several steps of the proof of the Coleman-Mandula theorem and, to the best of our knowledge, there is no generalisation of this theorem with this requirement lifted.
Still, it seems reasonable to expect that such a generalisation exists, because other no-go results do not rely on the finiteness of the spectrum.

 Another option is to relax the assumption of the Coleman-Mandula theorem on analyticity of the $S$-matrix and the requirement that it is non-vanishing almost everywhere. The $S$-matrix that violates these assumptions may still be non-trivial, e.g. if it is supported at some isolated points of the Mandelstam plane. 
However, it is  not clear how such an $S$-matrix can result from the application of the Feynman rules to the conventional Lagrangian field theories, so such a higher-spin theory should be rather exotic.

Irrespectively of the consequences for the higher-spin theory $S$-matrix and for the potential Lagrangian description,
one may wonder whether global higher-spin algebras non-trivially extending the Poincare algebra exist at all. The answer to this question is affirmative. In particular, the algebra 
\begin{equation}
\label{17jan17}
[T_2,T_2]=T_2, \qquad [T_s,T_2]=T_s,
\end{equation}
where the first commutator defines the Poincare algebra, the second one defines the appropriate transformations of the higher Killing tensors with respect to the Poincare algebra, while  other commutators are vanishing, satisfies the Jacobi identity. It is worth stressing that in (\ref{17jan17}) higher-spin generators do not generate internal symmetries, instead, they appear as a factor in a \emph{semidirect} product. For a recent discussion on the higher-spin algebras in flat space we refer the reader to \cite{Campoleoni:2021blr}.

Another possibility, which we did not cover here and which seems to be almost entirely unexplored is to consider theories that start with quartic or higher order vertices. The existing no-go results do not apply to these theories, at least, without substantial amendments.

Finally, another important option to avoid the aforementioned no-go results is to consider higher-spin gauge theories around the AdS background. 
As we will discuss in the next section, the existence of higher-spin gauge theories in the AdS space has strong support from holography. 
Despite the flat space no-go theorems are not directly applicable to the AdS case, higher-spin gauge theories in the AdS space have features suggested by these no-go results. In particular, an infinite spectrum of  fields plays an essential role in the construction, while the higher-spin scattering amplitudes display an exotic analytic behaviour.

\section{Elements of higher-spin holography}
\label{sec:11}

In the previous section we saw that the construction of interacting higher-spin gauge theories encounters substantial obstacles. In this and the following sections we will discuss different setups in which these no-go results are avoided and indications of consistent higher-spin theories exist. We will start from higher-spin gauge theories in the AdS space.
In this case the aforementioned no-go results are not directly applicable, while the
existence of higher-spin gauge theories has strong support from holography. 

We first remark that the  perturbative approach of section \ref{section:7} can be directly applied to 
gauge theories in  the AdS space. At the conceptual level this procedure
 remains identically the same, e.g. the key consistency condition  is still  gauge invariance of the action, global symmetries are generated by parameters valued in the kernel of the free gauge transformations etc. However, at the technical level the analysis becomes more tedious primarily due to the fact that the AdS space covariant derivatives do not commute.
 We will not review the results of this analysis in detail,
  assuming that it is more or less understood how it goes.
 An interested reader can find further details e. g. in \cite{Joung:2013nma}.

In a similar way the deformation procedure based on amplitudes can be extended to the AdS space. This extension, however, is somewhat less transparent as it leads to questions like: how to define the scattering amplitude in the AdS space or how to see whether an AdS amplitude corresponds to a local theory based on amplitude's analytic properties? The first of these questions will be answered below. The answer to the second question is also  well understood by now, but this discussion goes beyond the scope of the present course. 
Still, we will assume that it is understood that up to technicalities, an extension of the procedure from section \ref{sec:9} to the AdS case can be performed  and that it closely mimics its flat space counterpart.

Despite the perturbative approach to gauge theories in the AdS space remains very similar to that in flat space, for higher-spin gauge fields it leads to different results due to peculiarities of the representation theory of $so(D-1,2)$\footnote{In this section we will use the standard notation  used in the AdS/CFT literature: $D$ denotes the dimension of the AdS space, while $d\equiv D-1$ is the dimension of its boundary.}.
Namely, a distinguishing feature of the representation theory associated with free fields in the AdS space compared to that in flat space is that AdS isometry algebra admits singleton representations. These representations appear to be of outmost relevance in the  higher-spin gauge theory context. 

To be more precise, singleton representations feature the higher-spin theories in two major ways. Firstly, singleton symmetry algebras provide a suitable candidate for a global higher-spin algebra in the AdS space. To see that, we recall that in section \ref{sec:632} we showed that the singleton symmetry algebra under $so(3,2)$ decomposes into a direct sum of tensor representations
with symmetries characterised by  Young diagrams of shapes $\mathbb{Y}(s-1,s-1)$. This statement was shown only for $so(3,2)$, but it can be extended to $SO(D-1,2)$ with any $D$, see below. At the same time, by solving the following exercise, one can show that exactly these tensor representations of $SO(D-1,2)$ are carried by the AdS space Killing tensors, which are the parameters of global higher-spin symmetries. Therefore, the singleton symmetry algebra has the right spectrum to serve as the global higher-spin algebra in the AdS space. 
 Already at this level we can see that the AdS background is more suitable for higher-spin interactions, as in flat space no appropriate algebra exists\footnote{Strictly speaking, as we mentioned in the previous section, existence of a non-trivial Lie algebra with the spectrum consisting of flat space Killing tensors is not ruled out unless one makes additional assumptions. Still an important difference with the flat space is that in the AdS case the structure constants of the higher-spin algebra can, indeed, be induced from consistent cubic vertices, see e.g. \cite{Joung:2013nma}. In other words, the argument analogous to that of section \ref{sec:10.3} does not rule out interactions in AdS.}.

\begin{zadacha}
  Show that traceless AdS Killing tensors
\begin{equation}
\label{15nov1}
\nabla^\mu\xi^{\mu(s-1)}=0, \qquad g_{\mu\mu}\xi^{\mu(s-1)}=0,
\end{equation}
carry a tensor representation of $so(D-1,2)$ (AdS isometry algebra, not Lorentz algebra), which is traceless and has the symmetry type associated with the two-row rectangular Young diagram with $s-1$ columns. Hint: solve (\ref{15nov1}) using the ambient space approach. Hint: to extend $\xi$ into ambient space away from AdS, you may set $k$ in   $X\cdot \partial_X\xi = k\xi$ to any value, e.g. $k=s-1$.
\end{zadacha}

Secondly, as shown by Flato and Fronsdal, the whole multiplet of higher-spin gauge fields is nothing but a tensor product of two singleton representations. This, in particular, suggests that we may start from a theory of singletons and consider scattering of their bilinears. By the Flato-Fronsdal theorem, the resulting observables can be then reinterpreted as the AdS counterpart of the higher-spin $S$-matrix. By design, such an $S$-matrix is invariant with respect to higher-spin global symmetries, so 
 this construction gives a simple solution to the global symmetry consistency conditions on the higher-spin scattering in the AdS space.
 
 First ideas to generate higher-spin theories in terms of singleton constituents were expressed in \cite{Flato:1980zk} and then they acquired a more precise and modern form in the holographic context \cite{Sezgin:2002rt,Klebanov:2002ja}.
 Relevance of the AdS background for higher-spin interactions was also highlighted in \cite{Fradkin:1986qy} and the associated 
higher-spin algebra in the 4d case was identified in \cite{Fradkin:1986ka}. 

\subsection{Bulk-boundary dictionary}

In this section we make a brief introduction into the AdS/CFT correspondence. It is a very rich topic and we are not going to cover all of it. Instead, we will focus on certain basic concepts, which will be sufficient for explaining how holography allows one to construct higher-spin theories. For a more comprehensive review see \cite{Petersen:1999zh,Aharony:1999ti,DHoker:2002nbb,Nastase:2007kj}. For a detailed review focused on the higher-spin holography, see \cite{Sleight:2016hyl}.

The most basic observation that underlies the AdS/CFT correspondence is the fact that isometries of AdS${}_D$ and conformal isometries of the Minkowski space  M${}_{D-1}$ form the very same algebra $so(D-1,2)$. Accordingly, every field in AdS${}_D$ carries a representation of $so(D-1,2)$, which can be also reinterpreted as a conformal operator in M${}_{D-1}$.
In this context, by saying \emph{field} we mean that there is a Klein-Gordon-type equation of motion involved. In contrast, \emph{operators} do not have to satisfy any equations of motion. Note that the on-shell states carried by a field  in AdS${}_D$ can be labelled by functions of $D-1$ independent variables:
starting from a general  function on all coordinates the
 equation of motion allows one to eliminate dependence on one of the  variables.\footnote{Mathematically, this means that the associated representation has the Gelfand-Kirillov dimension equal to $D-1$. The Gelfand-Kirillov dimension is the standard way to characterise ''the size'' of infinite-dimensional representations.}  These very same representations can be alternatively realised by arbitrary functions on the space of one dimension lower, like M${}_{D-1}$, without any equations of motion imposed. This explains why fields in AdS${}_D$ correspond to operators in M${}_{D-1}$\footnote{Certainly, fields also become operators in quantum field theory. We will still use terminology of fields and operators depending on whether Klein-Gordon-like equations are involved.}.  

Analogously, any objects defined in terms of fields in AdS can be reinterpreted as the analogous objects defined in terms of CFT operators. In particular, mathematically speaking, one can view AdS scattering amplitudes as forms on $so(D-1,2)$ representations. These can also be reinterpreted as correlators of the associated operators on the conformal field theory side. This duality of AdS amplitudes and CFT correlators is the main ingredient of the AdS/CFT dictionary that we will use.

For simplicity, below we will be dealing with the Euclidean signature. In the Euclidean signature the analytic structure of correlators is simplified due to the fact that the only singularities of the correlators are at coincident points. Equivalently, Euclidean field theory is devoid of ambiguities related to the operator ordering.\footnote{Lorentz invariance requires correlators to be functions of $x^2_{ij}=(x_i-x_j)^2$. In the Euclidean case $x_{ij}^2\ge 0$ and the only singularities of $(x^2_{ij})^\delta$ occur when $x_i\to x_j$. Instead, in the Lorentzian signature $x_{ij}^2$ changes the sign when $x_i$ crosses the light cone with a vertex at $x_j$, so at these points the correlator develops additional  singularities. Depending on the way one analytically continues the correlator when crossing these singular light cones, one obtains correlators with different operator orderings.} This makes dealing with the Euclidean signature easier. At the same time,  according to the standard theorems, once  correlators are known in the Euclidean signature, they can always be analytically continued to correlators in the Lorentzian signature with any given ordering, see \cite{Osterwalder:1973dx,Osterwalder:1974tc,Mack:1974sa,Hartman:2015lfa}. 

\subsubsection{The boundary of AdS}

There is the standard geometric construction that relates the two aforementioned spaces involved in the holographic duality: namely, \emph{M${}_{d}$ is the boundary of the conformally compactified AdS${}_{d+1}$ space}.
We will now clarify this statement, but before that, let us introduce the AdS space in the Euclidean signature.

Euclidean AdS${}_{d+1}$ is defined as a sheet of a two-sheeted hyperboloid 
\begin{equation}
\label{1dec1}
X^2=-l^2, \qquad X^{d+1}>0
\end{equation}
embedded into a $(d+2)$-dimensional flat ambient space endowed with the metric
\begin{equation}
\label{17apr1}
\eta ={\rm diag}(+,\dots,+,-).
\end{equation}
  In the light-cone type coordinates
\begin{equation}
\label{20jan1}
X^+ \equiv X^d + X^{d+1}, \qquad X^- \equiv X^{d+1}-X^d
\end{equation}
the ambient metric becomes
\begin{equation}
\label{20jan2}
ds^2 = -dX^+ dX^- +dX^0 dX^0 +\dots +dX^{d-1} dX^{d-1}.
\end{equation}
In the holographic context it is convenient to use the Poincare coordinates, which are defined by\footnote{One often distributes powers of $l$ between $X^+$ and $X^-$ differently: $(X^+,X^-) = z^{-1}(l^2,z^2+x^2)$. We will use (\ref{1dec2}) as it allows us to get rid of the $l$ dependence in the coordinates for the AdS boundary.}
\begin{equation}
\label{1dec2}
X^A=(X^+,X^-,X^0,\dots, X^{d-1})=\frac{l}{z}(1,z^2+x^2,x^\mu).
\end{equation}
In these terms the AdS metric reads
\begin{equation}
\label{1dec3}
ds^2 = l^2 \frac{dz^2+dx^2}{z^2}.
\end{equation}
\begin{zadacha}
Show that (\ref{20jan2}), indeed, induces the metric (\ref{1dec3}) on the Euclidean AdS hyperboloid (\ref{1dec1}) in the Poincare coordinates (\ref{1dec2}).
\end{zadacha}

The coordinate transformation (\ref{1dec2}) maps infinite points of the AdS space, $X\to\infty$, to a finite value $z=0$. Thus, $z=0$ can be regarded as an infinite boundary of the AdS space.
The fact that these points are actually infinite, can also be seen from the Poincare metric (\ref{1dec3}): due to a divergent factor of $z^{-2}$ points with $z=0$ are infinitely separated from points inside the  AdS space. Moreover, upon dropping a divergent conformal factor, the AdS metric (\ref{1dec3}) induces metric $ds^2 =dx^2$ on $z=0$, which is just the metric of the Euclidean Minkowski space.
 Accordingly, in the holographic context one often refers to the AdS space as the \emph{bulk}, while the Minkowski space of one dimension lower is referred to as the conformally compactified boundary or just the  \emph{boundary} of AdS.

Similarly to the AdS space itself, there is an ambient space description of the AdS boundary, which has the advantage of making the $so(d+1,1)$, that is the Euclidean conformal symmetry, manifest. In the reminder of this section we briefly review this construction. For more comprehensive review on  conformal field theories and the associated  ambient space formalism  we refer the reader to  \cite{Rychkov:2016iqz,Simmons-Duffin:2016gjk}.

The AdS boundary  can be defined as a hypercone in the ambient space
\begin{equation}
\label{1dec4}
P^2=0
\end{equation}
quotiented by generatrices
\begin{equation}
\label{1dec5}
P\sim \lambda P, \qquad \lambda\ne 0.
\end{equation}
Indeed, by sending $z\to 0$ in (\ref{1dec2}) and dropping the divergent $z^{-1}$ factor we obtain
\begin{equation}
\label{21jan1}
z X^A \to P^A=(P^+,P^-,P^1,\dots, P^d)=l(1,x^2,x^\mu),
\end{equation}
which satisfies (\ref{1dec4}). In terms of the ambient space geometry in the limit $z\to 0$ the point of the AdS hyperboloid goes to infinity and at the same time approaches the cone (\ref{1dec4}) along some generatrix. To get a finite point on the generatrix we need to rescale the coordinates of the point  by an infinite factor of order $z^{-1}$. 

Due to equivalence (\ref{1dec5}) any choice of a representative point on each generatrix is equally eligible. One typically chooses 
 \begin{equation}
 \label{20jan1x1}
   P^+=1,
 \end{equation}
 which means that (\ref{21jan1}) should be further rescaled by $l$.
 The advantage of (\ref{20jan1x1}) is that it leads to the intrinsic boundary Minkowski metric in the standard form $ds^2=dx^2$.

Clearly, the choice (\ref{20jan1x1}) is not $so(d+1,1)$ invariant. Those $so(d+1,1)$ transformations that violate it should be supplemented with shifts along generatrices $P\to \lambda P$ with  $\lambda$, that restore (\ref{20jan1x1}).
In intrinsic coordinates, transformations that require non-trivial shifts along generatrices  act as conformal transformations with non-trivial conformal factors and consist of dilations and conformal boosts.

A scalar operator ${\cal O}$ of conformal dimension $\Delta$ can be realised in the ambient space formalism as a function ${\cal O}(P)$ on the cone (\ref{1dec4}), which satisfies the homogeneity condition
\begin{equation}
\label{20jan2x1}
P^M \frac{\partial}{\partial P^M}{\cal O}(P) = -\Delta {\cal O}(P).
\end{equation}
This means that ambient space $so(d+1,1)$ transformations 
 \begin{equation}
 \label{20jan2x2}
J^{MN}{\cal O}(P)\equiv - i \left(P^M \frac{\partial}{\partial P_N} - P^N \frac{\partial}{\partial P_M}\right){\cal O}(P)
 \end{equation}
upon going to the intrinsic boundary coordinates $x^i$ give the standard conformal transformations for the conformal primary of dimension $\Delta$.
 The ambient space approach to conformal field theories  can be extended to spinning operators, but for simplicity, we will limit ourselves with the scalar case only. Moreover, to avoid unnecessary clutter, we will  set $l=1$ in what follows.

\subsubsection{Bulk-to-boundary propagators}

An important ingredient of the AdS/CFT correspondence is the so-called \emph{bulk-to-boundary propagators}. These can be regarded as intertwining kernels between the AdS and CFT realisations of the same representation of $so(d+1,1)$ in the sense that these are two-point functions, which on one leg  transform as on-shell fields in the bulk, while on the other one -- as conformal operators on the boundary. Putting it differently, the bulk-to-boundary propagator is ''an infinite-dimensional transformation matrix'' between a representation of $so(d+1,1)$ in two bases: the one which is more adapted to the CFT language, in which the boundary coordinate takes definite values, and the one adapted to the bulk, in which the bulk coordinate is definite. As will be discussed below, this makes the bulk-to-boundary propagator the AdS counterpart of plane waves in flat space, which allow one to transition between bases with definite coordinate and definite momenta\footnote{Note, however, a subtlety: the coordinate of an on-shell field cannot be well-defined as $\delta(x-x_0)$ does not obey wave equations, so whether it is the AdS or the flat space case, an on-shell field is always a superposition of states with definite coordinates.}.

In the simplest case of the scalar conformal operator of dimension $\Delta$, the bulk-to-boundary propagator is given by
\begin{equation}
\label{1dec7}
G_{b\partial}(P;X) = \frac{C}{(-2P\cdot X)^\Delta},
\end{equation}
where $C$ is an arbitrary constant. Its $so(d+1,1)$ covariance is manifest due to the covariant contraction of $so(d+1,1)$ indices. In the Poincare coordinates it reads
\begin{equation}
\label{1dec8}
G_{b\partial}(y;z,x) = C \left(\frac{z}{z^2+(y-x)^2} \right)^\Delta,
\end{equation}
where we used $y$'s for the boundary coordinates instead of $x$'s. 

The fact that (\ref{1dec7}) has homogeneity degree $-\Delta$ in $P$ implies that it transforms as a conformal operator of dimension $\Delta$. At the same time, it is straightforward to check that (\ref{1dec4}) entails
\begin{equation}
\label{1dec9}
\frac{\partial}{\partial X^M}\frac{\partial}{\partial X_M}G_{b\partial}(P;X)=0,
\end{equation}
which, according to (\ref{30sep7}) with $\kappa=-\Delta$ 
implies 
\begin{equation}
\label{1dec10}
(\Box -M^2)G_{b\partial}(P;X)=0,
\end{equation}
where $\Box$ is the AdS Laplacian, while 
\begin{equation}
\label{1dec11}
M^2 = \Delta (\Delta-d).
\end{equation}
This allows one to establish that the conformal dimension $\Delta$ of the operator in the CFT is, in fact, equal to the energy of the lowest energy state in the construction of section  \ref{sec:4.2},
\begin{equation}
\label{1dec11x1}
E_0=\Delta.
\end{equation}
Considering spinning fields, one can show that spins of bulk fields and of the associated boundary operators are equal.

With some additional analysis it can be demonstrated that $G_{b\partial}(P;X)$ with different $P$'s on the AdS boundary, actually, form a basis of solutions to (\ref{1dec10}). From this perspective, boundary coordinates can be regarded as the AdS counterpart of on-shell momenta in flat space, while the bulk-to-boundary propagators play the role analogous to  flat space plane waves with momenta put on-shell. This point of view is advocated, e. g. in \cite{Fronsdal:1974ew}. For the following discussion, this role of the bulk-to-boundary propagators is the most important one: these allow one to label all the solutions of the free equations of motion with points of the boundary serving as labels. As a result, they allow us to \emph{identify $n$-point correlators in the boundary theory with $n$-point bulk amplitudes}.  

For completeness, let us review another property of bulk-to-boundary propagators, which is more famous and explains 
 their name. To this end, we will consider  their near-boundary limit. Starting from  (\ref{1dec8}) and sending $z$ to zero for $y\ne x$ we find
\begin{equation}
\label{1dec12}
G_{b\partial}(y;z,x) \sim C z^\Delta \frac{1}{(y-x)^{2\Delta}}, \qquad z\to 0, \qquad y\ne x.
\end{equation}
Thus, in the near-boundary  limit for $y\ne x$ the bulk-to-boundary propagator goes as $z^\Delta$, while the dependence on the boundary coordinate is that of a two-point function $\langle {\cal O}(x) {\cal O}(y)\rangle$ of operators with dimension $\Delta$.

 At the same time, for $x = y$ (\ref{1dec8}) becomes more singular due to the relevance of $z^2$ in the denominator. This means that $G_{b\partial}$
has a contribution of the form
\begin{equation}
\label{1dec13}
G_{b\partial} \propto \delta^d(x-y)
\end{equation}
in the near-boundary limit. To fix the $z$-dependence of this contribution and the numerical prefactor, we will integrate $G_{b\partial}$ over $x$. This gives
\begin{equation}
\label{1dec14}
\begin{split}
\int d^dx G_{b\partial}(y;z,x) &= \frac{C}{z^\Delta} \int d^dx \left(\frac{1}{1+\left(\frac{y-x}{z} \right)^2}\right)^\Delta \\
&= C z^{d-\Delta}\int d^d\bar x\frac{1}{(1+\bar x^2)^\Delta}=C z^{d-\Delta} \frac{\pi^{\frac{d}{2}}\Gamma\left(\Delta-\frac{d}{2} \right)}{\Gamma(\Delta)}.
\end{split}
\end{equation}
In the second equality we made a change of integration variables so that the integrand no longer depends on $z$, while the last integral is one of the standard conformal integrals, whose precise value for our purposes is not very important. Together with (\ref{1dec13}), this implies 
\begin{equation}
\label{1dec15}
\begin{split}
G_{b\partial} = C z^{d-\Delta} \frac{\pi^{\frac{d}{2}}\Gamma\left(\Delta-\frac{d}{2} \right)}{\Gamma(\Delta)} \delta^d(x-y) ,\qquad z\to 0, \qquad x\to y.
\end{split}
\end{equation}

Typically, for $\Delta$ above the unitarity bound one has $\Delta >d-\Delta$\footnote{In the higher-spin case we will be interested in $\Delta=d-2$, which is smaller than $d-\Delta=2$ for $d<4$.}, so the dominant behaviour of the bulk-to-boundary propagator  in the near-boundary limit is given by  (\ref{1dec15}), not by (\ref{1dec12}). The fact that
the dominant behaviour of the bulk-to-boundary propagator in the near-boundary limit is localized at some boundary point $x=y$ allows one to interpret $G_{b\partial}$ as a solution of the bulk equations of motion with a source at this point. We would like to stress that $G_{b\partial}$ satisfies the bulk equations of motion identically, so "a source" on the boundary does not refer to a delta-function on the right-hand side of the equations of motion, but to the behaviour of the propagator itself in the near-boundary limit\footnote{This is the same difference as between the Green function and the kernel of a differential equation.}.

\subsection{Boundary theory}

In previous sections we presented some formal representation theory constructions for the singleton module, its bilinear tensor product and the associated symmetry algebra in $d=3$. In this section these concepts and results will be rephrased in the conformal field theory language. Moreover, these constructions will be extended to arbitrary dimensions. Eventually,
our goal is to construct the CFT correlators for bilinears of singletons, which will be then
interpreted as higher-spin scattering amplitudes in the AdS space.

We will start from the singleton representation itself. As mentioned in section \ref{sec:323}, 
\begin{equation}
\label{30nov1}
\Box \varphi=0
\end{equation}
in the $d$-dimensional Minkowski space is a conformal equation, thereby it carries a representation of $so(d,2)$, which becomes $so(d+1,1)$ in the Euclidean signature. This is precisely the singleton representation, which sits at the unitarity bound (\ref{17dec2}).
For completeness, let us demonstrate this using the ambient space formalism.

In the ambient space formalism it is convenient to describe fields on the cone (\ref{1dec4}) via the equivalence relation
\begin{equation}
\label{24jan1}
\varphi(P)\sim \varphi(P)+P^2 \psi (P),
\end{equation}
where $\psi$ is an arbitrary function. This relation implies that values of $\varphi$ away from the cone are irrelevant, as these can be gauged away via a suitable choice of $\psi$. The ambient-space version of (\ref{30nov1}) reads
\begin{equation}
\label{24jan2}
\frac{\partial}{\partial P^M}\frac{\partial}{\partial P_M}\varphi(P)=0.
\end{equation}
\begin{zadacha}
Show that (\ref{24jan2}) is consistent with (\ref{24jan1}) only for 
\begin{equation}
\label{24jan3}
P^M\frac{\partial}{\partial P^M}\varphi(P) = -\frac{d-2}{2}\varphi(P).
\end{equation}
\end{zadacha}

From this exercise we find that $\varphi$ has 
\begin{equation}\label{30nov1x1}
\Delta_{\varphi}\equiv\frac{d-2}{2},
\end{equation}
see (\ref{20jan2x1}), which is the correct value of the conformal dimension for the massless scalar field. One can also compute the quadratic Casimir for $\varphi$ and ensure oneself that it matches that of the singleton in the lowest-weight construction (\ref{8apr7}).
As for the equivalence of (\ref{30nov1}) and  (\ref{24jan2}), it can be seen  directly  by going into the intrinsic CFT coordinates in (\ref{24jan2}).

\begin{zadacha}
Show that the quadratic Casimir operator of $so(d+1,1)$ for (\ref{24jan1}), (\ref{24jan2}) is
\begin{equation}\label{24jan4}
 {\cal C}_2(so(d+1,1)) \equiv \frac{1}{2}J^{MN}J_{MN}=\Delta_{\varphi} (\Delta_{\varphi}-d),
\end{equation}
where $J$ was defined in (\ref{20jan2x2}). Check that it matches (\ref{8apr7}), once (\ref{1dec11x1}) is taken into account.
\end{zadacha}

\subsubsection{Higher-spin algebra}

The higher-spin algebra was defined so far in section \ref{sec:632} in $d=3$ as an algebra of general linear maps of the singleton representation space to itself. The utility of the oscillator construction used there was that the oscillators, in effect, allow us to solve (\ref{30nov1}), thus providing an efficient parametrisation of the singleton states and a simple definition of the associated higher-spin algebra. 

Instead, for general $d$ no equally efficient construction exists. Still, the higher-spin algebra is defined as general linear transformations acting from solutions of (\ref{30nov1}) to solutions, and it is also called the symmetry algebra of (\ref{30nov1}). It is generated by differential operators $L$, that preserve (\ref{30nov1}), that is
\begin{equation}
\label{25jan1}
L: \qquad \Box \varphi=0,\qquad \Rightarrow \qquad \Box (L\varphi) =0.
\end{equation}
 It is not hard to see that such $L$ should satisfy
\begin{equation}
\label{30nov1x2}
\Box L = L' \Box,
\end{equation}
where $L'$ is any differential operator. Among  $L$ that satisfy (\ref{30nov1x2}) some can be written in the form $L=M\Box$. These act trivially on solutions of (\ref{30nov1}) and form an ideal in the algebra of all $L$ that satisfy (\ref{30nov1x2}).
 \emph{The higher-spin algebra is then generated by $L$'s that satisfy (\ref{30nov1x2}) quotiented by trivial symmetries} \cite{Eastwood:2002su}. In particular, conformal transformations are symmetries of (\ref{30nov1}) and thereby, they form a subalgebra of the higher-spin algebra. Below we will see how $L$'s can be defined in terms of the free theory conserved currents and conformal Killing tensors.

\begin{zadacha}
Make sure that (\ref{30nov1x2}) is necessary and sufficient condition for $L$ to be a symmetry of (\ref{30nov1}). In other words, $L$ gives the most general operator that maps solutions of (\ref{30nov1}) to solutions.
\end{zadacha}

\begin{zadacha}
Check that trivial symmetries form an ideal inside the algebra of symmetries. 
\end{zadacha}

Finally, we would like to remark that in the same way as in section \ref{sec:632}, the  higher-spin algebra as defined here is naturally an associative algebra. Within the framework we presented above the associative product is given by the composition of differential operators. Once the associative product is available, the associated commutator automatically satisfies the Jacobi identity and, therefore, defines a Lie algebra. Despite that symmetry is only required to be a Lie algebra, in many applications it is useful to keep in mind that higher-spin Lie algebras are naturally embedded into associative ones. In particular, this allows one to add color to the higher-spin algebra simply by promoting the coefficients of  the differential operators $L$ to matrices.

\subsubsection{The Flato-Fronsdal theorem}

The Flato-Fronsdal theorem tells us that a tensor product of two singleton representations equals the sum of representations that can be identified with massless higher-spin fields in the bulk. Now we would like to present a version of this statement using the dual language of the conformal field theory.

Since the energy of the lowest-energy state maps into the conformal dimension of the dual operator (\ref{1dec11x1}) and the spin of the conformal operator equals the spin of the bulk field, the operator dual to a massless spin-$s$ field in the bulk has conformal dimension $\Delta_J\equiv d+s-2$ and spin $s$. The Flato-Fronsdal theorem suggests that this operator should be constructed as a bilinear in $\varphi$. Its precise form can be found from the following simple considerations. 

  A pair of $\varphi$'s contributes $d-2$ to the dimension of $J$. The remaining dimension should be contributed by $s$ derivatives. Considering that $J$ has to be a traceless rank-$s$ tensor to carry spin $s$, we have the following ansatz
\begin{equation}
\label{30nov2}
J_{\mu(s)} \equiv \sum_{k=0}^s a_k \partial_{\mu_1}\dots \partial_{\mu_k}\varphi \partial_{\mu_{k+1}}\dots \partial_{\mu_s}\varphi - {\text{traces}}.
\end{equation}
Then, coefficients $a_k$ can be fixed by requiring that $J_{\mu(s)}$ is a primary operator, that is it should be annihilated by conformal boosts $K$. A somewhat tedious computation gives \cite{Craigie:1983fb,Anselmi:1999bb}
\begin{equation}
\label{30nov3}
a_k = (-1)^k \left(\begin{array}{c}
s\\
k
\end{array} \right)\frac{\left(\frac{d-2}{2}\right)_s}{\left(\frac{d-2}{2}\right)_k\left(\frac{d-2}{2}\right)_{s-k}}.
\end{equation}
Equations (\ref{30nov2}), (\ref{30nov3}) present the CFT version of the Flato-Fronsdal theorem, in which \emph{currents} (\ref{30nov2}) \emph{are the holographic duals of massless higher-spin fields} in the bulk.

A direct computation shows that currents (\ref{30nov2}) are conserved on-shell in the sense that
\begin{equation}
\label{30nov4}
\partial_\mu J^{\mu(s)}\propto \Box\varphi \approx 0.
\end{equation}
Because of this the space of states associated with $J_{\mu(s)}$ is smaller than that of a genuine spin-$s$ operator. This phenomenon is the CFT counterpart of bulk gauge invariance.

Besides serving as the holographic duals of bulk higher-spin fields,  currents (\ref{30nov2}) also generate higher-spin symmetries. This construction is a version of the standard relation between symmetries and conserved currents, except that in our case currents are higher-rank tensors. It works as follows.

One starts with  rank-$(s-1)$ conformal Killing tensors, which are defined by
\begin{equation}
\label{30nov5}
\text{trace free part of }(\partial_{\mu_1}K_{\mu_2\dots \mu_s})=0.
\end{equation}
Then one can construct  rank-one currents
\begin{equation}
\label{30nov6}
{\mathbb J}^\mu(K) \equiv J^{\mu(s)}K_{\mu_2\dots \mu_s},
\end{equation}
which are also conserved on-shell due to (\ref{30nov4}), (\ref{30nov5}). By virtue of the Noether theorem, each such current generates a symmetry of the theory. Therefore, (\ref{30nov1}) has symmetries which can be labelled by 
 conformal Killing tensors. Of course, these symmetries are nothing but higher-spin symmetries. 

\begin{zadacha}
Show that conformal Killing tensors (\ref{30nov5}) transform as traceless tensors of shape $\mathbb{Y}(s-1,s-1)$ of $so(d+1,1)$. In other words, these have the same spectrum as generators of higher-spin global symmetries in the bulk.
\end{zadacha}

\subsection{CFT correlators as the higher-spin scattering amplitudes}

We will now consider  correlators of higher-spin conserved currents in the free theory of massless scalars. As advocated above, these can be regarded as the scattering amplitudes of massless higher-spin fields  in AdS${}_{d+1}$.
To be more general, we will consider a multiplet of massless scalars (\ref{30nov1}), which carry an additional $O(N)$ index in the vector representation. Relevance of this amendment will become clear in a moment. 

The associated action is
\begin{equation}
\label{30novinfty}
S=\frac{1}{2}\int d^dx \varphi^a \Box \varphi_a,
\end{equation}
where $a$ is the $O(N)$ index. Expressions for currents (\ref{30nov2}) can be straightforwardly generalised to the $O(N)$ case. In the following we will focus on the $O(N)$-singlet sector, that is on currents, for which the $O(N)$ indices are contracted\footnote{In (\ref{30nov7}) all odd spin currents are vanishing. This can be avoided if we replace $O(N)$ with $U(N)$ and consider currents of the form $J\sim \bar \varphi^a \varphi_a$.} 
\begin{equation}
\label{30nov7}
J_{\mu(s)} \equiv \sum_{k=0}^s a_k :\partial_{\mu_1}\dots \partial_{\mu_k}\varphi^a \partial_{\mu_{k+1}}\dots \partial_{\mu_s}\varphi_a - {\text{traces}}:.
\end{equation}
Above we added the sign of normal ordering $:\dots:$, which prepares us to the computation of correlators.

We will be interested in computing 
\begin{equation}
\label{30nov8}
\langle J_{s_1} \dots J_{s_n}\rangle
\end{equation}
as \emph{these serve as holographic duals of higher-spin amplitudes}. Considering that the theory we are dealing with is free, these correlators are evaluated simply by performing the Wick contractions of $\varphi$ and then evaluating sums of their
derivatives according to the definition of $J$ (\ref{30nov7}).
For simplicity we will focus on correlators  of spin-0 currents $J_0(x)\equiv :\varphi^a(x)\varphi_a(x):$, for which  the Wick contractions already give the end result. 

To start, the two-point function of $\varphi$'s is
\begin{equation}
\label{30nov9}
\langle \varphi^{a_1}(x_1) \varphi^{a_2}(x_2) \rangle = \frac{\delta^{a_1a_2}}{(x_{12}^2)^{\frac{d-2}{2}}} \equiv D(x_1,x_2)\delta^{a_1a_2},
\end{equation}
where  $x_{ij}\equiv x_i-x_j$. Then, performing the Wick contractions in the two-point function of spin-0 currents, we obtain
\begin{equation}
\label{30nov10}
\langle J_0(x_1)J_0(x_2) \rangle \equiv \langle :\varphi^{a_1}(x_1)\varphi_{a_1}(x_1): :\varphi^{a_2}(x_2)\varphi_{a_2}(x_2): \rangle 
=2ND^2(x_1,x_2) = \frac{2N}{(x_{12}^2)^{d-2}}.
\end{equation}
Here $2$ comes from two different possibilities of making the Wick contraction, while $N$ originates from the trace in the $O(N)$ indices. 

Analogously, for three-point functions one finds
\begin{equation}
\label{30nov11}
\begin{split}
&\langle :\varphi^{a_1}(x_1)\varphi_{a_1}(x_1): :\varphi^{a_2}(x_2)\varphi_{a_2}(x_2): :\varphi^{a_3}(x_3)\varphi_{a_3}(x_3): \rangle\\
&\qquad= 
8N D(x_1,x_2)D(x_2,x_3)D(x_3,x_1) = \frac{8N}{(x_{12}^2)^{\frac{d-2}{2}}(x_{23}^2)^{\frac{d-2}{2}}(x_{31}^2)^{\frac{d-2}{2}}}.
\end{split}
\end{equation}
Here again, $8$ is the combinatorial factor that counts different ways to make equivalent contractions, while $N$ originates from the $O(N)$ trace.

In the four-point case we obtain
\begin{equation}
\label{30nov12}
\begin{split}
\quad&\langle :\varphi^{a_1}(x_1)\varphi_{a_1}(x_1): :\varphi^{a_2}(x_2)\varphi_{a_2}(x_2): :\varphi^{a_3}(x_3)\varphi_{a_3}(x_3): :\varphi^{a_4}(x_4)\varphi_{a_4}(x_4):  \rangle\\
&= 
16N \big( D(x_1,x_2)D(x_2,x_3)D(x_3,x_4) D(x_4,x_1) \\
&+ D(x_1,x_3)D(x_3,x_2)D(x_2,x_4) D(x_4,x_1)+ D(x_1,x_4)D(x_4,x_3)D(x_3,x_2) D(x_2,x_1) \big)\\
&+(2N)^2 \big(D^2(x_1,x_2)D^2(x_3,x_4)+D^2(x_1,x_3)D^2(x_2,x_4)+D^2(x_1,x_4)D^2(x_2,x_3)\big).
\end{split}
\end{equation}
Note that (\ref{30nov12}) contains two types of contributions: connected and disconnected ones, which come with different powers of $N$. 

\begin{zadacha}
Reproduce (\ref{30nov10})-(\ref{30nov12}).
\end{zadacha}

To better understand  the role of $N$ in this analysis, let us rescale our spin-0 current, so that its two-point function gets unit normalised
\begin{equation}
\label{30nov13}
{\cal O}(x) \equiv\frac{1}{\sqrt{2N}}:\varphi^{a}(x)\varphi_{a}(x):, 
\qquad 
\langle {\cal O}(x_1){\cal O}(x_2) \rangle 
=2ND^2(x_1,x_2) = \frac{1}{(x_{12}^2)^{d-2}}.
\end{equation}
In these terms,  for higher-point functions we obtain
\begin{equation}
\label{30nov15}
\begin{split}
\langle {\cal O}(x_1){\cal O}(x_2) {\cal O}(x_3)\rangle 
= \sqrt{\frac{8}{N}}\frac{1}{(x_{12}^2)^{\frac{d-2}{2}}(x_{23}^2)^{\frac{d-2}{2}}(x_{31}^2)^{\frac{d-2}{2}}}
\end{split}
\end{equation}
and 
\begin{equation}
\label{30nov16}
\begin{split}
&\langle {\cal O}(x_1){\cal O}(x_2) {\cal O}(x_3){\cal O}(x_4)\rangle \\
&\qquad= \frac{4}{N}\left(\frac{1}{(x_{12}^2)^{\frac{d-2}{2}}(x_{23}^2)^{\frac{d-2}{2}}(x_{34}^2)^{\frac{d-2}{2}}(x_{41}^2)^{\frac{d-2}{2}}}+{\text{2 terms}}\right)\\
& \qquad+\frac{1}{(x_{12}^2)^{d-2}}\frac{1}{(x_{34}^2)^{d-2}}+{\text{2 terms}}.
\end{split}
\end{equation}
In a similar way one can proceed with higher-spin currents and higher-order correlators. \emph{Via the holographic dictionary this defines the S-matrix for higher-spin gauge fields in the bulk}.

Now we will  clarify the role of $N$ in this analysis. If ${\cal O}$ was a fundamental field, the fact that its three-point correlator is non-vanishing would imply that ${\cal O}$ interacts with some effective cubic vertex. Since, (\ref{30nov15}) is of order $N^{-\frac{1}{2}}$, the associated coupling constant is of order $N^{-\frac{1}{2}}$. Similarly, the four-point correlator (\ref{30nov16}) has a non-trivial connected part, which corresponds to a theory of a fundamental field ${\cal O}$ with effective cubic couplings of order $N^{-\frac{1}{2}}$ and quartic couplings of order $N^{-1}$. Thus, by going from the fundamental free fields $\varphi^a$ to their bilinears, one effectively generates correlators of an interacting theory with $N^{-\frac{1}{2}}$ playing the role of an effective coupling constant. Moreover, once the CFT correlators are reinterpreted as the AdS scattering amplitudes, \emph{$N^{-\frac{1}{2}}$ also plays the role of the bulk coupling constant}, $g\sim N^{-\frac{1}{2}}$.

\subsection{Remarks and further reading}

Having defined the $S$-matrix of the higher-spin theory in AdS, one may ask a couple of natural questions, suggested by our flat space analysis, in particular: how one can define the action of the bulk theory once its $S$-matrix is known, whether the higher-spin $S$-matrix in AdS corresponds to a local theory, what is the analytic structure of this $S$-matrix etc.? Detailed answers to these questions go well beyond the scope of the present course. We will still briefly review the existing results. For a more detailed review, see \cite{Sleight:2016hyl} and references therein. 

Answering the first question, the bulk action can be reconstructed from the CFT correlators up to the freedom of field redefinitions. 
The associated procedure closely mimics the flat space discussion of section \ref{sec:9.2}, in which every flat-space object should be replaced with its AdS counterpart: for external lines one uses the bulk-to-boundary propagators instead of plane waves\footnote{Here by plane waves we mean, essentially, $e^{ipx}$, which in flat space contribute through the Fourier transform.}, the propagators get replaced with the AdS space Green functions with the properly set boundary conditions, etc. Despite the fact that conceptually the procedure of reconstruction of the action from amplitudes in the AdS space remains the same as in flat space, at the technical level it becomes more tedious. Because of that, in practice, it is hard to reconstruct the bulk action beyond the cubic order \cite{Bekaert:2014cea,Sleight:2016dba} with only some partial and implicit results available at higher orders \cite{Bekaert:2015tva}. Dealing with the action in the AdS space is further complicated by the fact that the AdS covariant derivative do not commute. Because of that the very same action can be presented in different forms, which are not manifestly equal to each other. This difficulty becomes especially challenging when one has to deal with infinitely many derivatives, which is what one encounters in the higher-spin case. 

Due to these technical difficulties, the direct analysis of locality of the holographic higher-spin theory, based on its action,  becomes a formidable task. A more feasible approach to address this question is based on  the analysis of the analytic structure of amplitudes, which are readily provided by the boundary theory. Pretty much like in flat space, see section \ref{sec:9.2}, analytic properties of AdS amplitudes for local contact diagrams and for exchanges are different. Using this knowledge, one may try to establish whether the quartic vertex in the holographic higher-spin theory is local or its amplitude, rather, behaves as it comes from the exchange diagram.
In practice, one finds that holographic higher-spin amplitudes are not analytic, but rather of a distributional type \cite{Taronna:2016ats,Bekaert:2016ezc,Sleight:2017pcz,Ponomarev:2017qab}. Due to that, it is not quite clear whether the usual connection between locality and analytic properties can be applied in the higher-spin case. Still, it does not seem possible that such amplitudes can be produced from a conventional local action \cite{Sleight:2017pcz}.

In this section we focused on the holographic duality, which features higher-spin theories in the bulk and  free scalars on the boundary. In the same way one can define a bulk higher-spin theory by replacing scalars with another singleton representation -- the spin-$\frac{1}{2}$ fermion.
 One may then wonder whether other conformal theories can be used to construct higher-spin amplitudes in a similar manner. In this respect an important result was proven in \cite{Maldacena:2011jn}, where it was shown that a 3d conformal field theory with at least one conserved higher-spin current is necessarily either a theory of a free scalar or a free spin-$\frac{1}{2}$ fermion. Its extension to higher dimensions can be found in  \cite{Boulanger:2013zza,Alba:2013yda}. These results are naturally regarded as the AdS counterparts of the Coleman-Mandula theorem with the important difference that higher-spin scattering in the AdS space is not ruled out completely, instead, it is almost uniquely fixed by symmetries. It is worth mentioning that these results can be extended to the case of the so-called slightly-broken higher-spin symmetry \cite{Maldacena:2012sf}, for which the boundary theories are interacting and conservation of higher-spin currents holds up to  terms of the special type. 
 
 Finally, we would like to mention an alternative version of the bulk reconstruction procedure, which works at the level of partition functions. Namely,  by changing the variables in the path integral from the boundary field $\varphi(x)$ to the bi-local field $\varphi(x)\varphi(y)$ and properly extending the latter to the bulk, one obtains a candidate partition function for the bulk higher-spin theory \cite{deMelloKoch:2010wdf}.

\section{Frame-like approach and Chern-Simons theories}
\label{sec:12}

We will start this section by reviewing another approach to higher-spin gauge fields. Unlike the description used above, which involved fields that should be rather regarded as higher-spin generalisations of the metric, 
the frame-like formalism utilises fields which are reminiscent of the Yang-Mills connections. This formalism can be viewed as a higher-spin extension of the geometric ideas behind the Yang-Mills theory and the Cartan formulation of gravity. The frame-like approach is extensively used in the higher-spin literature and, in particular, serves as the important ingredient of the Vasiliev theories. Besides that, the frame-like formalism allows one to rephrase the three-dimensional gravity as a Chern-Simons theory, which then has a natural extension to higher spins.

\subsection{The Cartan gravity}

We will start by briefly reviewing the Cartan\footnote{It is often referred to as the Cartan-Weyl gravity.} gravity. For more comprehensive review on the topic, see \cite{Ortin:2015hya,Didenko:2014dwa}.

\subsubsection{Geometry}
Instead of using the standard basis for tangent vectors, $\partial_\mu$, which one does in General Relativity, one can pick any other basis in the tangent space. Let the transition matrices between the new basis and the coordinate basis be $e_{\mu|}{}^a$. Then, for a vector $A$, its components in the coordinate basis $A^{\mu}$ are related to those in the new basis $A^a$ via
\begin{equation}
\label{8dec1}
A^a =e_{\mu|}{}^a A^{\mu}, \qquad A^{\mu} = (e^{-1})^{\mu|}{}_a A^a, 
\end{equation}
where $e^{-1}$ is inverse of $e$. Analogous manipulations can be done with forms and tensors involving both vector and form indices.

 In Cartan's approach, in addition, one requires that 
\begin{equation}
\label{8dec2}
g_{\mu\nu} = e_{\mu|}{}^a e_{\nu|}{}^b \eta_{ab}, \qquad \eta_{ab}= (e^{-1})^{\mu|}{}_a (e^{-1})^{\nu|}{}_b g_{\mu\nu},
\end{equation}
which means that the local basis is chosen so that the metric in it is Minkowskian. 
Condition (\ref{8dec2}) does not fix the basis $e$ uniquely. Indeed, if the metric is $\eta$ for some basis $e$, the same applies to other bases that differ from $e$ by Lorentz transformations. These Lorentz transformations can be performed independently at every point of space-time. Field-theoretically speaking, this means that when passing from $g$ to $e$ viewed as dynamical fields, we earn an additional local Lorentz symmetry. One can loosely describe (\ref{8dec2}) by saying that the  field $e$ is the square root of the metric with the local Lorentz symmetry being an irrelevant phase.

 One of the main motivations to choose the local basis with the Minkowskian metric is that it allows one to extend the defining relation of the Clifford algebra, $\{\gamma_b,\gamma_a\}=\eta_{ab}$, to the curved space. To this end, one simply replaces $\gamma_a \to \gamma_\mu \equiv e_{\mu|}{}^a\gamma_a$, which leads to $\{\gamma_\mu,\gamma_\nu \}=g_{\mu\nu}$.
 This, in turn, allows one to couple fermions to gravity. Strictly speaking, from this perspective one does not quite need $e$ alone as it appears only in combination $e\cdot \gamma$.

Each metric -- $g$ and $\eta$ -- can be used to raise/lower its own type of indices according to the standard rules. Moreover, it can be checked that 
\begin{equation}
\label{8dec3}
(e^{-1})^{\mu|}{}_a = \eta_{ab} g^{\mu\nu}e_{\nu|}{}^b.
\end{equation}
In other words, the inverse transformation matrix $e^{-1}$ can be obtained from $e$ by simply raising/lowering indices. Because of that, $e$ differs from $e^{-1}$ only by the position of indices and we will use the same symbol, $e$, for both of them. Transformation matrix $e$ is called the \emph{frame field}, the soldiering form, the tetrad or even the zweibein, the dreibein, the vierbein, etc. in special dimensions.

The covariant derivative can be straightforwardly  rewritten in the local Lorentz basis. For example, for a vector field one has
\begin{equation}
\label{8dec4}
 \partial_\mu v^b + \omega_{\mu|}{}^b{}_cv^c   \equiv \nabla_\mu v^b=  e_{\nu|}{}^b \nabla_\mu v^\nu=e_{\nu|}{}^b \nabla_\mu (e^{\nu|}{}_c v^c),
\end{equation}
where we introduced the \emph{spin-connection} $\omega$, which is just the standard Yang-Mills-like connection responsible for the parallel transport of tangent vectors in the local Lorentz basis. Equation (\ref{8dec4}) is just the requirement of consistency of the covariant derivatives in the coordinate and local Lorentz bases.
It is not hard to see that it is equivalent to the requirement that $e$ is covariantly constant, when both world and local Lorentz indices are parallel transported with the appropriate connection
\begin{equation}
\label{8dec5}
\nabla_\nu e_{\mu|}{}^a = \partial_\nu e_{\mu|}{}^a - \Gamma_\nu{}^\rho{}_\mu e_{\rho|}{}^a + \omega_{\nu|}{}^a{}_b e_{\mu|}{}^b=0.
\end{equation}
Besides that, covariant constancy of the metric in the local Lorentz basis -- which follows from the analogous property in the coordinate basis -- entails 
\begin{equation}
\label{8dec6}
0=\nabla_\nu \eta_{ab}=\partial_\nu\eta_{ab}-\omega_{\nu|}{}^c{}_a \eta_{cb}-\omega_{\nu|}{}^c{}_b=-(\omega_{\nu| b a}+\omega_{\nu| ab}),
\end{equation}
hence, the spin connection is antisymmetric in the last two indices. To highlight this property we will separate the fibre indices of $\omega$ with a comma, 
$\omega_{\nu| ab} \to \omega_{\nu| a,b}$. 

Considering that $\Gamma_\nu{}^\rho{}_\mu = \Gamma_\mu{}^\rho{}_\nu$, we find from (\ref{8dec5})
\begin{equation}
\label{8dec7}
T_{\nu\mu|}{}^a \equiv \nabla_\nu e_{\mu|}{}^a-\nabla_\mu e_{\nu|}{}^a = 
\partial_\nu e_{\mu|}{}^a - \partial_\mu e_{\nu|}{}^a + 
\omega_{\nu|}{}^{a,}{}_b e_{\mu|}{}^b - \omega_{\mu|}{}^{a,}{}_b e_{\nu|}{}^b=0.
\end{equation}
In this equation $\omega$ is a rank-three tensor which is antisymmetric in two indices. The same refers to $T$, which is called \emph{torsion}. This means that there are as many equations (\ref{8dec7}) as components of $\omega$, which allows one to express $\omega$ as a function of $e$ and its derivatives. In other words, in the same way as in the usual General Relativity the affine connection is completely fixed in terms of the metric, in the Cartan gravity the spin connection is also fixed once the frame field is specified. 

\begin{zadacha}
Denoting
\begin{equation}
\label{9dec1}
n_{\nu\mu|\rho}\equiv \partial_\nu g_{\mu\rho} - \partial_\mu g_{\nu\rho},
\end{equation}
(\ref{8dec7}) becomes
\begin{equation}
\label{9dec2}
n_{\nu\mu|\rho} = \omega_{\mu|\rho,\nu}-\omega_{\nu|\rho,\mu}.
\end{equation}
Solve (\ref{9dec2}) for $\omega$ keeping in mind (\ref{8dec6}).
\end{zadacha}

For the connection $\omega$ one naturally defines the associated Yang-Mills-like curvature tensor
\begin{equation}
\label{9dec3}
[\nabla_\mu,\nabla_\nu]v^a \equiv R_{\mu\nu|}{}^{a,}{}_{b}v^b
\end{equation}
and the explicit computation gives
\begin{equation}
\label{9dec4}
R_{\mu\nu|}{}^{a,}{}_{b}= \partial_\mu \omega_{\nu|}{}^a{}_b - \partial_\nu \omega_{\mu|}{}^a{}_b+
\omega_{\mu|}{}^a{}_c \omega_{\nu|}{}^c{}_b-\omega_{\nu|}{}^a{}_c \omega_{\mu|}{}^c{}_b.
\end{equation}
Considering consistency of covariant derivatives in the coordinate and the local Lorentz bases (\ref{8dec4}), defining relation (\ref{9dec3}) for $R$ and the analogous formula for the Riemann tensor, we find that $R$ is nothing but the Riemann curvature  tensor with the indices properly converted with the frame field
\begin{equation}
\label{9dec5}
R_{\mu\nu|}{}^{a,}{}_{b} = e_{\lambda|}{}^a e^{\rho|}{}_b R_{\mu\nu|}{}^{\lambda,}{}_{\rho}.
\end{equation}

The formalism of the Cartan gravity can be somewhat streamlined using the language of differential forms. For example, (\ref{8dec7}) and (\ref{9dec4}) can be rewritten as
\begin{equation}
\label{9dec6}
T = de^a - e_b \omega^{ab}, \qquad R^{ab}=d\omega^{ab}+\omega^{a}{}_c \omega^{cb},
\end{equation}
where $\omega$ and $e$ are 1-forms, $R$ and $T$ are 2-forms, all products are understood as wedge products and $d$ is the de Rham differential. In the reminder of this section we will use the language of differential forms extensively.

\subsubsection{Action}
Using the geometric objects that we have just introduced, the Einstein-Hilbert action
\begin{equation}
\label{9dec7}
S_{EH}=\int d^dx \sqrt{-g}R
\end{equation}
can be rewritten in the Cartan-Weyl form
\begin{equation}
\label{9dec8}
S_{CW}=\frac{1}{(d-2)!}\int R^{ab}e^{c_3}\dots e^{c_d}\epsilon_{abc_3\dots c_d},
\end{equation}
where $\epsilon$ is the totally antisymmetric Levi-Civita tensor, see appendix \ref{appa:conventions} for conventions.
To show this, let us first write the form indices and differentials explicitly
\begin{equation}
\label{27jan1x1}
\begin{split}
(d-2)!S_{CW}&= \frac{1}{2}\int R_{\mu\nu|}{}^{a,b} e_{\lambda_3|}{}^{c_3}\dots e_{\lambda_d|}{}^{c_d}\epsilon_{abc_3\dots c_d}dx^\mu dx^\nu dx^{\lambda_3}\dots dx^{\lambda_d}\\
&=\frac{1}{2}\int R_{kl|}{}^{a,b} e_{\mu|}{}^k e_{\nu|}{}^l e_{\lambda_3|}{}^{c_3}\dots e_{\lambda_d|}{}^{c_d}\epsilon_{abc_3\dots c_d}dx^\mu dx^\nu dx^{\lambda_3}\dots dx^{\lambda_d}\\
&=
\frac{\sigma}{2}\int R_{kl|}{}^{a,b} e_{\mu|}{}^k e_{\nu|}{}^l e_{\lambda_3|}{}^{c_3}\dots e_{\lambda_d|}{}^{c_d}\epsilon_{abc_3\dots c_d}\epsilon^{\mu\nu\lambda_3\dots \lambda_d}d^dx.
\end{split}
\end{equation}
In the last line the factor of $\sigma$ appeared due to differences between the Levi-Civita symbol with upper and lower indices in the Lorentzian signature, (\ref{27jan1}), (\ref{27jan3}). Next, we use the identity (\ref{27jan4}), which leads to
\begin{equation}
\label{9dec9}
(d-2)!S_{CW}= \frac{\sigma}{2}\int R_{kl|}{}^{a,b} 
{\rm det}[e]
\epsilon_{abc_3\dots c_d}\epsilon^{kl c_3\dots c_d}d^dx.
\end{equation}
Then, employing (\ref{9dec11}) we obtain
\begin{equation}
\label{9dec9x1}
S_{CW}= \int R_{kl|}{}^{a,b} 
{\rm det}[e]
\delta^{[k}_a\delta^{l]}_b
d^dx.
\end{equation}
Since
\begin{equation}
\label{9dec12}
{\rm det}[g]={\rm det}^2[e] {\rm det}[\eta],
\end{equation}
we have
\begin{equation}
\label{9dec12x1}
{\rm det}[e]=\sqrt{-g}.
\end{equation}
Finally, evaluating the trace in (\ref{9dec9x1}), we obtain
\begin{equation}
\label{9dec12x2}
S_{CW}= \int \sqrt{-g} R
d^dx.
\end{equation}

 In a similar manner one can introduce the cosmological term
\begin{equation}
\label{9dec13}
S_\Lambda=-\frac{2\Lambda}{d!} \int e^{a_1}\dots e^{a_d}\epsilon_{a_1\dots a_d}.
\end{equation}
Therefore, the Einstein-Hilbert action with the cosmological term when written in the Cartan-Weyl form reads
\begin{equation}
\label{1feb3}
S_{CW}+S_{\Lambda}= \frac{1}{(d-2)!}\int R^{ab}e^{c_3}\dots e^{c_d}\epsilon_{abc_3\dots c_d}-
\frac{2\Lambda}{d!} \int e^{a_1}\dots e^{a_d}\epsilon_{a_1\dots a_d}.
\end{equation}

 Being written in terms of differential forms, this action has manifest diffeomorphism invariance. Under infinitesimal local Lorentz transformations we have
\begin{equation}
\label{9dec14}
\delta e^a = -\lambda^{a,}{}_b e^b, \qquad \delta \omega^{a,b}=d\lambda^{a,b}+\omega^{a,}{}_c \lambda^{c,b}+\omega^{b,}{}_c \lambda^{a,c}\equiv \nabla\lambda^{a,b}.
\end{equation}
In other words, $e$ transforms as a vector, while $\omega$ transforms as a local Lorentz connection. The standard Yang-Mills analysis then implies that $R$ (\ref{9dec4}) transforms in the adjoint representation of the local Lorentz algebra or, equivalently, as the antisymmetric rank-two tensor, which is also consistent with its index structure. Considering that in (\ref{9dec9}) and (\ref{9dec13}) all local Lorentz indices are contracted covariantly, we conclude that these terms are manifestly invariant with respect to the local Lorentz symmetry.

Since (\ref{1feb3}) is  equivalent to the matter free Einstein-Hilbert action with the cosmological constant, it leads to the equivalent dynamics. Moreover, similarly to the Einstein-Hilbert action, (\ref{1feb3}) can be treated in two different ways. The more standard way is  to regard $\omega$ and $e$ as two independent dynamical fields.
Then, by varying with respect to $\omega$ one finds the zero-torsion condition (\ref{8dec7})
\begin{equation}
\label{2feb1}
\frac{\delta(S_{CW}+S_{\Lambda})}{\delta \omega}=0 \qquad \Leftrightarrow \qquad T=0.
\end{equation}
In turn, by varying with respect to $e$ one finds, 
\begin{equation}
\label{2feb2}
\frac{\delta(S_{CW}+S_{\Lambda})}{\delta e}=0 \qquad \Leftrightarrow \qquad \left( R_{\mu\nu|}{}^{a,b}+\frac{1}{l^2}e_{\mu|}{}^a e_{\nu|}{}^b\right) e^{\mu|}{}_a=0,
\end{equation}
where we assumed that the frame field is invertible. Then, (\ref{2feb1}) is to be used to eliminate $\omega$ in terms of $e$ and $\partial e$, while (\ref{2feb2}) gives the matter free Einstein equations in the AdS space.
 This approach is analogous to the Palatini approach to gravity, in which the metric and the affine connection are viewed as independent fields. Alternatively, one can immediately regard  $\omega$ as a function of $e$ and $\partial e$ such that the torsion is zero. Then, (\ref{1feb3}) is the two-derivative action for a single dynamical field $e$.
The approach of rewriting a second-derivative action as a first-derivative one by introducing auxiliary fields is analogous to the Hamiltonian formalism in classical mechanics. 

 Note  that in the first-derivative form action (\ref{9dec9}) is at most $d$-linear in fields. On the contrary, once $\omega$ is eliminated, the action features the inverse of $e$, which upon decomposition around a background leads to an infinite series in fluctuations. For this reason, the first order form of the action can be useful if one wants to avoid an infinite tail of vertices.
  Moreover, let us note that once $\omega$ is eliminated, (\ref{1feb3}) can no longer be written in terms of differential forms.

\subsubsection{Linearised Cartan gravity in flat space}

Before going to free higher-spin theories in the frame-like formulation, let us consider how linearised gravity works. 
We start from action (\ref{9dec8}) and consider its perturbations around the flat background. For simplicity we consider the Minkowski space in the Cartesian coordinates
\begin{equation}
\label{10dec1}
dh = 0, \qquad \upsilon=0,
\end{equation}
where $h$ and $\upsilon$ are the background values of $e$ and $\omega$ respectively. Substituting 
\begin{equation}
\label{10dec2}
e\to h+e, \qquad \omega \to \upsilon+\omega
\end{equation}
into (\ref{9dec9}) and keeping only the terms which are not more than quadratic in fluctuations, we find
\begin{equation}
\label{10dec3}
S=(d-2)\int d\omega^{a,b}e^c h^{n_4}\dots h^{n_d}\epsilon_{abcn_4\dots n_d}+
\int \omega^{a,c}\omega_c{}^{,b}h^{n_3}\dots h^{n_d}\epsilon_{abn_3\dots n_d},
\end{equation}
where we denoted $S=(d-2)! S_{CW}$.
\begin{zadacha}
Reproduce this.
\end{zadacha}

Our next goal is to bring this expression into a more convenient form. For the first term we will integrate $d$ by parts. For the second term we will use the fact that antisymmetrization over $d+1$ lower indices $a$, $b$, $c$, $n_3$, $\dots$, $n_d$ is zero, which allows us to write 
\begin{equation}
\label{10dec4}
S=(d-2)\int\left(de^a-\frac{1}{2}h_m\omega^{a,m}\right) \omega^{b,c} h^{n_4}\dots h^{n_d}\epsilon_{abcn_4\dots n_d}.
\end{equation}
\begin{zadacha}
Reproduce this.
\end{zadacha}

Action (\ref{10dec4}) has an obvious symmetry
\begin{equation}
\label{10dec5}
\delta e^a = d\xi^a,
\end{equation}
which follows from the fact that $e$ enters only via $de$. Besides that, it has the linearised Lorentz symmetry (\ref{9dec14})
\begin{equation}
\label{10dec6}
\delta e^a = -h_m\lambda^{a,m}, \qquad \delta \omega^{a,b}=d\lambda^{a,b}.
\end{equation}
Checking this explicitly, we find 
\begin{equation}
\label{10dec7}
\delta_\lambda S=\frac{d-2}{2}\int (h_md\lambda^{a,m} \omega^{b,c}-
h_m\omega^{a,m} d\lambda^{b,c}) h^{n_4}\dots h^{n_d}\epsilon_{abcn_4\dots n_d}.
\end{equation}
To see that this is zero use again the fact that antisymmetrization on all lower Lorentz indices vanishes. 

\begin{zadacha}
Check that  (\ref{10dec7}) is, indeed, zero.
\end{zadacha}

We already showed that the Cartan-Weyl action reproduces the Einstein-Hilbert one at full non-linear level. This, of course, implies that their linearisations agree. Still, to get acquainted with the features of the frame-like formalism before going to the higher-spin case, we would like to show that the Cartan-Weyl action upon linearisation is equivalent to the spin-2 Fronsdal theory. 

To see this, we first vary (\ref{10dec4}) with respect to $\omega$, which gives
\begin{equation}
\label{10dec8}
\frac{\delta S}{\delta \omega}=0 \qquad \Leftrightarrow \qquad T^a \equiv de^a -h_m\omega^{a,m}=0.
\end{equation}
This allows one to eliminate $\omega$ in terms of $de$ and leaves us with the second-derivative theory of $e$.

Next, one notices that out of the three irreducible Lorentz components of $e$
\begin{equation}
\label{10dec9}
\mathbb{Y}(1)\otimes \mathbb{Y}(1)= \mathbb{Y}(2)\oplus \mathbb{Y}(1,1)\oplus \mathbb{Y}(0),
\end{equation}
the antisymmetric one can be gauged away via the algebraic $\lambda$ symmetry (\ref{10dec6}). The remaining components of $e$ can be combined into a traceful symmetric rank-2 tensor, which is the same dynamical field as in the Fronsdal formalism
\begin{equation}
\label{1feb1}
\varphi_{aa}=e_{a|a}.
\end{equation}
Moreover, the remaining symmetry is given by (\ref{10dec5}), which reproduces the Fronsdal's gauge symmetry transformation. By uniqueness of the action with this set of fields and gauge symmetries, one argues equivalence of (\ref{10dec4}) and the spin-2 Fronsdal theory.

To finish the analysis of the action, we vary with respect to the frame field
\begin{equation}
\label{2feb4}
\frac{\delta S}{\delta e}=0 \qquad \Leftrightarrow \qquad d\omega^{a,b}\Big|_{\mathbb{Y}(2)\oplus \mathbb{Y}(0)}=0,
\end{equation}
where the projection is enforced by contraction with non-trivial components of $e$, (\ref{1feb1}). More explicitly, this projection amounts to the trace
\begin{equation}
\label{2feb5}
\partial_a\omega_{c|}{}^{a,b}(\partial e)-\partial_c\omega_{a|}{}^{a,b}(\partial e)=0.
\end{equation}
Here we use $\omega(\partial e)$ to emphasise that $\omega$ is understood to be expressed in terms of first derivatives of $e$ via (\ref{10dec8}).
Equation (\ref{2feb5})  is another form of the Fronsdal equation for spin 2. Equivalently, (\ref{2feb5}) says that the trace of the linearised Riemann tensor is zero.

Linearisation of the Cartan gravity with the cosmological term around the (A)dS background goes along the same lines.

\subsubsection{Cartan gravity vs Yang-Mills theory}

In the previous discussion we already noticed some similarities between the Cartan form of gravity and the Yang-Mills theories. 
Here we will make this similarity more manifest.

To start, let us redefine our generators of the $(A)dS$ algebra (\ref{7apr12}), so that $i$'s are avoided
\begin{equation}
\label{15dec1}
\begin{split}
[\hat P_a,\hat P_b] &=\frac{1}{l^2}\hat J_{ab}, \qquad [\hat P_a,\hat J_{bc}]=2\eta_{a[b}\hat P_{c]}\\
[\hat J_{ab},\hat J_{cd}] &=2\eta_{a[c}\hat J_{d]b}-2\eta_{b[c}\hat J_{d]a},
\end{split}
\end{equation}
where
\begin{equation}
\label{15dec1x1}
P=-i\hat P, \qquad 
J=-i\hat J.
\end{equation}
We would like to show that the Cartan gravity (\ref{1feb3})
has some features of the Yang-Mills theory with the  symmetry algebra (\ref{15dec1}). The cosmological constant $\Lambda$ in (\ref{1feb3}) and the AdS radius $l$ in (\ref{15dec1}) should be identified as in (\ref{7apr7}).

As the Yang-Mills analogy suggests, we combine the frame field and the spin connection into a single connection on a principal bundle with gauge algebra (\ref{15dec1})
\begin{equation}
\label{15dec2}
A=e+\omega \equiv  e^a \hat P_a +\frac{1}{2}\omega^{a,b}\hat J_{a,b}.
\end{equation}
In the same way, we combine gauge parameters 
\begin{equation}
\label{15dec3}
\varepsilon =\xi+\lambda \equiv \xi^a \hat P_a+\frac{1}{2}\lambda^{a,b}\hat J_{a,b}.
\end{equation}
Then, the Yang-Mills gauge transformation law for connections
\begin{equation}
\label{15dec4}
\delta A = d\varepsilon + [A,\varepsilon]
\end{equation}
after decomposition into $P$ and $J$ components leads to
\begin{equation}
\label{15dec5}
\begin{split}
\delta e &= \nabla \xi + [e,\lambda],\\
\delta \omega &= \nabla \lambda+[e,\xi].
\end{split}
\end{equation}
\begin{zadacha}
Check this.
\end{zadacha}
\begin{zadacha}
Starting from (\ref{15dec5}), substitute $e$, $\omega$, $\xi$ and $\lambda$ from (\ref{15dec2}), (\ref{15dec3}) in terms of components $e^a$, $\omega^{a,b}$, $\xi^a$ and $\lambda^{a,b}$. Evaluating the commutators with (\ref{15dec1}) find the transformations of the component fields. 
\end{zadacha}

It is straightforward to see that $\lambda$-transformations in (\ref{15dec5}) reproduce Lorentz transformations of the non-linear Cartan gravity (\ref{9dec14}). Moreover, considering the flat space case,  $l\to \infty$, and \emph{linearising} $\xi$-transformation around the flat background, we reproduce the $\xi$-symmetry of the \emph{linearised} Cartan gravity (\ref{10dec5}). 
The analogous statement is also true for linearisation around the AdS background.

Proceeding with the Yang-Mills analogy, we also define the Yang-Mills field strength
\begin{equation}
\label{15dec6}
F = dA+\frac{1}{2}[A,A] = T^{a}\hat P_a+\frac{1}{2}\left( R^{a,b}+\frac{1}{l^2}e^a e^b\right)\hat J_{a,b}.
\end{equation}
One can see that $F=0$ describes the AdS space of radius $l$. From this point of view, the AdS space can be regarded as the vacuum solution of (\ref{1feb3}).

\begin{zadacha}
Show that $T^a$ and $R^{a,b}$ appearing in (\ref{15dec6}) are, indeed, the torsion and the curvature defined earlier (\ref{9dec6}).
\end{zadacha}

 It is important to stress, however, that the $\xi$-transformations (\ref{15dec5}) \emph{do not} give symmetries of the non-linear Cartan gravity in $d \ge 4$. Indeed, considering the flat space case for simplicity, we find that
 \begin{equation}
 \label{15dec6x1}
 \begin{split}
 \delta_{\xi}S&=(d-2)\int R^{a,b}\nabla\xi^{c_3}\dots e^{c_d}\epsilon_{abc_3\dots c_d}\\
 &=-(d-2)\int \nabla R^{a,b}\xi^{c_3}e^{c_4}\dots e^{c_d}\epsilon_{abc_3\dots c_d}-(d-2)(d-3)
 \int R^{a,b}\xi^{c_3}T^{c_4}\dots e^{c_d}\epsilon_{abc_3\dots c_d}.
 \end{split}
 \end{equation}

\begin{zadacha}
Show that 
\begin{equation}
\label{12dec5}
\nabla \epsilon_{a[d]} =0.
\end{equation}
\end{zadacha}

Keeping in mind the Bianchi identity
\begin{equation}
\label{15dec7}
\nabla R^{a,b}=0,
\end{equation}
we find that 
 \begin{equation}
 \label{15dec8}
 \begin{split}
 \delta_{\xi}S=-(d-2)(d-3)
 \int R^{a,b}\xi^{c_3}T^{c_4}\dots e^{c_d}\epsilon_{abc_3\dots c_d}.
 \end{split}
 \end{equation}
Thus, indeed, for $d\ge 4$, local translations $\xi$ do not generate symmetries of the Cartan action.

To summarise, we can see that the Cartan formulation of gravity has some features of the  Yang-Mills theory with the symmetry algebra being the isometry of its vacuum solution. In particular, the dynamical fields can be naturally combined into a Yang-Mills connection gauging the respective isometry. Moreover, the associated Yang-Mills field strength naturally combines the torsion and the the curvature of the Cartan gravity. At the same time, only the Lorentz part of the Yang-Mills theory is a symmetry of the Cartan-Weyl action at the non-linear level, while local translations are symmetries only of the linearised theory. Accordingly, the Cartan gravity is not the Yang-Mills theory gauging the isometry of its vacuum. This quite obviously follows from the fact that the action (\ref{1feb3}) does not have the Yang-Mills form.

\subsection{Frame-like approach to higher-spin fields}
\label{12:2}
In this section we will explain how the Cartan formulation of gravity can be extended to free higher-spin theories in flat \cite{Vasiliev:1980as} and the AdS spaces \cite{Vasiliev:1986td,Lopatin:1987hz}.  This approach is usually referred to as the \emph{frame-like formalism}.

\subsubsection{Flat space}
As in the linearised Cartan gravity case, we will aim for a 1-derivative action, which involves two dynamical fields and which is formulated in terms of differential forms. The key feature of the Fronsdal action is that it has a differential gauge symmetry with the parameter $\xi^{a(s-1)}$ being traceless and symmetric. From the linearised gravity example we learned that this symmetry can be easily achieved if the action contains a 1-form field $e^{a(s-1)}$, which is also traceless and symmetric on the local Lorentz indices and enters into the action as $de$ only. The associated gauge symmetry reads
\begin{equation}
\label{11decx1}
\delta e^{a(s-1)}=d\xi^{a(s-1)}.
\end{equation}

The higher-spin frame field $e$ besides explicit Lorentz indices also has one implicit form index.
Converting the form index to the local Lorentz basis via the background frame field and decomposing the resulting representation of the local Lorentz algebra into irreducible components, we find\footnote{We would like to remind the reader that we use notation $\mathbb{Y}$ to highlight the fact that in addition to possessing the Young symmetry  a tensor is also traceless. Tensors possessing only Young symmetries and not satisfying any trace constraints are denoted $\mathbf{Y}$.}
\begin{equation}
\label{11dec1}
e: \qquad \mathbb{Y}(s-1)\otimes \mathbb{Y}(1) = \mathbb{Y}(s)\oplus \mathbb{Y}(s-1,1)\oplus \mathbb{Y}(s-2).
\end{equation}

Indeed, the form index can be either symmetrized with the Lorentz ones
\begin{equation}
\label{3feb1}
e^{a|a(s-1)} \qquad \to \qquad  \mathbf{Y}(s)
\end{equation}
or properly antisymmetrized so that the resulting tensor has the hook-type Young symmetry
\begin{equation}
\label{3feb2}
e^{b|a(s-1)}-e^{a|a(s-2)b} \qquad \to \qquad  \mathbf{Y}(s-1,1).
\end{equation}
These tensors, however, do not give irreducible representation of the Lorentz algebra, as these have non-vanishing traces. Projecting them out, we obtain the first two terms on the right-hand side of (\ref{11dec1}). Traces themselves, which are both proportional to 
\begin{equation}
\label{3feb3}
e_{m|}{}^{a(s-2)m}  \qquad \to \qquad  \mathbb{Y}(s-2)
\end{equation}
give the last term in (\ref{11dec1}). In summary, the extra box can be added to $\mathbb{Y}(s-1)$ either into the first or into the second row or, it can also cancel one box from the original diagram.

In the following we will assume that tensor products are computed as in this example, that is  by  studying various options of adding new boxes to a Young diagram and then taking all possible traces. A general algorithm for computing the tensor product  of two irreducible representations of $gl(d)$ associated with given Young diagrams is given by the so-called Littlewood-Richardson rule, which is hard to clearly formulate, let alone to prove. Moreover, to take into account trace constraints one needs additional considerations. We do not review the systematics of tensor product here and refer the interested reader to \cite{Bekaert:2006py} and references therein. Evaluation of  tensor products of Young diagrams in some simple cases can be found in \cite{Didenko:2014dwa}.

Returning to the discussion of the dynamical fields in the frame-like approach we note that the first and the third terms on the right-hand side of (\ref{11dec1}) can be naturally combined into a double-traceless rank-$s$ field $\varphi$ from the Fronsdal approach
\begin{equation}
\label{1feb4}
\varphi_{a(s)}=e_{a|a(s-1)}.
\end{equation}
The second component on the right-hand side of (\ref{11dec1}) is absent in the Fronsdal formalism. As in the case of the linearised gravity, we may get rid of this component by requiring the presence of an algebraic Lorentz-like symmetry that acts on $e$ as follows
\begin{equation}
\label{11dec2}
\delta e^{a(s-1)}=-h_m \lambda^{a(s-1),m},
\end{equation}
where $\lambda$ is traceless and has the Young symmetry of type $\mathbb{Y}(s-1,1)$.
The linearised gravity case suggests that this symmetry comes together with the spin-connection-like 1-form
\begin{equation}
\label{11dec3}
\delta \omega^{a(s-1),b} = d\lambda^{a(s-1),b}.
\end{equation}

Having settled with the set of dynamical fields and gauge symmetries, we proceed to the action. Its structure is fixed by a couple of considerations. To start, it should schematically be of the form $de\omega +\omega^2$, so that after solving for $\omega$ we have a second order action for $e$. Focusing on the $de\omega$ term, which is a 3-form, to obtain a d-form, which would make the result suitable for integration over the d-dimensional space, we need to multiply $de\omega$ with $d-3$ copies of the background frame field $h$. Since $h$'s are  1-forms, their wedge product is antisymmetric, which antisymmetrizes Lorentz indices that $h$'s carry. To form an action, which is a Lorentz scalar, these indices should be contracted in a Lorentz-covariant manner. The only way to do that is, similarly to the linearised gravity case, to contract the indices carried by $h$'s with the Levi-Civita tensor.

 The remaining three indices of the Levi-Civita tensor should be contracted with $de$ and $\omega$. Considering that these Lorentz indices are antisymmetric, there is the only way to make this contraction non-trivial: one index gets contracted with $e$, while the other two should be contracted with two indices in different rows of the two-row Young diagram carried by $\omega$. The remaining indices of $e$ and $\omega$ should be contracted between each other, which, can be done in a unique way. This fixes the form of the $de \omega$ term up to an overall factor. 
 
 In a similar manner one fixes the $\omega^2$ term.  The relative factor between the two terms is inessential and can be fixed from the requirement that the action has the symmetry as in (\ref{11dec2}), (\ref{11dec3}) without any prefactors. The resulting action reads
\begin{equation}
\label{11dec4}
S=\int\left(de^{af(s-2)}-\frac{1}{2}h_m\omega^{af(s-2),m}\right) \omega^{b}{}_{f(s-2),}{}^{c} h^{n_4}\dots h^{n_d}\epsilon_{abcn_4\dots n_d}.
\end{equation}

\begin{zadacha}
Show that (\ref{11dec4}) is gauge invariant with respect to  (\ref{11dec2}), (\ref{11dec3}).
\end{zadacha}

Now we would like to show the equivalence of (\ref{11dec4}) and the spin-s Fronsdal action. It was already explained that $e$ once its hook component is removed by an algebraic gauge shift (\ref{11dec2}) can be identified with the off-shell field of the Fronsdal formulation and, moreover, gauge symmetry (\ref{11decx1}) reproduces gauge transformation from the Fronsdal theory. It only remains to show that $\omega$ does not propagate additional degrees of freedom. 

To see that, we consider the action variation with respect to $\omega$
\begin{equation}
\label{11dec5}
\frac{\delta S}{\delta \omega}=0 \qquad \Leftrightarrow \qquad T^{a(s-1)}\Big|_{\Pi(\omega)}=0,
\end{equation}
where
\begin{equation}
\label{11dec6}
T^{a(s-1)}\equiv de^{a(s-1)}-h_m \omega^{a(s-1),m}
\end{equation}
is the higher-spin counterpart of the linearised torsion (\ref{10dec8}). In (\ref{11dec5}) a vertical bar followed by $\Pi(\omega)$ implies that the argument should be projected into Young symmetries present in $\omega$. This projection is enforced by the fact that (\ref{11dec5}) was obtained by varying the action with respect to $\omega$ and this way one may only  get  as many equations as components $\omega$ features. 

To be more precise, considering that $T$ is a two form, on top of $s-1$ symmetric Lorentz indices, it also has two antisymmetric form indices. Converting them to the same basis and performing the tensor product, we find that $T$ has the following Lorentz-irreducible components
\begin{equation}
\label{11dec7}
T:\qquad \mathbb{Y}(1,1)\otimes \mathbb{Y}(s-1)=\mathbb{Y}(s,1)\oplus \mathbb{Y}(s-1,1,1)\oplus 
\mathbb{Y}(s-1)\oplus\mathbb{Y}(s-2,1).
\end{equation}
Similarly, for $\omega$, being a 1-form, we obtain
\begin{equation}
\label{11dec8}
\omega:\qquad \mathbb{Y}(1)\otimes \mathbb{Y}(s-1,1)=\mathbb{Y}(s,1)\oplus \mathbb{Y}(s-1,2)\oplus\mathbb{Y}(s-1,1,1)\oplus 
\mathbb{Y}(s-1)\oplus\mathbb{Y}(s-2,1).
\end{equation}
Comparing (\ref{11dec7}) and (\ref{11dec8}), we find that all components of $T$ are present in $\omega$. This implies that projection in (\ref{11dec5}) is, in fact, trivial and one has
\begin{equation}
\label{11dec9}
T\equiv de^{a(s-1)}-h_m \omega^{a(s-1),m}=0.
\end{equation}

Zero-torsion constraint is now to be used to express $\omega$ in terms of $de$. This is, however, possible only for those components of $\omega$ in (\ref{11dec8}) that are also present in $T$, (\ref{11dec7}). By comparing these two decompositions, one can see that $\omega$ has an extra component of shape $\mathbb{Y}(s,2)$, which, therefore is not an auxiliary field in the sense that it cannot be expressed in terms of $de$ via (\ref{11dec9}). To see this more explicitly, we note that this extra component can be written as
\begin{equation}
\label{11dec10}
\omega^{a(s-1),b}_{\rm e}=-h_n \tau^{a(s-1),bn}.
\end{equation}
Plugging this into (\ref{11dec9}), one can see that $\omega_e$ is, in fact, annihilated by contraction with $h_m$ or, in other words, it does not contribute to (\ref{11dec9}).

This issue is resolved by noting that 
\begin{equation}
\label{11dec11}
\delta\omega^{a(s-1),b}=-h_n \tau^{a(s-1),bn}
\end{equation}
is a symmetry of the action (\ref{11dec4}), so $\omega_{\rm e}$ can be gauged away. 
To see this we can vary the action with respect to (\ref{11dec11}) and find that the variation is vanishing. Indeed, this variation contains two terms: with $\omega$ varied inside the bracket and outside of it. The first term vanishes because, as above, (\ref{11dec11}) is annihilated by contraction with $h_m$ inside the bracket. Analogously, the second term with $\omega$ varied outside the bracket vanishes due to the fact that similarly to $T$ the bracket does not have the $\mathbb{Y}(s,2)$ component, hence, it cannot be contracted with $\tau$ Lorentz covariantly and non-trivially.

Therefore, we found that all components of the spin-connection-like field $\omega$ can be either gauged away or are auxiliary fields, in the sense that these can be expressed in terms of other fields via equations of motion. The remaining field, the frame-like field $e$, after gauging away its hook-type component can be identified with the off-shell Fronsdal field and transforms in the same way with respect to the remaining $\xi$ symmetry (\ref{11decx1}). By the uniqueness of the Fronsdal action we conclude its equivalence to (\ref{11dec4}).

For completeness, we give the schematic form of the variation of the action with respect to $e$
\begin{equation}
\label{2feb6}
\frac{\delta S}{\delta e}=0 \qquad \Leftrightarrow \qquad d\omega^{a(s-1),b}\Big|_{\mathbb{Y}(s)\oplus \mathbb{Y}(s-2)}=0.
\end{equation}

\subsubsection{AdS space}

In the AdS space one has
\begin{equation}
\label{12dec1}
\nabla h=0, \qquad \nabla^2 n^a = -\frac{1}{l^2}h^a h_b n^b,
\end{equation}
where $\nabla$ is the AdS covariant derivative, carrying a form index and $n$ is an arbitrary differential form with one Lorentz index. Free fields in the AdS space can be described similarly to those in flat space with the only difference that derivatives no longer commute, $\nabla^2\ne 0$. This leads to the deformation of the action (\ref{11dec4}), which we will now present.

To start, we just replace the de Rham differential in (\ref{11dec4}) with the AdS covariant derivative
\begin{equation}
\label{12dec2}
S_0=\int\left(\nabla e^{af(s-2)}-\frac{1}{2}h_m\omega^{af(s-2),m}\right) \omega^{b}{}_{f(s-2),}{}^{c} h^{n_4}\dots h^{n_d}\epsilon_{abcn_4\dots n_d}.
\end{equation}
The action of Lorentz-like gauge symmetries is naturally deformed into 
\begin{equation}
\label{12dec3}
\delta e^{a(s-1)}=-h_m\lambda^{a(s-1),m}, \qquad \delta\omega^{a(s-1),b}=\nabla \lambda^{a(s-1),b}.
\end{equation}
By varying (\ref{12dec2}) with respect to (\ref{12dec3}), we find that part of the variation vanishes as in flat space, while 
\begin{equation}
\label{12dec4}
\delta_{\lambda}S_0 =\int \nabla e^{a f(s-2)}\nabla\lambda^b{}_{f(s-2),}{}^c h^{n_4}\dots h^{n_d}\epsilon_{abc n_4 \dots n_d}
\end{equation}
remains. Integrating $\nabla$ by parts to make $\nabla^2\lambda$, using (\ref{12dec5})
and (\ref{12dec1}), we arrive to 
\begin{equation}
\label{12dec6}
\begin{split}
\delta_{\lambda}S_0 &= -\frac{1}{R^2}\int e^{af(s-2)} \big( h^b h_m \lambda^m{}_{f(s-2),}{}^c+(s-2)h_f h^m\lambda^b{}_{mf(s-3),}{}^c\\
&\qquad\qquad\qquad\qquad\qquad\qquad+
h^c h_m\lambda^b{}_{f(s-2),}{}^m\big)h^{n_4}\dots h^{n_d}\epsilon_{abcn_4\dots n_d}.
\end{split}
\end{equation}
\begin{zadacha}
Show (\ref{12dec6}).
\end{zadacha}

To compensate this non-invariance, we need to add other terms to the action. By inspecting different options, that would suit natural requirements e.g. presence of not more than two derivatives, we find the only suitable term to be 
\begin{equation}
\label{12dec7}
S_{\Lambda}= \alpha \int e^{af(s-2)}e^b{}_{f(s-2)}h^{n_3}\dots h^{n_d}\epsilon_{abn_3\dots n_d},
\end{equation}
where $\alpha$ remains to be fixed. 
Its variation with respect to (\ref{12dec3}) gives
\begin{equation}
\label{12dec8}
\delta_\lambda S_{\Lambda}=- 2\alpha \int h_m \lambda^{af(s-2),m}e^b{}_{f(s-2)}h^{n_3}\dots h^{n_d}\epsilon_{abn_3\dots n_d},
\end{equation}
Our goal now is to choose $\alpha$ so that (\ref{12dec6}) and (\ref{12dec8}) cancel each other. 

To see that (\ref{12dec6}) and (\ref{12dec8}) can, indeed, cancel each other, one needs to do some rather tedious manipulations that use Young symmetries of the fields and parameters as well as the fact that antisymmetrization of $d+1$ indices gives zero. To simplify this analysis a bit, one may notice that only the component of $e$ of the form
\begin{equation}
\label{12dec9}
e^{a(s-1)}=h_p \sigma^{a(s-1),p},
\end{equation}
can contribute to (\ref{12dec6}), (\ref{12dec8}). With $e$ in the form (\ref{12dec9}) one can systematically simplify (\ref{12dec6}), (\ref{12dec8}) using (\ref{27jan4}), (\ref{9dec11}) and reduce these expressions to contractions between $\lambda$ and $\sigma$.
\begin{zadacha}
Use Young symmetry to show that
\begin{equation}
\label{12dec10}
\lambda^{a(s-1),b}\sigma_{a(s-2)b,a}=-\frac{1}{s-1}\lambda^{a(s-1),b}\sigma_{a(s-1),b}.
\end{equation}
\end{zadacha}

Eventually, one finds that the total action is gauge invariant for
\begin{equation}
\label{12dec11}
\alpha=\frac{1}{l^2}\frac{s(d+s-4)}{2(d-2)}.
\end{equation}
\begin{zadacha}
Check this.
\end{zadacha}

So far, we found that the action 
\begin{equation}
\label{12dec12}
\begin{split}
S& =\int\left(\nabla e^{af(s-2)}-\frac{1}{2}h_m\omega^{af(s-2),m}\right) \omega^{b}{}_{f(s-2),}{}^{c} h^{n_4}\dots h^{n_d}\epsilon_{abcn_4\dots n_d}\\
& \qquad\qquad\qquad\qquad\qquad+\frac{1}{l^2}\frac{s(d+s-4)}{2(d-2)} \int e^{af(s-2)}e^b{}_{f(s-2)}h^{n_3}\dots h^{n_d}\epsilon_{abn_3\dots n_d}
\end{split}
\end{equation}
is invariant with respect to Lorentz-like higher-spin symmetries (\ref{12dec3}). However, due to the fact that $\nabla^2\ne 0$
it is not invariant under Fronsdal-like transformations
\begin{equation}
\label{12dec13}
\delta e^{a(s-1)}=\nabla \xi^{a(s-1)}.
\end{equation}
To restore gauge invariance, we should require that $\omega$ transforms as well. Considering the differential form degree of $\omega$ and its symmetry and trace properties on the local Lorentz indices, this transformation should be of the form
\begin{equation}
\label{12dec14x1}
\delta \omega = \beta (\sigma_+ \xi),
\end{equation}
where
\begin{equation}
\label{12dec14}
(\sigma_+\xi)^{a(s-1),b}\equiv \Big(h^b \xi^{a(s-1)}-h^a \xi^{a(s-2)b}- \frac{s-2}{d+s-4}[h_m \xi^{a(s-2)m} 
\eta^{ab}-h_m\xi^{a(s-3)mb}\eta^{aa}
]\Big), 
\end{equation}
and $\beta$ is the proportionality coefficient to be fixed. 
\begin{zadacha}
Show that the bracket on the right-hand side of (\ref{12dec14}) is traceless on any pair of indices and has the Young symmetry associated with the index structure on the left-hand side.
\end{zadacha}

After a rather tedious computation one finds that gauge invariance of (\ref{12dec12}) requires
\begin{equation}
\label{12dec15}
\beta=-\frac{1}{l^2}\frac{(s-1)(d+s-4)}{d-2}.
\end{equation}

Thus, by properly deforming the action and gauge symmetries, we managed to make the former gauge invariant under Lorentz-like symmetries and under Fronsdal-like gauge transformations. Besides that, (\ref{12dec12}) is invariant under (\ref{11dec11}), which can be seen in the exactly the same way as in flat space. 

The analysis of the resulting theory also follows the same steps as in flat space. Namely, by varying the action with respect to $\omega$, we obtain the torsion-free constraint
\begin{equation}
\label{12dec16}
T\equiv \nabla e^{a(s-1)}-h_m \omega^{a(s-1),m}=0.
\end{equation}
It allows one to solve for all the components of $\omega$ in terms of $\nabla e$ except for those that can be gauge away by (\ref{11dec11}). The Lorentz-like symmetry (\ref{12dec3}) is still used to gauge away the hook-type part of $e$, which leaves us with the double-traceless and symmetric field of the Fronsdal theory. Gauge symmetry (\ref{12dec13}) then reproduces Fronsdal's gauge transformations in the AdS space. By uniqueness of a theory with this field content and symmetries we establish equivalence of (\ref{12dec12}) and the Fronsdal theory in the AdS space.

For completeness, we give the schematic form of the variation of the action with respect to $e$
\begin{equation}
\label{2feb7}
\frac{\delta S}{\delta e}=0 \qquad \Leftrightarrow \qquad \left(\nabla\omega^{a(s-1),b}+\beta(\sigma_+e)\right)\Big|_{\mathbb{Y}(s)\oplus \mathbb{Y}(s-2)}=0.
\end{equation}

\subsection{Chern-Simons theories}

In this section we consider a particular case of $d=3$ gravity and its extension to higher-spin theories. This section follows \cite{Campoleoni:2011tn}, to which we refer the reader for further details. 

In three dimensions the Cartan-Weyl action for gravity reads
\begin{equation}
\label{15dec9}
S_{CW}=\frac{1}{16\pi G}\int \left(e^a R^{bc}+\frac{1}{3l^2}e^ae^be^c \right)\epsilon_{abc}.
\end{equation}
A first major difference with the case of arbitrary dimension is that in $d=3$ the local translations are symmetries of the action, see (\ref{15dec8}). Thus, unlike in general dimensions,  in three dimensions, gravity can, indeed, be regarded as a gauge theory for the full isometry of the vacuum. 

Besides that, varying (\ref{15dec9}) we find
\begin{equation}
\label{15dec10}
\begin{split}
\frac{\delta S}{\delta e}&=0 \qquad \Leftrightarrow \qquad \tilde R^{a,b}\equiv R^{a,b}+\frac{1}{l^2}e^ae^b=0,\\
\frac{\delta S}{\delta \omega}&=0 \qquad \Leftrightarrow \qquad T^a=0.
\end{split}
\end{equation}
In general dimensions the first line of (\ref{15dec10}) contains a projector into the irreducible components, carried by $e$, which results in the equation that only sets the trace of $\tilde R$ to zero and thus gives the matter-free Einstein equations, see (\ref{2feb2}). However, in three dimensions, the remaining -- traceless -- components carried by $\tilde R$ are automatically zero, due to the fact, that these are described by not allowed Young diagrams. Therefore, it is only in $d=3$ that equations of motion for the Cartan gravity give zero curvature conditions.

The latter fact suggests that the Cartan gravity in three dimensions can be rewritten as the Chern-Simons theory. This is, indeed, the case and one has
\begin{equation}
\label{15dec11}
S_{CW}=\frac{1}{16\pi G}\int {\rm tr}\left( AdA + \frac{2}{3}A A A\right),
\end{equation}
where the invariant trace is chosen to be\footnote{This trace is not unique, which follows from the fact that $so(2,2)\sim sl(2,\mathbb{R})_L\oplus sl(2,\mathbb{R})_L$, see below.}
\begin{equation}
\label{15dec12}
{\rm tr}(\hat P_a\hat P_b)=0, \qquad {\rm tr}(\hat P_a\hat M_{bc})=\epsilon_{abc}, \quad {\rm tr}(\hat M_{ab}\hat M_{cd})=0
\end{equation}
and $A$ was defined in (\ref{15dec2}).

To further simplify the form of this theory, one may change variables as
\begin{equation}
\label{15dec13}
\omega^a\equiv \frac{1}{2}\epsilon^{abc}\omega_{b,c}, \qquad \hat J_a\equiv -\frac{1}{2}\epsilon_{abc}\hat J^{bc}.
\end{equation}
In these terms, the $so(2,2)$ commutation relations read
\begin{equation}
\label{15dec14}
[\hat P_a,\hat P_b]=\frac{1}{l^2}\epsilon_{abc}\hat J^c, \qquad [\hat J_a,\hat P_b]=\epsilon_{abc}\hat P^c,\qquad [\hat J_a,\hat J_b]=\epsilon_{abc}\hat J^c.
\end{equation}
In other words, in $3d$ translations $\hat P$ and Lorentz transformations $\hat J$ have the same number of components and transform in the equivalent representations of the Lorentz algebra. 

Furthermore, by going to 
\begin{equation}
\label{15dec15}
A_L^a = \omega^a +\frac{1}{l}e^a, \qquad A_R^a = \omega^a-\frac{1}{l}e^a,
\end{equation}
which corresponds to changing the generators to
\begin{equation}
\label{15dec16}
T_{La}\equiv \frac{1}{2}(\hat J_a+l \hat P_a), \qquad T_{Ra}\equiv \frac{1}{2}(\hat J_a-l \hat P_a),
\end{equation}
we find that the commutation relations read
\begin{equation}
\label{15dec17}
[T_{La},T_{Lb}]=\epsilon_{abc}T_L^c, \qquad [T_{Ra},T_{Rb}]=\epsilon_{abc}T_R^c, \qquad [T_L,T_R]=0.
\end{equation}
This change of variables manifests the fact that $so(2,2)\sim so(2,1)_L\oplus so(2,1)_R\sim sl(2,\mathbb{R})_L\oplus sl(2,\mathbb{R})_R$.
In terms of these new variables the action reads 
\begin{equation}
\label{15dec18}
S=\frac{1}{16\pi G}\int {\rm tr}\left( A_LdA_L + \frac{2}{3}A_L A_L A_L\right)-
\frac{1}{16\pi G}\int {\rm tr}\left( A_RdA_R + \frac{2}{3}A_R A_R A_R\right).
\end{equation}

\subsection{Higher-spin extensions}

More generally, we can consider an action of the form (\ref{15dec18}) in which $A_L$ and $A_R$ are connections of some algebra $g_L\oplus g_R$ instead of $ sl(2,\mathbb{R})_L\oplus sl(2,\mathbb{R})_R$. We would like to show that this leads to some three-dimensional higher-spin gauge theories coupled to gravity.

 Focusing first  on $g_L$ we pick any of its $sl(2,\mathbb{R})$ subalgebras and identify it with the gravitational $sl(2,\mathbb{R})_L$. This $sl(2,\mathbb{R})_L$ acts on the remaining generators of $g_L$, $T_L$, schematically, as follows
\begin{equation}
\label{15dec19}
[T_{La},T_L] = T_L.
\end{equation}
The Jacobi identity for $g_L$ then implies that $g_L$ forms a representation of $sl(2,\mathbb{R})_L$ with respect to the adjoint action (\ref{15dec19}). This representation can be decomposed into irreducible ones.

 We will assume that these representations are finite-dimensional. All finite-dimensional representations of $so(2,1)\sim sl(2,\mathbb{R})$ are classified by $s$ with $2s\in \mathbb{N}$ and the dimension being $2s+1$. For simplicity we will focus on representations with integer $s$. These can be realised as traceless symmetric rank-$(s-1)$ tensors $T_{L a(s-1)}$ of $so(2,1)$. The associated commutation relations read
\begin{equation}
\label{15dec20}
[T_{La},T_{Lb(s-1)}]=(s-1)\epsilon_{abc}T_{Lb(s-2)}{}^c.
\end{equation}

We will now treat the components of the connection $A^{a(s-1)}_L$, associated with $T_{a(s-1)}$, as small fields over the gravitational background. The equations of motion for $A^{a(s-1)}_L$ are just the zero-curvature conditions, contributing to the  $T_{a(s-1)}$ sector. More specifically, we find
\begin{equation}
\label{15dec21}
\begin{split}
0=F_{L}\Big|_{T_s}&=\left(dA_L +\frac{1}{2}[A_L,A_L]\right)\Big|_{T_s}\\
&=dA_L^{a(s-1)}T_{La(s-1)}+[A_L^bT_{Lb},A_L^{c(s-1)}T_{Lc(s-1)}]+\dots,
\end{split}
\end{equation}
where $\dots$ refer to terms of the form $A^{a(s_1-1)} A^{a(s_2-1)}$, with both $s_1$ and $s_2$ different from two, which in the free field approximation are irrelevant. Applying (\ref{15dec20}) for the commutator, we find the linearised equations to be
\begin{equation}
\label{15dec22}
dA_L^{a(s-1)}+(s-1)A_L^bA_L^{ca(s-2)}\epsilon_{bc}{}^a=0.
\end{equation}
Analogous considerations can be applied to the $g_R$ sector of the theory, which leads to 
\begin{equation}
\label{15dec23}
dA_R^{a(s-1)}+(s-1)A_R^bA_R^{ca(s-2)}\epsilon_{bc}{}^a=0.
\end{equation}

Changing the variables as in the gravity case to
\begin{equation}
\label{15dec24}
A_L^{a(s-1)}=\omega^{a(s-1)}+\frac{1}{l}e^{a(s-1)}, \qquad 
A_R^{a(s-1)}=\omega^{a(s-1)}-\frac{1}{l}e^{a(s-1)}
\end{equation}
we can rewrite (\ref{15dec22}), (\ref{15dec23}) as
\begin{equation}
\label{15dec25}
\begin{split}
d\omega^{a(s-1)}+(s-1)v^b\omega^{ca(s-2)}\epsilon_{bc}{}^a+\frac{s-1}{l^2}
h^be^{ca(s-2)}\epsilon_{bc}{}^a &=0,\\
d e^{a(s-1)}+(s-1)h^b\omega^{ca(s-2)}\epsilon_{bc}{}^a+(s-1)
v^be^{ca(s-2)}\epsilon_{bc}{}^a&=0.
\end{split}
\end{equation}
Note that since we are interested only in the linearised equations, we replaced gravitational fields with their background values
(\ref{10dec2}).

Finally, we dualise $\omega$'s in the Lorentz indices as
\begin{equation}
\label{15dec26}
\omega^{a(s-1)}=\frac{1}{s}\epsilon^{abc}\omega^{a(s-2)}{}_{b,c}.
\end{equation}
New $\omega$'s are traceless tensors with the Young symmetry $\mathbb{Y}(s-1,1)$. Thus, $\omega$ and $e$ reproduce the field content of the frame-like formulation of linearised massless higher-spin fields discussed in section (\ref{12:2}). 
\begin{zadacha}
Show that (\ref{15dec26}) can be inverted as
\begin{equation}
\label{1feb10}
\omega^{a(s-1),b}=-(s-1)\epsilon^{dab}\omega^{a(s-2)}{}_d.
\end{equation}
\end{zadacha}

It can be shown that equations (\ref{15dec25}) are equivalent to those following from (\ref{12dec12}). 
\begin{zadacha}
Show that
\begin{equation}
\label{1feb6}
v^be^{ca(s-2)}\epsilon_{bc}{}^a=v^{a,}{}_ce^{a(s-2)c},
\end{equation}
where $v^a$ is related to $v^{a,b}$ as different forms of the spin connection in (\ref{15dec13}). 
\end{zadacha}
This exercise implies that the first and the third terms on the left-hand side of the second equation in (\ref{15dec25}) combine into $\nabla e^{a(s-1)}$.
Similarly, the first two terms on the left-hand side of the first equation in (\ref{15dec26}) combine into $\nabla \omega^{a(s-1)}$.
\begin{zadacha}
By substituting (\ref{15dec26}) show that 
\begin{equation}
\label{1feb7}
h^b\omega^{ca(s-2)}\epsilon_{bc}{}^a=-\frac{1}{s-1}h_b\omega^{a(s-1),b}.
\end{equation}
\end{zadacha}
To get a differential equation for $\omega^{a(s-1),b}$ from the first equation in (\ref{15dec25}), we need to apply to the latter the operator on the right-hand side of (\ref{1feb10}). The third term then leads to a contraction, that we evaluate in the following exercise.
\begin{zadacha}
Show that 
\begin{equation}
\label{1feb11}
\begin{split}
&\epsilon^{dab}h^m \big((s-2)e^{ca(s-3)}{}_d\epsilon_{mc}{}^a + e^{ca(s-2)}\epsilon_{mcd} \big)\\
&\;=(s-1)(h^b e^{a(s-1)}-h^a e^{ba(s-2)})-(s-2)(h^me^{a(s-2)}{}_m\eta^{ba}-h^m e^{ba(s-2)}{}_m \eta^{aa}).
\end{split}
\end{equation}
\end{zadacha}
Summing up, we find that (\ref{15dec25}) amounts to
\begin{equation}
\label{1feb12}
\begin{split}
\nabla\omega^{a(s-1),b}-\frac{(s-1)^2}{R^2}
(\sigma_+ e)^{a(s-1),b}&=0,\\
\nabla e^{a(s-1)}-h_b\omega^{a(s-1),b}&=0.
\end{split}
\end{equation}
where $\sigma_+$ for general dimension was defined in (\ref{12dec14}). As one can check, (\ref{1feb12}), indeed, agrees with (\ref{12dec16}), (\ref{2feb7}) for general dimensions. It should be kept in mind, that the projector present in (\ref{2feb7})  becomes trivial in three dimensions and, therefore, can be dropped.

To summarise, we found that the action (\ref{15dec18}), for rather general assumptions on $g$, upon linearisation over the AdS background reproduces the direct sum of Fronsdal theories in $3d$. For this reason, theory (\ref{15dec18}) is naturally regarded as the higher-spin extension of $3d$ gravity. One should, however, keep in mind that neither graviton nor higher-spin fields propagate in $3d$. This follows from our general discussions on the representation theory of the Poincare and $so(d-1,2)$ groups: in $3d$ massless representations with $s\ge 2$ carry no degrees of freedom. This property is also a general feature of the Chern-Simons theories, which in the topologically trivial case describe only pure gauge solutions. 

Finally, we would like to make a general remark about the field choice. Namely, it may be interesting to attempt to rewrite non-linear action (\ref{15dec18}) in terms of the Fronsdal fields at the interacting level. Then, the torsion constraint becomes non-linear. As a result, solving for $\omega$ becomes feasible only in perturbations. In practice, rewriting of  (\ref{15dec18}) in terms of Fronsdal fields has been carried out only to leading orders  in interactions \cite{Campoleoni:2012hp} and the result does not display any apparent structures. In other words, it is only with the Chern-Simons setting that one can write the action (\ref{15dec18}) in a closed and concise form. In principle, one should expect that this phenomenon is pretty general: to be able to write an interacting theory in a closed form one should be fortunate with the choice of fields.

\subsection{Further reading}

For gravity with the cosmological constant, one can strengthen its similarity with the Yang-Mills theory by rewriting the gravity action in the form that involves the shifted curvature (\ref{15dec10}) quadratically \cite{MacDowell:1977jt}. Moreover, one can make the local $so(D-1,2)$ symmetry more manifest by employing $so(D-1,2)$ tensors \cite{Stelle:1979aj}.

The frame-like formalism for massless higher-spin theories was suggested in \cite{Vasiliev:1980as,Vasiliev:1986td,Lopatin:1987hz} and then extended to many other setups. These include, for example, partially-massless \cite{Skvortsov:2006at}, massive and continuous spin symmetric fields \cite{Zinoviev:2008ze,Ponomarev:2010st,Khabarov:2017lth} as well as massless mixed-symmetry fields in flat space \cite{Skvortsov:2008sh}.  For massive fields of arbitrary symmetry the frame-like action is not known, see, however, \cite{Zinoviev:2008ve} for some particular cases. 
To the best of our knowledge, in the AdS space the frame-like formulation is not known even for arbitrary massless fields, see \cite{Alkalaev:2003qv,Boulanger:2008up,Skvortsov:2009zu} for partial results. 
The frame-like formalism was also used to construct cubic vertices for higher-spin fields of different types, see e.g. \cite{Fradkin:1986qy,Alkalaev:2010af,Vasiliev:2011knf,Boulanger:2011se,Boulanger:2012dx,Zinoviev:2014zka}.

The Chern-Simons description for massless higher-spin theories in 3d was suggested in \cite{Blencowe:1988gj}. These theories attracted considerable attention recently in the holographic context \cite{Henneaux:2010xg,Campoleoni:2010zq}. Besides that, there is a proposal for a massless higher-spin theory in 3d extended with propagating scalar degrees of freedom \cite{Prokushkin:1998bq}. A holographic conjecture involving this theory was suggested in \cite{Gaberdiel:2010pz}, see \cite{Gaberdiel:2012uj} for review.

\section{Light-cone deformation procedure and chiral theories}
\label{sec:13}

At the very beginning of section \ref{sec:3}, when adapting UIR's of the Poincare group to the field theory setting
we  made a choice to make the Lorentz symmetry manifest by employing Lorentz tensors. This lead us to gauge symmetry for massless fields: it was necessary to argue away the excessive degrees of freedom that Lorentz tensors carry. In the following sections gauge symmetry served as a key principle that constrained interactions of massless fields. 

There are, however, alternative ways to deal with massless fields. Namely, to the best of our knowledge, there is no argument that would require that Lorentz symmetry should be realised in field theory in a manifest manner by employing  tensor fields.
For massless theories, this implies that we do not have to introduce unwanted degrees of freedom that come with Lorentz tensors and thereby we may avoid the difficulties related to gauge invariance and maintaining it at non-linear level.
 There is, however, a price to pay: due to the fact that Lorentz invariance is no longer manifest, it should be verified explicitly. 

The process of going from the manifestly Lorentz-covariant gauge description  to the non-covariant description without gauge degrees of freedom can be, in general, described as gauge fixing. When fixing a gauge in a known theory we arrive at an equivalent theory. Similarly, one may expect that consistent interactions in a gauge theory are in one-to-one correspondence with interactions in its gauge fixed counterpart. However,  it turns out that the relation between the two approaches to interactions is more subtle. The reason is that field theory comes together with various additional requirements, such as locality, and these may not map  into each  other  identically upon fixing a gauge. 
 
Below we will consider one example of such a non-manifestly Lorentz covariant approach, the light-cone gauge formalism. It is the simplest and most developed in the four-dimensional Minkowski space, so we will focus on this case only. As we will see below, the light-cone formalism allows one to construct additional consistent interactions of massless fields already at the cubic order.  Moreover, going further, it enables the construction of chiral higher-spin theories, which can be regarded as higher-spin counterparts of the self-dual Yang-Mills theory and self-dual gravity. This section mostly follows the review part of \cite{Ponomarev:2016cwi}.

\subsection{Massless UIR's in the light-cone gauge}
\label{sec:13.1}

For the purpose of this section, it will be convenient to define the Poincare algebra without $i$ factors, see (\ref{15dec1x1})
\begin{equation}
\label{15feb1}
\begin{split}
[\hat P^a,\hat P^b]&=0,\\
[\hat J^{ab},\hat P^c]&=\hat P^a \eta^{bc}-\hat P^b \eta^{ac},\\
[\hat J^{ab},\hat J^{cd}]&=\hat J^{ad}\eta^{bc}-\hat J^{bd}\eta^{ac}-\hat J^{ac}\eta^{bd}+\hat J^{bc}\eta^{ad}.
\end{split}
\end{equation}
In the rest of this section, hats will be systematically omitted.

In  the light-cone coordinates
\begin{equation}
\label{15feb2}
\begin{split}
x^{+}&=\frac{1}{\sqrt{2}}(x^3+x^0), \qquad x^{-}=\frac{1}{\sqrt{2}}(x^3-x^0),\\
x&=\frac{1}{\sqrt{2}}(x^1-ix^2), \qquad \bar x =\frac{1}{\sqrt{2}}(x^1 + ix^2),
\end{split}
\end{equation}
 the metric reads
\begin{equation}
\label{15feb3}
ds^2 = 2dx^+ dx^- + 2dx d\bar x.
\end{equation}
Accordingly, for the derivatives one has
\begin{equation}
\label{15feb4}
\begin{split}
\partial^- &= \frac{1}{\sqrt{2}}(\partial^3 - \partial^0), \qquad \partial^+ = \frac{1}{\sqrt{2}}(\partial^3+\partial^0), \\
\bar\partial&\equiv \partial^{\bar x} =\frac{1}{\sqrt{2}}(\partial^1+i\partial^2) ,\qquad 
\partial\equiv \partial^{ x} =\frac{1}{\sqrt{2}}(\partial^1-i\partial^2).
\end{split}
\end{equation}

We start by introducing an off-shell field, that transforms as
\begin{equation}
\label{15feb5}
\begin{split}
P^a \Phi^{\lambda} &= \partial^a \Phi^\lambda,\\
J^{ab} \Phi^\lambda &= (x^a\partial^b-x^b \partial^a+S^{ab})\Phi^\lambda,
\end{split}
\end{equation}
where $S^{ab}$ is the spin part of the angular momentum. In the light-cone formalism one requires
\begin{equation}
\label{15feb6}
S^{+a} \Phi^\lambda=0, \qquad S^{ab} \partial_a \Phi^\lambda=0.
\end{equation}
Then, the first condition in (\ref{15feb6}) implies that the only non-trivial components of $S$ are given by $S^{x\bar x}$, $S^{x-}$ and $S^{\bar x -}$. The second condition, in turn, allows one to express all these components in terms of $S^{x\bar x}$
\begin{equation}
\label{15feb7}
S^{x-}\Phi^\lambda=-S^{x\bar x} \frac{\partial}{\partial^+}\Phi^\lambda, \qquad S^{\bar x-}\Phi^\lambda = S^{x\bar x}\frac{\bar\partial}{\partial^+}\Phi^\lambda
\end{equation}
with the latter given by
\begin{equation}
\label{15feb8}
S^{x\bar x}\Phi^\lambda=-\lambda \Phi^\lambda.
\end{equation}
To put the field on-shell one imposes the standard equation of motion
\begin{equation}
\label{15feb9}
\partial_a \partial^a \Phi^\lambda=0.
\end{equation}

It can be shown that  $J$ and $P$ defined above, indeed, generate the Poincare algebra. Moreover, $\lambda$ appearing in  (\ref{15feb9}) is helicity. 
In appendix \ref{app:2x1} we review how helicity is defined and show that (\ref{15feb8}) is consistent with this definition. 
For the following discussion it will be sufficient to know that the spin-$s$ field is equivalent to a pair of fields with helicities $\pm s$ with  some extra  reality conditions imposed. 

\subsection{Light-cone gauge for the Fronsdal theory}
\label{sec:13.2}

Above we demonstrated how massless representations in $4d$ flat space can be realised in the light-cone formalism. This formalism is 
completely self-sufficient and does not require connection to the Fronsdal approach to justify its consistency. Still, for completeness, in this section we
will demonstrate that the Lorentz-covariant approach of the Fronsdal theory leads to the light-cone theory after gauge fixing. A basic and detailed discussion of the light-cone gauge fixing for point particles, strings and lower-spin massless fields can be found in \cite{Zwiebach:789942}. In the following sections this connection will be used to compare the results for higher-spin interactions in two formalisms.

The gauge fixing procedure starts by imposing 
\begin{equation}
\label{15feb10}
\varphi^{+ a(s-1)}=0
\end{equation}
or, in other words, all components of the Fronsdal field with at least one ''$+$'' index are set to zero. It is not hard to see that this gauge can be achieved for all $\varphi$, except those that satisfy $\partial^+\varphi =0$. Strictly speaking, solutions of $\partial^+\varphi=0$ 
make the original and the gauged fixed theories inequivalent. Additional subtleties with these modes occur when we use $x^+$ as the time variable in the quantization context. For our purposes this issue will be irrelevant. An interested reader can find related discussions in \cite{Perry:1994kp,Burkardt:1995ct,6db1d5c6aea346a68475d79dd4ab667e,Harindranath:1996hq,Heinzl:2000ht,Mannheim:2020rod}.

\begin{zadacha}
Consider a massless spin-1 field in the Fourier space. Show that $A^+(p)=0$ can be achieved for all $A$ except those with $p^+=0$. Analogously, consider the spin-2 case and show that $h^{++}$ can be set to zero by properly adjusting $\xi^+$ in the gauge transformation. Then show that $h^{+a}=0$ can be achieved by adjusting the remaining components of $\xi^{a}$.
\end{zadacha}

Next, we consider the Fronsdal equations (\ref{14sep19}) with two indices along ''+'' direction, while other indices being arbitrary. Due to the gauge choice (\ref{15feb10}), the only term that survives is
\begin{equation}
\label{15feb11}
\partial^+\partial^+\varphi_m{}^{ma(s-2)}\approx 0.
\end{equation}
Again, we use the fact that $\partial^+$ can be inverted, so (\ref{15feb11}) entails
\begin{equation}
\label{15feb12}
\varphi_m{}^{ma(s-2)}\approx 0.
\end{equation}
Let us introduce $I$, $J$, $\dots$ that will only run over $\{x,\bar x\}$. Then, (\ref{15feb12}) implies
\begin{equation}
\label{15feb13}
\varphi_I{}^{IJ(s-2)}\approx 0.
\end{equation}
It is not hard to see that traceless symmetric two-dimensional tensor $\varphi^{I(s)}$ has only two non-vanishing components, which will be denoted as
\begin{equation}
\label{15feb14}
\Phi^{s} \equiv \varphi^{x(s)}, \qquad \Phi^{-s} \equiv \varphi^{\bar{x}(s)}.
\end{equation}
For real Fronsdal fields these are complex conjugate to each other. 

Next, we consider the Fronsdal equation with one of the indices taking value ''+''. Keeping in mind (\ref{15feb12}), we find
\begin{equation}
\label{15feb15}
\partial^+\partial_m\varphi^{m a(s-1)}\approx 0 \qquad \Rightarrow \qquad \partial_m\varphi^{m a(s-1)}\approx 0.
\end{equation}
Focusing on $a$'s with values in  $\{x,\bar x\}$, we obtain
\begin{equation}
\label{15feb16}
\partial^+ \varphi^{- I(s-1)}+ \partial_J \varphi^{J I(s-1)}\approx 0 \qquad \Rightarrow \qquad \varphi^{-I(s-1)}\approx -\frac{\partial_J}{\partial^+}\varphi^{JI(s-1)}.
\end{equation}
This implies that $\varphi$ with a single index taking value ''-'' can be expressed in terms of the two fields we introduced in (\ref{15feb14}).

Analogously, considering (\ref{15feb15}) with one index taking value ''$-$'', while the remaining being from the set $\{x,\bar x\}$, we find 
\begin{equation}
\label{15feb17}
\partial^+ \varphi^{-- I(s-2)}+ \partial_J \varphi^{-J I(s-2)}\approx 0 \qquad \Rightarrow \qquad \varphi^{--I(s-2)}\approx -\frac{\partial_J}{\partial^+}\varphi^{-JI(s-2)}.
\end{equation}
By employing (\ref{15feb16}) we obtain
\begin{equation}
\label{15feb18}
 \varphi^{--I(s-2)}\approx \left(\frac{\partial_J}{\partial^+}\right)^2\varphi^{JJI(s-2)},
\end{equation}
so $\varphi$'s with two indices taking value ''$-$'' can also be expressed in terms of $\Phi^{\pm s}$.

Proceeding iteratively, we find that
\begin{equation}
\label{15feb19}
 \varphi^{-(n)I(s-n)}\approx \left(-\frac{\partial_J}{\partial^+}\right)^n\varphi^{J(n)I(s-n)}.
\end{equation}
This shows that all components of the Fronsdal field that were not gauged away by (\ref{15feb10}) can be expressed in terms of the two dynamical fields $\Phi^{s}$ and $\Phi^{-s}$.

Finally, we substitute $\varphi$ expressed in terms of $\Phi^{\pm s}$ into the Fronsdal action (\ref{14sep13}). Since in the light-cone gauge the Fronsdal field is traceless (\ref{15feb12})
and divergenceless (\ref{15feb15}), only the first term in (\ref{14sep13}) survives. This leads to
\begin{equation}
\label{15feb20}
S=-\int d^4x \partial_m \Phi^{-s} \partial^m \Phi^s.
\end{equation}
Summing over spins, we find the action in a more symmetric form
\begin{equation}
\label{15feb21}
S_2 \equiv \int d^4x L_2, \qquad L_2 = -\frac{1}{2}\sum_{\lambda \in {\mathbb{Z}}} \partial_a\Phi^{-\lambda}\partial^a \Phi^\lambda.
\end{equation}

We claim that fields $\Phi^{\pm s}$ that we introduced in (\ref{15feb14}) define representations of helicities $\pm s$ from the previous section. In particular, it is trivial to see that upon variation (\ref{15feb20}), indeed, gives (\ref{15feb9}). It is also straightforward to see that the tensor transformation law for $\varphi$ leads to 
\begin{equation}
\label{27feb1x1}
S^{x\bar x}\varphi^{x(s)} = -s \varphi^{x(s)} , \qquad S^{x\bar x}\varphi^{\bar x(s)} = s\varphi^{\bar x(s)},
\end{equation}
which reproduces (\ref{15feb8}) with helicities $\pm s$  (\ref{15feb14}). The remaining conditions are somewhat harder to obtain, as some of the Lorentz transformations break the light-cone gauge, hence, these should be accompanied with the compensating gauge transformations. This computation can be carried out in a straightforward manner. Alternatively,  one can notice that the light-cone gauge (\ref{15feb10}) together with the vanishing divergence of the potentials (\ref{15feb15}) entails (\ref{15feb6}), which, in turn, allows one to derive (\ref{15feb7}).

\begin{zadacha}
Derive (\ref{15feb6})-(\ref{15feb8}) for the spin-1 case from gauge fixing the spin-1 potential as explained above.
\end{zadacha}

In this section we reviewed how the light-cone gauge fixing procedure works for the Fronsdal theory. Eventually, it results in a very simple action (\ref{15feb20}) and involves only two off-shell fields $\Phi^{\pm s}$, reminiscent of scalar fields. It is worth stressing, however, that only $\Phi^0$ actually transforms as a scalar field, while fields with non-vanishing helicities have a more non-trivial transformation rule given in the previous section.

\subsection{Poincare charges in free theory}
\label{sec:13.3}
Poincare symmetry of (\ref{15feb20}) despite being present is not manifest. For free fields, at least, we know how $\Phi^{\lambda}$ transform and invariance of (\ref{15feb20})  can be explicitly checked. However, when we proceed to the non-linear level, the transformation law may get deformed. Thus, when constructing interactions, not only do we need to check that the action is Poincare invariant, we also need to find a suitable deformation of  field transformations under the Poincare algebra. In this regard, the perturbative procedure for construction of interactions in the light-cone gauge is reminiscent of the covariant deformation procedure, for which, we  had to find how the gauge transformations get deformed as well as to check that the action is gauge invariant.

In practice Poincare invariance of the action will be implemented following the approach suggested in \cite{Dirac:1949cp}. Namely, using the Hamiltonian formalism to dynamics, we will first identify the phase space of the theory -- fields and the associated canonical momenta at fixed time. The action of symmetries on this phase space can be realised via the Poisson/Dirac bracket with charges -- suitably chosen functions on the phase space. The fact that these charges generate the Poincare algebra when acting on fields translates into the requirement that the charges commute with the Poisson/Dirac bracket in the same way as the associated generators of the Poincare algebra. At the non-linear level the charges may get deformed, but Poincare invariance of the deformed theory implies that commutation relations between them should remain the same.

 Before proceeding to the non-linear level, we first need to identify the necessary structures in  the free theory. 
To start, as we are going to use the Hamiltonian formalism we need to define a space-time variable, which will be regarded as the Hamiltonian time. This can be done in many different ways, though, satisfying the requirement that equal-time surfaces serve as the Cauchy surfaces, in the sense that once the initial data is defined on these surfaces, one can  reproduce the fields in the whole space-time via the Hamilton equations. 

A notorious feature of the Hamiltonian formalism when applied to field theories is that it breaks manifest Lorentz covariance. 
As we have already broken manifest Lorentz covariance by imposing the light-cone gauge (\ref{15feb10}) and
since it seems reasonable to reduce manifest covariance breaking to the minimum, we will choose the Hamiltonian time as
\begin{equation}
\label{15feb22}
t=x^+.
\end{equation}
The remaining coordinates $x^\perp=\{ x,\bar x, x^-\}$ will be regarded as space coordinates. Note that equal-time surfaces are then hyperplanes with one light-like direction, but not light-cones. Due to that terminology ''light-cone approach'' is somewhat unfortunate and some authors refer to this formalism as the ''light-front'' one. We will stick to ''light-cone'' as it is more typical in the higher-spin literature. 

Considering that $x^+$ is time, time derivatives are given by $\partial^-$ and, accordingly, the canonical momenta are defined as
\begin{equation}
\label{15feb23}
\Pi^\lambda \equiv \frac{\delta L_2}{\delta (\partial^-\Phi^\lambda)}= -\partial^+\Phi^{-\lambda}.
\end{equation}
The associated Poisson bracket is
\begin{equation}
\label{15feb24}
[\partial^+ \Phi^\lambda(x^\perp,x^+),\Phi^\mu(y^\perp,x^+)]_{P}=\delta^{\lambda+\mu,0}\delta^3(x^\perp,y^\perp),
\end{equation}
where $\delta^{\lambda+\mu,0}$ is the Kronecker delta,
and the Hamiltonian is given by
\begin{equation}
\label{15feb25}
H_2\equiv \sum_{\lambda\in\mathbb{Z}}\int d^3x^\perp (\Pi^\lambda \partial^-\Phi^\lambda-L_2)=\sum_{\lambda\in\mathbb{Z}}\int d^3x^\perp \partial \Phi^{-\lambda}\bar\partial\Phi^\lambda
\end{equation}
with the integral going over an equal-time surface.

The right-hand side of  (\ref{15feb23}) does not involve time derivatives $\partial^-\Phi$ and, thereby, presents a constraint. It is analogous to a constraint for a point particle, that expresses momentum as a function of coordinates only. This situation is a consequence of a theory being linear in time derivatives. We will not review here the analysis of constraints, as it is very standard. The end result in a given case is that the Poisson bracket (\ref{15feb24}) gets replaced with the Dirac bracket
\begin{equation}
\label{15feb26}
[\partial^+ \Phi^\lambda(x^\perp,x^+),\Phi^\mu(y^\perp,x^+)]_D=\frac{1}{2}\delta^{\lambda+\mu,0}\delta^3(x^\perp,y^\perp),
\end{equation}
which can be written more conveniently as
\begin{equation}
\label{17feb1}
[ \Phi^\lambda(x^\perp,x^+),\Phi^\mu(y^\perp,x^+)]_D=\frac{1}{\partial_x^+ - \partial_y^+}\delta^{\lambda+\mu,0}\delta^3(x^\perp,y^\perp).
\end{equation}
The remaining aspects of the Hamiltonian formalism remain intact. In particular, the time evolution is generated via the Dirac bracket with the Hamiltonian
\begin{equation}
\label{17feb2}
\partial^-F(\Phi) = [F(\Phi),H_2]_D.
\end{equation}
For a field itself it gives
\begin{equation}
\label{17feb3}
\partial^-\Phi^\lambda(x) = -\frac{\partial\bar\partial}{\partial^+}\Phi^\lambda(x),
\end{equation}
which, as it is not hard to see, is consistent with the variation of (\ref{15feb21}).
Note that if the Poisson bracket was not replaced with the Dirac one, in (\ref{17feb3}) we would have gotten an extra factor of two.
Below the lower index "D" for the Dirac bracket will be often omitted.

The action (\ref{15feb21}) is invariant with respect to the Poincare symmetry with the action of generators defined in (\ref{15feb5})-(\ref{15feb8}). Following the standard procedure, one identifies the Noether currents
\begin{equation}
\label{17feb4}
\begin{split}
P^i \qquad & \to \qquad T^{i,j}=\sum_{\lambda}\frac{\delta L_2}{\delta (\partial_j \Phi^\lambda)} \partial^i \Phi^\lambda - \eta^{ij}L_2,\\
J^{ij} \qquad  & \to \qquad L^{ij,k}=x^i T^{j,k}-x^j T^{i,k}+R^{ij,k},
\end{split}
\end{equation}
where $R^{ij,k}$  is the spin current 
\begin{equation}
\label{17feb5}
R^{ij,k}=\sum_{\lambda} \frac{\delta L}{\delta (\partial_k \Phi^\lambda)} S^{ij} \Phi^\lambda.
\end{equation}
Currents (\ref{17feb4}) define the associated Noether charges 
\begin{equation}
\label{17feb6}
Q_2[P^i]=\int d^3x^\perp T^{i,+}, \qquad Q_2[J^{ij}]=\int d^3x^\perp L^{ij,+},
\end{equation}
where the integral  goes over an equal-time surface. For simplicity, we will choose this surface to be $x^+=0$. 
Explicitly, for charges (\ref{17feb6}) we find
 \begin{equation}
  \label{17feb7}
  Q_2[P^i] = -\sum_{\lambda}\int d^3x^\perp \partial^+\Phi^{-\lambda}  p_2^i \Phi^\lambda, \qquad  Q_2[J^{ij}] = -\sum_{\lambda}\int d^3x^\perp \partial^+\Phi^{-\lambda}  j_2^{ij} \Phi^\lambda,
  \end{equation}
where 
 \begin{equation}
\begin{split}
p_2^+ &=\partial^+, \qquad  p_2^- = -\frac{\partial \bar\partial}{\partial^+}, \qquad\qquad \quad\;  p_2 = \partial,  \qquad\qquad\quad\;    \bar p_2=
\bar\partial,\\
j_2^{+-} &=  - x^-\partial^+,\qquad   j_2^{x\bar x} = x\bar\partial - \bar x \partial -\lambda, \\
j_2^{x+}& = x\partial^+, \qquad   j_2^{x-} = -x\frac{\partial \bar\partial}{\partial^+} - x^- \partial+\lambda\frac{\partial}{\partial^+}, \\
j_2^{\bar x+}& = \bar x\partial^+, \qquad
   j_2^{\bar x-} = -\bar x\frac{\partial \bar\partial}{\partial^+} - x^- \bar\partial
-\lambda\frac{\bar\partial}{\partial^+}.
\end{split}
\label{17feb8}
\end{equation}
To arrive to (\ref{17feb8}) we dropped terms that involve $x^+$, as these anyway vanish on $x^+=0$.
\begin{zadacha}
Derive $Q_2[P^+]$ and $Q_2[P^-]$ starting from (\ref{17feb4}) for the respective Noether currents.
\end{zadacha}

These charges, when commuted with the field from the right, generate the Poincare algebra action
  \begin{equation}
  \label{17feb9}
 [\Phi^\lambda,Q_2 [P^i]] = p_2^i \Phi^\lambda, \qquad  [\Phi^\lambda, Q_2[J^{ij}]] = j_2^{ij} \Phi^\lambda.
  \end{equation} 
  This, in particular, implies that they commute with each other the same way as the respective symmetry generators. More explicitly,   for two symmetry generators $T_1$ and $T_2$ one has
  \begin{equation}
  \label{17feb10}
  [Q_2[T_1],Q_2[T_2]]_D=Q_2[[T_1,T_2]_L],
  \end{equation}
where $[,]_{L}$ is the Lie algebra commutator.
Note, however, that the action generated by $Q_2[T]$ is different from (\ref{15feb5})-(\ref{15feb8}). This happens because we realised symmetries on a phase space of the theory, given by $\Phi$ and $\Pi$ at $x^+=0$. This is why $x^+$ dependence dropped out. Besides that, generators that involve time derivatives are now realised on the equal-time surface phase space with the two actions related by the Hamiltonian equations of motion.

\subsection{Interactions via the light-cone deformation procedure}
\label{sec:13.4}

In an interacting theory charges that generate symmetries may get deformed. For a symmetry generator $T$ one has
\begin{equation}
\label{17feb10o1}
Q[T] \equiv Q_2[T] + \delta Q[T] = Q_2[T]+g Q_3[T]+ g^2 Q_4[T]+\dots ,
\end{equation}
where $Q_3$ is cubic in fields, $Q_4$ is quartic in fields, etc. We require that the Poincare symmetry of the theory survives at the non-linear level. This implies that non-linear charges (\ref{17feb10o1}) still generate the Poincare algebra or, in other words, that $Q[T]$ commute  the same way as $T$
\begin{equation}
\label{17feb10x1}
[Q[T_1],Q[T_2]]=Q[[T_1,T_2]].
\end{equation}

From the Hamiltonian formalism perspective, when deforming a theory with interaction terms, we deform the Hamiltonian which, in turn, changes the time evolution. This change has no effect on the way symmetries act on the phase space variables, unless these symmetries leave the Cauchy surface and changes in the dynamics becomes relevant.
An example of a generator that leaves the Cauchy surface is time translations, which are generated by the Hamiltonian itself, which is, clearly, deformed once interactions are added. Generators of this type are called Hamiltonians or  \emph{dynamical} generators and they all may get deformed in the interacting theory. The remaining generators -- those that leave the Cauchy surface invariant -- are called \emph{kinematical} and do not receive any corrections\footnote{There are some exotic theories in which kinematic generators receive corrections as well \cite{Weinberg}.}.

\begin{zadacha}
Following the same steps that we used in the previous section to construct the Poincare charges for the free higher-spin fields, compute $Q[P^0]$, $Q[P^i]$ and $Q[J^{0i}]$ in scalar theory \begin{equation}
\label{17feb11}
S = \int d^dx \left(\frac{1}{2}\varphi \Box \varphi + \frac{g}{6}\varphi^3\right)
\end{equation}
with $x^0$ taken as time. Make sure that only  dynamical charges $Q[P^0]$ and $Q[J^{0i}]$ receive non-linear corrections. 
\end{zadacha}

With our choice of the Cauchy surface $x^+=0$,  the only generators transverse to it are $P^-$, $J^{x-}$ and $J^{\bar x -}$. Therefore, these generators are dynamical and they will be collectively denoted as $D$. The remaining generators are  kinematical and they will be denoted $K$. Accordingly, at the non-linear level one has
\begin{equation}
\label{17feb10x2}
Q[K]=Q_2[K].
\end{equation}

The commutators of the Poincare algebra can  be broken into classes depending on types of generators they feature. The first type of commutators is of the form
\begin{equation}
\label{17feb12}
[K_1,K_2]=K_3.
\end{equation}
Since the kinematic generators do not receive corrections at the non-linear level, for these generators (\ref{17feb10x1}) is automatically satisfied as a consequence of  (\ref{17feb10}).
Other two groups of commutators are of the form
\begin{equation}
\label{17feb13}
[K,D] = K, \qquad [K,D_1]=D_2.
\end{equation}
For them (\ref{17feb10x1}) leads to
\begin{equation}
\label{17deb14}
[Q_2[K],\delta Q[D]]=0, \qquad [Q_2[K],\delta Q[D_1]]=\delta Q_2[D_2].
\end{equation}
These equations feature $Q_2[K]$, which are known, while unknown $\delta Q[D]$ enter linearly. For this reason, these constraints, called \emph{kinematical}, can be immediately solved to all orders. 
The most non-trivial type of commutators is of the form
\begin{equation}
\label{17feb15}
[D_1,D_2]=0
\end{equation}
as it involves terms quadratic in  $\delta Q[D]$.

Eventually, once the solution to the consistency conditions (\ref{17feb10x1}) is found the action of the theory is obtained from its Hamiltonian 
\begin{equation}
\label{20apr1x1}
H\equiv Q[P^-]
\end{equation}
in the standard way
\begin{equation}
\label{17feb15xx1}
S=\sum_\lambda\Big[ \int d^4 x \Pi^\lambda \partial^-\Phi^\lambda\Big]-\int dx^+ H.
\end{equation}

\subsubsection{Kinematical constraints}
\label{sec:13.4.1}

With the general procedure to construct interactions settled, let us be more specific and see what it leads to in practice. We will start by considering kinematical constraints. 

To proceed further it will be convenient to Fourier transform transverse coordinates
  \begin{align}
\notag
  \Phi(x,x^+) &= (2\pi)^{-\frac{3}{2}} \int{e^{+i (x^- p^+ + \bar x p +x\bar p)}\Phi(p,x^+)d^{3}p^\perp},\\
    \label{21sep17}
   \Phi(p,x^+) &= (2\pi)^{-\frac{3}{2}} \int{e^{-i (x^- p^+ + \bar x p +x\bar p)}\Phi(x,x^+)d^{3}x^\perp}
  \end{align}
and then follow this with the change
of variables $p=iq$ to avoid $i$ factors. In effect this amounts to the replacement 
\begin{equation}
\label{17feb16}
\frac{\partial}{\partial x^i} \to q_i,\qquad x^i \to -\frac{\partial}{\partial q_i}.
\end{equation}
Some of the useful formulas derived previously in the Fourier transformed version are presented in appendix \ref{app:2}. It is also conventional to denote $\beta \equiv q^+$.

To evaluate commutators with the quadratic charges of the kinematical generators one can use
\begin{equation}
\label{18feb1}
\begin{split}
[F(\Phi),Q_2[P^{i}]] &= [\Phi, Q_2[P^{i}]]\frac{\delta F(\Phi)}{\delta \Phi}=p_2^{i}\Phi \frac{\delta F(\Phi)}{\delta \Phi}\\
[F(\Phi),Q_2[J^{ij}]] &= [\Phi, Q_2[J^{ij}]]\frac{\delta F(\Phi)}{\delta \Phi}=j_2^{ij}\Phi \frac{\delta F(\Phi)}{\delta \Phi}
\end{split}
\end{equation} 
with $p_2$ and $j_2$ given in (\ref{29sep6}). 

Now, we consider dynamical generator $P^-$ with the associated charge being the Hamiltonian $H$. For the charge $H_n$ which is $n$-linear in fields we make an ansatz, which is just the most general functional on the $n$-particle phase space
\begin{equation}
\label{18feb2}
H_n =   \frac{1}{n!}\sum_{\lambda_i}\int d^{3n}q^\perp   \tilde h^{\lambda_1 \dots \lambda_n}_n\left(q_i^\perp;\frac{\partial}{\partial q_i^\perp}\right) \prod_{i=1}^n\Phi^{\lambda_i}(q^\perp_i).
\end{equation}
Here we assume that derivatives in $\tilde h$ are to the right of $q$, so they act on $\Phi$ only. By requiring 
\begin{equation}
\label{18feb3}
[H_n, Q_2[P^x]]=0, \qquad [H_n, Q_2[P^{\bar x}]]=0, \qquad 
[H_n, Q_2[P^+]]=0, 
\end{equation}
we find that $\tilde h$ does not depend on these derivatives, moreover, it has a factor with a momentum-conserving delta function in spatial directions. In other words, we find that 
\begin{equation}
\label{18feb4}
  H_n=  \frac{1}{n!}\sum_{\lambda_i}\int d^{3n}q^\perp \delta^3 (\sum_{i=1}^n q^\perp_i) h^{\lambda_1 \dots \lambda_n}_n(q_i^\perp) \prod_{i=1}^n\Phi^{\lambda_i}(q^\perp_i).
\end{equation}
This result could have been expected: indeed, translation invariance leads to the absence of manifest dependence on coordinates $x\sim \frac{\partial}{\partial q}$ as well as to momentum conservation.

Similar analysis for commutators of 
\begin{equation}
\label{18feb5}
  J_n \equiv Q_n[J^{x-}] , \qquad   \bar J_n  \equiv Q_n[J^{\bar x-}]
\end{equation}
leads to
 \begin{equation}
\begin{split}
  J_n 
  &=
    \frac{1}{n!}\sum_{\lambda_i}\int d^{3n}q^\perp \delta^3 (\sum_{i=1}^n q^\perp_i) \Big[j_n^{\lambda_1 \dots \lambda_n} (q_i^\perp)-\frac{1}{n}
  h_n^{\lambda_1\dots \lambda_n}(q_i^\perp)\big( \sum_j \frac{\partial}{\partial \bar q_j}\big)\Big]\prod_{i=1}^n\Phi^{\lambda_i}(q^\perp_i),\\
    \label{22sep1o1}
  \bar J_n 
 & =  \frac{1}{n!}\sum_{\lambda_i}\int d^{3n}q^\perp \delta^3 (\sum_{i=1}^n q^\perp_i) \Big[\bar j_n^{\lambda_1 \dots \lambda_n}(q_i^\perp) -\frac{1}{n}
  h_n^{\lambda_1\dots \lambda_n}(q_i^\perp)\big( \sum_j \frac{\partial}{\partial  q_j}\big)\Big]\prod_{i=1}^n\Phi^{\lambda_i}(q^\perp_i).
  \end{split}
\end{equation}
 The shift of the integral kernels in (\ref{22sep1o1}) by $h\partial_q$ was made for convenience, more precisely, to isolate dependence on $q$ derivatives into a separate term and then deal with $j$'s that only depend on $q$.

Next, we proceed to the kinematical commutators with $Q_2[J]$. After rather straightforward computations one finds
\begin{equation}
\label{18feb6}
\begin{split}
[Q_2[J^{x+}],H_n]&=0\qquad \Rightarrow \qquad \delta^3 (\sum_{i=1}^n q^\perp_i)\sum_{i=1}^n \beta_i \frac{\partial}{\partial \bar q_i} h_n^{\lambda_1\dots \lambda_n}=0,\\
[Q_2[J^{\bar x+}],H_n]&=0\qquad \Rightarrow \qquad \delta^3 (\sum_{i=1}^n q^\perp_i)\sum_{i=1}^n \beta_i \frac{\partial}{\partial  q_i} h_n^{\lambda_1\dots \lambda_n}=0,\\
[Q_2[J^{x\bar x}],H_n]&=0\qquad \Rightarrow \qquad \delta^3 (\sum_{i=1}^n q^\perp_i)\sum_{i=1}^n (N_{q_i}- N_{\bar{q}_i}+\lambda_i)h_n^{\lambda_1\dots \lambda_n}=0,\\
[Q_2[J^{+-}],H_n]+H_n&=0\qquad \Rightarrow \qquad \delta^3 (\sum_{i=1}^n q^\perp_i) \sum_{i=1}^n \beta_i \frac{\partial}{\partial \beta_i} h_n^{\lambda_1\dots \lambda_n}=0.
\end{split}
\end{equation}
Analogous equations can be found for $J$ and $\bar J$.
\begin{zadacha}
Derive the constraint on $h$ in the first line of (\ref{18feb6})
\end{zadacha}

Equations (\ref{18feb6}) can be easily solved. In particular, the first two equations imply that $h$ can depend on $q$ and $\bar q$ only via their combinations with $\beta $\footnote{These equations have more solutions when understood in the sense of distributions. We will not discuss distributional solutions here.}
\begin{equation}
\label{18feb7}
\bar{\mathbb{P}}_{ij} \equiv \bar q_i\beta_j -\bar q_j \beta_i , \qquad \mathbb{P}_{ij} \equiv q_i\beta_j -q_j \beta_i.
\end{equation}
It is worth remarking that not all $\bar{\mathbb{P}}_{ij}$ and $\mathbb{P}_{ij}$ are independent. In the three-point case this will be discussed below.

The remaining two conditions (\ref{30sep4}), (\ref{30sep5})  specify the homogeneity degrees of 
$h_n$ on its arguments 
\begin{equation}
\label{18feb7x1}
\begin{split}
\sum_{i,j=1}^n \left(\mathbb{P}_{ij}\frac{\partial}{\partial \mathbb{P}_{ij}}-\bar{\mathbb{P}}_{ij}\frac{\partial}{\partial \bar{\mathbb{P}}_{ij}}\right)h_n^{\lambda_1\dots \lambda_n} +\sum_{i=1}^n \lambda_i h_n^{\lambda_1\dots \lambda_n} &=0,\\
\sum_{i,j=1}^n \left(\mathbb{P}_{ij}\frac{\partial}{\partial \mathbb{P}_{ij}}+\bar{\mathbb{P}}_{ij}\frac{\partial}{\partial \bar{\mathbb{P}}_{ij}}\right)h_n^{\lambda_1\dots \lambda_n} +\sum_{i=1}^n \beta_i\frac{\partial}{\partial \beta_i} h_n^{\lambda_1\dots \lambda_n} &=0.
\end{split}
\end{equation}
Constraints for $j_n$ and $\bar j_n$ are solved analogously.
This finishes the discussion of the kinematical constraints to all orders. 

\subsubsection{Dynamical constraints}
\label{sec:13.4.2}
There are three dynamical constraints to satisfy
\begin{equation}
\label{19feb1}
[H,J]=0, \qquad [H,\bar J]=0, \qquad [J,\bar J]=0.
\end{equation}
Focusing on the first commutator and expanding it in the number of fields at $n$-th order we obtain
\begin{equation}
\label{19feb2}
[H_2,J_n]+[H_3,J_{n-1}]+\dots +[H_{n-1},J_3]+[H_n,J_2]=0.
\end{equation}

As in the covariant deformation procedure, one starts by analysing the lowest order in deformations
\begin{equation}
\label{19feb2x1}
[H_2,J_3]+[H_3,J_2]=0.
\end{equation}
At this order we expect to find conditions on first non-linear perturbations $H_3$ and $J_3$. More precisely, as we will see below, this condition allows us to solve for $J_3$ in terms of $H_3$ for any $H_3$. With extra requirements, namely, locality of $J_3$, we will find non-trivial constraints on $H_3$.
Since (\ref{19feb2x1}) is linear in deformations, as a result we obtain a set of independent solutions for $H_3$, which can be multiplied by yet to be fixed cubic coupling constants.

At the next order  one considers 
\begin{equation}
\label{19feb2x2}
[H_2,J_4]+[H_3,J_3] +[H_4,J_2]=0.
\end{equation}
Similarly, it allows us to solve for $J_4$ in terms of $H_4$ and $H_3$. With additional locality requirements on  $J_4$ this constraint becomes by far more non-trivial to satisfy.
Analogously, one proceeds to higher orders.

Since $H_2$ and $J_2$ are fixed by the free theory, we can evaluate the first and the last term in (\ref{19feb2}) immediately. An explicit computation gives
  \begin{equation}
  [H_2, J_n]  
  =  -\frac{1}{n!}\sum_{\lambda_i}\int d^{3n}q^\perp \delta (\sum_{i=1}^n q^\perp_i)  
  {\cal H}
 \Big[j_n^{\lambda_1 \dots \lambda_n} +\frac{1}{n}
 \big( \sum_j \frac{\partial}{\partial \bar q_j}\big) h_n^{\lambda_1\dots \lambda_n}\Big]
 \prod_{i=1}^n\Phi^{\lambda_i}(q^\perp_i),
  \label{19feb3}
  \end{equation}
where 
  \begin{equation}
  \label{19feb4}
  {\cal H}\equiv \sum_{i=1}^n p^-_2(q^\perp_i).
  \end{equation}
Similarly, for the last commutator in (\ref{19feb2}) we find
 \begin{align}
  \label{19feb5}
 [H_n,J_2] = \frac{1}{n!}\sum_{\lambda_i}\int d^{3n}q^\perp \delta (\sum_{i=1}^n q^\perp_i){\cal J} h_n^{\lambda_1 \dots \lambda_n} 
 \prod_{i=1}^n\Phi^{\lambda_i}(q^\perp_i),
 \end{align}
 where
 \begin{equation}
 \label{19feb6}
{\cal J}=\sum_{i=1}^n \Big(-\frac{q_i\bar q_i}{\beta_i}\frac{\partial}{\partial \bar q_i}-q_i \frac{\partial}{\partial \beta_i}+\lambda_i \frac{q_i}{\beta_i}\Big).
 \end{equation}
The remaining terms are evaluated using that
\begin{equation}
\label{19feb6x1}
[F(\Phi),G(\Phi)]=[\Phi^i,\Phi^j]\frac{\delta F(\Phi)}{\delta\Phi^i} \frac{\delta G(\Phi)}{\delta\Phi^j}.
\end{equation}

The second commutator in (\ref{19feb1}) does not require a separate computation as it is just the complex conjugate to the first one. Besides that, it can be shown that once $J_n$ and $\bar J_n$ are solved for in terms of $H_n$, the third commutator 
 in (\ref{19feb1}) at the relevant order is satisfied identically, see \cite{Ponomarev:2016lrm} for details.
 
 Eventually, the action (\ref{17feb15xx1}) acquires the form
 \begin{equation}
\label{17feb15x1}
S
=S_2 - \sum_{n=3}^\infty \frac{1}{n!}\int d^4x  h_n \prod_{i=1}^n\Phi^{\lambda_i}.
\end{equation}
Therefore, up to numerical factors $h_n$ give vertices of the theory in the light-cone gauge.

\subsection{Applications of the light-cone deformation procedure}
\label{sec:13.5}

We will now apply the light-cone deformation procedure to the construction of interactions for massless higher-spin fields. At the first step we consider cubic interactions.

Before doing that, we would like to comment on how locality is defined in the light-cone gauge formalism.
First, note that, vertices in the light-cone deformation procedure do not involve $\partial^-\sim q^-$, as it is the time derivative from the Hamiltonian formalism perspective, while the Hamiltonian is a function of the phase space variables only. From the field theory perspective, dependence on $\partial^-$ can be always eliminated by field redefinitions\footnote{As explained in section \ref{section:7}, once a vertex is trivial on-the-free-shell, it can be eliminated by field redefinitions. Vertices of the form $(\partial_i^- + \partial_i\bar\partial_i /\partial_i^+)(\dots)$ clearly vanish on the free shell, therefore, field redefinitions allow one to trade $\partial_i^-$ for $- \partial_i\bar\partial_i /\partial_i^+$ and eliminate $\partial^-$ entirely.}.
 As a result,  in the light-cone formalism vertices  only involve derivatives $\partial^+$, $\partial^x$ and $\partial^{\bar x}$.
 Secondly, in section \ref{sec:13.2} we could see 
that elimination of tensor components with ''-'' indices brings $\partial^+$'s in the denominator, (\ref{15feb19}). Because of that, local covariant interactions when written in the light-cone gauge inevitably involve negative powers of $\partial^+$. This happens, for example, for the light-cone gauge version of the Yang-Mills theory. Due to this, vertices with negative powers of $\partial^+$ are still regarded as \emph{local}. In contrast, non-analytic dependence on $\partial$ and $\bar\partial$ will be regarded as \emph{non-locality}. As in the covariant formalism one can consider stronger versions of locality, e.g. by demanding that vertices only feature finitely many derivatives $\partial$ and $\bar\partial$.

\subsubsection{Cubic vertices}
\label{sec:13.5.1}

According to the general results of the previous section, at the cubic order we need to solve
  \begin{equation}
\begin{split}
   & \sum_{\lambda_i}\int d^{3n}q^\perp \delta (\sum_{i=1}^3 q^\perp_i)  
    \prod_{i=1}^3\Phi^{\lambda_i}(q^\perp_i)\\
&\qquad\qquad \qquad\left(  -{\cal H}
 \Big[j_3^{\lambda_1 \lambda_2 \lambda_3} +\frac{1}{3}
 \big( \sum_j \frac{\partial}{\partial \bar q_j}\big) h_3^{\lambda_1\lambda_2 \lambda_3}\Big]
 +{\cal J}h_3^{\lambda_1\lambda_2\lambda_3}
 \right)=0
 \end{split}
  \label{19feb7}
  \end{equation}
and its complex conjugate, where $h$ is a function of $\mathbb{P}_{ij}$, $\bar{\mathbb{P}}_{ij}$ and $\beta_i$, so that (\ref{18feb7x1}) is satisfied. An immediate observation that one can make is that ${\cal H}$, see (\ref{19feb4}), is a purely algebraic operator in momentum space. This implies that $j$ can be solved from (\ref{19feb7}) for any $h$. 
However, in general this way of solving (\ref{19feb7})  is not satisfactory as it leads to non-local $j$ in the sense defined above. Indeed, ${\cal H}$ involves $\partial$ and $\bar\partial$ in the numerator, so ${\cal H}^{-1}$ is a non-local factor.
Instead, for (\ref{19feb7}) to give a local $j$ the ${\cal J}h$ term should be proportional to ${\cal H}$.

As in the covariant analysis we should factor out on-shell trivial vertices, as these lead to fake interactions. One has
\begin{equation}
\label{19feb8}
0\approx q_1\cdot q_2 = -\frac{\mathbb{P}_{12}\bar{\mathbb{P}}_{12}}{\beta_1\beta_2}
\end{equation}
and similarly for other pairs of fields. Note also that momentum conservation leads to
\begin{equation}
\label{19feb9}
\mathbb{P}\equiv \mathbb{P}_{12}=\mathbb{P}_{23}=\mathbb{P}_{31}, \qquad
\bar{\mathbb{P}}\equiv \bar{\mathbb{P}}_{12}=\bar{\mathbb{P}}_{23}=\bar{\mathbb{P}}_{31},
\end{equation}
so at three-point level there is only one independent $\mathbb{P}$ and one independent $\bar{\mathbb{P}}$. We will also need that
\begin{equation}
\label{19feb10}
{\cal H}= \frac{\mathbb{P}\bar{\mathbb{P}}}{\beta_1\beta_2\beta_3}.
\end{equation}
Finally, a simple computation shows that one can use momentum conservation for $h$ in (\ref{19feb7}) in the sense that this does not change its contribution to the equation and, hence, leads to the same $j$. 

With these auxiliary results at hand, we proceed to solving the kinematical constraints (\ref{18feb7x1}). Both $\mathbb{P}$ and $\bar{\mathbb{P}}$ should enter $h_3$ with non-negative powers, otherwise, the respective $h_3$ is non-local. Moreover, at least one power -- either of $\mathbb{P}$ or of $\bar{\mathbb{P}}$ -- should be vanishing for $h_3$ to give a non-trivial interaction, see (\ref{19feb8}). Then, the first equation in (\ref{18feb7x1}) entails that $h$ has the following dependence on $\mathbb{P}$ and $\bar{\mathbb{P}}$ depending on the value of the total helicity
\begin{equation}
\label{19feb11}
\begin{split}
h_3 &\sim \bar{\mathbb{P}}^{\lambda_1+\lambda_2+\lambda_3}, \qquad \lambda_1+\lambda_2+\lambda_3 >0, \\
h_3& \sim {\mathbb{P}}^{-\lambda_1-\lambda_2-\lambda_3}, \qquad \lambda_1+\lambda_2+\lambda_3 <0, \\
h_3& \sim 1, \qquad \lambda_1+\lambda_2+\lambda_3 =0.
\end{split}
\end{equation}

By requiring, in addition, the second equation in (\ref{18feb7x1}) to hold, we find the general solution to the kinematical constraints to be
\begin{equation}
\label{19feb12}
\begin{split}
h_3 & =\frac{\bar{\mathbb{P}}^{\lambda_1+\lambda_2+\lambda_3}}{\beta_1^{\lambda_1}\beta_2^{\lambda_2}\beta_3^{\lambda_3}}f\left(\frac{\beta_1}{\beta_2}\right), \qquad \lambda_1+\lambda_2+\lambda_3 >0, \\
h_3 &= \frac{{\mathbb{P}}^{-\lambda_1-\lambda_2-\lambda_3}}{\beta_1^{-\lambda_1}\beta_2^{-\lambda_2}\beta_3^{-\lambda_3}}f\left(\frac{\beta_1}{\beta_2}\right), \qquad \lambda_1+\lambda_2+\lambda_3 <0, \\
h_3 &=\frac{1}{\beta_1^{\lambda_1}\beta_2^{\lambda_2}\beta_3^{\lambda_3}} f\left(\frac{\beta_1}{\beta_2}\right), \qquad \lambda_1+\lambda_2+\lambda_3 =0,
\end{split}
\end{equation}
where $f$ is a general function of its argument. Indeed, in each case the prefactor of $f$ has the right total homogeneity degree in $\beta$, while $\beta_1/\beta_2$ is the only independent homogeneity degree zero variable in $\beta$ once conservation of momentum is taken into account. 

Focusing on the case $\lambda_1+\lambda_2+\lambda_3>0$, from (\ref{19feb7}) and its complex conjugate we find
\begin{equation}
\label{19feb13}
\begin{split}
j_3 &= -\frac{2}{3}\frac{h_3}{\bar{\mathbb{P}}}\big((\lambda_3-\lambda_2)\beta_1+(\lambda_1-\lambda_3)\beta_2+(\lambda_2-\lambda_1)\beta_3\big)-\frac{h_3 \mathbb{P}}{{\cal H}}\frac{f' }{f\beta_2^2},\\
\bar j_3& = -\frac{h_3 \bar{\mathbb{P}}}{{\cal H}}\frac{f' }{f\beta_2^2}.
\end{split}
\end{equation}
We, thus, obtain that $j_3$ is always local. At the same time, $\bar{j}_3$ features $\mathbb{P}$ in the denominator contributed by ${\cal H}$, so it is non-local unless we require $f'=0$.

Analogously, for $\lambda_1+\lambda_2+\lambda_3<0$ we obtain
\begin{equation}
\label{19feb14}
\begin{split}
\bar j_3& = \frac{2}{3}\frac{h_3}{{\mathbb{P}}}\big((\lambda_3-\lambda_2)\beta_1+(\lambda_1-\lambda_3)\beta_2+(\lambda_2-\lambda_1)\beta_3\big)-\frac{h_3 \bar{\mathbb{P}}}{{\cal H}}\frac{f' }{f\beta_2^2},\\
 j_3 &= -\frac{h_3 {\mathbb{P}}}{{\cal H}}\frac{f' }{f\beta_2^2}.
\end{split}
\end{equation}
Similarly, both $j$ and $\bar j$ are local only for $f'=0$.

Finally,  $\lambda_1+\lambda_2+\lambda_3=0$ can be considered as a particular case of (\ref{19feb13}). Thus, we find again that $f'=0$. However, in addition to that,  $h_3$ no longer depends on $\bar{\mathbb{P}}$, so the first term in the expression for $j_3$ is non-local unless
\begin{equation}
\label{19feb15}
(\lambda_3-\lambda_2)\beta_1+(\lambda_1-\lambda_3)\beta_2+(\lambda_2-\lambda_1)\beta_3=0.
\end{equation}
For general $\beta$ satisfying momentum conservation this leads to
\begin{equation}
\label{19feb16}
\lambda_1=\lambda_2=\lambda_3=0.
\end{equation}
Therefore, in the total helicity zero case the light-cone deformation procedure admits non-trivial local solutions at the cubic level only when all fields are scalars. 

Summarising our results so far, we found that the light-cone deformation procedure allows us to construct the following local cubic vertices
\begin{equation}
\label{21feb1}
\begin{split}
S_3  &=C^{\lambda_1\lambda_2\lambda_3}\int d^4x \frac{\bar{\mathbb{P}}^{\lambda_1+\lambda_2+\lambda_3}}{\beta_1^{\lambda_1}\beta_2^{\lambda_2}\beta_3^{\lambda_3}}
\Phi^{\lambda_1}\Phi^{\lambda_2}\Phi^{\lambda_3}
, \qquad \lambda_1+\lambda_2+\lambda_3 >0, \\
S_3 &=C^{\lambda_1\lambda_2\lambda_3}\int d^4x\frac{{\mathbb{P}}^{-\lambda_1-\lambda_2-\lambda_3}}{\beta_1^{-\lambda_1}\beta_2^{-\lambda_2}\beta_3^{-\lambda_3}}
\Phi^{\lambda_1}\Phi^{\lambda_2}\Phi^{\lambda_3}
, \qquad \lambda_1+\lambda_2+\lambda_3 <0, \\
S_3 & =C^{000}\int d^4x\Phi^{0}\Phi^{0}\Phi^{0}, \qquad \lambda_1=\lambda_2=\lambda_3 =0,
\end{split}
\end{equation}
where $C$ are undetermined coupling constants. Equation (\ref{21feb1}) makes it manifest that $h_3$ splits into two pieces: the piece that depends only on $\mathbb{P}$ and the piece that depends only on $\bar{\mathbb{P}}$. These will be referred to as holomorphic and antiholomorphic vertices correspondingly. Besides that holomorphic $h_3(\mathbb{P})$ gives rise to trivial $j_3$ while the associated $\bar{j}_3$ is also holomorphic. Similarly, antiholomorphic $h_3(\bar{\mathbb{P}})$ leads to trivial $\bar{j}_3$, while the associated $j_3$ is antiholomorphic. This structure of the light-cone deformation procedure at the cubic level will be important at the next order. 

\subsection{Comparison with the covariant classification}
\label{sec:13.6}

It is instructive to compare these results with those found in the covariant classification of section \ref{sec:8.3}. We remind the reader that for a general triplet of spins in 4d one can construct \emph{two} independent parity-preserving vertices: the one with $s_1+s_2+s_3$ derivatives and the one with $s_1+s_2-s_3$ derivatives, where $s_3$ is the lowest spin. 

To compare this result with the light-cone one, we recall that a field of spin $s$ corresponds to a pair of fields with helicities $s$ and $-s$. Thus, for a triplet of spins $(s_1,s_2,s_3)$, we have the following helicity configurations
\begin{equation}
\label{21feb2}
\begin{split}
&(s_1,s_2,s_3), \quad (s_1,s_2,-s_3), \quad (s_1,-s_2,s_3), \quad  (-s_1,s_2,s_3),\\
&(s_1,-s_2,-s_3), \quad (-s_1,s_2,-s_3), \quad (-s_1,-s_2,s_3), \quad  (-s_1,-s_2,-s_3).
\end{split}
\end{equation}
For general spins the total helicity for any of the eight configurations in (\ref{21feb2}) is non-vanishing and the respective light-cone vertex is given by one of the first two lines in (\ref{21feb1}). Therefore, in total we obtain eight independent vertices. These are, however, not real and do not have definite parity. As it will be discussed below, once vertices with opposite helicities enter with the same coefficients, the resulting vertex is real and parity-invariant. In other words, parity-invariant vertices have the form
\begin{equation}
\label{21feb3}
\begin{split}
S_3  =\int d^4x \frac{\bar{\mathbb{P}}^{\lambda_1+\lambda_2+\lambda_3}}{\beta_1^{\lambda_1}\beta_2^{\lambda_2}\beta_3^{\lambda_3}}
\Phi^{\lambda_1}\Phi^{\lambda_2}\Phi^{\lambda_3}+
\int d^4x\frac{{\mathbb{P}}^{\lambda_1+\lambda_2+\lambda_3}}{\beta_1^{\lambda_1}\beta_2^{\lambda_2}\beta_3^{\lambda_3}}
\Phi^{-\lambda_1}\Phi^{-\lambda_2}\Phi^{-\lambda_3}.
\end{split}
\end{equation}
This implies that for a general triplet of spins, the light-cone deformation procedure allows one to construct \emph{four} parity-invariant cubic vertices.

 Comparing this with the conclusions from the covariant classification, we see that these results do not match: \emph{the light-cone deformation procedure allows one to construct additional cubic vertices, which are not available in the manifestly covariant formalism.}

To see how this discrepancy occurs, one may gauge fix the covariant  cubic vertices. To this end, we first note that at the non-linear level the light-cone gauge condition (\ref{15feb10}) remains intact. This follows from the fact that it corresponds to the first condition in (\ref{15feb6}), which only involves kinematical generators and, hence, stays undeformed. On the other hand, tracelessness (\ref{15feb12}) and divergencelessness (\ref{15feb15}) are only valid in the free theory and will get corrected in the non-linear theory
\begin{equation}
\label{21feb4}
\varphi_{m}{}^{ma(s-2)}=O(g)+\dots,  \qquad \partial_m\varphi^{m a(s-1)}=O(g)+\dots.
\end{equation}
The same refers to other equations, e.g. to (\ref{15feb19}), which expresses components of the Fronsdal field in terms of its physical light-cone components $\Phi^{s}$ and $\Phi^{-s}$.

These corrections, however, are irrelevant for the comparison of cubic vertices in the covariant and light-cone approaches. Indeed, divergence and trace, once appear in the Fronsdal action, appear quadratically. Thus, in non-linear theory, gauge fixed Fronsdal action will have corrections that start at order $O(g^2)$. Cubic vertices, in turn, already feature a coupling constant, so to get order $O(g)$  light-cone action one can use undeformed relations of section \ref{sec:13.2} for covariant cubic vertices.
In summary, for the purpose of gauge fixing the cubic vertices, we can use relations between Fronsdal fields and the light-cone fields as they appear in the free theory, see section \ref{sec:13.2}.

It is not hard to see by explicit evaluation that upon gauge fixing building blocks of covariant vertices give different result depending on the signs of helicities of fields involved
\begin{equation}
\label{21feb5}
\begin{split}
\partial_{x_i}\cdot \partial_{u_j} &= \frac{\bar{\mathbb{P}}_{ij}}{\beta_j}, \qquad \lambda_j >0,\\
\partial_{x_i}\cdot \partial_{u_j} &= \frac{{\mathbb{P}}_{ij}}{\beta_j}, \qquad \lambda_j <0,\\
\partial_{u_i}\cdot \partial_{u_j}&=1, \qquad \lambda_i\lambda_j<0,\\
\partial_{u_i}\cdot \partial_{u_j}&=0, \qquad \lambda_i\lambda_j>0.
\end{split}
\end{equation}
\begin{zadacha}
Show (\ref{21feb5}).
\end{zadacha}

With these auxiliary results we are ready to gauge fix covariant cubic vertices. The covariant vertex with the maximal number of derivatives is given by
\begin{equation}
\label{21sep6}
{ Y}_1^{s_1}{ Y}_2^{s_2}{ Y}_3^{s_3},
\end{equation}
see (\ref{5nov1}), (\ref{6nov5}) for definitions. If $\lambda_i$ is positive, then ${ Y}_i$ produces $\bar{\mathbb{P}}$, see the first line of (\ref{21feb5}). Analogously, ${ Y}_i$ produces ${\mathbb{P}}$ for negative $\lambda_i$. Considering that $\mathbb{P}\bar{\mathbb{P}}$ leads to fake interactions, we find that (\ref{21sep6}) gives non-trivial light-cone vertices only when all helicities are of the same sign.
 The explicit computation gives
\begin{equation}
\label{21sep7}
{ Y}_1^{s_1}{ Y}_2^{s_2}{ Y}_3^{s_3} \to \frac{(-2\bar{\mathbb{P}})^{s_1+s_2+s_3}}{\beta_1^{s_1}\beta_2^{s_2}\beta_3^{s_3}}+
\frac{(-2\mathbb{P})^{s_1+s_2+s_3}}{\beta_1^{s_1}\beta_2^{s_2}\beta_3^{s_3}}.
\end{equation}
Here the first term corresponds to $(s_1,s_2,s_3)$ helicity configuration, while the second one corresponds to $(-s_1,-s_2,-s_3)$.

In a similar manner we proceed with vertex
\begin{equation}
\label{21sep8}
G^{s_3}{ Y}_1^{s_1-s_3}{ Y}_2^{s_2-s_3},
\end{equation}
where it is assumed that $s_3\le s_1$ and $s_3\le s_2$. 
Again, to have a non-trivial light-cone interaction, we need to demand that signs of helicities of the first and the second fields are the same.
Indeed, otherwise ${ Y}_1{ Y}_2$ produce a $\mathbb{P}\bar{\mathbb{P}}$ contribution.
 Helicity of the third field should have the opposite sign, otherwise traces vanish (\ref{21feb5}) and, hence, $G$ is equal to zero. Proceeding to the computation in these non-trivial cases, we find
\begin{equation}
\label{21sep9}
G^{s_3}{ Y}_1^{s_1-s_3}{ Y}_2^{s_2-s_3}
\to (-1)^{s_3}\frac{(-2\bar{\mathbb{P}})^{s_1+s_2-s_3}}{\beta_1^{s_1}\beta_2^{s_2}\beta_3^{-s_3}}+
(-1)^{s_3}\frac{(-2\mathbb{P})^{s_1+s_2-s_3}}{\beta_1^{s_1}\beta_2^{s_2}\beta_3^{-s_3}}.
\end{equation}
These vertices correspond to helicity configurations $(s_1,s_2,-s_3)$ and $(-s_1,-s_2,s_3)$.

\begin{zadacha}
Derive (\ref{21sep7}) and (\ref{21sep9}).
\end{zadacha}

This analysis allows one to make a couple of qualitative conclusions. First, is that parity-invariant vertices of the covariant formalism upon gauge fixing, indeed, produce pairs of light-cone vertices, that come in parity-invariant configurations. Second, we can see that the number of derivatives in the covariant vertex naturally translates into the power of $\mathbb{P}$ or $\bar{\mathbb{P}}$ upon the light-cone gauge fixing. This conclusion can also be achieved by simple scaling arguments. 

As for the comparison of two classifications, one can see that the covariant approach does not contain vertices that correspond to the light-cone vertices with the following helicity configurations
\begin{equation}
\label{21feb2xx1}
\begin{split}
 (s_1,-s_2,s_3), \quad  (-s_1,s_2,s_3),\quad
(s_1,-s_2,-s_3), \quad (-s_1,s_2,-s_3).
\end{split}
\end{equation}
Formally, one can reproduce these from 
\begin{equation}
\label{22feb1}
G^{s_2}{ Y}_1^{s_1-s_2}{ Y}_3^{s_3-s_2}, \qquad G^{s_1}{ Y}_2^{s_2-s_1}{ Y}_3^{s_3-s_1}
\end{equation}
by following the same steps as before. Despite (\ref{22feb1}) are formally gauge invariant -- they are of the form (\ref{6nov4}) -- they feature negative powers of ${ Y}_3$. Not only such vertices are non-local from the covariant perspective, as they contain inverse space-time derivatives, these also involve negative numbers of contractions of tensor indices, which is meaningless. Still, in the light-cone approach the associated vertices are meaningful and local in the sense that these do not involve negative powers of $\partial$ and $\bar\partial$.

In this section we assumed that spins $s_1$, $s_2$ and $s_3$ are general, in particular, none of them is vanishing, the sum of  helicities can never be zero. For particular configurations the analysis is a bit different, but conceptually goes along the same lines. We will not give it here.  

Finally, we would like to note that the additional vertices of the light-cone deformation procedure correspond to lower-derivative vertices in the covariant approach. For example, according to our rule for counting derivatives from powers of $\mathbb{P}$ and $\bar{\mathbb{P}}$, 
\begin{equation}
\label{22feb2}
\frac{\bar{\mathbb{P}}^{2}}{\beta_1^{s}\beta_2^{2}\beta_3^{-s}}
\end{equation}
is a two-derivative vertex that corresponds to the helicity configuration $(s,2,-s)$. It can be regarded as a  generalisation of \emph{the minimal gravitational coupling} to higher-spin fields.
Thus, unlike in the covariant approach to higher-spin interactions, \emph{in the light-cone gauge higher-spin interactions do not necessarily require higher derivatives, moreover, the light-cone deformation procedure allows one to couple higher-spin fields to gravity minimally, at least, at the cubic level}.

\subsection{Chiral theories}
\label{sec:13.7}

In this section we will proceed with the analysis of the light-cone deformation procedure beyond the leading order in the coupling constants. Since the treatment  at this level becomes very technical, we will present only its qualitative conclusions  here. 

As we learned from the analysis of higher-spin interactions in the covariant form,  the level of quartic vertices is where the consistency conditions get hard to satisfy without relaxing requirements on locality. However, in the previous section we found that the light-cone deformation procedure is not equivalent to the covariant one: it gives different results already at the level of cubic vertices. Accordingly, one may expect that the light-cone approach may give new possibilities for higher-spin theories at higher orders in perturbations and, hopefully, construct consistent theories without violating locality. As we will see, partially, this is the case.

At the order $g^2$ we are lead to the dynamical condition (\ref{19feb2x2})
\begin{equation}
\label{22feb3}
[H_2,J_4]+[H_3,J_3] +[H_4,J_2]=0.
\end{equation}
From (\ref{19feb3}) and (\ref{19feb5}) one concludes that the first and the last terms do not contribute to the $q$-independent sector. 
Therefore, by setting $q$ to zero from (\ref{22feb3}) we find
\begin{equation}
\label{22feb4}
[H_3,J_3]\big|_{q=0}=0.
\end{equation}
Next, one can see that only the antiholomorphic part $H_3(\bar{\mathbb{P}})$ and the associated $J_3(\bar{\mathbb{P}})$ contribute to (\ref{22feb4}).
 Moreover,  $[H_3(\bar{\mathbb{P}}),J_3(\bar{\mathbb{P}})]$ is, actually, $q$-independent, so (\ref{22feb4}) is equivalent to 
\begin{equation}
\label{22feb4x1}
[H_3(\bar{\mathbb{P}}),J_3(\bar{\mathbb{P}})]=0.
\end{equation}
This results in some non-linear constraints on the cubic coupling constants $C^{\lambda_1\lambda_2\lambda_3}$ with $\lambda_1+\lambda_2+\lambda_3>0$.

Analogously, from the complex conjugate dynamical constraint
\begin{equation}
\label{22feb3x1}
[H_2,\bar J_4]+[H_3,\bar J_3] +[H_4,\bar J_2]=0.
\end{equation}
 we find 
\begin{equation}
\label{22feb5}
[H_3,\bar J_3]\big|_{\bar q=0}=0,
\end{equation}
which leads to 
\begin{equation}
\label{22feb6}
[H_3({\mathbb{P}}),\bar J_3({\mathbb{P}})]=0
\end{equation}
and puts constraints on $C^{\lambda_1\lambda_2\lambda_3}$ with $\lambda_1+\lambda_2+\lambda_3<0$.

Obviously, 
each of equations (\ref{22feb4x1}), (\ref{22feb6}) can be solved in a trivial manner that is by setting all the coupling constants to zero.
Non-trivial solutions to  (\ref{22feb4x1}), (\ref{22feb6}) are much harder to find, but this still can be done systematically, see  \cite{Ponomarev:2016lrm}. We will consider here solution 
\begin{equation}
\label{22feb7}
\begin{split}
C^{\lambda_1\lambda_2\lambda_3}=\frac{g\ell^{\lambda_1+\lambda_2+\lambda_3-1}}{(\lambda_1+\lambda_2+\lambda_3-1)!}, \qquad \lambda_1+\lambda_2+\lambda_3 > 0,
\end{split}
\end{equation}
where $\ell$ is a dimensionful coupling constant. Consistency conditions feature other solutions, which are related to (\ref{22feb7}) by adding color, doing contractions in $\ell$ and making truncations in the spectrum. All these solutions are qualitatively very similar, so below we will only focus on (\ref{22feb7}).

Normally, one wants to obtain a theory with the action, which is real and parity-invariant. As mentioned before, for the cubic action in the light-cone approach this implies 
\begin{equation}
\label{22feb8}
C^{\lambda_1\lambda_2\lambda_3}=C^{-\lambda_1-\lambda_2-\lambda_3}.
\end{equation}
This case, however, is rather complicated to analyse. We will briefly mention  the results of this analysis towards the end of this section. For now, we will not require (\ref{22feb8}), instead, we will consider the chiral setup in which the light-cone consistency conditions simplify significantly. 

Namely, let the antiholomorphic coupling constants be as in (\ref{22feb7}), while the holomorphic action will be vanishing
\begin{equation}
\label{22feb9}
H_3 (\mathbb{P})=0.
\end{equation}
In this case, $\bar{J}_3$ vanishes and the only non-vanishing component of $J_3$ is the antiholomorphic one
\begin{equation}
\label{22feb10}
\bar{J}_3=0, \qquad J_3=J_3(\bar{\mathbb{P}}).
\end{equation}
This means that the second commutator in (\ref{22feb3x1}) equals zero, so this equation can be solved simply by setting
\begin{equation}
\label{22feb11}
H_4=0, \qquad \bar J_4=0.
\end{equation}
By setting in addition $J_4=0$, the first and the third commutators in (\ref{22feb3}) are also trivially vanishing. The remaining contribution is as on the left-hand side of (\ref{22feb6}) and it is vanishing for the chosen set of coupling constants (\ref{22feb7}). Thus, we managed to solve all the consistency conditions at  order $g^2$. It is trivial to see that higher-order consistency conditions are satisfied identically.

As a short summary, we found that the light-cone consistency conditions at order $g^2$ split into a couple of sectors. The first one, (\ref{22feb4x1}), features only the cubic antiholomorphic vertices and the second one, (\ref{22feb6}), only contains the cubic holomorphic vertices. Other consistency conditions mix both types of cubic vertices, as well as higher order corrections. However,  due to a particular structure of this last type of consistency conditions, these can be trivially solved by setting cubic action of one holomorphicity to zero together with all higher-order corrections. In this chiral setting it only remains to solve (\ref{22feb4x1}) or (\ref{22feb6}) depending on whether holomorphic or antiholomorphic part is kept non-trivial and these are known how to solve systematically.

This leads us to the so-called chiral higher-spin theory, which in the antiholomorphic case is given by
\begin{equation}
\label{22feb12}
\begin{split}
S &= 
 -\frac{1}{2}\int d^4x \sum_{\lambda} \partial_a\Phi^{-\lambda}\partial^a \Phi^\lambda\\
 &+
\frac{g\ell^{-1}}{(\lambda_1+\lambda_2+\lambda_3-1)!}\int d^4x \sum_{\lambda_i}\frac{(\ell\bar{\mathbb{P}})^{\lambda_1+\lambda_2+\lambda_3}}{\beta_1^{\lambda_1}\beta_2^{\lambda_2}\beta_3^{\lambda_3}}
\Phi^{\lambda_1}\Phi^{\lambda_2}\Phi^{\lambda_3}.
\end{split}
\end{equation}
By referring to it as ''theory'' we mean that this action is consistent to all orders, despite featuring only cubic interactions.
Below we will briefly discuss its properties.

First of all, action (\ref{22feb12}) is not real. In this regard chiral theories are reminiscent of self-dual Yang-Mills and self-dual gravity theories. The latter theories admit a light-cone gauge action, which also involves only cubic interactions of one holomorphicity \cite{Chalmers:1996rq}. Similarity of (\ref{22feb12}) with self-dual lower-spin theories does not end here. Namely, it can be shown that both self-dual gravity and the chiral higher-spin theory  can be brought to the form of the self-dual Yang-Mills theory with a gauge algebra which involves space-time derivatives. Moreover, all these three theories after a sequence of changes of variables can be written in the form of some integrable  2d sigma models, which allows one to argue that these theories are integrable as well. We refer the reader to \cite{Ponomarev:2017nrr} for details.  

Integrability of self-dual Yang-Mills theory and self-dual gravity is well known and leads to a certain list of properties, such as the vanishing of the scattering amplitudes. These properties carry over to chiral higher-spin theories. The vanishing of $n$-point tree-level amplitudes for the chiral higher-spin theory with $n>3$ was verified explicitly\footnote{Kinematics of massless 3-point scattering is singular and requires a separate discussion.}. Moreover, it was found that (\ref{22feb12}) is one-loop finite. Part of these cancellations can be attributed to the self-dual nature of the theory, while some cancellations are specific to the higher-spin case. 
The twistor-geometric constructions for self-dual Yang-Mills and gravity can also be extended to the chiral higher-spin case. For this and other related results see, e.g. \cite{Skvortsov:2018jea,Krasnov:2021nsq}

Note that triviality of scattering in the chiral higher-spin theory resolves the apparent contradiction between the non-linearity of its action  and the no-go theorems, such as those we discussed in sections \ref{sec:10.1} and \ref{sec:10.2}. 
Considering that scattering in chiral theories is  trivial, one may wonder whether these can be obtained as field redefinitions of a free theory. This is  not the case: field redefinitions of free theories have been explicitly factored out in our analysis.
 Instead, cancellation of tree-level amplitudes in chiral higher-spin theories appears in a more subtle manner and is related to integrability.

Finally, it is natural to ask whether there is any parity-invariant completion of (\ref{22feb12}), which would be analogous to the completion of the self-dual Yang-Mills to the usual Yang-Mills theory. To answer this question, one has to impose (\ref{22feb8}), thus keeping both chiral parts of the cubic action, and proceed with solving the light-cone consistency conditions. Then, the direct analysis of  (\ref{22feb3}) and (\ref{22feb3x1}) indicates that the deformation procedure requires non-trivial quartic interactions. A more refined analysis shows that no matter what local $H_4$, $J_4$ and $\bar J_4$ we try, these consistency conditions cannot be satisfied. A summary of this analysis can be found in \cite{Ponomarev:2017nrr}. Therefore, when requiring the theory to be parity-invariant, we encounter a familiar obstacle to local higher-spin interactions. It is not clear whether this problem can be overcome. Still, simplicity and uniformity of (\ref{22feb12}) with the lower-spin self-dual theories suggests that it provides an important step towards understanding higher-spin theories in flat space.

\subsection{Further reading}

The analysis of cubic interactions for massless higher-spin fields in the light-cone gauge formalism was initiated in \cite{Bengtsson:1983pd,Bengtsson:1986kh}. The  next order analysis was carried out in \cite{Metsaev:1991mt,Metsaev:1991nb}
and, in particular, (\ref{22feb7}) was obtained. The light-cone formalism has various extensions:  it was extended to higher dimensions, to massive and mixed-symmetry fields \cite{Metsaev:2005ar,Metsaev:2007rn} as well as to massless fields in the AdS space \cite{Metsaev:2018xip,Skvortsov:2018uru}. Let us note, that despite the cubic vertices both in the covariant formalism and in the light-cone approach are known for some time, the fact that these classifications do not match was emphasised only recently \cite{Bengtsson:2014qza,Conde:2016izb}. The light-cone gauge fixing of cubic vertices is relatively straightforward and it was carried out by different authors, see e.g. \cite{Ponomarev:2016jqk,Conde:2016izb,Sleight:2016xqq}.
 Besides that the light-cone gauge approach is intimately connected to the spinor-helicity formalism \cite{Ananth:2012un,Ponomarev:2016cwi}, which facilitates the contact with the amplitude literature, such as \cite{Benincasa:2007xk}.

\section{Conclusions and further reading}

In these introductory lectures we focused on interactions of massless higher-spin fields. These interactions -- whether in flat or the AdS spaces -- appear to be severely constrained. The no-go theorems as well as other results indicate that scattering of massless higher-spin fields is either trivial or rather degenerate. We illustrated these general conclusions with some toy models of massless higher-spin theories. 

A  part of this course was devoted to perturbative approaches to higher-spin interactions. Their power lies in their complete generality, which, in particular, allows one to show absence of interactions of a particular class. In these constructions an important role is played by locality of the theory in question. At the same time, the example of the Chern-Simons theories shows that these general approaches are unlikely to produce a theory in the closed form, unless one is fortunate with the convenient choice of the field variables. In fact, only chiral higher-spin theories in the light-cone gauge were constructed in a general perturbative fashion. In contrast, the Chern-Simons theories and the holographic theories rely on the additional input. 

In agreement with the expectations from the no-go theorems, all toy models that we considered appear to be degenerate in one or another sense. In particular, 3d higher-spin theories are of the Chern-Simons type, hence, these are topological. Chiral theories have trivial scattering as a consequence of being integrable theories of the self-dual type. Despite this, it is rather remarkable that at least in these sectors, higher-spin theories do exist and generalise their lower-spin counterparts in a very natural manner.
Finally, degeneracy of holographic higher-spin theories, that we considered, manifests itself in their duality with free boundary theories. Despite the fact that holographic higher-spin theories are interacting from the bulk perspective, scattering in these theories is of a rather degenerate type. Moreover, as we briefly mentioned, holographic theories do not have a conventional type local bulk action, once formulated in terms of Fronsdal fields. 

In this course we touched upon some developments in the higher-spin literature. One important development that was left out is Vasiliev theories. These are based on the frame-like formalism, reviewed in the course, which is further extended with an infinite set of auxiliary fields, each carrying derivatives of the dynamical fields of unbounded order. This extension allows one to achieve certain nice properties, e.g. Vasiliev theories have the higher-spin symmetries manifestly built in. At the same time, an infinite set of auxiliary fields presents a difficulty as it requires a scheme that would allow one to eliminate them, thus, making contact with conventional field-theoretic approaches. In addition, since these auxiliary fields carry derivatives of unbounded order, locality in the Vasiliev theories gets somewhat obscured. Whether Vasiliev theories are local is a subject of current research. For review of the Vasiliev theories, see \cite{Vasiliev:1999ba,Bekaert:2004qos,Didenko:2014dwa}. For recent results on Vasiliev theories in the context of holography and locality, see \cite{Giombi:2009wh,Giombi:2010vg,Giombi:2012ms,Boulanger:2015ova,Skvortsov:2015lja,Didenko:2017lsn,Didenko:2018fgx}.

Another interesting development is conformal higher spin theories. These generalise conformal gravity to higher-spin case. Quite remarkably, these theories can be formulated in the closed form and, moreover, are manifestly local. Unfortunately, similarly to conformal gravity, these theories are non-unitary. Still, these can be interesting in many ways. 
Here we would like to mention, that amplitudes in conformal higher-spin theories were computed \cite{Joung:2015eny,Beccaria:2016syk} and, similarly to holographic amplitudes, they display distributional structure. For review on conformal higher-spin theories, see \cite{Segal:2002gd}.

Finally, we would like to mention recent proposals for conformal higher-spin theories based on twistor geometry \cite{Hahnel:2016ihf,Adamo:2016ple} as well as proposed higher-spin theories based on matrix models \cite{Sperling:2017dts,Steinacker:2019fcb}.

 \acknowledgments

I would like to thank the students participating  in the course for their feedback, which helped me improving these lectures. I would also like to thank A. Alkalaev, V. Didenko, M.~Grigoriev, S.~Pekar and, especially, E. Skvortsov, as well as all participants of the higher-spin course at UMONS  -- A. Bedhotiya, S. Dhasmana, J. O'Connor, M. Serrani, R.~Van~Dongen,   -- for their valuable comments on the preliminary versions of the text.

\appendix

\section{Conventions}
\label{appa:conventions}

We use the mostly plus convention $\eta = {\rm diag}(-,+,\dots,+)$.

To deal with symmetric tensors we use the following notation
\begin{equation}
\label{a14sep1}
\varphi^{a(s)}=\varphi^{a_1\dots a_s}.
\end{equation}
Moreover, we will use identical indices to indicate that indices have to be symmetrized. Symmetrization is defined with the normalization that makes it a projector. For example,  
\begin{equation}
\label{a14sep2}
\partial^a\varphi^{a(s)}=\frac{1}{s+1}\left(\partial^{a_1}\varphi^{a_2\dots a_{s+1}}+\dots + \partial^{a_{s+1}}\varphi^{a_1\dots a_{s}}\right),
\end{equation}
where the r.h.s. contains $s+1$ non-trivial permutations of indices in the expression on the l.h.s. Alternatively, one can sum over all $(s+1)!$ permutations of indices and then the overall factor in front of the bracket would be $[(s+1)!]^{-1}$.

\paragraph{Levi-Civita tensor.}
We will introduce the Levi-Civita tensor with lower indices so that
\begin{equation}
\label{27jan1}
\epsilon_{1 \dots d}= 1.
\end{equation}
By raising indices we find the Levi-Civita tensor with upper indices
\begin{equation}
\label{27jan2}
\epsilon^{a_1 \dots a_d} = \eta^{a_1b_1}\dots \eta^{a_d b_d} \epsilon_{b_1\dots b_d}.
\end{equation}
It should be remarked that 
\begin{equation}
\label{27jan3}
\epsilon^{1 \dots d} =\sigma, \qquad \sigma\equiv  {\rm det}[\eta],
\end{equation}
so for the Minkowski metric $\sigma =-1$.

There are a couple of useful formulas that involve the Levi-Civita tensor
\begin{equation}
\label{27jan4}
A_{a_1}{}^{b_1}\dots A_{a_d}{}^{b_d} \epsilon_{b_1\dots b_d}={\rm det}[A]\epsilon_{a_1\dots a_d}
\end{equation}
and
\begin{equation}
\label{9dec11}
\epsilon^{k[n]l[d-n]}\epsilon_{k[n]m[d-n]}=\sigma n!(d-n)!\delta^{[l_1}_{m_1}\dots \delta^{l_{d-n}]}_{m_{d-n}},
\end{equation}
where, as usual, antisymmetrization is understood as a projection.

\section{The Coleman-Mandual theorem}
\label{app:cm}

In this appendix we sketch the proof of the Coleman-Mandula theorem stated in section \ref{sec:10.2}.
 For a more rigorous discussion we refer the reader to the original paper \cite{Coleman:1967ad} and to \cite{Weinberg} for shortcuts and clarifications. Here we mostly follow \cite{Weinberg}.

\subsection{Step 1: generators that commute with translations}
The proof goes in several steps. We will start by considering only those symmetry transformations $B_\alpha$ that commute with the translation generator
\begin{equation}
\label{25nov1}
[B_\alpha,P_\mu]=0.
\end{equation}
It is convenient to choose the basis for the states in the theory so that they have  definite momenta. Then, the action of $B$ on single-particle states is of the form\footnote{Here $|p\rangle_m$ is what was denoted $\varphi_{p,\sigma}$ in section \ref{sec:2}.}
\begin{equation}
\label{25nov2}
B_{\alpha}|p\rangle^m = (b_\alpha(p))^m{}_{m'} |p\rangle^{m'},
\end{equation}
where $m, m'$ label the states with given momentum $p$, which also includes spin labels. Note that according to our assumptions -- more precisely, the assumption of having finitely many mass shells with the mass below any fixed value --  the space of states with fixed $p$ is finite-dimensional.

The argument  then works differently for $B_{\alpha}$ that map into the pure trace $b_\alpha$ -- that is those  proportional to the unit matrix $\delta^m{}_{m'}$ -- and for those that map into traceless $b_{\alpha}$. We will first consider the pure trace part of $b$.

By one of our assumptions,  $B$ act on two-particle states by acting on each single-particle state separately. Together with (\ref{25nov2}), this gives
\begin{equation}
\label{25nov3}
(b_\alpha(p,q))^m{}_{m',}{}^{n}{}_{n'}=  (b_\alpha(p))^m{}_{m'} \delta^n{}_{n'}+(b_\alpha(q))^n{}_{n'} \delta^m{}_{m'}.
\end{equation}
Next, we consider a two-particle scattering process of particles with momenta $p$ and $q$ into $p'$ and $q'$, 
\begin{equation}
\label{14apr1}
S(p,m;q,n \to p',m';q',n')\equiv \delta^d(p'+q'-p-q) S(p',q';p,q)^{m'n'}{}_{mn},
\end{equation}
where $p^2 = (p')^2$ and $q^2 = (q')^2$.
Invariance of the $S$-matrix with respect to $B$  implies
\begin{equation}
\label{25nov4}
b_\alpha(p',q')S(p',q';p,q)=S(p',q';p,q)b_{\alpha}(p,q).
\end{equation}
Assuming that $S$ is non-vanishing, we multiply  both sides of (\ref{25nov4}) with  $S^{-1}$
and take the trace.
Due to the cyclicity of the trace, we find
\begin{equation}
\label{25nov4x1}
{\rm tr}\, b_\alpha(p',q') = {\rm tr}\, b_\alpha(p,q).
\end{equation}

The trace of the tensor product of two matrices is the product of their traces. Together with   (\ref{25nov3}) and (\ref{25nov4x1}) 
this leads to
\begin{equation}
\label{25nov5}
N(q^2){\rm tr}\, b_{\alpha}(p')+N(p^2){\rm tr}\, b_{\alpha}(q')=N(q^2){\rm tr}\, b_{\alpha}(p)+N(p^2){\rm tr}\, b_{\alpha}(q),
\end{equation}
where $N(q^2)$ and $N(p^2)$ result from taking the traces of the Kronecker delta's and count the numbers of states on each mass shell. 
Equivalently, we have
\begin{equation}
\label{25nov5x1}
\frac{{\rm tr}\, b_{\alpha}(p')}{N(p^2)}+\frac{{\rm tr}\, b_{\alpha}(q')}{N(q^2)}=\frac{{\rm tr}\, b_{\alpha}(p)}{N(p^2)}+\frac{{\rm tr}\, b_{\alpha}(q)}{N(q^2)}.
\end{equation}
Its general solution reads
\begin{equation}
\label{25nov6}
\frac{{\rm tr}\, b_\alpha(p)}{N(p^2)} =  c^\mu_{\alpha} p_\mu +d_\alpha,
\end{equation}
where both $c$ and $d$ are $p$-independent. Indeed, the $d$-part on both sides of (\ref{25nov5x1}) cancels out identically, while the $c$-part cancels out due to momentum conservation. We, therefore, find that the pure trace part of $b_\alpha$ is either proportional to momenta or is an internal symmetry. 

We now proceed with $B_{\alpha}$ that map into traceless $b_\alpha$, which we will denote 
\begin{equation}
\label{14apr2}
(b^{\sharp}_\alpha)^{n'}{}_n\equiv (b_\alpha)^{n'}{}_n - \frac{{\rm tr}\, b_\alpha(p)}{N(p^2)}\delta^{n'}{}_n.
\end{equation}
From (\ref{25nov6}) it follows that the generators represented by $b^{\sharp}_\alpha$ are
\begin{equation}
\label{14apr3}
B^{\sharp}_{\alpha} \equiv B_\alpha - c^\mu_{\alpha} P_\mu -d_\alpha.
\end{equation}
On the two-particle states these act by 
\begin{equation}
\label{14apr4}
(b^{\sharp}_\alpha(p,q))^m{}_{m',}{}^{n}{}_{n'}=  (b^{\sharp}_\alpha(p))^m{}_{m'} \delta^n{}_{n'}+(b^{\sharp}_\alpha(q))^n{}_{n'} \delta^m{}_{m'}.
\end{equation}
Our next goal is to show that $B_\alpha^{\sharp} \to b^{\sharp}_\alpha(p,q)$ is an isomorphism of Lie algebras. To this end, we need to show that 
\begin{equation}
\label{15aprx1}
l^\alpha b^{\sharp}_\alpha(p,q)=0 \quad \Rightarrow \quad l^\alpha B_\alpha^{\sharp}=0,
\end{equation}
where $l^\alpha$ are some coefficients. Equation (\ref{15aprx1}) implies that $B_\alpha^{\sharp} \to b^{\sharp}_\alpha(p,q)$ is invertible, so it is an isomorphism.

Since $B^{\sharp}$ are symmetries of the $S$-matrix, we have
\begin{equation}
\label{14apr5}
b^{\sharp}_\alpha(p',q')S(p',q';p,q)=S(p',q';p,q)b^{\sharp}_{\alpha}(p,q).
\end{equation}
Again, we assume that the $S$-matrix is non-vanishing. Then, (\ref{14apr5})  implies that $b^{\sharp}_\alpha(p',q')$ and $b^{\sharp}_\alpha(p,q)$ are related by the similarity transformation. This, in turn, implies that 
\begin{equation}
\label{14apr6}
 l^\alpha b^{\sharp}_{\alpha}(p,q) = 0 \qquad \Rightarrow \qquad l^\alpha b^{\sharp}_{\alpha}(p',q') = 0.
\end{equation}
Considering  (\ref{14apr4}) and that $b^{\sharp}_\alpha(p)$ are traceless,
one finds
\begin{equation}
\label{14apr7}
 l^\alpha b^{\sharp}_{\alpha}(p',q') = 0 \qquad \Rightarrow \qquad l^\alpha b^{\sharp}_{\alpha}(p') = 0.
\end{equation}

We, thus, have shown that $ l^\alpha b^{\sharp}_{\alpha}(p,q) = 0$ implies $ l^\alpha b^{\sharp}_{\alpha}(p') = 0$,
where $p'$ is constrained to be on the same mass shell as $p$ and, moreover, there should exist $q'$ on the same mass shell with $q$, so that $p+q=p'+q'$. 
With some extra work, this limitation on $p'$ can be lifted, that is one can prove that
\begin{equation}
\label{14apr8}
 l^\alpha b^{\sharp}_{\alpha}(p,q) = 0 \qquad \Rightarrow \qquad  l^\alpha b^{\sharp}_{\alpha}(k) = 0, 
\end{equation}
where $k$ is an arbitrary on-shell momentum. The fact that $ l^\alpha b^{\sharp}_{\alpha}(k)$ vanishes for any $k$ means that $ l^\alpha B^{\sharp}_{\alpha}=0$. Thus, we managed to show (\ref{15aprx1}), which implies that 
the Lie algebra generated by $B^{\sharp}_\alpha$ is isomorphic to its representation on  two-particle states with the trace part removed, $ b^{\sharp}_{\alpha}(p,q)$.

Next, we apply the standard theorem, see e.g. \cite{Weinberg}, which  tells us that a Lie algebra of finite-dimensional Hermitian matrices -- like $ b^{\sharp}_{\alpha}(p,q)$ for fixed $p$ and $q$ -- is at most the direct sum of a semi-simple Lie algebra and some number of $U(1)$ Lie algebras. We will now explore the consequence of this theorem focusing on the semi-simple part. The associated symmetry generators will be denoted  $B^{\flat}_\alpha$.

The Lorentz group acts on these generators in the standard way\footnote{See section \ref{sec:2} for notations.}
\begin{equation}
\label{15apr1}
G(\Lambda,0): \qquad B^{\flat}_{\alpha} \quad \to \quad B^{\flat}_{\alpha}(\Lambda)\equiv U(G(\Lambda,0))B^{\flat}_{\alpha}U^{-1}(G(\Lambda,0)).
\end{equation}
Since $B^{\flat}_{\alpha}$ commute with $P_\mu$ it follows that $B^{\flat}_{\alpha}(\Lambda)$ commute with $\Lambda_\mu{}^\nu P_\nu$. Considering that  $\Lambda_\mu{}^\nu P_\nu$ is just a linear combination of translations, we conclude that $B^{\flat}_{\alpha}(\Lambda)$  commutes with $P_\mu$. This, in turn, entails that $B^{\flat}_{\alpha}(\Lambda)$ is some linear combination of $B^{\flat}_{\alpha}$ 
\begin{equation}
\label{15apr2}
 U(G(\Lambda,0))B^{\flat}_{\alpha}U^{-1}(G(\Lambda,0)) = D^{\beta}{}_{\alpha}(\Lambda)B^{\flat}_{\beta}. 
\end{equation}
Therefore, $B^{\flat}_{\alpha}$ realise a representation of the Lorentz group. We would like to show that it is unitary. 

To do that one notices that $B^{\flat}_{\alpha}(\Lambda)$ commute the same way as $B^{\flat}_{\alpha}$ -- $U$ and $U^{-1}$ factors cancel out. 
This  means that the structure constants of the algebra generated by $B^{\flat}_{\alpha}$ are invariant under Lorentz transformations, that is
\begin{equation}
\label{15apr3}
f^{\gamma}{}_{\alpha\beta} = D^{\alpha'}{}_\alpha(\Lambda)D^{\beta'}{}_\beta(\Lambda)D^{\gamma}{}_{\gamma'}(\Lambda)f^{\gamma'}{}_{\alpha'\beta'}.
\end{equation}
As a consequence, the Lie algebra metric
\begin{equation}
\label{15apr4}
g_{\beta\delta}\equiv f^{\gamma}{}_{\alpha\beta}f^{\alpha}{}_{\gamma\delta}
\end{equation}
is also Lorentz invariant. Moreover, since the Lie algebra generated by $B^{\flat}_{\alpha}$ is semi-simple, metric (\ref{15apr4}) is positive-definite. Altogether, this implies that $B^{\flat}_{\alpha}$ realise a unitary  finite-dimensional representation of the Lorentz group. As we mentioned in section \ref{sec:2}, 
for finite-dimensional representations, this is only possible if the representation  carried by  $B^{\flat}_{\alpha}$ is trivial. Thus, $B^{\flat}_{\alpha}$ generate internal symmetries. 

 With some extra arguments, one can show that  the $U(1)$ part of $B^{\sharp}_{\alpha}$ also commutes with the Lorentz algebra.
 
Summarising the results of the first part of the proof, we found that symmetries of the $S$-matrix that commute with momenta are either momenta -- the first term on the right hand side of (\ref{25nov6}) -- or internal symmetries -- the second term in (\ref{25nov6}) and all generators $B^{\sharp}_{\alpha}$.

\subsection{Step 2: locality in momentum space}
On the next step, we take up the possibility of symmetry generators that do not commute with translations. In general, the symmetry generator in the momentum basis reads
\begin{equation}
\label{14jan1}
A_\alpha|p\rangle^n =\int d^dp' {\cal A}_\alpha(p',p)^n{}_{n'}|p'\rangle^{n'}.
\end{equation}
Since the kernel ${\cal A}$ maps physical states to physical states, it should vanish unless both $p$ and $p'$ are on the mass shell. Our goal is to show that  ${\cal A}$  vanishes for any $p\ne p'$.

To achieve this, one considers a generator
\begin{equation}
\label{14jan2}
A_\alpha^f = \int d^dx e^{iPx}A_\alpha e^{-iPx}f(x),
\end{equation}
where $f$ is an arbitrary function.
It is a symmetry generator as it is defined via a composition of symmetry generators $P$ and $A_\alpha$. It is straightforward to show that $A^f$ acts on the single-particle states as
\begin{equation}
\label{14jan3}
A^f_\alpha|p\rangle^n =\int d^dp' \tilde f(p'-p){\cal A}_\alpha(p',p)^n{}_{n'}|p'\rangle^{n'},
\end{equation}
where $\tilde f$ is the Fourier transform of $f$
\begin{equation}
\label{14jan4}
\tilde f(p)\equiv \int d^d xe^{ixp}f(x).
\end{equation}

Next, we return to the analysis of the  2-to-2 scattering, $p+q=p'+q'$. Let $\Delta$ be such that $p+\Delta$ is still on-shell, while all $q+\Delta$, $p'+\Delta$ and $q'+\Delta$ are off-shell. Then, picking $\tilde f$ in (\ref{14jan3}) with the support in the vicinity of $\Delta$, we find that 
\begin{equation}
\label{14jan5}
A_\alpha^f|q\rangle^m=0 ,\qquad A_\alpha^f|p'\rangle^{n'}=0,\qquad A_\alpha^f|q'\rangle^{m'}=0.
\end{equation}
Indeed, outside the support of $\tilde f$, the $\tilde f$ factor vanishes in (\ref{14jan3}), while inside the support of $\tilde f$, the ${\cal A}$ factor vanishes, as ${\cal A}$ only relates physical states. The condition of invariance of the $S$-matrix with respect to $A^f$ reads
\begin{equation}
\label{14jan6}
\langle p',q'| S |A^f p,q\rangle + \langle p',q'| S | p,A^f q\rangle =
\langle A^f p',q'| S | p,q\rangle + \langle p',A^f q'| S | p, q\rangle,
\end{equation}
where we again used our assumption about the action of symmetries on multi-particle states. Due to (\ref{14jan5}) this reduces to
\begin{equation}
\label{14jan7}
\langle p',q'| S |A^f p,q\rangle=0.
\end{equation}
This can happen for two reasons: either $S$ or $A^f$ is vanishing. By invoking some geometric considerations, it is not hard to see, that one can change on-shell $p$, $q$, $p'$ and $q'$, so that momentum conservation is still satisfied, moreover, 
 $p+\Delta$ remains on-shell, while 
$q+\Delta$, $p'+\Delta$ and $q'+\Delta$ remain off-shell. In other words, the argument presented above holds in a certain continuous range of the Mandelstam variables. Keeping in mind our assumption that  the $S$-matrix can only  have isolated zeros, 
we conclude that 
 (\ref{14jan7}) entails $A^f=0$. This, in turn, implies that 
\begin{equation}
\label{14jan8}
A(p',p)=0,\qquad  \text{for} \qquad  p-p'=\Delta.
\end{equation}

The same argument can be applied to other $\Delta$ that shift an on-shell momentum $p$ to an on-shell momentum. Typically, one can choose the remaining three momenta so that the momentum conservation is still satisfied, while after a shift by $\Delta$ they all go off-shell. The value of $\Delta$ which is excluded by these arguments is $\Delta=0$ as, clearly, all on-shell states remain on-shell. Accordingly, we find that $A$ is supported only on $p=p'$ or
\begin{equation}
\label{14jan9}
A(p',p)=0, \qquad \forall \quad p'\ne p.
\end{equation}

\subsection{Step 3: constraining derivatives in momenta}
With some technical assumptions the fact that the integral kernel $A(p,p')$ is only supported on $p=p'$ implies that it is given by $\delta^d(p-p')$ and its derivatives of \emph{finite} order.
In other words, the action of $A$ on one particle states is given by
\begin{equation}
\label{17jan1}
A_\alpha |p\rangle^n = \sum_{i=0}^k A_\alpha^{(i)|\mu_1\dots \mu_i |n}{}_{n'}(p)\frac{\partial}{\partial p^{\mu_1}}\dots  \frac{\partial}{\partial p^{\mu_i}}|p\rangle^{n'}.
\end{equation}
The last step of the proof is to reduce the analysis of symmetries of this type to those that commute with momenta, discussed in the first part of the proof.

To achieve this one considers a $k$-fold commutator of (\ref{17jan1}) with momenta, which is also a symmetry of the $S$-matrix
\begin{equation}
\label{17jan2}
[P^{\mu_1},[P^{\mu_2}, \dots [P^{\mu_k},A_\alpha]\dots ]]= (-1)^k A_\alpha^{(k)|\mu_1\dots \mu_k}(p),
\end{equation}
where $A$ is symmetric in $\mu$ indices.
It no longer contains derivatives of momenta, hence, it commutes with the translation generators. Therefore, the results of the first step of the proof can be applied and we have
\begin{equation}
\label{17jan3}
A_\alpha^{(k)|\mu_1\dots \mu_k|n}{}_{n'}(p) = b_\alpha^{\mu_1\dots \mu_k|n}{}_{n'}+c_{\alpha}^{\mu| \mu_1\dots \mu_k}p_\mu \delta^n{}_{n'}.
\end{equation}
Here $c$ is just the pure trace $c$ term from (\ref{25nov6}), while $b$ combines $p$-independent internal traceless and pure-trace symmetries. Note that we dropped $N(p^2)$ from (\ref{25nov6}) for the pure trace part. This can be done for the following reason. First, by our assumption, there are finitely many mass shells in the system, so $p^2$ takes discrete eigenvalues. Moreover, as we showed, $A$ acts locally in momentum space. Altogether, this implies that $A$ acts within a single mass shell, so $N(p^2)$ can be replaced with a number $N(-m^2)$.

We will first focus on the case with $m^2\ne 0$ and take into account that $A$ may only act within a single mass shell. In general, invariance of the mass shell $p^2+m^2=0$ with respect to transformation $O$ implies 
\begin{equation}
\label{17jan7}
(P^2+m^2)O = O'(P^2+m^2),
\end{equation}
where $O'$ is an arbitrary operator. Condition (\ref{17jan7}) can be rewritten as 
\begin{equation}
\label{17jan8}
[P^2,O] = (O'-O)(P^2+m^2).
\end{equation}
We would like to apply this conclusion to $O$ defined by
\begin{equation}
\label{17jan9}
O=[P^{\mu_2},[P^{\mu_3}, \dots [P^{\mu_k},A_\alpha]\dots]], \qquad k\ge 1.
\end{equation}
By evaluating $[P^2,O]$, we find
\begin{equation}
\label{17jan9x1}
[P^2,O]= (-1)^k 2p_{\mu_1}(b_\alpha^{\mu_1\dots \mu_k}+c_{\alpha}^{\mu| \mu_1\dots \mu_k}p_\mu ).
\end{equation}
This should be compared with the admissible form for $[P^2,O]$ on the right-hand side of (\ref{17jan8}). We find that this requires 
\begin{equation}
\label{17jan4}
(-1)^k 2p_{\mu_1}(b_\alpha^{\mu_1\dots \mu_k}+c_{\alpha}^{\mu| \mu_1\dots \mu_k}p_\mu )=0.
\end{equation}
This constraint is applicable for $k\ge 1$, because otherwise $O$ does not exist, see (\ref{17jan9}).

Equation (\ref{17jan4}) should be satisfied for any $p$ on the mass shell with $m^2>0$. This  leads to
\begin{equation}
\label{17jan5}
b_\alpha^{\mu_1\dots \mu_k}=0, \qquad c_{\alpha}^{\mu| \mu_1\dots \mu_k}=-c_{\alpha}^{\mu_1| \mu\dots \mu_k}.
\end{equation}
The symmetry condition on $c$ can be solved non-trivially only for $k=1$, see exercise \ref{z1} for $k=2$ case. 
We, thus, find the only non-trivial solution to be
\begin{equation}
\label{17jan6}
c^{\mu|\mu_1}=-c^{\mu_1|\mu},
\end{equation}
which generates Lorentz transformations.

In summary, we are left with the following possibilities for symmetries of the $S$-matrix in massive theories: for $k=1$ these may only contain Lorentz transformations, while for $k=0$ $A$'s commute with momenta and,  as was shown on previous steps of the proof, may be either internal symmetries or momenta themselves. This finishes the proof of the Coleman-Mandula theorem for massive particles. 

This argument can be naturally extended to include massless particles. For massless particles the right-hand side of (\ref{17jan8}) does not have the mass term, so we find
\begin{equation}
\label{17jan10}
(-1)^k 2p_{\mu_1}(b_\alpha^{\mu_1\dots \mu_k}+c_{\alpha}^{\mu| \mu_1\dots \mu_k}p_\mu ) = (O'-O)p^2.
\end{equation}
In addition to the solutions that we have already discussed in the massive case, (\ref{17jan10}) can be solved as
\begin{equation}
\label{17jan11}
c_{\alpha}^{\mu| \mu_1\dots \mu_k} = \eta^{\mu ( \mu_1} d_\alpha^{\mu_2 \dots \mu_k)}
\end{equation}
for some $d$. Solutions to (\ref{17jan11}) with $k=1$ correspond to conformal symmetries. Solutions with $k\ge 2$ can be argued away, e.g. by noticing that the associated $A$ under commutator generate an unbounded number of derivatives in $p$, which contradicts (\ref{17jan1}). Summarising, we find that for massless particles, the symmetry of the $S$-matrix may consist of a direct product of the conformal algebra and the algebra of internal symmetries.

\section{Helicity}
\label{app:2x1}

For the 4d Poincare group it is conventional to introduce the Pauli-Lubanski pseudovector
\begin{equation}
\label{27feb1}
W_{\mu}\equiv \frac{1}{2}\varepsilon_{\mu\nu\rho\sigma}J^{\nu\rho}P^\sigma.
\end{equation}
It is straightforward to compute that $[P,W]=0$, so one can pick a basis in the space of states so that both $P$ and $W$ take definite values. We will use this basis in the following. 

Focusing on massless fields, we take the standard momentum as in (\ref{15sep5}). Then, the only non-vanishing components  of $P$ are $p_3=p_0$ and the only component of $J$, that is non-trivially realised is $J^{12}$. It then straightforward to see that the only non-vanishing components of $W$ are given by
\begin{equation}
\label{27feb2}
W_0 = J^{12}p_3, \qquad W_3 = J^{12}p_0.
\end{equation}
This implies that for the chosen basis, for the states with the standard momentum, $J^{12}$ also takes a definite value
\begin{equation}
\label{27feb3}
J^{12}|p,\lambda\rangle = \lambda|p,\lambda\rangle.
\end{equation}

Equation (\ref{27feb3}) holds for the frame, in which momentum takes the standard form. To write it in the Lorentz-covariant form we note that (\ref{27feb2}) and (\ref{27feb3}) together entail
\begin{equation}
\label{27feb4}
W_\mu |p,\lambda\rangle=\lambda P_\mu |p,\lambda\rangle.
\end{equation}
Since both $W_\mu$ and $P_\mu$ transform as vectors under parity-preserving Lorentz transformations, by keeping $\lambda$ Lorentz invariant, (\ref{27feb4}) takes a manifestly covariant form. Together with the Wigner approach of induced representations, this implies that $\lambda$ defined as the proportionality coefficient between $W$ and $P$ is the same for all the states in the representation and, hence, can be used to label different massless representations in the same way as spin does. 

To understand the connection between helicity $\lambda$ and spin, let us return to representations of the Wigner little group. In a given case it is $SO(2)$, which for the standard momentum is generated by $J^{12}$. In the main body of the text irreducible representations of $SO(2)$ were given as traceless symmetric tensors of $SO(2)$ with spin being the rank of a tensor. By counting the number of independent components of such a tensor, it is not hard to see that it is two for $s>0$ and one for $s=0$. This may seem to be in  contradiction with (\ref{27feb3}), which suggests that a representation space of the Wigner little group is generated by a single vector $|p,\lambda\rangle$, so it is one-dimensional.

To clarify what actually happens, we consider a simple example of the $SO(2)$ vector representation. In this case $J^{12}$ acts via a matrix 
\begin{equation}
\label{26feb4}
J_{12}=
\left( \begin{array}{cc}
0 & i\\
-i & 0
\end{array} \right). 
\end{equation}
It is straightforward to find that its eigenvalues are $\lambda=\pm 1$ and the associated eigenvectors are
\begin{equation}
\label{26feb5}
v_{\pm}=\left(\begin{array}{c}
1\\
\pm i
\end{array}\right)
. 
\end{equation}
Obviously, these are not real, so there is no contradiction with irreducibility of the real vector representation of $SO(2)$. 

Still, when we are dealing with a real vector representation, it may be convenient to use basis (\ref{26feb5}) in which $J^{12}$ acts diagonally.  At the same time, the coordinates of a vector in this basis should satisfy certain reality conditions to ensure that the associated vector is, indeed, real. The same refers to representations of the Poincare algebra obtained from these by the Wigner induced representation technique.
A similar result holds for fields of any spin $s>0$: a symmetric traceless tensor of rank $s$ has two eigenvectors with respect to $J^{12}$ with eigenvalues being $\pm s$ and both eigenvectors corresponding to complex tensors. 

\subsection{Helicity in the light-cone gauge}

With the necessary background reviewed, let us demonstrate that the representation given in (\ref{15feb5})-(\ref{15feb8}), indeed, has helicity $\lambda$. To this end, we evaluate $W_0$ for the standard momentum, see (\ref{27feb2}). With some simple algebra one finds that only the spin part of $J$ contributes, moreover,
\begin{equation}
\label{27feb5}
S^{12} = -i S^{x\bar x}.
\end{equation}
Taking into account, in addition, the relative factors due to the definition of generators (\ref{15dec1x1}), we find that (\ref{15feb8}) leads to
\begin{equation}
\label{27feb6}
W_0 = \lambda p^3 = \lambda p_0.
\end{equation}
Therefore, the proportionality coefficient between $W$ and $P$ is, indeed, given by $\lambda$.

Finally, let us mention the intuitive meaning of helicity. Remembering that only the spin part of $J$ contributes to $W$, we have
\begin{equation}
\label{27feb7}
W_0 = \frac{1}{2}\varepsilon_{0\nu\rho \sigma}S^{\nu\rho}P^\sigma = \lambda P_0.
\end{equation}
For the expression to be non-vanishing, indices $\nu$, $\rho$ and  $\sigma$ may take only spatial values. Moreover, the Levi-Civita tensor reduces to the spatial one
\begin{equation}
\label{27feb8}
 \frac{1}{2}\varepsilon_{ij k}S^{ij}P^k = \lambda P_0.
\end{equation}
Then, helicity becomes
\begin{equation}
\label{27feb9}
\lambda = \frac{S_k p^k}{p_0}, \qquad S_{k} \equiv  \frac{1}{2}\varepsilon_{ij k}S^{ij}.
\end{equation}
Since $p_0=\pm\sqrt{p^k p_k}$, (\ref{27feb9}) implies that helicity, up to a sign, is a projection of spin on  the spatial part of momentum.

\section{Fourier transform for the light-cone approach}
\label{app:2}
In the light-cone deformation procedure it is convenient to make the Fourier transform with respect to spatial coordinates (\ref{21sep17}), which is then followed by the change of variables $p=iq$. 
For readers convenience, we present here some of the useful formulas in the Fourier transformed form.

In these terms the canonical commutator reads
\begin{equation}
\label{29sep5}
[\Phi^{\lambda_1}(q^\perp_1,x^+),\Phi^{\lambda_2}(q^\perp_2,x^+)] = \frac{\delta^{\lambda_1+\lambda_2,0}\delta^3(q^\perp_1+q^\perp_2)}{\beta_1-\beta_2}
\end{equation}
and the Noether charges are 
\begin{align}
\notag
P_2^i &= \sum_{\lambda}\int d^3q^\perp_1 d^3q^\perp_2 \delta^3(q^\perp_1+q^\perp_2)\beta_1 \Phi^{-\lambda}(q^\perp_1,x^+) p^i_2(q_2,\partial_2) \Phi^\lambda(q^\perp_2,x^+),\\
\label{29sep6}
J_2^{ij} &= \sum_{\lambda}\int d^3q^\perp_1 d^3q^\perp_2 \delta^3(q^\perp_1+q^\perp_2)\beta_1 \Phi^{-\lambda}(q^\perp_1,x^+) j^{ij}_2(q_2,\partial_2) \Phi^\lambda(q^\perp_2,x^+),
\end{align}
where
  \begin{equation}
  \label{29sep7}
\begin{split}
p_2^+ &=q^+,  \qquad p_2^-  = -\frac{q \bar q}{\beta}, \qquad p_2 = q, \qquad \bar p_2=
\bar q,\\
j_2^{+-} &=  \frac{\partial}{\partial \beta}\beta, \qquad  j_2^{x\bar x} =  N_q- N_{\bar q}  -\lambda,\\
j_2^{x+}& = -\beta \frac{\partial}{\partial \bar q},  \qquad j_2^{x-} = \frac{\partial}{\partial \bar q}\frac{q \bar q}{\beta} + q \frac{\partial}{\partial \beta}+\lambda\frac{q}{\beta},\\
j_2^{\bar x+}& = -\beta \frac{\partial}{\partial q},\qquad
  j_2^{\bar x-} = \frac{\partial}{\partial q}\frac{q \bar q}{\beta} +  \bar q \frac{\partial}{\partial \beta}
-\lambda\frac{\bar q}{\beta}
\end{split}
\end{equation}
and 
\begin{equation}
N_q \equiv q\frac{\partial}{\partial q}, \qquad N_{\bar q} \equiv \bar q\frac{\partial}{\partial \bar q}.
\end{equation}

\bibliography{hs}
\bibliographystyle{JHEP}

\end{document}